\documentclass[twocolumn]{aastex62}
\newcommand \versnum {18}

\usepackage{amsmath}
\usepackage{grffile}          
\usepackage{color}

\tightenlines

\newcommand{\galname}{DDO053}

\newcommand     \beq    {\begin{equation}}
\newcommand     \beqa   {\begin{eqnarray}}
\newcommand     \Cdust  {C_{\rm dust}}
\newcommand     \cm     {\,{\rm cm}}

\newcommand     \dust   {{\rm d}}
\newcommand     \eeq    {\end{equation}}
\newcommand     \eeqa   {\end{eqnarray}}

\newcommand     \fpdr   {f_{\rm PDR}}

\newcommand     \gtsim  {\gtrsim}                
\newcommand     \Ha     {{\rm H}}
\newcommand     \HI     {{\rm H\,I}}

\newcommand     \Herschel {{\it Herschel}}
\newcommand     \hers   {{\it Herschel}}
\newcommand     \HH     {{\rm H}_2}

\newcommand     \IRAC   {{IRAC}}

\newcommand     \Jy     {\,{\rm Jy}}

\newcommand     \K              {\,{\rm K}}
\newcommand     \kms    {\,{\rm km~s}^{-1}}
\newcommand     \kpc    {\,{\rm kpc}}
\newcommand     \Ldust  {L_{\rm d}}

\newcommand     \Lstar {L_\star}     

\newcommand     \Lsol   {L_{\odot}}
\newcommand     \ltsim  {\lesssim}               

\newcommand     \mm     {\,{\rm mm}}

\newcommand     \Mpc    {\,{\rm Mpc}}
\newcommand     \Msol   {M_{\odot}}

\newcommand     {\msun} {$M_{\odot}$}
\newcommand     \Mdust  {M_{\rm d}}
\newcommand     \MH     {M_{\rm H}}
\newcommand     \MIPS   {{MIPS}}

\newcommand     \nH     {n_{\rm H}}

\newcommand     \PDR    {{\rm PDR}}
\newcommand     \fPDR    {f_{\rm PDR}}
\newcommand     \fPDRj    {f_{{\rm PDR},j}}
\newcommand     \pc     {\,{\rm pc}}
\newcommand     \qpah   {q_{\rm PAH}}
\newcommand     \qPAH   {q_{\rm PAH}}
\newcommand     \qpahj   {q_{{\rm PAH},j}}

\newcommand     \SigLd  {{\Sigma_{Ld}}}
\newcommand     \SigLdmin {{\Sigma_{Ld,{\rm min}}}}
\newcommand     \SigMd  {{\Sigma_{Md}}}

\newcommand     \Spitzer {{\it Spitzer}}
\newcommand     \spit   {{\it Spitzer}}

\newcommand     \Ubar   {\overline{U}}

\newcommand     \Umax   {U_{\rm max}}

\newcommand     \Umin   {U_{\rm min}}

\newcommand     \yr     {\,{\rm yr}}

\newcommand     \logoh  {$12+\log_{10}$(O/H)}
\newcommand     \logohpt  {$12+\log_{10}$(O/H)$_{\rm PT}$}
\newcommand     \logohpp  {$12+\log_{10}$(O/H)$_{\rm PP04N2}$}

\newcommand	\nulummipsshort	{$\nu\,L_{24\micron}$}
\newcommand	\nulummipsmid	{$\nu\,L_{70\micron}$}

\newcommand	\lumdust	{$L_{\rm d}$}

\newcommand	\mdust	{$M_{\rm d}$}

\newcommand{\btdnote}[1]{}

\newcommand{\newtext}[1]{{#1}}

\newcommand{\omittext}[1]{}
\newcommand{\omittextr}[1]{}

\newlength{\figwidth}
\newlength{\figwidthw}
\newlength{\figwidthww}
\newlength{\figwidthd}
\countdef\decade=200
\advance\decade by \year
\countdef\hours=201
\advance\hours by \time
\divide\hours by 60
\countdef\mins=202
\advance\mins by \hours
\multiply\mins by 60
\multiply\hours by 100
\countdef\miltime=203
\advance\miltime by \hours
\advance\miltime by \time
\advance\miltime by -\mins
\renewcommand\today{\number\decade.\number\month.\number\day.\number\miltime}
\begin{document}

\title{ {\bf Modeling Dust and Starlight in Galaxies Observed by {\it Spitzer} and {\it Herschel}:
       The KINGFISH Sample}
}

\author[0000-0001-9015-2654]{G.~Aniano}
\affiliation{Dept.\ of Astrophysical Sciences, Peyton Hall,
  Princeton University, Princeton NJ 08544, USA}
\affiliation{LinkedIn Corporation, 700 E Middlefield Rd., Mountain View, CA 94043}

\author[0000-0002-0846-936X]{B.~T.~Draine}
\affiliation{Dept.\ of Astrophysical Sciences, Peyton Hall,
  Princeton University, Princeton NJ 08544, USA}

\author[0000-0001-9162-2371]{L.~K.~Hunt}
\affiliation{INAF - Osservatorio Astrofisico di Arcetri,
  Largo E. Fermi 5, 50125 Firenze, Italy}

\author[0000-0002-4378-8534]{K.~Sandstrom}
\affiliation{Center for Astrophysics and Space Sciences,
  University of California, 9500 Gilman Drive, San Diego CA 92093, USA}

\author[0000-0002-5189-8004]{D.~Calzetti}
\affiliation{Dept.\ of Astronomy, University of Massachusetts,
  Amherst MA 01003, USA}

\author[0000-0001-5448-1821]{R.~C.~Kennicutt}
\affiliation{Steward Observatory, University of Arizona,
  Tucson AZ 85721-0065, USA}
\affiliation{Dept.\ of Physics \& Astronomy, Texas A\&M University,
  College Station TX 77843-4242, USA}

\author[0000-0002-5782-9093]{D.~A.~Dale}
\affiliation{Physics \& Astronomy Dept., University of Wyoming,
  Laramie WY 82071, USA}

\author[0000-0002-0283-8689]{M.~Galametz}
\affiliation{AIM, CEA, CNRS, Universit\'e Paris-Saclay,
  Universit\'e Paris Diderot,
  Universit\'e Paris Cit\'e,
  F-91191 Gif-sur-Yvette, France}
  
\author[0000-0001-5340-6774]{K.~D.~Gordon}
\affiliation{Space Telescope Science Institute, 3700 San Martin Drive,
  Baltimore MD 21218, USA}

\author[0000-0002-2545-1700]{A.~K.~Leroy}
\affiliation{Ohio State University, Columbus, OH 43210, USA}

\author[0000-0003-1545-5078]{J.-D.~T.~Smith}
\affiliation{Dept.\ of Physics \& Astronomy, University of Toledo,
  Toledo, OH 43606, USA}

\author[0000-0001-5617-7129]{H.~Roussel}
\affiliation{Institut d'Astrophysique de Paris,
  Universit\'{e} Pierre et Marie Curie (UPMC),
  Sorbonne Universit\'e, CNRS, 75014 Paris, France}

\author[0000-0002-0809-2574]{M.~Sauvage}
\affiliation{AIM, CEA, CNRS, Universit\'e Paris-Saclay,
  Universit\'e Paris Diderot,
  Universit\'e Paris Cit\'e,
  F-91191 Gif-sur-Yvette, France}
  
\author[0000-0003-4793-7880]{F.~Walter}
\affiliation{Max-Planck-Institut fur Astronomie, Konigstuhl 17,
  D-69117 Heidelberg, Germany}

\author[0000-0003-3498-2973]{L.~Armus}
\affiliation{Spitzer Science Center,
  California Institute of Technology, MC 314-6, Pasadena CA 91125, USA}

\author[0000-0002-5480-5686]{A.~D.~Bolatto}
\affiliation{Dept.\ of Astronomy, University of Maryland,
  College Park MD 20742, USA}

\author[0000-0003-0946-6176]{M.~Boquien}
\affiliation{Centro de Astronom\'ia (CITEVA), Universidad de Antofagasta,
  Avenida Angamos 601, Antofagasta, Chile}

\author[0000-0001-8513-4945]{A.~Crocker}
\affiliation{Dept.\ of Physics, Reed College, Portland OR 97202, USA}

\author[0000-0001-9419-6355]{I.~De~Looze}
\affiliation{Sterrenkundig Observatorium, Ghent University,
  Krijgslaan 281 S9, B-9000 Gent, Belgium}
\affiliation{Department of Physics \& Astronomy,
  University College London, Gower St., London WC1E6BT, UK}

\author[0000-0002-3106-7676]{J.~Donovan~Meyer}
\affiliation{National Radio Astronomy Observatory,
  Charlottesville VA 22903, USA}

\author[0000-0003-3367-3415]{G.~Helou}
\affiliation{Spitzer Science Center,
  California Institute of Technology, MC 314-6, Pasadena CA 91125, USA}

\author[0000-0003-3339-0546]{J.~Hinz}
\affiliation{Steward Observatory, University of Arizona,
  Tucson AZ 85721-0065, USA}

\author[0000-0002-9280-7594]{B.~D.~Johnson}
\affiliation{Harvard-Smithsonian Center for Astrophysics,
  Cambridge MA 02138, USA}
  
\author[0000-0002-8762-7863]{J.~Koda}
\affiliation{Department of Physics and Astronomy, SUNY Stony Brook,
  Stony Brook NY 11794-3800, USA}

\author{A.~Miller}
\affiliation{Physics \& Astronomy Dept., University of Wyoming,
  Laramie WY 82071, USA}
\affiliation{Dept.\ of Physics, Montana State University,
  Bozeman MT 59717, USA}
  
\author{E.~Montiel}
\affiliation{University of California, Davis, Davis CA 95616, USA}
\affiliation{SOFIA Science Center, NASA Ames Research Center,
  Moffett Field CA 94036, USA}

\author[0000-0001-7089-7325]{E.~J.~Murphy}
\affiliation{National Radio Astronomy Observatory,
  Charlottesville VA 22903 USA}

\author[0000-0003-1682-1148]{M.~Rela\~no}
\affiliation{Dept.\ F\'isica Te\'orica y del Cosmos, Universidad de Granada,
  Granada, Spain}
\affiliation{Instituto Universitario Carlos I de F\'isica Te\'orica
  y Computacional, Universidad de Granada, 18071, Granada, Spain}
  
\author[0000-0003-4996-9069]{H.-W.~Rix}
\affiliation{Max-Planck-Institut fur Astronomie,
  Konigstuhl 17, D-69117 Heidelberg, Germany}

\author[0000-0002-3933-7677]{E.~Schinnerer}
\affiliation{Max-Planck-Institut fur Astronomie,
  Konigstuhl 17, D-69117 Heidelberg, Germany}

\author{R.~Skibba}
\affiliation{Center for Astrophysics and Space Sciences,
  University of California, 9500 Gilman Drive, San Diego CA 92093, USA}
\affiliation{Freelance science journalist, San Diego CA, USA}

\author[0000-0003-0030-9510]{M.~G.~Wolfire}
\affiliation{Dept.\ of Astronomy, University of Maryland,
  College Park MD 20742, USA}

\author{C.~W.~Engelbracht}
\affiliation{deceased}

\correspondingauthor{B.~T.~Draine}
\email{draine@astro.princeton.edu}

\begin{abstract}
Dust and starlight are modeled
for the KINGFISH project galaxies.
With data from
3.6$\micron$ to 500$\micron$, models are strongly constrained.
For each pixel in each galaxy we estimate 
(1) dust surface density; 
(2) $\qpah$, the dust mass fraction in PAHs; 
(3) distribution of starlight intensities heating the dust; 
(4) luminosity emitted by the dust; 
and (5) dust luminosity from regions with high starlight intensity.
The models successfully reproduce both
global and resolved spectral energy distributions. 
We provide well-resolved maps for the dust properties.
As in previous studies, we find 
$\qpah$ to be an increasing function of metallicity,
above a threshold $Z/Z_\odot\approx 0.15$.
Dust masses are obtained
by summing the dust mass over
the map pixels;
these ``resolved'' dust masses are consistent with the
masses inferred from model fits to the global photometry.
The global dust-to-gas ratios obtained from this study 
correlate with galaxy metallicities. Systems with $Z/Z_\odot\gtsim 0.5$
have most of their refractory elements locked up in dust, whereas
when $Z/Z_\odot\ltsim 0.3$ most of these elements tend to
remain in the gas phase. 
Within galaxies, we find that $\qpah$ is suppressed in regions with
unusually warm dust with $\nu L_\nu(70\micron)\gtsim 0.4L_{\rm dust}$.
With knowledge of one long-wavelength flux density ratio
(e.g., $f_{160}/f_{500}$), the minimum starlight intensity heating the
dust ($\Umin$) can be estimated to within $\sim$50\%.
For the adopted dust model, dust
masses can be estimated to within 
$\sim$0.07\,dex accuracy using
the 500$\micron$ luminosity $\nu L_\nu(500)$ alone.
There are additional systematic errors arising from the choice of
dust model, but these are hard to estimate.
These calibrated prescriptions
may be useful for studies of high-redshift galaxies.
\end{abstract}


\section{Introduction
\label{sec:intro}
}

\btdnote{This uses results from\\
/scratch/draine/work\_2017/dustmaps/galname/\\
Resolution\_M160\_111\_SSS\_111
and\\
Resolution\_S250\_100\_SSS\_100
copied to\\ DRAINE2TB/ganiano/KINGFISH/Run\_2017/galname/
}

Interstellar dust affects the appearance of galaxies
by attenuating short-wavelength radiation from stars and ionized gas, and
contributing IR, 
submm, mm, and microwave emission.
Dust is also an important agent in the
fluid dynamics, chemistry, heating, cooling, and even ionization balance
in some interstellar regions,
with a major role in the process of star formation.
Despite the importance of dust, determination of the physical properties of
interstellar dust grains has been a challenging task
[for a review, see \citet{Draine_2003a}].
Even the overall amount
of dust present in other galaxies has often been very uncertain.

The ``Key Insights on Nearby Galaxies: a Far-Infrared Survey with
Herschel'' (KINGFISH) \citep{Kennicutt+Calzetti+Aniano+etal_2011}
project is an imaging and spectroscopic survey of 61 nearby (distance
$D<30\Mpc$) galaxies with the {\it Herschel Space Observatory}.  The
KINGFISH galaxy sample was chosen to cover a wide range of integrated
properties and local interstellar medium (ISM) environments found in
the nearby Universe.  KINGFISH is a direct descendant of the ``{\it
  Spitzer} Infrared Nearby Galaxies Survey'' (SINGS)
\citep{Kennicutt+Armus+Bendo+etal_2003} which produced complete {\it
  Spitzer} imaging with the Infrared Array Camera (IRAC)
\citep{Fazio+Hora+Allen+etal_2004} and the Multiband Imaging
Photometer for Spitzer (MIPS)
\citep{Rieke+Young+Engelbracht+etal_2004} instruments on {\it Spitzer
  Space Telescope} \citep{Werner+Roellig+Low+etal_2004}.  The new {\it
  Herschel} observations include a complete mapping of the galaxies
with the Photodetector Array Camera and Spectrometer (PACS)
\citep{Poglitsch+Waelkens+Geis+etal_2010} and the Spectral and
Photometric Imaging Receiver (SPIRE)
\citep{Griffin+Abergel+Abreu+etal_2010} instruments.  The merged
KINGFISH and SINGS data-set provides panchromatic mapping of the
galaxies, across a wide range of local extragalactic ISM environments.
In addition, we have KINGFISH and SINGS data for 9 additional
galaxies that fell within the 61 KINGFISH target fields.
The photometric maps cover wavelengths from 3.6$\micron$ to
500$\micron$, allowing us to produce well-resolved maps of the dust in
nearby galaxies.

\citet{Skibba+Engelbracht+Dale+etal_2011} modeled the dust in the
KINGFISH galaxy sample using ``modified blackbody'' models.  In the
present work we employ a physically-motivated dust model based on a
mixture of amorphous silicate grains and carbonaceous grains, each
with a distribution of grain sizes 
\citep[][hereafter DL07]{Draine+Li_2007}.  
The dust grains are heated by starlight, and
the model allows for a distribution of intensities for the
starlight heating the dust.  With a small number
of adjustable parameters, the DL07 model reproduces the observed 
spectral energy distribution (SED)
of the dust emission for a variety of astrophysical systems, giving
some confidence in the reliability of dust masses estimated using the
model.  The DL07 model has been found to be consistent with the
3.6$\micron$--500$\micron$ emission from the dust in the star-forming
galaxies NGC~628 and NGC~6946
\citep{Aniano+Draine+Calzetti+etal_2012}, the dust across M31
\citep{Draine+Aniano+Krause+etal_2014}, the emission from annular
rings in the KINGFISH galaxy sample
\citep{Hunt+Draine+Bianchi+etal_2015},
and the overall dust SEDs from KINGFISH galaxies
\citep{Dale+Cook+Roussel+etal_2017}.

The present work is a sequel to the KINGFISH study of 
NGC~628 and NGC~6946 \citep[][hereafter
  AD12]{Aniano+Draine+Calzetti+etal_2012}.  AD12 developed the 
image processing and dust modeling techniques employed 
here, using the spiral galaxies NGC~628 and NGC~6946
as examples.  
The present work takes into account
a recent ``recalibration'' of the DL07 model made possible by {\it Planck} 
observations of diffuse Galactic emission
\citep{Planck_DL07_2016}.
We expand the spatially-resolved
dust modeling to the full KINGFISH galaxy sample, producing maps of
dust mass surface density, PAH fraction, and intensities of the
starlight heating the dust.
Dependences of  
dust/gas ratio and PAH abundance on galaxy metallicity are examined, 
and resolved trends within galaxies are studied.
While the present results are undoubtedly model-dependent, 
comparison of different dust models is beyond the scope of the present work.

The paper is organized as follows.  
A brief overview of the KINGFISH sample is given in Section \ref{sec:sample},
and
in Section \ref{sec:datasources} we discuss the data sources.  
Background subtraction and data
processing are described in Section \ref{sec:dataproc},
and the dust
model is summarized in Section \ref{sec:dustmodel},
including the {\it Planck-}based dust mass ``recalibration''
(Section \ref{sec:renormalization}).
Results are reported in Section \ref{sec:results} with 
a comparison of dust parameter estimates 
based on different dust modeling strategies
given in Section \ref{sec:goldstandard};
global trends with metallicity are described in 
Sections \ref{sec:pahglobaloh} and \ref{sec:globaloh};
and resolved trends of DL07 fitted parameters are discussed 
in Sect. \ref{sec:resolveddl07}.
We summarize the main results in Section \ref{sec:summary}.
Appendix \ref{app:maps} (on-line version) displays maps of selected
dust parameters for each of the
62 galaxies where we have reliable dust detections,
at both MIPS160 and SPIRE250 resolution.
Appendix \ref{app:upper limits} describes the method used to obtain
upper limits for the dust mass for the eight galaxies
(5 dwarfs, 3 ellipticals)
where we were unable to measure the dust mass reliably.
The on-line data set with the KINGFISH data and dust models is
described in Appendix \ref{app:online data}.
In Appendix \ref{app:dependence on psf} we examine the robustness of
the results as the PSF is reduced, precluding use of the lower-resolution
cameras (e.g., MIPS160 and SPIRE500).

\section{Galaxy Sample}\label{sec:sample}

The observational program was designed to cover the 61 galaxies
in the KINGFISH galaxy sample.  Because we will also be discussing the
9 extra galaxies, and the statistical properties of various subsamples,
we list these for clarity in Table \ref{tab:subsamples}.
For each galaxy, we list in Table \ref{tab:geom} the type,
adopted distance, and major and minor optical radii
(corresponding to $\sim$25th mag arcsec$^{-2}$ isophotes),
all taken from \citet[][Table 1]{Kennicutt+Calzetti+Aniano+etal_2011}.

The galaxies IC\,3583, NGC\,586, NGC\,1317, NGC\,1481, NGC\,1510, NGC\,3187,
NGC\,4533, NGC\,7335, and NGC\,7337 were not part of the
KINGFISH sample, but were observed because
each happened
to be in the field of view of
a KINGFISH galaxy.  For these galaxies, we have our standard
imaging with PACS and SPIRE, as well as prior observations
with IRAC and MIPS, so we are able to measure and model their SEDs with the
same techniques as the KINGFISH galaxies.
Information for these 9 ``extra'' galaxies is appended
to many of the tables below. 

\begin{table}
  \footnotesize
  \begin{center}
    \caption{\label{tab:subsamples} Subsamples}
    \begin{tabular}{l c c c c}
      \hline
      Sample & name & KF galaxies & Extra galaxies & Total\\
      \hline
      Full sample   & KF70 & 61 & 9 & 70\\
      Dust detected & KF62 & 53 & 9 & 62\\
      \ion{H}{1} detected & KF57 & 57 & 0 & 57\\
      CO detected   & ---  & 35 & 0 & 35\\
      CO upper limits & ---& 5 & 0 & 5\\
      \hline
      \end{tabular}
  \end{center}
\end{table}

\subsection{Metallicities}

\begin{figure}
\begin{center}
\includegraphics[angle=0,width=\linewidth,
                 clip=true,trim=0.0cm 5.0cm 0.0cm 2.5cm]
                {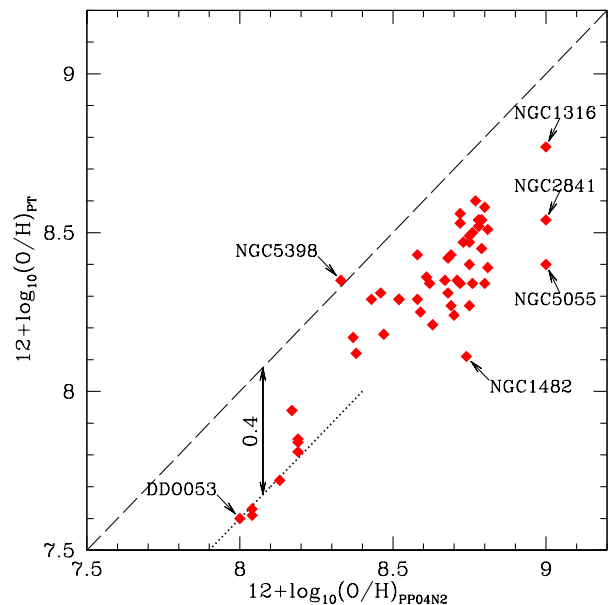} 
\caption{%
\logoh\ for 55 KINGFISH galaxies using two different
estimators (see text).
The long-dashed line corresponds to identity between the two quantities.
The PT estimate for O/H is systematically below the
PP04N2 estimate (by $\sim$0.4 dex at low O/H: dotted line), 
and there is considerable additional
scatter between the two estimates.
\btdnote{f1.pdf}
}
\label{fig:oxyoxy}
\end{center}
\end{figure}

Table \ref{tab:geom} also lists the oxygen abundance 
$12 + \log_{10}({\rm O/H})$ for the galaxies in our sample.
These are ``characteristic'' abundances, which
\citet{Moustakas+Kennicutt+Tremonti+etal_2010} take to be the values
at galactocentric radius $R=0.4R_{25}$.
For 6 of the KINGFISH galaxies 
(DDO154, 
IC\,342, NGC\,628, NGC\,2146, NGC\,3077, and
NGC\,5457) we use metallicities based on
observations of weak lines (specifically, [NII]5726 and [OIII]4364) 
that allow ``direct'' determination of the electron temperature
in the \ion{H}{1} regions responsible for the line emission
\citep{vanZee+Haynes+Salzer_1997,
    Pilyugin+Thuan+Vilchez_2007,
    Engelbracht+Rieke+Gordon+etal_2008,
    Storchi-Bergmann+Calzetti+Kinney_1994,
    Li+Bresolin+Kennicutt_2013,Croxall+Pogge+Berg+etal_2016}.
For these galaxies we list the preferred weak line metallicities in the
PP04N2 column.

\begin{table*}
\vspace*{-0.5cm}

  \footnotesize
\begin{center}
  \caption{\label{tab:geom}The 61 KINGFISH galaxies + 9 extra galaxies}
    \begin{tabular}{| c | c c c c | c c c | c c | c |}
      \hline
       & & & & & \multicolumn{3}{c|}{optical} & \multicolumn{2}{c|}{M160 mask} & \multicolumn{1}{c|}{S250 mask} \\
Galaxy & Type & \multicolumn{2}{c}{$12+\log_{10}({\rm O/H})$} & $D$ & $R_{\rm maj}$ & $R_{\rm min}$ & $\Omega$ &$\Sigma_{Ld,{\rm min}}$ & $\Omega$ & $\Omega$ \\
 && PT$^a$ & PP04N2$^b$ & (Mpc) & (kpc) & (kpc) & (arcmin$^2$) & ($L_\odot\pc^{-2}$) & (arcmin$^2$) &  (arcmin$^2$) \\
\hline
DDO053   & Im    &$7.60\pm0.11$         & \,  8.00$^c$ &  3.61 &0.81                  &0.70                  &1.61                  &--- &--- &--- \\
DDO154   & IBm   &$7.54\pm0.09$         & \,  7.67$^c$ &  4.30 &1.89                  &1.36                  &5.15                  &--- &--- &--- \\
DDO165   & Im    &$7.63\pm0.08$         & \,  8.04$^c$ &  4.57 &2.30                  &1.22                  &4.98                  &--- &--- &--- \\
Hol1     & IABm  &$7.61\pm0.11$         & \,  8.04$^c$ &  3.90 &2.06                  &2.06                  &10.4                  &--- &--- &--- \\
Hol2     & Im    &$7.72\pm0.14$         & \,  8.13 &  3.05 &3.52                  &2.78                  &39.0                  &0.72 &25.2                  &23.0                  \\
IC342    & SABcd &$8.70\pm0.20$        $^d$ & \,  8.85$^e$ &  3.28 &10.5                  &9.18                  &332.                  &3.47 &417.                  &398.                  \\
IC2574   & SABm  &$7.85\pm0.14$         & \,  8.19 &  3.79 &7.27                  &2.72                  &51.1                  &0.66 &54.5                  &52.7                  \\
M81dwB   & Im    &$7.84\pm0.13$         & \,  8.19$^c$ &  3.60 &0.46                  &0.30                  &0.40                  &--- &--- &--- \\
NGC0337  & SBd   &$8.18\pm0.07$         & \,  8.47 & 19.30 &8.08                  &4.98                  &4.01                  &3.55 &11.1                  &6.77                  \\
NGC0584  & E4    &$8.43\pm0.20$         & \,  8.69$^c$ & 20.80 &12.6                  &6.67                  &7.20                  &--- &--- &--- \\
NGC0628  & SAc   &$8.35\pm0.01$         & \,  8.64$^f$ &  7.20 &11.0                  &10.0                  &78.8                  &1.70 &76.1                  &70.3                  \\
NGC0855  & E     &$8.29\pm0.10$         & \,  8.43$^c$ &  9.73 &2.83                  &2.79                  &3.09                  &--- &--- &--- \\
NGC0925  & SABd  &$8.25\pm0.01$         & \,  8.59 &  9.12 &13.9                  &7.57                  &47.0                  &1.48 &46.3                  &44.8                  \\
NGC1097  & SBb   &$8.47\pm0.02$         & \,  8.75 & 14.20 &19.3                  &12.9                  &45.8                  &2.19 &66.3                  &55.7                  \\
NGC1266  & SB0   &$8.29\pm0.20$         & \,  8.52 & 30.60 &6.85                  &6.75                  &1.83                  &2.63 &9.09                  &8.37                  \\
NGC1291  & SB0/a &$8.52\pm0.20$         & \,  8.78 & 10.40 &14.8                  &14.8                  &75.1                  &0.65 &123.                  &119.                  \\
NGC1316  & SAB0  &$8.77\pm0.20$         & \,  9.00$^g$ & 21.00 &12.2                  &8.49                  &8.73                  &1.02 &17.5                  &14.2                  \\
NGC1377  & S0    &$8.29\pm0.20$         & \,  8.52 & 24.60 &6.37                  &6.37                  &2.49                  &2.57 &7.20                  &8.46                  \\
NGC1404  & E1    &$8.54\pm0.20$         & \,  8.78$^c$ & 20.20 &9.75                  &8.69                  &7.71                  &--- &--- &--- \\
NGC1482  & SA0   &$8.11\pm0.13$         & \,  8.74 & 22.60 &8.09                  &4.40                  &2.59                  &2.19 &24.4                  &13.0                  \\
NGC1512  & SBab  &$8.56\pm0.12$         & \,  8.72 & 11.60 &15.0                  &9.27                  &38.5                  &0.89 &34.4                  &15.4                  \\
NGC2146  & Sbab  &$8.68\pm0.10$        $^e$ & \,  8.68$^e$ &  17.20 &16.8                  &7.35                  &15.5                  & 4.8 &39.2                  &18.5                  \\
NGC2798  & SBa   &$8.34\pm0.08$         & \,  8.72 & 25.80 &9.61                  &9.61                  &5.15                  &2.63 &13.0                  &10.2                  \\
NGC2841  & SAb   &$8.54\pm0.03$         & \,  9.00$^g$ & 14.10 &16.7                  &6.77                  &21.1                  &0.66 &60.9                  &55.1                  \\
NGC2915  & I0    &$7.94\pm0.13$         & \,  8.17 &  3.78 &1.04                  &0.51                  &1.37                  &0.58 &5.58                  &5.51                  \\
NGC2976  & SAc   &$8.36\pm0.06$         & \,  8.61 &  3.55 &3.04                  &1.28                  &11.5                  &2.57 &25.6                  &21.3                  \\
NGC3049  & SBab  &$8.53\pm0.01$         & \,  8.72 & 19.20 &6.09                  &3.99                  &2.45                  &1.48 &7.92                  &7.99                  \\
NGC3077  & I0pec &--- & \,  8.64$^h$ &  3.83 &3.34                  &3.29                  &27.8                  &2.14 &39.2                  &32.6                  \\
NGC3184  & SABcd &$8.51\pm0.01$         & \,  8.81 & 11.70 &12.6                  &11.8                  &40.4                  &0.85 &64.0                  &60.5                  \\
NGC3190  & SAap  &$8.49\pm0.20$         & \,  8.75 & 19.30 &12.2                  &3.78                  &4.61                  &1.17 &12.3                  &11.2                  \\
NGC3198  & SBc   &$8.34\pm0.02$         & \,  8.76 & 14.10 &17.5                  &5.98                  &19.5                  &1.23 &36.9                  &30.0                  \\
NGC3265  & E     &$8.27\pm0.14$         & \,  8.69 & 19.60 &3.65                  &2.80                  &0.99                  &2.04 &6.39                  &6.10                  \\
NGC3351  & SBb   &$8.60\pm0.01$         & \,  8.77 &  9.33 &10.1                  &6.74                  &28.9                  &1.35 &49.1                  &43.3                  \\
NGC3521  & SABbc &$8.39\pm0.02$         & \,  8.81 & 11.20 &17.9                  &7.83                  &41.4                  &2.19 &91.4                  &76.6                  \\
NGC3621  & SAd   &$8.27\pm0.02$         & \,  8.75 &  6.55 &11.7                  &6.55                  &66.4                  &1.07 &103.                  &85.7                  \\
NGC3627  & SABb  &$8.34\pm0.24$         & \,  8.62 &  9.38 &12.4                  &5.26                  &27.6                  &2.00 &87.2                  &54.9                  \\
NGC3773  & SA0   &$8.43\pm0.03$         & \,  8.58 & 12.40 &2.13                  &1.80                  &0.93                  &1.86 &5.85                  &5.81                  \\
NGC3938  & SAc   &$8.42\pm0.20$         & \,  8.68 & 17.90 &14.0                  &12.8                  &20.8                  &1.15 &39.4                  &34.5                  \\
NGC4236  & SBdm  &$8.17\pm0.20$         & \,  8.37 &  4.45 &10.4                  &2.85                  &55.4                  &0.71 &64.7                  &63.2                  \\
NGC4254  & SAc   &$8.45\pm0.01$         & \,  8.79 & 14.40 &11.3                  &9.76                  &19.7                  &2.40 &45.9                  &33.2                  \\
NGC4321  & SABbc &$8.50\pm0.03$         & \,  8.76 & 14.30 &15.4                  &13.1                  &36.7                  &1.66 &60.6                  &40.4                  \\
NGC4536  & SABbc &$8.21\pm0.08$         & \,  8.63 & 14.50 &16.0                  &6.25                  &17.6                  &2.09 &37.0                  &25.2                  \\
NGC4559  & SABcd &$8.29\pm0.01$         & \,  8.58 &  6.98 &10.9                  &4.08                  &33.8                  &1.17 &43.2                  &40.3                  \\
NGC4569  & SABab &$8.58\pm0.20$         & \,  8.80 &  9.86 &11.5                  &4.85                  &21.2                  &1.95 &18.3                  &17.2                  \\
NGC4579  & SABb  &$8.54\pm0.20$         & \,  8.79 & 16.40 &14.0                  &11.1                  &21.4                  &2.24 &22.4                  &18.0                  \\
NGC4594  & SAa   &$8.54\pm0.20$         & \,  8.79 &  9.08 &11.5                  &2.98                  &15.5                  &1.55 &31.1                  &27.8                  \\
NGC4625  & SABmp &$8.35\pm0.17$         & \,  8.67 &  9.30 &2.95                  &2.55                  &3.23                  &1.58 &7.47                  &6.47                  \\
NGC4631  & SBd   &$8.12\pm0.11$         & \,  8.38 &  7.62 &17.2                  &0.30                  &3.28                  &3.16 &83.4                  &43.3                  \\
NGC4725  & SABab &$8.35\pm0.13$         & \,  8.71 & 11.90 &18.6                  &12.9                  &62.7                  &1.15 &64.1                  &55.0                  \\
NGC4736  & SAab  &$8.31\pm0.03$         & \,  8.68 &  4.66 &7.60                  &6.15                  &80.0                  &1.38 &124.                  &120.                  \\
NGC4826  & SAab  &$8.54\pm0.10$         & \,  8.78 &  5.27 &6.13                  &4.70                  &38.5                  &2.88 &37.3                  &17.3                  \\
NGC5055  & SAbc  &$8.40\pm0.03$         & \,  9.00$^g$ &  7.94 &14.6                  &8.14                  &69.7                  &2.75 &87.6                  &67.6                  \\
NGC5398  & SBdm  &$8.35\pm0.05$         & \,  8.33 &  7.66 &3.14                  &1.85                  &3.67                  &1.62 &7.29                  &5.72                  \\
NGC5408  & IBm   &$7.81\pm0.09$         & \,  8.19 &  4.80 &1.13                  &0.53                  &0.97                  &2.95 &6.66                  &5.42                  \\
NGC5457  & SABcd &$8.46\pm0.10$        $^h$ & \,  8.38$^i$ &   6.70 &17.1                  &16.9                  &238.                  &0.74 &398.                  &385.                  \\
\hline

    \multicolumn{4}{l}{$a$ From \citet{Moustakas+Kennicutt+Tremonti+etal_2010}
                            except as noted} &
    \multicolumn{7}{l}{$b$ Derived from KK metallicities from
                            \citet{Moustakas+Kennicutt+Tremonti+etal_2010}
                            except as noted} \\
    \multicolumn{4}{l}{$c$ \citet{vanZee+Haynes+Salzer_1997}} &
    \multicolumn{4}{l}{$d$ \citet{Pilyugin+Thuan+Vilchez_2006}} &
    \multicolumn{3}{l}{$e$ \citet{Engelbracht+Rieke+Gordon+etal_2008}} \\
    \multicolumn{4}{l}{$f$ \citet{Berg+Skillman+Croxall+etal_2015}} &
    \multicolumn{4}{l}{$g$ see text} &
    \multicolumn{3}{l}{$h$ \citet{Storchi-Bergmann+Calzetti+Kinney_1994}} \\
    \multicolumn{4}{l}{$i$ \citet{Li+Bresolin+Kennicutt_2013}}\\
    \end{tabular}
  \end{center}
\vspace*{-1.0em}
\end{table*}
\addtocounter{table}{-1}
\begin{table*}[t]
  \footnotesize
  \begin{center}
    \caption{\label{tab:geom2} continued.}
    \begin{tabular}{|c | c c c c | c c c | c c | c |}
      \hline
       & & & & & \multicolumn{3}{c|}{optical} & \multicolumn{2}{c|}{M160 mask} & \multicolumn{1}{c|}{S250 mask} \\
Galaxy & Type & \multicolumn{2}{c}{$12+\log_{10}({\rm O/H})$} & $D$ & $R_{\rm maj}$ & $R_{\rm min}$ & $\Omega$ &$\Sigma_{Ld,{\rm min}}$ & $\Omega$ & $\Omega$ \\
 && PT & PP04N2$^a$ & (Mpc) & (kpc) & (kpc) & (arcmin$^2$) & ($L_\odot\pc^{-2}$) & (arcmin$^2$) &  (arcmin$^2$) \\
\hline
NGC5474  & SAcd  &$8.31\pm0.22$         & \,  8.46 &  6.80 &4.73                  &4.21                  &16.0                  &1.10 &21.1                  &17.9                  \\
NGC5713  & SABbc &$8.24\pm0.06$         & \,  8.70 & 21.40 &8.59                  &7.65                  &5.33                  &1.51 &35.6                  &14.3                  \\
NGC5866  & S0    &$8.47\pm0.20$         & \,  8.73 & 15.30 &10.4                  &3.90                  &6.44                  &1.91 &8.64                  &8.77                  \\
NGC6946  & SABcd &$8.400\pm0.030$       & \,  8.75 &  6.80 &11.4                  &9.63                  &87.8                  &7.08 &102.                  &100.                  \\
NGC7331  & SAb   &$8.340\pm0.020$       & \,  8.80 & 14.50 &22.1                  &6.83                  &26.7                  &3.02 &47.1                  &32.3                  \\
NGC7793  & SAd   &$8.310\pm0.020$       & \,  8.64 &  3.91 &5.31                  &3.55                  &45.8                  &1.35 &76.2                  &58.9                  \\
\hline
EIC3583  & IBm   &--- & \,--- & 14.20 &5.16                  &2.96                  &2.82                  &1.32 &6.57                  &5.96                  \\
ENGC0586 & SAa   &--- & \,--- & 20.80 &5.63                  &1.92                  &0.93                  &0.78 &5.31                  &4.93                  \\
ENGC1317 & SABa  &--- & \,--- & 21.00 &11.9                  &11.6                  &11.6                  &1.05 &10.8                  &10.7                  \\
ENGC1481 & SA0   &--- & \,--- & 22.60 &3.68                  &2.90                  &0.78                  &0.98 &5.40                  &2.00                  \\
ENGC1510 & SA0   &---& \,  8.38$^b$ &  11.60 &2.80                  &2.15                  &1.66                  &0.95 &7.74                  &1.76                  \\
ENGC3187 & SBc   &--- & \,--- & 19.30 &7.19                  &2.69                  &1.93                  &1.74 &7.92                  &5.53                  \\
ENGC4533 & SAd   &--- & \,--- & 14.50 &5.48                  &0.38                  &0.37                  &1.12 &6.12                  &3.83                  \\
ENGC7335 & SAO   &--- & \,--- & 83.40 &27.9                  &26.9                  &4.01                  &3.16 &2.61                  &2.60                  \\
ENGC7337 & SBb   &--- & \,--- & 87.20 &25.1                  &21.7                  &2.67                  &2.00 &2.16                  &1.98                  \\
\hline

    \multicolumn{7}{l}{$a$ Derived from KK metallicities from
                            \citet{Moustakas+Kennicutt+Tremonti+etal_2010}
                            except as noted} &
    \multicolumn{4}{l}{$b$ \citet{Marble+Engelbracht+vanZee+etal_2010}}\\
    \end{tabular}
  \end{center}
\vspace*{-1.0em}
\end{table*}

For the remaining 55
KINGFISH galaxies, we consider two popular
``strong line'' estimators:
the ``PT'' \citet{Pilyugin+Thuan_2005} method, 
taken from \citet{Moustakas+Kennicutt+Tremonti+etal_2010}, 
and the
``PP04N2'' method based on [NII]/H$\alpha$ \citep{Pettini+Pagel_2004}.
Abundance measurements by 
\citet[][``characteristic'' values from their Table 8]{Moustakas+Kennicutt+Tremonti+etal_2010} with the ``KK04'' 
\citep{Kobulnicky+Kewley_2004}
calibration were converted to PP04N2 values, 
according to the parameters recommended by  \citet{Kewley+Ellison_2008}.
This procedure is described in detail by 
\citet{Hunt+Dayal+Magrini+Ferrara_2016} who use the same metallicities
in their analysis; they preferred 
the PP04N2 calibration because it shows tighter scaling relations overall than other calibrations, and because
its behavior in the mass-metallicity relation is quite 
similar to weak-line electron-temperature determinations 
\citep[e.g.,][]{Andrews+Martini_2013}.

For NGC\,1316, NGC\,2841, and NGC\,5055 the original KK04 O/H values ($\sim$9.4)
from \citet{Moustakas+Kennicutt+Tremonti+etal_2010}
exceeded the range of applicability for the transformations
formulated by \citet{Kewley+Ellison_2008}.
Thus we have (somewhat arbitrarily) given these three galaxies a
maximum metallicity of $12+\log_{10}({\rm O/H}) = 9.0$,
consistent with what is advocated by
\citet{Pilyugin+Thuan+Vilchez_2007}.
Ultimately, the metallicities for these three galaxies are uncertain, but
toward the high end of the observed range.

Figure \ref{fig:oxyoxy} compares the PT and PP04N2 metallicity
estimates for the 55 KINGFISH
galaxies where ``direct'' method estimates are unavailable.
Note that the PT and PP04N2 metallicities differ by as much as
0.5 dex (e.g., DDO154, type IBm) or even 0.63 dex (e.g., NGC1482, type SA0).
It is evident that the metallicity
estimates have significant uncertainties, and that there are systematic
differences between the two methods \citep[see also][]{Kewley+Ellison_2008}.
Below we will argue, by comparing
PAH abundances estimated from infrared observations with
these two metallicity estimates, 
that the PP04N2 estimate appears to be more reliable, at least for
the galaxies in the KINGFISH sample.

\section{Observations and Data Reduction}
\label{sec:datasources}

\subsection{Infrared, Far-Infrared, and Submm}

Most of the galaxies in the KINGFISH sample are part of the SINGS
galaxy sample and were imaged
by {\it Spitzer Space Telescope}
as part of the SINGS observing program
\citep{Kennicutt+Armus+Bendo+etal_2003}.  IRAC and MIPS imaging
obtained by other {\it Spitzer Space Telescope} observing programs 
was available for the remaining KINGFISH galaxies.

The KINGFISH project imaged the
galaxies with the {\it Herschel Space Observatory}
\citep{Pilbratt+Riedinger+Passvogel+etal_2010}, following the
observing strategy described by
\citet{Kennicutt+Calzetti+Aniano+etal_2011}, using the 70,
100, and $160\micron$ PACS filters, and the
250, 350, and 500$\micron$ SPIRE filters.
The maps were designed
to cover a region out to $\gtsim 1.5$ times the optical radius
$R_{25}$, with good signal to noise (S/N) and redundancy.

Following AD12, we will use ``camera'' to identify each optical
configuration of the observing instruments, i.e., each different
channel or filter arrangement of the instruments will be referred to
as a different ``camera''.  With this nomenclature, each ``camera'' has
a characteristic spectral response and
point-spread function (PSF).  We will refer to the IRAC, MIPS, PACS,
and SPIRE cameras using their nominal wavelengths in microns: IRAC3.6,
IRAC4.5, IRAC5.8, IRAC8.0, MIPS24, MIPS70, MIPS160, PACS70, PACS100,
PACS160, SPIRE250, SPIRE350, and SPIRE500.

IRAC imaged the galaxies in four
bands, centered at 3.6$\micron$, 4.5$\micron$, 5.8$\micron$, and 8.0$\micron$,
as described by
\citet{Kennicutt+Armus+Bendo+etal_2003}.  The images were processed by
the SINGS Fifth Data Delivery pipeline.\footnote{Details can be found
  in the data release documentation:
\url{https://irsa.ipac.caltech.edu/data/SPITZER/SINGS/doc/sings_fifth_delivery_v2.pdf}}
The IRAC images are calibrated for point sources.  Photometry of
extended sources requires so-called ``aperture corrections''.  We
multiply the intensities in each pixel by the asymptotic (infinite
radii) value of the aperture correction (i.e., the aperture correction
corresponding to an infinite radius aperture).  We use the factors
0.91, 0.94, 0.66 and 0.74 for the 3.6$\micron$, 4.5$\micron$, 5.8$\micron$,
and 8.0$\micron$
bands, respectively, as described in the IRAC Instrument Handbook
(V2.0.1)\footnote{
\url{http://irsa.ipac.caltech.edu/data/SPITZER/docs/irac/iracinstrumenthandbook/IRAC\_Instrument\_Handbook.pdf}}.

Imaging with MIPS at 24, 70, and 160$\micron$
was carried out following the observing strategy
described in \citet{Kennicutt+Armus+Bendo+etal_2003}.  The data were
reduced using the LVL (Local Volume Legacy) project
pipeline.\footnote{Details can be found in the data release
  documentation:
\url{https://irsa.ipac.caltech.edu/data/SPITZER/LVL/LVL_DR5_v5.pdf}}.
A correction for nonlinearities in the MIPS70 camera was applied, as
described by \citet{Dale+Cohen+Johnson+etal_2009} and
\citet{Gordon+Meixner+Meade+etal_2011}.

The galaxies were observed with the PACS and SPIRE
instruments on {\it Herschel}, using the ``Scan Map''
observing mode.
Both PACS and SPIRE images were first reduced to ``level 1''
(flux-calibrated brightness time series, with attached sky
coordinates) using
HIPE v11.1.0 \citep{Ott_2010}, and maps
(``level 2'') were created using the Scanamorphos data reduction
pipeline \citep{Roussel_2013}, v24.0.  This reduction strategy
used the latest available PACS and SPIRE calibrations
(as of 2014 July),
and was designed to preserve the low
surface brightness diffuse emission.

The assumed beam sizes are 465.4, 822.6, and 1769$\,{\rm arcsec}^2$
for SPIRE250, SPIRE350, and SPIRE500, respectively.  Additionally, we
excluded discrepant bolometers from the map and adjusted the pointing
to match the MIPS24 map.

\subsection{\ion{H}{1} Observations}

To measure the \ion{H}{1} gas mass we use 
\ion{H}{1}\,21\,cm line observations made with the
NSF's NRAO\footnote{The National Radio Astronomy
  Observatory is a facility of the National Science Foundation
  operated under cooperative agreement by Associated Universities,
  Inc.} 
Karl G.\ Jansky Very Large Array (VLA).

For 23 of our galaxies we have data from
The \ion{H}{1} Nearby Galaxies Survey
\citep[THINGS][]{Walter+Brinks+deBlok+etal_2008} 
and for four galaxies we use data from the LittleTHINGS survey
\citep{Hunter+Ficut-Vicas+Ashley+etal_2012}.
For 10 galaxies without 
THINGS or LittleTHINGS observations, we obtained VLA 21-cm maps in
programs AL731 and AL735, in some cases also incorporating
archival VLA observations. 
For 8 targets, we reduced and incorporated VLA archival
observations of the 21 cm line.    
For one galaxy, NGC~4559, 
we use archival WSRT observations. 
These observations are described in
\citet{Leroy+Walter+Sandstrom+etal_2013}.  
For each galaxy, the source of the \ion{H}{1} map is listed in
Table~\ref{tab:gas}.
The dominant uncertainty on the measured \ion{H}{1} masses
comes from the calibration uncertainties of $\sim10$\%.

For NGC\,1266 
\citet{Alatalo+Blitz+Young+etal_2011} estimated
$M(\HI)= 9.5\times10^{6}\Msol$ based on
21\,cm absorption of
the radio continuum from the nucleus
(for an assumed $T_{\rm spin}=100\K$).
However, we estimate that \ion{H}{1}\,21cm line emission from
as much as $\sim$$2\times10^9\Msol$ of \ion{H}{1} could have gone undetected
because of the strong continuum ($0.1\Jy$ at 1.4GHz),
hence the \ion{H}{1} mass must be considered highly uncertain.

Thus we have \ion{H}{1}
data for 57 of the 61 KINGFISH galaxies.
\ion{H}{1}\,21\,cm observations were not available for 
NGC\,1316 (SAB0),
NGC\,1377 (S0),
NGC\,1404 (E1),
NGC\,5866 (S0), 
nor for the nine extra galaxies.

\subsection{CO Observations}

To estimate H$_2$ masses,
we use observations of CO line emission together with an assumed
ratio of $\HH$ mass to CO luminosity.  
The adopted CO-to-$\HH$ ``conversion factors'' are discussed in
\S\ref{sec:gas masses}.

For 38 KINGFISH galaxies
we use $^{12}$CO\,$J\!=\!2-\!1$ maps from the 
the HERA CO Line Emission Survey (HERACLES)
\citep{Leroy+Walter+Bigiel+etal_2009,Leroy+Walter+Sandstrom+etal_2013}.

For NGC~4826, we use $^{12}$CO\,$J\!=\!1\!-\!0$ mapping from the
Nobeyama Radio Observatory (Koda et al. in prep). 
We propagate uncertainties on the CO
integrated intensities from the spectra through the gridding and
masking of the cube as described in
\citet{Leroy+Walter+Sandstrom+etal_2013}.

For NGC~1266 we use $M(\HH)=1.6\times10^9\Msol$
from
\citet{Alatalo+Blitz+Young+etal_2011}.
We arbitrarily adopt a $\pm$50\% uncertainty.
For NGC~3190 we use CO line fluxes from 
\citet{Martinez-Badenes+Lisenfeld+Espada+etal_2012}.

Thus we have CO data for 41 of the 61 galaxies in the KINGFISH sample.
CO observations are not available for any of the nine extra galaxies.

\begin{table}
  {\scriptsize 
    \begin{center}
\caption{\label{tab:gas}
\ion{H}{1} \& CO Observation Summary}
\begin{tabular}{lll}
\hline 
\multicolumn{1}{l}{Galaxy} & \multicolumn{1}{l}{\ion{H}{1} Source} & \multicolumn{1}{l}{CO Source} \\
\hline 
DDO~053   &  LittleTHINGS                   & ... \\
DDO~154   &  THINGS              & HERACLES \\
DDO~165   &  LittleTHINGS                   & ...  \\
Holmberg~I     &  THINGS         & HERACLES \\
Holmberg~II    &  THINGS         & HERACLES \\
IC~342    &  Hyperleda                   & ... \\
IC~2574   &  THINGS              & HERACLES \\
M81~dwB   &  THINGS              & HERACLES \\
NGC~0337  &  Archival            & HERACLES \\
NGC~0584  &  Hyperleda                   & ... \\
NGC~0628  &  THINGS              & HERACLES       \\
NGC~0855  &  Hyperleda                   & ... \\
NGC~0925  &  THINGS              & HERACLES \\
NGC~1097  &  Hyperleda                   & ... \\
NGC~1266  &  ABY11               & ABY11 \\
NGC~1291  &  Hyperleda                   & ... \\
NGC~1482  &  Hyperleda                   & ... \\
NGC~1512  &  Hyperleda                   & ... \\
NGC~2146  &  AL735               & HERACLES \\
NGC~2798  &  AL735               & HERACLES \\
NGC~2841  &  THINGS              & HERACLES \\
NGC~2915  &  Hyperleda                   & ... \\
NGC~2976  &  THINGS              & HERACLES \\
NGC~3049  &  AL735               & HERACLES \\
NGC~3077  &  THINGS              & HERACLES \\
NGC~3184  &  THINGS              & HERACLES \\
NGC~3190  &  AL735               & MBLE12 \\
NGC~3198  &  THINGS              & HERACLES \\
NGC~3265  &  Hyperleda                   & ... \\
NGC~3351  &  THINGS              & HERACLES \\
NGC~3521  &  THINGS              & HERACLES \\
NGC~3621  &  THINGS              & ... \\
NGC~3627  &  THINGS              & HERACLES \\
NGC~3773  &  Hyperleda                   & ... \\
NGC~3938  &  Archival, AL731     & HERACLES \\
NGC~4236  &  AL731, AL735        & HERACLES \\
NGC~4254  &  Archival, AL731     & HERACLES \\
NGC~4321  &  Archival            & HERACLES \\
NGC~4536  &  Archival, AL731, AL735   & HERACLES \\
NGC~4559  &  Archival(WSRT)      & HERACLES \\
NGC~4569  &  Archival            & HERACLES \\
NGC~4579  &  Archival            & HERACLES \\
NGC~4594  &  Archival, AL735     & HERACLES \\
NGC~4625  &  Archival            & HERACLES \\
NGC~4631  &  Archival            & HERACLES \\
NGC~4725  &  AL735               & HERACLES \\
NGC~4736  &  THINGS              & HERACLES \\
NGC~4826  &  THINGS              & CANON \\
NGC~5055  &  THINGS              & HERACLES \\
NGC~5398  &  Hyperleda                   & ... \\
NGC~5408  &  Hyperleda                   & ... \\
NGC~5457  &  THINGS              & HERACLES \\
NGC~5474  &  Archival            & HERACLES \\
NGC~5713  &  Archival                & HERACLES \\
NGC~6946  &  THINGS              & HERACLES \\
NGC~7331  &  THINGS              & HERACLES \\
NGC~7793  &  THINGS              & ... \\
\hline
\multicolumn{3}{l}{THINGS = \citet{Walter+Brinks+deBlok+etal_2008}}\\
\multicolumn{3}{l}{HERACLES = \citet{Leroy+Walter+Bigiel+etal_2009,Leroy+Walter+Sandstrom+etal_2013}}\\
\multicolumn{3}{l}{ABY11 = \citet{Alatalo+Blitz+Young+etal_2011}}\\
\multicolumn{3}{l}{LittleTHINGS = \citet{Hunter+Ficut-Vicas+Ashley+etal_2012}}\\
\multicolumn{3}{l}{MBLE12 = \citet{Martinez-Badenes+Lisenfeld+Espada+etal_2012}}\\
\multicolumn{3}{l}{Hyperleda = \citet{Makarov+Prugniel+Terekhova+etal_2014}}\\
\multicolumn{3}{l}{CANON = \citet{Donovan_Meyer+Koda+Momose+etal_2013}}\\
\end{tabular}
    \end{center}
    }
\end{table}
\begin{table*}
{\footnotesize
\begin{center}
\caption{\label{tab:resolutions}Image Resolutions}
\begin{tabular}{c c c c c}
\hline 
                    & FWHM$^a$         & 50\% power$^a$                    &  Final grid                & Compatible\\
Camera       & ($\arcsec$) & diameter ($\arcsec$)   & pixel$^b$ ($\arcsec$) & cameras$^c$\\
\hline 
IRAC3.6  &  1.90 &  2.38 &  --- & not used as a final-map PSF\\
IRAC4.5  &  1.81 &  2.48 &  --- & not used as a final-map PSF\\
IRAC6  &  2.11 &  3.94 &  --- & not used as a final-map PSF\\
IRAC8  &  2.82 &  4.42 &  --- & not used as a final-map PSF\\
PACS70  &  5.67 &  8.46 &  --- & not used as a final-map PSF\\
MIPS24   &  6.43 &  9.86 & --- & not used as a final-map PSF\\
PACS100  &  7.04 &  9.74 & --- & not used as a final-map PSF\\
PACS160  & 11.2  & 15.3  & 5.0 & IRAC; MIPS24; PACS\\
SPIRE250 (S250) & 18.2  & 20.4  & 6.0 & IRAC; MIPS24; PACS; SPIRE250 \\
MIPS70   & 18.7  & 28.8  & 10.0 & IRAC; MIPS24,70; PACS; SPIRE250 \\
SPIRE350 & 24.9  & 26.8  & 10.0 & IRAC; MIPS24,70; PACS; SPIRE250,350\\
SPIRE500 & 36.1  & 39.0  & 15.0 & IRAC; MIPS24,70; PACS; SPIRE\\
MIPS160 (M160) & 38.8  & 58.0  & 18.0 & IRAC; MIPS; PACS; SPIRE\\
\hline
\multicolumn{5}{l}{$^a$ Values from 
\citet{Aniano+Draine+Gordon+Sandstrom_2011} for the circularized PSFs.}\\
\multicolumn{5}{l}{$^b$ The pixel size in the final-map grids is chosen to 
Nyquist-sample the PSFs.}\\
\multicolumn{5}{l}{$^c$ Other cameras that can be convolved into the camera 
PSF (see text for details).}
\end{tabular}
\end{center}
}
\end{table*}

\section{Image Analysis}
\label{sec:dataproc}

\subsection{Background Subtraction}

All camera images are first rotated to RA/Dec coordinates, and then
trimmed to a common sky region.  For each image we estimate the
best-fit ``tilted plane'' background (consisting of instrumental
background, Galactic foreground emission, and cosmic infrared
background emission) using an iterative procedure described in AD12.
The procedure uses multiple cameras to identify regions in the image
where only background emission is present.  Regions where excess
emission is detected at more than one wavelength are not used for
background estimation.

\subsection{Convolution to Common Resolution}

After background subtraction, the images are convolved to a common
point spread function (PSF), and resampled on a common final-map grid,
with pixel sizes for each final-map PSF as given in
Table \ref{tab:resolutions}.
Finally, the dispersion in intensities of
the background pixels (which includes noise coming from unresolved
undetected background sources) is used to estimate the pixel flux
uncertainties.  By comparing the MIPS and PACS images, we can also
estimate a calibration uncertainty.  
The procedures used are
fully described in AD12.

As discussed in AD12, multiwavelength observations must be degraded to
a common PSF before dust models are fit to the observed intensities.
The convolution to a common PSF is carried out using the methods
described by \citet{Aniano+Draine+Gordon+Sandstrom_2011}.  In the
present work, we present, for each galaxy, resolved results at two
final-map PSFs: SPIRE250 and MIPS160, henceforth abbreviated as S250
and M160.  S250 is the PSF with smallest FWHM (full width at half
maximum) that allows use of enough cameras to adequately constrain the
dust SED (IRAC, MIPS24, and PACS70, 100, 160, SPIRE250).  The M160 PSF
allows inclusion of {\it all} the cameras (IRAC, MIPS, PACS, SPIRE),
therefore producing the most reliable maps; this will be our ``gold
standard''.  Table \ref{tab:resolutions} 
lists the resolutions of the
cameras, the pixel size in the final-map grids used, and the other
cameras that can be used at this resolution.
In Appendix \ref{app:dependence on psf} we 
compare dust mass estimates obtained with different final-map PSFs.

\subsection{\label{sec:image segmentation}
            Image Segmentation}

After convolution to a common ``final-map''
PSF and background subtraction,
we next fit a dust model
to the observed SED of each pixel in the field.  
In order for dust mass estimation
to be reliable, the pixel's SED must be measured in a number of bands with
a reasonable signal/noise ratio.  
However, estimates of the total dust infrared luminosity
per unit area from a single pixel, $\SigLd$,
are reliable so long as there is a significant
detection of far-infrared emission after background subtraction.

The procedure used for automatically identifying ``galaxy'' pixels
is described in Appendices A and B of AD12.
For purposes of dust mass estimation, we need
to limit the modeling to a ``galaxy mask'' consisting of
pixels where the emission from the galaxy of interest has
sufficiently high surface brightness for dust mass estimation
(via SED fitting)
to be reasonably reliable.

A simple criterion for ``sufficiently high surface brightness'' is that
the total dust luminosity/projected area $\SigLd$
exceed a specified threshold value, $\SigLdmin$.
The value chosen for $\SigLdmin$ will depend on the noisiness of the
data [which may depend on the brightness of the (subtracted)
Galactic foreground emission, as well as on the presence of other
extragalactic objects in the field, stars, 
or even small-scale structure in the Galactic
foreground, which may compromise background estimation and subtraction].
The choice of $\SigLdmin$ will also depend on the choice of
final-map PSF: use of a larger PSF improves the signal/noise
in each pixel
by smoothing, and also enables use of more cameras to constrain the
dust modeling, and thus may allow use of a lower threshold $\SigLdmin$.
In the present study, $\SigLdmin$ was chosen subjectively for each galaxy.

In this paper we report results for two final-map
PSFs: S250, and M160,
with the PSF FWHM corresponding to \omittext{a} linear scales
${\rm FWHM} = 0.88 \kpc$ for S250 and \omittext{$1.75 \kpc$ for S500}
\newtext{$1.88\kpc$ for M160} at
the median distance $D=10\Mpc$ of the KINGFISH galaxy sample. 
Table \ref{tab:geom} lists $\SigLdmin$ used to define
the galaxy masks for
the \omittext{S250 and} M160 resolution studies,
\newtext{for the 62 galaxies where we detect dust emission}.  
\omittext{For this study we have chosen to adopt a 
common value of $\SigLdmin$ for both S250 and M160.}
Our adopted values of $\SigLdmin$ vary from galaxy to galaxy,
ranging from values
as low as \omittext{0.38}
$\newtext{0.58}\Lsol\pc^{-2}$ (\omittext{DDO\,165}\newtext{NGC\,2915}) 
to values as high as
\omittext{25}$\newtext{7}\Lsol\pc^{-2}$ 
(\omittext{NGC\,2146}\newtext{NGC\,6946}).
The median $\SigLdmin=$\omittext{1.9}$\newtext{1.6}\Lsol\pc^{-2}$ 
(e.g., \omittext{NGC\,4725}\newtext{NGC\,4625}). 
For each galaxy where dust is reliably detected
\newtext{at M160 resolution, we also generate a S250 resolution mask,
intended to comprise the region where the S250 resolution 
data permit reliable estimation of the dust surface density.
Our S250 masks are often similar in size to the M160 mask, 
but for some galaxies the S250 mask is considerably smaller 
than the M160 mask -- the most extreme example is NGC\,1481, 
where the S250 mask area is only 37\% of M160 mask area.}
The M160 and S250 masks
are shown in Figures
17.1-17.62.
The solid angle of each mask is listed in Table \ref{tab:geom}.
\omittext{They are generally quite similar, 
although in some cases the M160 mask is somewhat
larger (e.g., DDO053 or NGC1512, 
where the M160 mask is twice as large as the S250 mask).}
Because of the improved signal-to-noise ratio (S/N) in each pixel,
most of the analysis in this paper will be done with the M160
resolution images and masks.

\omittext{However, for} \newtext{For} 5 dwarf
galaxies where dust detection is uncertain
(DDO053, DDO154, DDO165, Hol1, and M81dwB)
we choose instead to use
masks defined by \ion{H}{1} observations.
For 3 elliptical galaxies where dust detection is
uncertain (NGC0584, NGC0855, and NGC1404)
we use \omittext{the} $\SigLdmin$-based masks. \omittext{discussed above}.
\newtext{We do not detect dust in any of these 8 galaxies.}
See Appendix \ref{app:upper limits} for further details.

\begin{table*}
  \scriptsize
  \begin{center}
    \caption{\label{tab:spitzer}Spitzer Photometry of KINGFISH
             Galaxies with Dust Detections}
\setlength{\tabcolsep}{2pt}
    \begin{tabular}{| c c | c c c c | c c c |}

      \hline
      &&\multicolumn{4}{c|}{IRAC $F_\nu(\Jy)$}&\multicolumn{3}{c|}{MIPS $F_\nu(\Jy)$}\\
Galaxy & mask & $3.6\micron$ & $4.5\micron$ & $5.8\micron$ & $8.0\micron$ & $24\micron$ & $70\micron$ & $160\micron$ \\
\hline
Hol2     & S250 &$0.090\pm0.018$       & $0.065\pm0.014$       & $0.041\pm0.015$       & $0.047\pm0.013$       & $0.187\pm0.033$       & ---                   & ---                  \\
" & M160 & $0.097\pm0.024$       & $0.070\pm0.019$       & $0.042\pm0.015$       & $0.047\pm0.014$       & $0.178\pm0.027$       & $ 3.1\pm 1.4$         & $ 3.3\pm 1.9$        \\
IC342    & S250 &$14.3\pm2.3$          & $ 8.6\pm 1.6$         & $12.0\pm3.6$          & $ 30.\pm  6.$         & $36.2\pm4.3$          & ---                   & ---                  \\
" & M160 & $14.3\pm2.5$          & $ 8.6\pm 1.5$         & $12.0\pm3.4$          & $ 30.\pm  6.$         & $36.3\pm4.1$          & $338.\pm 161.$        & $913.\pm 302.$       \\
IC2574   & S250 &$0.136\pm0.028$       & $0.101\pm0.021$       & $0.058\pm0.023$       & $0.061\pm0.035$       & $0.27\pm0.06$         & ---                   & ---                  \\
" & M160 & $0.135\pm0.035$       & $0.100\pm0.027$       & $0.057\pm0.021$       & $0.059\pm0.029$       & $0.246\pm0.044$       & $ 4.8\pm 2.1$         & $ 9.5\pm 5.0$        \\
NGC0337  & S250 &$0.092\pm0.010$       & $0.065\pm0.006$       & $0.139\pm0.035$       & $0.37\pm0.07$         & $0.76\pm0.08$         & ---                   & ---                  \\
" & M160 & $0.092\pm0.015$       & $0.066\pm0.008$       & $0.137\pm0.036$       & $0.36\pm0.07$         & $0.72\pm0.08$         & $10.1\pm3.1$          & $16.5\pm3.9$         \\
NGC0628  & S250 &$0.87\pm0.10$         & $0.60\pm0.06$         & $1.03\pm0.25$         & $ 2.7\pm 0.5$         & $3.18\pm0.38$         & ---                   & ---                  \\
" & M160 & $0.86\pm0.11$         & $0.59\pm0.06$         & $1.00\pm0.24$         & $2.67\pm0.48$         & $3.10\pm0.34$         & $ 32.\pm 12.$         & $106.\pm  16.$       \\
NGC0925  & S250 &$0.330\pm0.048$       & $0.226\pm0.044$       & $0.29\pm0.08$         & $0.63\pm0.12$         & $0.85\pm0.12$         & ---                   & ---                  \\
" & M160 & $0.33\pm0.06$         & $0.226\pm0.048$       & $0.28\pm0.07$         & $0.61\pm0.12$         & $0.82\pm0.10$         & $13.0\pm3.8$          & $ 37.\pm  7.$        \\
NGC1097  & S250 &$1.19\pm0.12$         & $0.79\pm0.07$         & $1.22\pm0.29$         & $ 3.1\pm 0.5$         & $ 6.5\pm 0.7$         & ---                   & ---                  \\
" & M160 & $1.17\pm0.13$         & $0.79\pm0.08$         & $1.19\pm0.28$         & $ 3.0\pm 0.5$         & $ 6.4\pm 0.7$         & $ 57.\pm 26.$         & $143.\pm  34.$       \\
NGC1266  & S250 &$0.062\pm0.007$       & $0.048\pm0.005$       & $0.053\pm0.014$       & $0.099\pm0.020$       & $0.88\pm0.09$         & ---                   & ---                  \\
" & M160 & $0.059\pm0.008$       & $0.046\pm0.006$       & $0.051\pm0.013$       & $0.095\pm0.019$       & $0.83\pm0.09$         & $11.5\pm3.8$          & $ 9.2\pm 4.1$        \\
NGC1291  & S250 &$1.99\pm0.19$         & $1.24\pm0.12$         & $0.84\pm0.23$         & $0.73\pm0.14$         & $0.53\pm0.11$         & ---                   & ---                  \\
" & M160 & $1.97\pm0.20$         & $1.23\pm0.13$         & $0.82\pm0.21$         & $0.72\pm0.14$         & $0.51\pm0.08$         & $ 5.7\pm 9.6$         & $ 25.\pm  9.$        \\
NGC1316  & S250 &$1.55\pm0.14$         & $0.98\pm0.08$         & $0.65\pm0.15$         & $0.50\pm0.09$         & $0.353\pm0.042$       & ---                   & ---                  \\
" & M160 & $1.47\pm0.14$         & $0.93\pm0.08$         & $0.62\pm0.14$         & $0.47\pm0.09$         & $0.334\pm0.038$       & $ 4.8\pm 1.3$         & $ 9.6\pm 2.3$        \\
NGC1377  & S250 &$0.065\pm0.007$       & $0.092\pm0.009$       & $0.25\pm0.06$         & $0.43\pm0.07$         & $1.79\pm0.19$         & ---                   & ---                  \\
" & M160 & $0.061\pm0.008$       & $0.087\pm0.010$       & $0.24\pm0.05$         & $0.41\pm0.07$         & $1.66\pm0.17$         & $ 6.0\pm 1.3$         & $ 2.9\pm 1.0$        \\
NGC1482  & S250 &$0.380\pm0.035$       & $0.271\pm0.022$       & $0.63\pm0.14$         & $1.60\pm0.27$         & $2.33\pm0.24$         & ---                   & ---                  \\
" & M160 & $0.372\pm0.042$       & $0.283\pm0.029$       & $0.60\pm0.14$         & $1.58\pm0.27$         & $2.07\pm0.22$         & $ 31.\pm 12.$         & $ 37.\pm 10.$        \\
NGC1512  & S250 &$0.321\pm0.030$       & $0.207\pm0.018$       & $0.209\pm0.049$       & $0.39\pm0.07$         & $0.42\pm0.05$         & ---                   & ---                  \\
" & M160 & $0.350\pm0.042$       & $0.228\pm0.030$       & $0.22\pm0.05$         & $0.41\pm0.07$         & $0.44\pm0.06$         & $ 6.1\pm 1.8$         & $20.1\pm3.8$         \\
NGC2146  & S250 & --                   & $2.11\pm0.16$         &  --                   & $ 8.9\pm 1.5$         & $10.4\pm1.1$          & ---                   & ---                  \\
" & M160 &  --                   & $2.17\pm0.21$         &  --                   & $ 8.8\pm 1.6$         & $10.8\pm1.2$          & $202.\pm  42.$        & $112.\pm  75.$       \\
NGC2798  & S250 &$0.126\pm0.014$       & $0.091\pm0.008$       & $0.19\pm0.06$         & $0.63\pm0.11$         & $1.54\pm0.16$         & ---                   & ---                  \\
" & M160 & $0.123\pm0.018$       & $0.088\pm0.010$       & $0.18\pm0.06$         & $0.61\pm0.11$         & $1.36\pm0.14$         & $21.8\pm3.7$          & $ 21.\pm  5.$        \\
NGC2841  & S250 &$1.25\pm0.12$         & $0.77\pm0.14$         & $0.69\pm0.18$         & $1.14\pm0.34$         & $0.96\pm0.12$         & ---                   & ---                  \\
" & M160 & $1.23\pm0.12$         & $0.77\pm0.11$         & $0.68\pm0.17$         & $1.13\pm0.25$         & $0.94\pm0.12$         & $ 9.9\pm 4.1$         & $ 56.\pm 12.$        \\
NGC2915  & S250 &$0.050\pm0.008$       & $0.033\pm0.006$       & $0.022\pm0.008$       & $0.027\pm0.009$       & $0.056\pm0.012$       & ---                   & ---                  \\
" & M160 & $0.045\pm0.010$       & $0.030\pm0.008$       & $0.019\pm0.008$       & $0.024\pm0.007$       & $0.047\pm0.010$       & $1.12\pm0.34$         & $0.72\pm1.21$        \\
NGC2976  & S250 &$0.392\pm0.038$       & $0.267\pm0.023$       & $0.43\pm0.10$         & $1.00\pm0.18$         & $1.40\pm0.16$         & ---                   & ---                  \\
" & M160 & $0.377\pm0.042$       & $0.257\pm0.026$       & $0.41\pm0.10$         & $0.96\pm0.17$         & $1.34\pm0.15$         & $19.0\pm3.2$          & $ 46.\pm  9.$        \\
NGC3049  & S250 &$0.043\pm0.006$       & $0.0291\pm0.0049$     & $0.048\pm0.014$       & $0.120\pm0.024$       & $0.44\pm0.05$         & ---                   & ---                  \\
" & M160 & $0.044\pm0.008$       & $0.031\pm0.007$       & $0.047\pm0.013$       & $0.115\pm0.022$       & $0.404\pm0.044$       & $ 2.6\pm 1.3$         & $ 4.2\pm 1.5$        \\
NGC3077  & S250 &$0.52\pm0.06$         & $0.35\pm0.05$         & $0.39\pm0.10$         & $0.85\pm0.17$         & $1.49\pm0.17$         & ---                   & ---                  \\
" & M160 & $0.52\pm0.07$         & $0.35\pm0.06$         & $0.38\pm0.10$         & $0.84\pm0.17$         & $1.36\pm0.15$         & $18.0\pm4.3$          & $ 32.\pm  8.$        \\
NGC3184  & S250 &$0.52\pm0.06$         & $0.348\pm0.040$       & $0.54\pm0.14$         & $1.37\pm0.25$         & $1.46\pm0.18$         & ---                   & ---                  \\
" & M160 & $0.51\pm0.07$         & $0.341\pm0.048$       & $0.53\pm0.14$         & $1.33\pm0.24$         & $1.42\pm0.16$         & $ 15.\pm  5.$         & $ 63.\pm 13.$        \\
NGC3190  & S250 &$0.340\pm0.038$       & $0.215\pm0.028$       & $0.188\pm0.049$       & $0.29\pm0.06$         & $0.263\pm0.033$       & ---                   & ---                  \\
" & M160 & $0.319\pm0.038$       & $0.203\pm0.029$       & $0.175\pm0.045$       & $0.28\pm0.06$         & $0.245\pm0.030$       & $ 4.9\pm 1.5$         & $13.7\pm2.6$         \\
NGC3198  & S250 &$0.273\pm0.031$       & $0.184\pm0.022$       & $0.21\pm0.10$         & $0.67\pm0.12$         & $1.07\pm0.12$         & ---                   & ---                  \\
" & M160 & $0.274\pm0.041$       & $0.185\pm0.029$       & $0.21\pm0.09$         & $0.66\pm0.12$         & $1.03\pm0.12$         & $ 9.8\pm 2.8$         & $ 35.\pm  8.$        \\
NGC3265  & S250 &$0.0288\pm0.0041$     & $0.0198\pm0.0025$     & $0.036\pm0.010$       & $0.095\pm0.019$       & $0.291\pm0.033$       & ---                   & ---                  \\
" & M160 & $0.0273\pm0.0040$     & $0.0189\pm0.0031$     & $0.034\pm0.010$       & $0.090\pm0.017$       & $0.269\pm0.029$       & $ 2.4\pm 0.9$         & $ 2.5\pm 0.9$        \\
NGC3351  & S250 &$0.75\pm0.07$         & $0.488\pm0.042$       & $0.58\pm0.14$         & $1.26\pm0.23$         & $2.52\pm0.29$         & ---                   & ---                  \\
" & M160 & $0.74\pm0.08$         & $0.48\pm0.05$         & $0.56\pm0.13$         & $1.23\pm0.22$         & $2.44\pm0.26$         & $ 21.\pm  8.$         & $ 58.\pm 14.$        \\
NGC3521  & S250 &$1.88\pm0.18$         & $1.28\pm0.11$         & $2.10\pm0.49$         & $ 5.8\pm 1.0$         & $ 5.6\pm 0.6$         & ---                   & ---                  \\
" & M160 & $1.86\pm0.19$         & $1.28\pm0.12$         & $2.07\pm0.48$         & $ 5.7\pm 1.0$         & $ 5.5\pm 0.6$         & $ 64.\pm 22.$         & $203.\pm  39.$       \\
NGC3621  & S250 &$1.62\pm0.24$         & $1.18\pm0.13$         & $ 2.2\pm 0.6$         & $ 4.0\pm 0.7$         & $3.63\pm0.43$         & ---                   & ---                  \\
" & M160 & $1.62\pm0.27$         & $1.18\pm0.15$         & $ 2.1\pm 0.6$         & $ 4.0\pm 0.7$         & $3.60\pm0.41$         & $ 46.\pm 13.$         & $122.\pm  25.$       \\
NGC3627  & S250 &$1.86\pm0.17$         & $1.28\pm0.10$         & $2.06\pm0.47$         & $ 5.3\pm 0.9$         & $ 7.5\pm 0.8$         & ---                   & ---                  \\
" & M160 & $1.88\pm0.20$         & $1.29\pm0.13$         & $2.04\pm0.48$         & $ 5.2\pm 0.9$         & $ 7.4\pm 0.8$         & $ 88.\pm 27.$         & $218.\pm  35.$       \\
NGC3773  & S250 &$0.0236\pm0.0031$     & $0.0157\pm0.0021$     & $0.020\pm0.006$       & $0.046\pm0.011$       & $0.143\pm0.019$       & ---                   & ---                  \\
" & M160 & $0.0218\pm0.0036$     & $0.0146\pm0.0027$     & $0.019\pm0.005$       & $0.043\pm0.009$       & $0.129\pm0.015$       & $ 1.4\pm 0.5$         & $ 1.8\pm 0.9$        \\
NGC3938  & S250 &$0.313\pm0.033$       & $0.213\pm0.021$       & $0.38\pm0.09$         & $0.99\pm0.17$         & $1.10\pm0.13$         & ---                   & ---                  \\
" & M160 & $0.308\pm0.040$       & $0.211\pm0.026$       & $0.37\pm0.09$         & $0.96\pm0.17$         & $1.06\pm0.12$         & $13.2\pm4.3$          & $ 46.\pm 10.$        \\
NGC4236  & S250 &$0.22\pm0.06$         & $0.155\pm0.047$       & $0.092\pm0.098$       & $0.15\pm0.06$         & $0.51\pm0.08$         & ---                   & ---                  \\
" & M160 & $0.21\pm0.06$         & $0.150\pm0.047$       & $0.087\pm0.072$       & $0.145\pm0.048$       & $0.48\pm0.07$         & $ 7.8\pm 5.2$         & $ 16.\pm  5.$        \\

      \hline
    \end{tabular}
  \end{center}
\btdnote{uses tab\_spitzer\_guts1.tex created by tablemaker\_v9}
\end{table*}
\addtocounter{table}{-1}
\begin{table*}
  \scriptsize
  \begin{center}
    \caption{Spitzer Photometry of KINGFISH Galaxies with Dust Detections,
             contd.}
\setlength{\tabcolsep}{2pt}
    \begin{tabular}{| c c | c c c c | c c c |}
      \hline
      &&\multicolumn{4}{c|}{IRAC $F_\nu(\Jy)$}&\multicolumn{3}{c|}{MIPS $F_\nu(\Jy)$}\\
Galaxy & mask & $3.6\micron$ & $4.5\micron$ &$5.8\micron$ & $8.0\micron$ & $24\micron$ &$70\micron$ & $160\micron$ \\
\hline
NGC4254  & S250 & $0.68\pm0.06$         & $0.48\pm0.09$         & $1.29\pm0.30$         & $ 3.9\pm 0.7$         & $4.23\pm0.45$         & ---                   & ---                   \\
" & M160 & $0.68\pm0.07$         & $0.47\pm0.11$         & $1.26\pm0.29$         & $ 3.8\pm 0.8$         & $4.11\pm0.45$         & $ 46.\pm 14.$         & $127.\pm  23.$        \\
NGC4321  & S250 & $0.88\pm0.09$         & $0.589\pm0.048$       & $1.04\pm0.24$         & $2.86\pm0.49$         & $3.39\pm0.36$         & ---                   & ---                   \\
" & M160 & $0.89\pm0.11$         & $0.60\pm0.06$         & $1.03\pm0.25$         & $2.81\pm0.49$         & $3.33\pm0.35$         & $ 38.\pm 11.$         & $126.\pm  26.$        \\
NGC4536  & S250 & $0.394\pm0.040$       & $0.293\pm0.026$       & $0.55\pm0.14$         & $1.61\pm0.29$         & $3.43\pm0.37$         & ---                   & ---                   \\
" & M160 & $0.389\pm0.048$       & $0.295\pm0.039$       & $0.53\pm0.13$         & $1.59\pm0.29$         & $3.30\pm0.35$         & $ 30.\pm 13.$         & $ 54.\pm 11.$         \\
NGC4559  & S250 & $0.402\pm0.043$       & $0.280\pm0.029$       & $0.38\pm0.09$         & $0.84\pm0.15$         & $1.13\pm0.14$         & ---                   & ---                   \\
" & M160 & $0.388\pm0.047$       & $0.271\pm0.033$       & $0.37\pm0.09$         & $0.81\pm0.14$         & $1.07\pm0.12$         & $15.7\pm4.0$          & $ 46.\pm  9.$         \\
NGC4569  & S250 & $0.65\pm0.06$         & $0.421\pm0.034$       & $0.47\pm0.11$         & $0.98\pm0.17$         & $1.41\pm0.16$         & ---                   & ---                   \\
" & M160 & $0.61\pm0.06$         & $0.395\pm0.034$       & $0.44\pm0.10$         & $0.92\pm0.16$         & $1.32\pm0.14$         & $10.8\pm3.9$          & $ 37.\pm  7.$         \\
NGC4579  & S250 & $0.76\pm0.07$         & $0.490\pm0.041$       & $0.44\pm0.11$         & $0.70\pm0.13$         & $0.79\pm0.09$         & ---                   & ---                   \\
" & M160 & $0.74\pm0.08$         & $0.475\pm0.043$       & $0.43\pm0.10$         & $0.68\pm0.13$         & $0.75\pm0.08$         & $ 8.9\pm 2.1$         & $ 36.\pm  6.$         \\
NGC4594  & S250 & $3.26\pm0.31$         & $2.04\pm0.17$         & $1.38\pm0.33$         & $1.22\pm0.24$         & $0.71\pm0.09$         & ---                   & ---                   \\
" & M160 & $3.14\pm0.30$         & $1.96\pm0.17$         & $1.33\pm0.31$         & $1.17\pm0.23$         & $0.68\pm0.08$         & $ 7.4\pm 2.6$         & $ 36.\pm  7.$         \\
NGC4625  & S250 & $0.045\pm0.005$       & $0.0296\pm0.0034$     & $0.050\pm0.015$       & $0.126\pm0.023$       & $0.132\pm0.017$       & ---                   & ---                   \\
" & M160 & $0.044\pm0.007$       & $0.0291\pm0.0048$     & $0.049\pm0.015$       & $0.119\pm0.022$       & $0.122\pm0.015$       & $1.83\pm0.48$         & $ 4.7\pm 1.2$         \\
NGC4631  & S250 & $1.20\pm0.11$         & $0.85\pm0.07$         & $ 2.2\pm 0.5$         & $ 5.8\pm 1.0$         & $ 8.0\pm 0.8$         & ---                   & ---                   \\
" & M160 & $1.22\pm0.14$         & $0.88\pm0.08$         & $ 2.2\pm 0.5$         & $ 5.9\pm 1.0$         & $ 8.0\pm 0.8$         & $134.\pm  21.$        & $265.\pm  47.$        \\
NGC4725  & S250 & $1.03\pm0.10$         & $0.66\pm0.06$         & $0.60\pm0.16$         & $1.02\pm0.18$         & $0.85\pm0.12$         & ---                   & ---                   \\
" & M160 & $1.03\pm0.11$         & $0.65\pm0.07$         & $0.60\pm0.15$         & $1.00\pm0.18$         & $0.83\pm0.10$         & $ 8.5\pm 4.3$         & $ 51.\pm  9.$         \\
NGC4736  & S250 & $3.29\pm0.36$         & $2.14\pm0.21$         & $ 2.3\pm 0.6$         & $ 4.8\pm 0.9$         & $ 5.6\pm 0.6$         & ---                   & ---                   \\
" & M160 & $3.26\pm0.36$         & $2.12\pm0.22$         & $ 2.3\pm 0.6$         & $ 4.8\pm 0.8$         & $ 5.5\pm 0.6$         & $ 92.\pm 29.$         & $163.\pm  29.$        \\
NGC4826  & S250 & $2.05\pm0.18$         & $1.31\pm0.10$         & $1.22\pm0.28$         & $2.04\pm0.35$         & $2.46\pm0.26$         & ---                   & ---                   \\
" & M160 & $2.19\pm0.21$         & $1.40\pm0.11$         & $1.25\pm0.29$         & $2.05\pm0.35$         & $2.43\pm0.26$         & $ 52.\pm  9.$         & $ 85.\pm 22.$         \\
NGC5055  & S250 & $2.25\pm0.20$         & $1.51\pm0.12$         & $ 2.3\pm 0.6$         & $ 5.7\pm 1.0$         & $ 5.7\pm 0.6$         & ---                   & ---                   \\
" & M160 & $2.28\pm0.22$         & $1.52\pm0.13$         & $ 2.3\pm 0.6$         & $ 5.7\pm 1.0$         & $ 5.7\pm 0.6$         & $ 73.\pm 14.$         & $269.\pm  44.$        \\
NGC5398  & S250 & $0.038\pm0.007$       & $0.027\pm0.005$       & $0.029\pm0.009$       & $0.058\pm0.011$       & $0.260\pm0.029$       & ---                   & ---                   \\
" & M160 & $0.039\pm0.010$       & $0.027\pm0.008$       & $0.029\pm0.009$       & $0.055\pm0.012$       & $0.240\pm0.027$       & $ 1.8\pm 0.7$         & $ 2.6\pm 1.0$         \\
NGC5408  & S250 & $0.082\pm0.021$       & $0.064\pm0.012$       & $0.041\pm0.021$       & $0.039\pm0.010$       & $0.403\pm0.044$       & ---                   & ---                   \\
" & M160 & $0.082\pm0.038$       & $0.064\pm0.021$       & $0.042\pm0.036$       & $0.039\pm0.014$       & $0.370\pm0.043$       & $ 3.0\pm 0.6$         & $ 1.8\pm 0.8$         \\
NGC5457  & S250 & $2.73\pm0.39$         & $1.83\pm0.26$         & $ 3.1\pm 0.9$         & $ 7.6\pm 1.4$         & $10.9\pm1.3$          & ---                   & ---                   \\
" & M160 & $2.74\pm0.41$         & $1.84\pm0.27$         & $ 3.1\pm 0.9$         & $ 7.6\pm 1.4$         & $10.9\pm1.2$          & $123.\pm  33.$        & $410.\pm  82.$        \\
NGC5474  & S250 & $0.101\pm0.012$       & $0.069\pm0.008$       & $0.082\pm0.026$       & $0.103\pm0.025$       & $0.160\pm0.026$       & ---                   & ---                   \\
" & M160 & $0.101\pm0.015$       & $0.067\pm0.010$       & $0.082\pm0.024$       & $0.101\pm0.022$       & $0.151\pm0.021$       & $ 3.4\pm 1.4$         & $ 8.7\pm 2.5$         \\
NGC5713  & S250 & $0.198\pm0.020$       & $0.136\pm0.018$       & $0.39\pm0.09$         & $1.13\pm0.20$         & $2.38\pm0.25$         & ---                   & ---                   \\
" & M160 & $0.205\pm0.029$       & $0.137\pm0.025$       & $0.40\pm0.10$         & $1.09\pm0.19$         & $2.35\pm0.25$         & $ 23.\pm  8.$         & $ 38.\pm 10.$         \\
NGC5866  & S250 & $0.61\pm0.05$         & $0.390\pm0.030$       & $0.26\pm0.06$         & $0.285\pm0.050$       & $0.214\pm0.026$       & ---                   & ---                   \\
" & M160 & $0.55\pm0.05$         & $0.355\pm0.029$       & $0.24\pm0.06$         & $0.265\pm0.046$       & $0.195\pm0.022$       & $ 7.8\pm 1.4$         & $16.5\pm4.0$          \\
NGC6946  & S250 & $ 3.4\pm 0.5$         & $2.44\pm0.32$         & $ 5.2\pm 1.3$         & $13.8\pm2.5$          & $19.7\pm2.1$          & ---                   & ---                   \\
" & M160 & $ 3.4\pm 0.6$         & $2.41\pm0.38$         & $ 5.1\pm 1.3$         & $13.6\pm2.4$          & $19.4\pm2.1$          & $202.\pm  59.$        & $438.\pm  98.$        \\
NGC7331  & S250 & $1.51\pm0.16$         & $1.00\pm0.09$         & $1.59\pm0.37$         & $ 4.0\pm 0.7$         & $4.04\pm0.43$         & ---                   & ---                   \\
" & M160 & $1.52\pm0.20$         & $1.01\pm0.11$         & $1.54\pm0.37$         & $ 3.8\pm 0.7$         & $3.91\pm0.41$         & $ 58.\pm 13.$         & $155.\pm  35.$        \\
NGC7793  & S250 & $0.74\pm0.09$         & $0.503\pm0.045$       & $0.81\pm0.21$         & $1.90\pm0.34$         & $2.12\pm0.25$         & ---                   & ---                   \\
" & M160 & $0.74\pm0.10$         & $0.50\pm0.06$         & $0.80\pm0.21$         & $1.88\pm0.33$         & $2.09\pm0.23$         & $ 33.\pm  8.$         & $108.\pm  22.$        \\
\hline
IC3583  & S250 & $0.037\pm0.005$       & $0.0237\pm0.0030$     & $0.019\pm0.007$       & $0.034\pm0.009$       & $0.048\pm0.010$       & ---                   & ---                   \\
" & M160 & $0.034\pm0.007$       & $0.0224\pm0.0040$     & $0.017\pm0.005$       & $0.033\pm0.008$       & $0.045\pm0.007$       & $0.84\pm0.31$         & $ 1.8\pm 0.7$         \\
NGC0586 & S250 & $0.00063\pm0.00221$   & $0.00031\pm0.00132$   & $-.0006\pm0.0031$     & $-.0001\pm0.0032$     & $0.029\pm0.006$       & ---                   & ---                   \\
" & M160 & $0.00061\pm0.00563$   & $0.00051\pm0.00264$   & $-.0002\pm0.0026$     & $-.0000\pm0.0021$     & $0.0254\pm0.0041$     & $0.36\pm0.33$         & $ 1.3\pm 0.6$         \\
NGC1317 & S250 & $0.262\pm0.028$       & $0.164\pm0.017$       & $0.159\pm0.040$       & $0.269\pm0.049$       & $0.253\pm0.030$       & ---                   & ---                   \\
" & M160 & $0.248\pm0.030$       & $0.156\pm0.020$       & $0.149\pm0.037$       & $0.253\pm0.046$       & $0.234\pm0.027$       & $ 5.1\pm 1.1$         & $10.6\pm2.6$          \\
NGC1481 & S250 & $0.0148\pm0.0025$     &  --                   & $0.0132\pm0.0041$     &  --                   & $0.050\pm0.007$       & ---                   & ---                   \\
" & M160 & $0.019\pm0.007$       &  --                   & $0.016\pm0.007$       &  --                   & $0.052\pm0.008$       & $0.62\pm0.74$         & $0.73\pm0.46$         \\
NGC1510 & S250 & $0.0149\pm0.0017$     & $0.0106\pm0.0011$     & $0.0102\pm0.0027$     & $0.0193\pm0.0037$     & $0.127\pm0.014$       & ---                   & ---                   \\
" & M160 & $0.024\pm0.005$       & $0.0172\pm0.0046$     & $0.0133\pm0.0043$     & $0.024\pm0.006$       & $0.134\pm0.016$       & $1.08\pm0.48$         & $ 1.3\pm 0.5$         \\
NGC3187 & S250 & $0.0249\pm0.0048$     & $0.019\pm0.005$       & $0.030\pm0.009$       & $0.064\pm0.014$       & $0.101\pm0.014$       & ---                   & ---                   \\
" & M160 & $0.027\pm0.010$       & $0.021\pm0.010$       & $0.029\pm0.010$       & $0.061\pm0.018$       & $0.093\pm0.013$       & $1.25\pm0.46$         & $ 3.8\pm 1.1$         \\
NGC4533 & S250 & $0.0158\pm0.0023$     & $0.0100\pm0.0017$     & $0.0088\pm0.0047$     & $0.022\pm0.006$       & $0.030\pm0.007$       & ---                   & ---                   \\
" & M160 & $0.0165\pm0.0042$     & $0.0098\pm0.0026$     & $0.0067\pm0.0047$     & $0.021\pm0.007$       & $0.029\pm0.007$       & $0.51\pm0.35$         & $ 1.1\pm 0.8$         \\
NGC7335 & S250 & $0.047\pm0.007$       & $0.0305\pm0.0034$     & $0.021\pm0.006$       & $0.027\pm0.006$       & $0.0299\pm0.0049$     & ---                   & ---                   \\
" & M160 & $0.040\pm0.008$       & $0.0260\pm0.0043$     & $0.020\pm0.006$       & $0.027\pm0.006$       & $0.0294\pm0.0046$     & $0.54\pm0.17$         & $ 1.6\pm 0.5$         \\
NGC7337 & S250 & $0.0278\pm0.0040$     & $0.0179\pm0.0022$     & $0.0135\pm0.0040$     & $0.0215\pm0.0045$     & $0.0201\pm0.0035$     & ---                   & ---                   \\
" & M160 & $0.024\pm0.006$       & $0.0156\pm0.0036$     & $0.0117\pm0.0042$     & $0.019\pm0.005$       & $0.0171\pm0.0031$     & $0.33\pm0.25$         & $0.89\pm0.40$         \\

      \hline
    \end{tabular}
  \end{center}
\btdnote{uses tab\_spitzer\_guts2.tex created by tablemaker\_v9}
\end{table*}
\begin{table*}
  \scriptsize
  \begin{center}
    \caption{\label{tab:herschel}Herschel Photometry of KINGFISH Galaxies
             with Dust Detections}
    \begin{tabular}{| c c | c c c | c c c |}
      \hline
      &&\multicolumn{3}{c|}{PACS $F_\nu(\Jy)$}&\multicolumn{3}{c|}{SPIRE $F_\nu(\Jy)$}\\
Galaxy & mask & $70\micron$ & $100\micron$ & $160\micron$& $250\micron$ & $350\micron$ & $500\micron$ \\
\hline
Hol2     & S250 & $ 4.3\pm 3.1$         & $ 4.2\pm 3.7$         & $ 3.2\pm 2.5$         & $ 1.8\pm 0.5$         &  --                   &  --                  \\
" & M160 & $ 4.1\pm 2.4$         & $ 4.0\pm 2.9$         & $ 2.9\pm 2.0$         & $1.70\pm0.39$         & $0.96\pm0.25$         & $0.44\pm0.14$        \\
IC342    & S250 & $465.\pm 230.$        & $904.\pm 402.$        & $1108.\pm 325.$       & $587.\pm  81.$        &  --                   &  --                  \\
" & M160 & $462.\pm 196.$        & $901.\pm 336.$        & $1101.\pm 293.$       & $584.\pm  81.$        & $262.\pm  37.$        & $ 97.\pm 14.$        \\
IC2574   & S250 & $ 5.8\pm 5.9$         & $ 6.6\pm 7.0$         & $ 8.5\pm 5.7$         & $ 6.4\pm 1.7$         &  --                   &  --                  \\
" & M160 & $ 5.4\pm 4.2$         & $ 6.0\pm 4.8$         & $ 7.9\pm 4.7$         & $ 6.0\pm 1.3$         & $ 3.9\pm 0.8$         & $1.79\pm0.44$        \\
NGC0337  & S250 & $13.8\pm3.9$          & $ 21.\pm  6.$         & $19.8\pm4.0$          & $ 8.4\pm 1.1$         &  --                   &  --                  \\
" & M160 & $12.9\pm3.7$          & $ 20.\pm  6.$         & $18.6\pm3.6$          & $ 7.8\pm 1.1$         & $ 3.6\pm 0.6$         & $1.42\pm0.23$        \\
NGC0628  & S250 & $ 41.\pm 18.$         & $ 81.\pm 27.$         & $113.\pm  19.$        & $ 61.\pm  8.$         &  --                   &  --                  \\
" & M160 & $ 40.\pm 15.$         & $ 79.\pm 22.$         & $109.\pm  17.$        & $ 59.\pm  7.$         & $27.5\pm3.3$          & $10.6\pm1.4$         \\
NGC0925  & S250 & $ 15.\pm  8.$         & $ 29.\pm 11.$         & $ 37.\pm  8.$         & $24.4\pm3.2$          &  --                   &  --                  \\
" & M160 & $ 15.\pm  6.$         & $ 27.\pm  9.$         & $ 36.\pm  7.$         & $23.1\pm2.8$          & $13.1\pm1.6$          & $ 5.9\pm 0.8$        \\
NGC1097  & S250 & $ 79.\pm 36.$         & $126.\pm  56.$        & $134.\pm  36.$        & $ 67.\pm  8.$         &  --                   &  --                  \\
" & M160 & $ 78.\pm 29.$         & $124.\pm  48.$        & $131.\pm  35.$        & $ 65.\pm  7.$         & $28.5\pm3.2$          & $10.3\pm1.3$         \\
NGC1266  & S250 & $ 16.\pm  6.$         & $ 18.\pm  8.$         & $ 12.\pm  5.$         & $ 4.4\pm 0.6$         &  --                   &  --                  \\
" & M160 & $14.5\pm4.4$          & $ 17.\pm  7.$         & $11.2\pm4.1$          & $ 4.1\pm 0.5$         & $1.59\pm0.23$         & $0.52\pm0.10$        \\
NGC1291  & S250 & $ 4.6\pm24.9$         & $ 8.4\pm27.6$         & $ 24.\pm 18.$         & $16.3\pm3.9$          &  --                   &  --                  \\
" & M160 & $ 4.4\pm18.2$         & $ 8.3\pm18.1$         & $ 23.\pm 13.$         & $15.7\pm2.5$          & $ 8.6\pm 1.5$         & $ 3.4\pm 0.9$        \\
NGC1316  & S250 & $ 5.3\pm 2.4$         & $ 9.9\pm 3.9$         & $10.8\pm2.7$          & $ 4.8\pm 0.7$         &  --                   &  --                  \\
" & M160 & $ 5.0\pm 1.9$         & $ 9.3\pm 3.4$         & $10.2\pm2.5$          & $ 4.6\pm 0.6$         & $1.88\pm0.28$         & $0.66\pm0.13$        \\
NGC1377  & S250 & $ 7.6\pm 2.6$         & $ 6.9\pm 3.0$         & $ 3.9\pm 1.5$         & $1.37\pm0.29$         &  --                   &  --                  \\
" & M160 & $ 7.0\pm 1.7$         & $ 6.3\pm 2.2$         & $ 3.6\pm 1.2$         & $1.22\pm0.18$         & $0.50\pm0.09$         & $0.171\pm0.050$      \\
NGC1482  & S250 & $ 43.\pm 17.$         & $ 54.\pm 23.$         & $ 43.\pm 10.$         & $15.4\pm1.8$          &  --                   &  --                  \\
" & M160 & $ 41.\pm 13.$         & $ 52.\pm 19.$         & $ 41.\pm 10.$         & $14.6\pm1.6$          & $ 5.5\pm 0.7$         & $1.64\pm0.29$        \\
NGC1512  & S250 & $ 7.0\pm 2.2$         & $13.7\pm3.8$          & $18.3\pm3.2$          & $10.2\pm1.3$          &  --                   &  --                  \\
" & M160 & $ 7.2\pm 3.1$         & $12.9\pm4.1$          & $18.6\pm4.0$          & $11.2\pm1.4$          & $ 5.6\pm 0.8$         & $2.22\pm0.36$        \\
NGC2146  & S250 & $200.\pm  37.$        & $242.\pm 140.$        & $182.\pm  96.$        & $ 63.\pm  7.$         &  --                   &  --                  \\
" & M160 & $194.\pm  42.$        & $235.\pm 105.$        & $176.\pm  75.$        & $ 62.\pm  7.$         & $22.6\pm2.5$          & $ 7.2\pm 0.9$        \\
NGC2798  & S250 & $ 26.\pm  7.$         & $ 30.\pm 10.$         & $21.9\pm5.0$          & $ 8.1\pm 1.0$         &  --                   &  --                  \\
" & M160 & $24.6\pm4.5$          & $ 28.\pm  8.$         & $ 20.\pm  5.$         & $ 7.6\pm 0.9$         & $2.88\pm0.36$         & $0.89\pm0.16$        \\
NGC2841  & S250 & $ 11.\pm 10.$         & $ 29.\pm 15.$         & $ 50.\pm 13.$         & $33.9\pm4.5$          &  --                   &  --                  \\
" & M160 & $ 11.\pm  8.$         & $ 28.\pm 11.$         & $ 48.\pm 13.$         & $32.9\pm3.8$          & $15.9\pm1.9$          & $ 6.3\pm 0.9$        \\
NGC2915  & S250 & $ 1.1\pm 0.7$         & $ 2.0\pm 1.4$         & $ 1.8\pm 1.2$         & $0.62\pm0.23$         &  --                   &  --                  \\
" & M160 & $0.98\pm0.52$         & $ 1.7\pm 1.2$         & $ 1.5\pm 1.1$         & $0.46\pm0.24$         & $0.24\pm0.13$         & $0.098\pm0.062$      \\
NGC2976  & S250 & $ 22.\pm  6.$         & $ 38.\pm  9.$         & $ 47.\pm  9.$         & $24.6\pm3.1$          &  --                   &  --                  \\
" & M160 & $20.6\pm4.5$          & $ 36.\pm  8.$         & $ 45.\pm  9.$         & $23.6\pm3.1$          & $11.4\pm1.5$          & $ 4.4\pm 0.6$        \\
NGC3049  & S250 & $ 4.0\pm 2.2$         & $ 5.7\pm 3.1$         & $ 5.4\pm 2.0$         & $2.78\pm0.44$         &  --                   &  --                  \\
" & M160 & $ 3.7\pm 1.7$         & $ 5.1\pm 2.4$         & $ 4.9\pm 1.5$         & $2.45\pm0.32$         & $1.32\pm0.18$         & $0.67\pm0.10$        \\
NGC3077  & S250 & $ 21.\pm  7.$         & $ 30.\pm 11.$         & $ 30.\pm  8.$         & $15.3\pm2.2$          &  --                   &  --                  \\
" & M160 & $ 20.\pm  6.$         & $ 29.\pm  9.$         & $ 29.\pm  8.$         & $15.0\pm2.0$          & $ 7.3\pm 1.0$         & $2.94\pm0.46$        \\
NGC3184  & S250 & $ 18.\pm 13.$         & $ 39.\pm 17.$         & $ 55.\pm 15.$         & $33.4\pm4.6$          &  --                   &  --                  \\
" & M160 & $ 18.\pm  9.$         & $ 38.\pm 13.$         & $ 54.\pm 13.$         & $32.3\pm3.8$          & $15.3\pm1.9$          & $ 5.8\pm 0.9$        \\
NGC3190  & S250 & $ 6.5\pm 2.9$         & $12.0\pm4.1$          & $15.7\pm3.7$          & $ 8.5\pm 1.1$         &  --                   &  --                  \\
" & M160 & $ 5.9\pm 2.3$         & $11.2\pm3.3$          & $14.6\pm2.8$          & $ 7.9\pm 0.9$         & $3.39\pm0.41$         & $1.16\pm0.18$        \\
NGC3198  & S250 & $11.7\pm4.4$          & $ 24.\pm  9.$         & $ 31.\pm  8.$         & $18.6\pm2.3$          &  --                   &  --                  \\
" & M160 & $11.4\pm3.8$          & $ 23.\pm  9.$         & $ 29.\pm  8.$         & $17.8\pm2.1$          & $ 9.3\pm 1.1$         & $ 3.9\pm 0.5$        \\
NGC3265  & S250 & $ 3.5\pm 1.6$         & $ 3.7\pm 2.0$         & $ 2.9\pm 1.2$         & $1.26\pm0.25$         &  --                   &  --                  \\
" & M160 & $ 3.2\pm 1.2$         & $ 3.4\pm 1.7$         & $ 2.6\pm 1.0$         & $1.15\pm0.16$         & $0.51\pm0.09$         & $0.199\pm0.047$      \\
NGC3351  & S250 & $ 27.\pm 14.$         & $ 50.\pm 23.$         & $ 55.\pm 17.$         & $31.9\pm4.0$          &  --                   &  --                  \\
" & M160 & $ 26.\pm 11.$         & $ 48.\pm 18.$         & $ 53.\pm 15.$         & $30.8\pm3.5$          & $13.8\pm1.6$          & $ 4.8\pm 0.7$        \\
NGC3521  & S250 & $ 83.\pm 28.$         & $168.\pm  51.$        & $210.\pm  36.$        & $108.\pm  12.$        &  --                   &  --                  \\
" & M160 & $ 81.\pm 26.$         & $165.\pm  48.$        & $206.\pm  38.$        & $107.\pm  12.$        & $ 46.\pm  5.$         & $16.6\pm2.1$         \\
NGC3621  & S250 & $ 52.\pm 20.$         & $100.\pm  32.$        & $130.\pm  27.$        & $ 67.\pm  9.$         &  --                   &  --                  \\
" & M160 & $ 51.\pm 18.$         & $ 98.\pm 27.$         & $128.\pm  26.$        & $ 67.\pm  8.$         & $31.7\pm3.8$          & $12.8\pm1.7$         \\
NGC3627  & S250 & $109.\pm  32.$        & $192.\pm  56.$        & $201.\pm  30.$        & $ 92.\pm 10.$         &  --                   &  --                  \\
" & M160 & $107.\pm  31.$        & $190.\pm  55.$        & $198.\pm  37.$        & $ 92.\pm 10.$         & $37.0\pm4.2$          & $12.4\pm1.6$         \\
NGC3773  & S250 & $ 1.5\pm 1.1$         & $ 2.2\pm 1.5$         & $ 2.4\pm 1.2$         & $1.06\pm0.22$         &  --                   &  --                  \\
" & M160 & $ 1.3\pm 0.8$         & $ 2.0\pm 1.2$         & $ 2.1\pm 0.9$         & $0.92\pm0.14$         & $0.42\pm0.07$         & $0.148\pm0.043$      \\
NGC3938  & S250 & $ 16.\pm  8.$         & $ 30.\pm 12.$         & $ 41.\pm 10.$         & $22.9\pm3.0$          &  --                   &  --                  \\
" & M160 & $ 15.\pm  6.$         & $ 29.\pm 10.$         & $ 39.\pm 10.$         & $22.1\pm2.6$          & $10.1\pm1.2$          & $ 3.8\pm 0.5$        \\
NGC4236  & S250 & $ 7.4\pm10.2$         & $ 11.\pm 13.$         & $ 16.\pm  8.$         & $10.8\pm2.3$          &  --                   &  --                  \\
" & M160 & $ 6.7\pm 8.0$         & $ 10.\pm 10.$         & $ 14.\pm  6.$         & $10.2\pm1.5$          & $ 6.5\pm 1.0$         & $ 3.3\pm 0.6$        \\

      \hline
    \end{tabular}
  \end{center}
\btdnote{uses tab\_herschel\_guts1.tex created by tablemaker\_v9}
\end{table*}
\addtocounter{table}{-1}
\begin{table*}
  \scriptsize
  \begin{center}
    \caption{Herschel Photometry of KINGFISH Galaxies with Dust Detections,
             contd.}
    \begin{tabular}{|c c | c c c | c c c|}
      \hline
      &&\multicolumn{3}{c|}{PACS $F_\nu(\Jy)$}&\multicolumn{3}{c|}{SPIRE $F_\nu(\Jy)$}\\
Galaxy & mask & $70\micron$ & $100\micron$ & $160\micron$ & $250\micron$ & $350\micron$ & $500\micron$ \\
\hline
NGC4254  & S250 & $ 61.\pm 18.$         & $113.\pm  30.$        & $129.\pm  21.$        & $ 62.\pm  7.$         &  --                   &  --                   \\
" & M160 & $ 59.\pm 17.$         & $110.\pm  30.$        & $125.\pm  23.$        & $ 61.\pm  7.$         & $24.7\pm2.8$          & $ 8.3\pm 1.0$         \\
NGC4321  & S250 & $ 45.\pm 15.$         & $ 90.\pm 27.$         & $118.\pm  25.$        & $ 63.\pm  7.$         &  --                   &  --                   \\
" & M160 & $ 45.\pm 14.$         & $ 88.\pm 26.$         & $115.\pm  27.$        & $ 62.\pm  7.$         & $26.8\pm3.1$          & $ 9.2\pm 1.2$         \\
NGC4536  & S250 & $ 41.\pm 17.$         & $ 57.\pm 22.$         & $ 56.\pm 11.$         & $26.8\pm3.1$          &  --                   &  --                   \\
" & M160 & $ 40.\pm 15.$         & $ 55.\pm 19.$         & $ 54.\pm 11.$         & $25.9\pm2.9$          & $11.6\pm1.4$          & $ 4.5\pm 0.6$         \\
NGC4559  & S250 & $ 19.\pm  6.$         & $ 35.\pm 11.$         & $ 43.\pm  9.$         & $25.0\pm3.1$          &  --                   &  --                   \\
" & M160 & $18.3\pm4.9$          & $ 34.\pm  9.$         & $ 41.\pm  9.$         & $24.0\pm2.9$          & $12.6\pm1.5$          & $ 5.5\pm 0.7$         \\
NGC4569  & S250 & $ 15.\pm  6.$         & $ 32.\pm 11.$         & $ 40.\pm  7.$         & $20.8\pm2.4$          &  --                   &  --                   \\
" & M160 & $14.0\pm4.5$          & $ 30.\pm  9.$         & $ 38.\pm  7.$         & $19.5\pm2.1$          & $ 8.2\pm 0.9$         & $2.81\pm0.36$         \\
NGC4579  & S250 & $10.6\pm3.9$          & $ 27.\pm  7.$         & $ 35.\pm  6.$         & $19.7\pm2.3$          &  --                   &  --                   \\
" & M160 & $10.2\pm3.1$          & $ 25.\pm  6.$         & $ 33.\pm  6.$         & $18.7\pm2.1$          & $ 8.2\pm 0.9$         & $2.88\pm0.39$         \\
NGC4594  & S250 & $ 8.4\pm 4.4$         & $ 26.\pm  9.$         & $ 38.\pm  7.$         & $23.5\pm2.8$          &  --                   &  --                   \\
" & M160 & $ 7.9\pm 3.5$         & $ 25.\pm  8.$         & $ 36.\pm  7.$         & $22.4\pm2.6$          & $10.8\pm1.3$          & $ 4.3\pm 0.6$         \\
NGC4625  & S250 & $ 1.7\pm 1.0$         & $ 3.8\pm 2.0$         & $ 4.9\pm 1.4$         & $2.49\pm0.38$         &  --                   &  --                   \\
" & M160 & $ 1.6\pm 0.8$         & $ 3.5\pm 1.6$         & $ 4.4\pm 1.4$         & $2.28\pm0.28$         & $1.11\pm0.15$         & $0.47\pm0.09$         \\
NGC4631  & S250 & $141.\pm  27.$        & $235.\pm  48.$        & $244.\pm  37.$        & $116.\pm  12.$        &  --                   &  --                   \\
" & M160 & $139.\pm  25.$        & $232.\pm  49.$        & $243.\pm  47.$        & $117.\pm  12.$        & $ 53.\pm  6.$         & $20.6\pm2.4$          \\
NGC4725  & S250 & $ 11.\pm  7.$         & $ 27.\pm 12.$         & $ 47.\pm 10.$         & $30.5\pm3.9$          &  --                   &  --                   \\
" & M160 & $ 11.\pm  6.$         & $ 26.\pm 10.$         & $ 45.\pm 10.$         & $29.9\pm3.6$          & $15.5\pm1.9$          & $ 6.3\pm 0.9$         \\
NGC4736  & S250 & $109.\pm  43.$        & $170.\pm  60.$        & $151.\pm  30.$        & $ 65.\pm  9.$         &  --                   &  --                   \\
" & M160 & $108.\pm  35.$        & $168.\pm  52.$        & $148.\pm  30.$        & $ 64.\pm  7.$         & $26.2\pm3.4$          & $ 9.1\pm 1.4$         \\
NGC4826  & S250 & $ 57.\pm 13.$         & $ 97.\pm 28.$         & $ 92.\pm 24.$         & $38.2\pm4.1$          &  --                   &  --                   \\
" & M160 & $ 55.\pm 11.$         & $ 95.\pm 24.$         & $ 91.\pm 22.$         & $38.1\pm4.1$          & $15.2\pm1.8$          & $ 5.1\pm 0.7$         \\
NGC5055  & S250 & $ 82.\pm 21.$         & $183.\pm  42.$        & $249.\pm  42.$        & $138.\pm  15.$        &  --                   &  --                   \\
" & M160 & $ 82.\pm 18.$         & $180.\pm  39.$        & $245.\pm  46.$        & $137.\pm  15.$        & $ 60.\pm  7.$         & $21.7\pm2.5$          \\
NGC5398  & S250 & $ 2.6\pm 1.4$         & $ 3.4\pm 1.8$         & $ 2.8\pm 1.1$         & $1.74\pm0.30$         &  --                   &  --                   \\
" & M160 & $ 2.3\pm 1.1$         & $ 3.2\pm 1.7$         & $ 2.5\pm 1.1$         & $1.59\pm0.22$         & $0.87\pm0.13$         & $0.41\pm0.08$         \\
NGC5408  & S250 & $ 3.5\pm 1.2$         & $ 3.3\pm 1.5$         & $ 2.2\pm 0.9$         & $0.80\pm0.21$         &  --                   &  --                   \\
" & M160 & $ 3.1\pm 1.0$         & $ 2.9\pm 1.2$         & $ 2.0\pm 0.8$         & $0.71\pm0.14$         & $0.37\pm0.08$         & $0.123\pm0.049$       \\
NGC5457  & S250 & $136.\pm  56.$        & $268.\pm  93.$        & $348.\pm  82.$        & $203.\pm  26.$        &  --                   &  --                   \\
" & M160 & $136.\pm  45.$        & $267.\pm  80.$        & $347.\pm  82.$        & $202.\pm  23.$        & $100.\pm 12.$         & $ 41.\pm  5.$         \\
NGC5474  & S250 & $ 4.4\pm 3.3$         & $ 8.0\pm 4.9$         & $ 7.7\pm 3.2$         & $ 4.8\pm 0.8$         &  --                   &  --                   \\
" & M160 & $ 4.3\pm 2.6$         & $ 7.9\pm 4.4$         & $ 7.1\pm 2.8$         & $ 4.6\pm 0.6$         & $2.64\pm0.38$         & $1.23\pm0.21$         \\
NGC5713  & S250 & $ 30.\pm  9.$         & $ 44.\pm 15.$         & $ 41.\pm 10.$         & $16.3\pm1.9$          &  --                   &  --                   \\
" & M160 & $ 28.\pm 10.$         & $ 42.\pm 14.$         & $ 40.\pm 10.$         & $16.6\pm2.0$          & $ 6.8\pm 0.9$         & $2.28\pm0.39$         \\
NGC5866  & S250 & $ 9.2\pm 3.2$         & $ 18.\pm  6.$         & $18.2\pm4.5$          & $ 7.8\pm 0.9$         &  --                   &  --                   \\
" & M160 & $ 8.4\pm 2.0$         & $16.7\pm4.5$          & $16.7\pm4.1$          & $ 7.1\pm 0.8$         & $2.89\pm0.34$         & $0.94\pm0.14$         \\
NGC6946  & S250 & $260.\pm  75.$        & $465.\pm 136.$        & $531.\pm 112.$        & $249.\pm  29.$        &  --                   &  --                   \\
" & M160 & $255.\pm  63.$        & $455.\pm 116.$        & $519.\pm 100.$        & $244.\pm  28.$        & $101.\pm  12.$        & $35.2\pm4.3$          \\
NGC7331  & S250 & $ 68.\pm 16.$         & $138.\pm  35.$        & $175.\pm  35.$        & $ 89.\pm 10.$         &  --                   &  --                   \\
" & M160 & $ 65.\pm 15.$         & $133.\pm  33.$        & $169.\pm  35.$        & $ 87.\pm  9.$         & $38.3\pm4.2$          & $14.2\pm1.6$          \\
NGC7793  & S250 & $ 36.\pm 14.$         & $ 71.\pm 24.$         & $ 91.\pm 23.$         & $ 54.\pm  7.$         &  --                   &  --                   \\
" & M160 & $ 36.\pm 13.$         & $ 70.\pm 23.$         & $ 89.\pm 23.$         & $ 53.\pm  6.$         & $27.9\pm3.2$          & $11.9\pm1.5$          \\
\hline
IC3583  & S250 & $ 1.1\pm 0.7$         & $ 2.0\pm 1.1$         & $ 2.1\pm 0.7$         & $1.27\pm0.24$         &  --                   &  --                   \\
" & M160 & $0.96\pm0.54$         & $ 1.8\pm 0.9$         & $ 2.0\pm 0.7$         & $1.16\pm0.19$         & $0.62\pm0.10$         & $0.25\pm0.06$         \\
NGC0586 & S250 & $0.29\pm0.62$         & $0.85\pm0.91$         & $ 1.4\pm 0.6$         & $0.92\pm0.18$         &  --                   &  --                   \\
" & M160 & $0.27\pm0.46$         & $0.76\pm0.72$         & $ 1.2\pm 0.6$         & $0.80\pm0.12$         & $0.38\pm0.07$         & $0.130\pm0.035$       \\
NGC1317 & S250 & $ 6.1\pm 1.9$         & $10.9\pm3.3$          & $11.1\pm2.8$          & $ 5.0\pm 0.7$         &  --                   &  --                   \\
" & M160 & $ 5.6\pm 1.4$         & $10.2\pm3.0$          & $10.3\pm2.7$          & $ 4.6\pm 0.6$         & $1.88\pm0.24$         & $0.64\pm0.10$         \\
NGC1481 & S250 & $ 1.0\pm 0.5$         & $ 1.0\pm 0.5$         & $0.88\pm0.33$         & $0.41\pm0.08$         &  --                   &  --                   \\
" & M160 & $ 1.3\pm 0.9$         & $0.93\pm0.81$         & $0.73\pm0.54$         & $0.42\pm0.08$         & $0.183\pm0.049$       & $0.024\pm0.034$       \\
NGC1510 & S250 & $1.12\pm0.27$         & $1.19\pm0.32$         & $0.97\pm0.25$         & $0.43\pm0.07$         &  --                   &  --                   \\
" & M160 & $ 1.5\pm 0.8$         & $ 1.3\pm 0.8$         & $ 1.3\pm 0.6$         & $0.78\pm0.15$         & $0.44\pm0.09$         & $0.22\pm0.06$         \\
NGC3187 & S250 & $ 1.5\pm 1.0$         & $ 2.6\pm 1.3$         & $ 3.9\pm 1.1$         & $2.47\pm0.36$         &  --                   &  --                   \\
" & M160 & $ 1.4\pm 0.9$         & $ 2.2\pm 1.2$         & $ 3.6\pm 1.2$         & $2.32\pm0.31$         & $1.33\pm0.18$         & $0.63\pm0.11$         \\
NGC4533 & S250 & $0.43\pm0.49$         & $ 1.4\pm 0.8$         & $ 1.6\pm 0.6$         & $0.95\pm0.16$         &  --                   &  --                   \\
" & M160 & $0.50\pm0.61$         & $ 1.3\pm 0.7$         & $ 1.5\pm 0.7$         & $0.82\pm0.14$         & $0.40\pm0.08$         & $0.190\pm0.043$       \\
NGC7335 & S250 & $0.60\pm0.44$         & $ 1.5\pm 0.7$         & $ 1.7\pm 0.5$         & $1.03\pm0.17$         &  --                   &  --                   \\
" & M160 & $0.56\pm0.29$         & $ 1.3\pm 0.6$         & $ 1.5\pm 0.5$         & $0.91\pm0.15$         & $0.42\pm0.07$         & $0.171\pm0.034$       \\
NGC7337 & S250 & $0.17\pm0.33$         & $0.62\pm0.62$         & $0.81\pm0.39$         & $0.69\pm0.11$         &  --                   &  --                   \\
" & M160 & $0.14\pm0.35$         & $0.46\pm0.51$         & $0.67\pm0.39$         & $0.57\pm0.10$         & $0.27\pm0.05$         & $0.097\pm0.022$       \\

      \hline
    \end{tabular}
  \end{center}
\btdnote{uses tab\_herschel\_guts2.tex created by tablemaker\_v9}
\end{table*}

\subsection{Integrated Fluxes}

The \Spitzer\ and \Herschel\ band surface brightnesses are
integrated over the M160 and S250 resolution galaxy masks to
obtain integrated flux densities.
The \IRAC\ and \MIPS\ flux densities are given in Table
\ref{tab:spitzer}, and the PACS and SPIRE flux densities are
given in Table \ref{tab:herschel}.
Note that MIPS70, MIPS160, SPIRE350, and SPIRE500 are not used at S250
resolution.

The uncertainties given in Tables \ref{tab:spitzer}
and \ref{tab:herschel} include uncertainties associated with
background subtraction, as well as calibration uncertainties.
As discussed in \S\ref{sec:mipspacs}, the fluxes measured by PACS
and MIPS sometimes differ by considerably more than the estimated
uncertainties: we know that some of the uncertainties
have been underestimated, although it is not clear how to improve on
our estimates.
  
\omittext{Our {\it Spitzer} and {\it Herschel} photometry is generally in good
agreement with the global fluxes measured by}
\omittextr{\citet{Dale+Aniano+Engelbracht+etal_2012}} 
\omittext{and}
\omittextr{\citet{Dale+Cook+Roussel+etal_2017},}
\omittext{particularly for the well-detected luminous galaxies.  
The masking utilized here works well for resolved dust modeling;}
\citet{Dale+Cook+Roussel+etal_2017} carried out careful foreground star
and background galaxy removal tailored for globally integrated
photometry.
\newtext{For 44 of the 53 KINGFISH galaxies where we claim dust detections, 
the SPIRE500 flux for our M160 galaxy mask is
within 10\%
of the global SPIRE500 photometry from \citet{Dale+Cook+Roussel+etal_2017}.
Thus, we are not missing
a significant reservoir of dust in the outer parts of the disk.\footnote{%
For NGC\,1512 \citet{Dale+Cook+Roussel+etal_2017} find a SPIRE 500 
flux that is 39\% larger than our value, but part of the difference
is because they included the companion
galaxy NGC\,1510, which we have treated separately.}}

\subsection{Gas Masses
\label{sec:gas masses}
}

For galaxies observed by THINGS, 
\ion{H}{1}\,21\,cm line intensities were extracted over 
the area of the
M160 resolution galaxy mask for each galaxy.
The \ion{H}{1} column density $N({\rm \HI})$ was estimated assuming
the 21\,cm emission to be optically thin.

For the 38 galaxies in the HERACLES sample, $^{12}$CO(2--1) 
line fluxes were
obtained by integrating over the M160 resolution galaxy mask, and the
H$_2$ mass was estimated from the CO\,$2\!\rightarrow\!1$ line flux
assuming $T_B(2\!\rightarrow\!1)/T_B(1\!\rightarrow\!0)=0.7$ and a
standard conversion factor\footnote{$X_{\rm CO,1-0}=2\times10^{20}{\rm
    H}_2\cm^{-2}(\K\kms)^{-1}$ corresponds to $\alpha_{\rm
    CO,1-0}=4.35\Msol\pc^{-2}(\K\kms)^{-1}$ if a factor 1.36 is
  assumed to allow for Helium and heavier elements.}  $X_{\rm
  CO,1-0}=2\times10^{20}{\rm H}_2\cm^{-2}(\K\kms)^{-1}$.  The adopted
$X_{\rm CO}$ value is representative of the values found in 26 nearby
star-forming galaxies by \citet{Sandstrom+Leroy+Walter+etal_2013}.

{For NGC\,1266 we use the integrated CO emission and the lower bound
on the \ion{H}{1} mass from \citet{Alatalo+Blitz+Young+etal_2011}.}

\section{Dust Modeling}
\label{sec:dustmodel}

\subsection{DL07 Dust Model}
We employ the DL07 dust model,
using ``Milky Way'' grain size distributions \citep{Weingartner+Draine_2001a}.
DL07 described the construction of the dust model, and AD12 described its usage 
in the context of the KINGFISH galaxies.
The DL07 dust model has a mixture of amorphous silicate grains and carbonaceous
grains, with a distribution of grain sizes.
The distribution of grain sizes was chosen to reproduce the wavelength
dependence of interstellar extinction within a few kpc of the Sun
\citep{Weingartner+Draine_2001a}. 
The silicate and carbonaceous content of the dust grains was constrained by 
observations of the gas phase depletions in the ISM.
It is assumed that the radiation field heating the dust
has a universal spectrum, taken to be 
that of the local interstellar radiation field estimated by
\citet{Mathis+Mezger+Panagia_1983}, scaled by a dimensionless factor $U$.
Following DL07, we assume that in each pixel there is dust 
exposed to radiation with
a single intensity $\Umin$, and also dust heated by 
a power-law distribution of starlight intensities with $\Umin<U<\Umax$:
\beq \label{eq:dMdU}
\frac{d\Mdust}{dU} = (1-\gamma)\Mdust \delta(U-\Umin) + 
\gamma M_d \frac{(\alpha-1)U^{-\alpha}}{\Umin^{1-\alpha}-\Umax^{1-\alpha}}
~~~,\eeq
where $\Mdust$ is the total dust mass in the pixel,
and $\gamma$ is the fraction of the dust mass that is 
heated by the power-law distribution of starlight intensities.

The DL07 model has 6 adjustable parameters
pertaining to the dust and the starlight heating the dust:
\begin{enumerate}
\item $\qpah$:  the fraction of the total grain mass contributed by 
  polycyclic aromatic hydrocarbons (PAHs) containing fewer than
  $10^3$ carbon atoms.
\item $\Umin$: the intensity of the
  diffuse ISM radiation field heating the dust,
  relative to the solar neighborhood.
\item $\alpha$: the exponent of the power-law distribution of 
heating starlight intensities between
$\Umin$ and $\Umax$.  The case $\alpha=2$ corresponds
to constant dust heating
power per logarithmic interval in starlight intensity $U$;
many galaxies seem to be characterized by $\alpha\approx 2$.
\item $\Umax$: the maximum heating starlight intensity of the 
power-law distribution of heating starlight intensities.
\item $\gamma$: the fraction of the dust mass exposed to the 
power-law distribution of starlight intensities.
 \item $\Mdust$: the dust mass in the pixel.
\smallskip

In addition, for modeling the observed fluxes in the IRAC bands,
we have an additional adjustable parameter (see AD12):
\smallskip
\item $\Omega_\star$: the solid angle subtended by stars within the pixel,
determined from the ``direct'' starlight intensity in the infrared, 
i.e., starlight that directly contributes to the IRAC photometry, 
without warming the dust. 
\end{enumerate}
The mean starlight intensity seen by the dust is
\beqa
\bar{U} &\!=\!& (1-\gamma)\Umin+\gamma\frac{(\alpha-1)}{(2-\alpha)}
\frac{\Umax^{2-\alpha}-\Umin^{2-\alpha}}{\Umin^{1-\alpha}-\Umax^{1-\alpha}}
~~{\rm if}~\alpha\neq 2,~
\\
&\!=\!& (1-\gamma)\Umin + \gamma \Umin 
\frac{\ln(\Umax/\Umin)}{1-(\Umin/\Umax)}
~~{\rm if}~\alpha=2.~
\eeqa
The parameter $\gamma$ is directly related to $\fPDR$, 
defined to be the fraction of  the total dust luminosity \lumdust\ 
that is radiated by dust in regions where
$U>10^2$:
\beqa \nonumber
&\fPDR = \hspace*{4.0cm}&
\\
\nonumber
&\frac{\gamma
\left[1 -\left(\frac{100}{\Umax}\right)^{2-\alpha}\right]}
{(1\!-\!\gamma)\frac{(2-\alpha)}{(\alpha-1)}
\left(\frac{\Umin}{\Umax}\right)^{2-\alpha}
\left[1\!-\!\left(\frac{\Umin}{\Umax}\right)^{\alpha-1}\right]
\!+\! \gamma
\left[1\!-\!\left(\frac{\Umin}{\Umax}\right)^{2-\alpha}\right]}
&
\eeqa
if $\alpha\neq2$, or
\beq
\fPDR=\frac{\gamma\ln\left(\frac{\Umax}{100}\right)}
{(1-\gamma)\left(1-\frac{\Umin}{\Umax}\right)+\gamma\ln\left(\frac{\Umax}{\Umin}\right)}
~~~~{\rm if}~\alpha=2 ~.~
\eeq
For each set of dust parameters $(\Mdust,\qpah,\gamma,\Umin,
\Umax,\alpha)$, and the adopted grain size
distribution and grain properties, the dust emission spectrum is
computed from first principles.  The observed SEDs are consistent with
models having $\Umax=10^7$, and we therefore fix $\Umax\equiv 10^7$.
Moreover, the model emission is linear in $\Mdust$, $\Lstar$, and
$\gamma$ (or, equivalently, $\fPDR$), so in the dust fitting
algorithms we only need to explore a three dimensional parameter space
($\qpah$, $\Umin$, and $\alpha$).  The limits on adjustable parameters
are given in Table \ref{tab:param limits}.  The allowed range for
$\Umin$ is determined by the wavelength coverage of the data used in
the fit.

\begin{table*}
{\footnotesize
\begin{center}
\caption{\label{tab:param limits}Allowed Ranges for Adjustable Parameters}
\begin{tabular}{c c c c c}
\hline
Parameter & min & max & & Parameter grid used\\
\hline
$\Lstar$ & 0 & $\infty$ & &continuous fit \\
$\Mdust$ & 0 & $\infty$ & &continuous fit \\
$\qpah$ & 0.00 & 0.10 & &in steps $\Delta\qpah=0.001$ \\
$\fPDR$ & 0.0 & $<1.00^a$ & &continuous fit \\
$\Umin$ & 0.7 & 30 & when $\lambda_{\rm max}=160\micron$ & in steps $\Delta\Umin=0.01$$^b$ \\
        & 0.07 & 30 & when $\lambda_{\rm max}=250\micron$ & in steps $\Delta\Umin=0.01$$^b$ \\
        & 0.01 & 30 & when $\lambda_{\rm max}=350\micron$& in steps $\Delta\Umin=0.01$$^b$ \\
        & 0.01 & 30 & when $\lambda_{\rm max}\geq 500\micron$& in steps $\Delta\Umin=0.01$$^b$\\
$\alpha$ &    1.0      & 3.0        & &in steps $\Delta\alpha=0.1$\\
$\Umax$ & $10^7$ & $10^7$ & &not adjusted \\
\hline
\multicolumn{5}{l}
{$^a$ For each set of $\Umin, \,\Umax,$ and $\alpha$ there is maximum value of $\fPDR$ possible.}\\
\multicolumn{5}{l}
{$^b$ The fitting procedure uses pre-calculated spectra for $\Umin \in \{$0.01, 0.015, 0.02, 0.03,}\\
\multicolumn{5}{l}
{\hspace*{0.5em}  0.05, 0.07, 0.1, 0.15, 0.2, 0.3, 0.4, 0.5, 0.6, 0.7, 0.8, 1.0, 1.2, 1.5, 2, 2.5, 3.0, 4.0, 5.0,}\\
\multicolumn{5}{l}
{\hspace*{0.5em} 6.0, 7.0, 8.0, 10, 12, 15, 20, 25, 30$\}$
interpolated onto a grid with $\Delta\Umin=0.01$}.
\end{tabular}
\end{center}
}
\end{table*}

The region observed is at a distance $D$ from the observer and
$\Omega_j$ is the solid angle of pixel $j$.  For each pixel $j$, the
best-fit model vector $\{\Lstar,\Mdust,\qpah,\gamma,\Umin,\alpha\}_j$
corresponds to a dust mass surface density: 
\beq \Sigma_{\Mdust,j}
\equiv \frac{1}{D^2\Omega_j} M_{\dust,j} ~.  
\eeq 
Similarly, we can
compute the infrared luminosity surface density $\Sigma_{\Ldust,j}$
and $\Sigma_{L_\PDR,j}$, 
the surface density of dust luminosity from regions with
$U>U_\PDR$, as:
\beq
\Sigma_{\Ldust,j} \equiv \frac{1}{D^2\Omega_j} L_{\dust,j}, 
\quad\quad \Sigma_{L_\PDR,j} \equiv \frac{1}{D^2\Omega_j} f_{\PDR,j}L_{\dust,j}
~,
\label{eq:total}
\eeq
where $L_{{\rm d},j}$ is the model luminosity radiated by
mass $M_{{\rm d},j}$ of dust heated by starlight characterized by
$(U_{{\rm min},j},\gamma_j,\alpha_j)$.

For each pixel $j$, we find the best-fit model parameters
$\{ U_{{\rm min},j}, \gamma_j, \alpha_j, M_{{\rm d},j}, q_{{\rm PAH},j}\}$
by minimizing $\chi^2$, as described by AD12.
After the resolved (pixel-by-pixel) modeling of the galaxy is
performed, we compute a set of global quantities by adding or taking
weighted means (denoted as $\langle...\rangle$) of the quantities in
each individual pixel of the map.  
The total dust mass
$M_{\dust}$, total dust luminosity $L_{\dust}$, and total dust
luminosity radiated by dust in regions with $U>10^2$, $L_{\dust,\rm tot}$,
are given by:
\beq
M_{\dust} \equiv \sum_{j=1}^N M_{\dust,j} \,,\quad L_{\PDR}
\equiv \sum_{j=1}^N L_{\dust,j} \,,\nonumber
\eeq 
\beq
L_{\PDR}\equiv
\sum_{j=1}^N L_{\PDR,j}=\sum_{j=1}^N L_{\dust,j}\fPDRj \,, 
\label{eq:mean_mdust}
\eeq where
the sums extend over all the pixels $j$ that correspond to the target
galaxy (i.e., the ``galaxy mask'' pixels, as described in AD12).  The
dust-mass weighted PAH mass fraction $\langle \qpah \rangle$, and mean
starlight intensity $\langle \overline{U} \rangle$, are given by: 
\beq
\langle \qpah\rangle \equiv \frac{\sum_{j=1}^N \qpahj \,
  M_{\dust,j}}{\sum_{j=1}^N M_{\dust,j}} \,,\quad
  \langle U\rangle \equiv \frac{\sum_{j=1}^N \overline{U}_j \,
  M_{\dust,j}}{\sum_{j=1}^N M_{\dust,j}} \,.  
\label{eq:mean_1}
\eeq 
We similarly define the dust mass-weighted minimum starlight intensity
\beq
\langle \Umin \rangle \equiv
\frac{\sum_{j=1}^N U_{{\rm min},j} M_{{\rm d},j}}{\sum_{j=1}^N M_{{\rm d},j}}
\eeq  
The dust-luminosity weighted value of $\fpdr$ is:
\beq
\label{eq:mean_fpdr}
\langle \fpdr\rangle  \equiv 
\frac{L_{\PDR}}{L_{\dust}}
~~.
\eeq
While the average value of $\alpha$ is of little physical significance
(the sum of two power laws is not a power law), for purposes of
discussion we define a representative value
\beq
\langle\alpha\rangle\equiv
\frac{\sum_{j=1}^N \gamma_j M_{{\rm d},j}U_{{\rm min},j}\alpha_j}
{\sum_{j=1}^N \gamma_j M_{{\rm d},j}U_{{\rm min},j}}
~~.
\eeq
We also fit a dust model to the global photometry of each galaxy
(i.e., a single-pixel dust model).  Below we will compare the result
of this single-pixel global model with summing over the fits to individual
pixels.

\begin{figure*}  
\centering 
\includegraphics[width=8.5cm,clip=true,trim=0.5cm 5.0cm 0.5cm 2.5cm]
{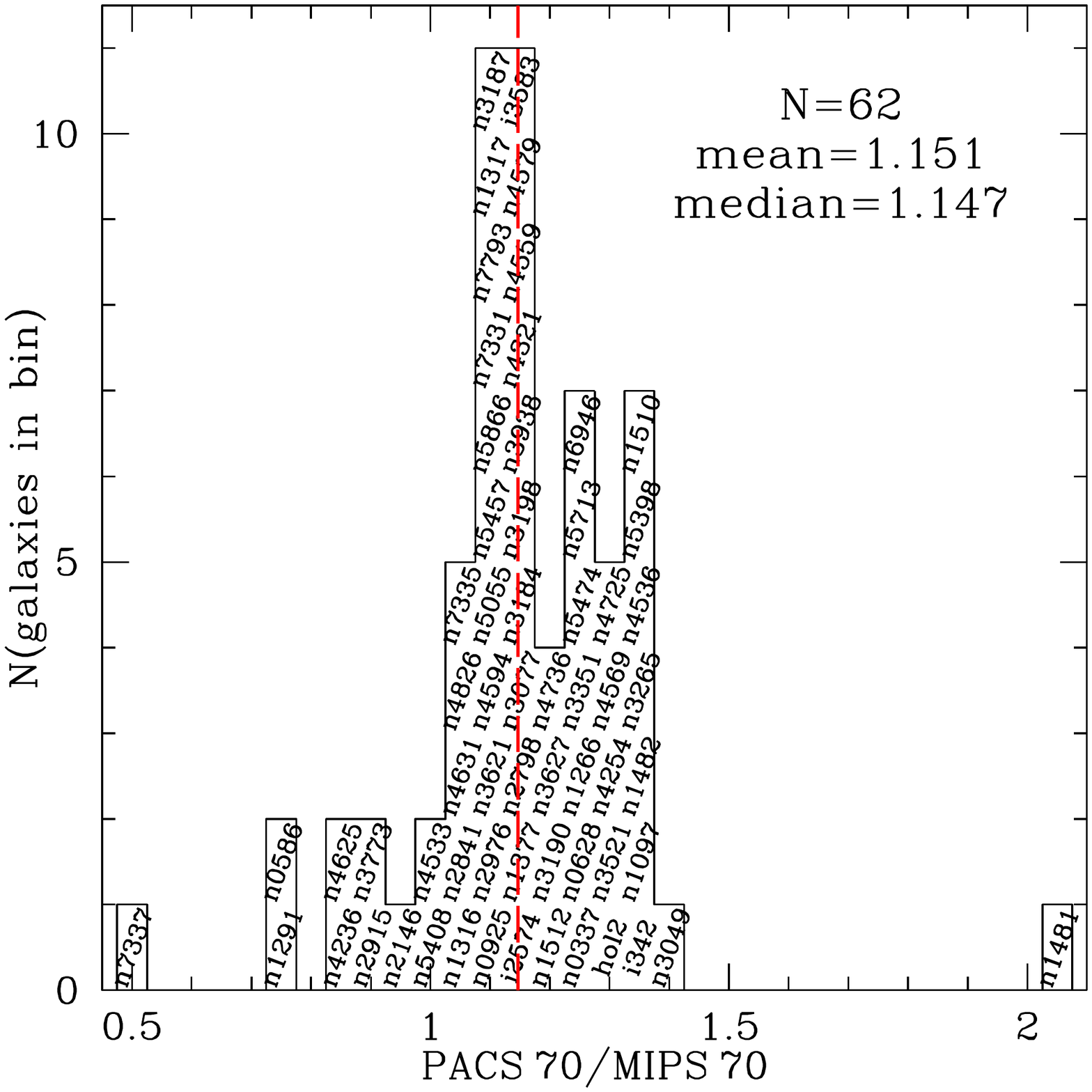} 
\includegraphics[width=8.5cm,clip=true,trim=0.5cm 5.0cm 0.5cm 2.5cm]
{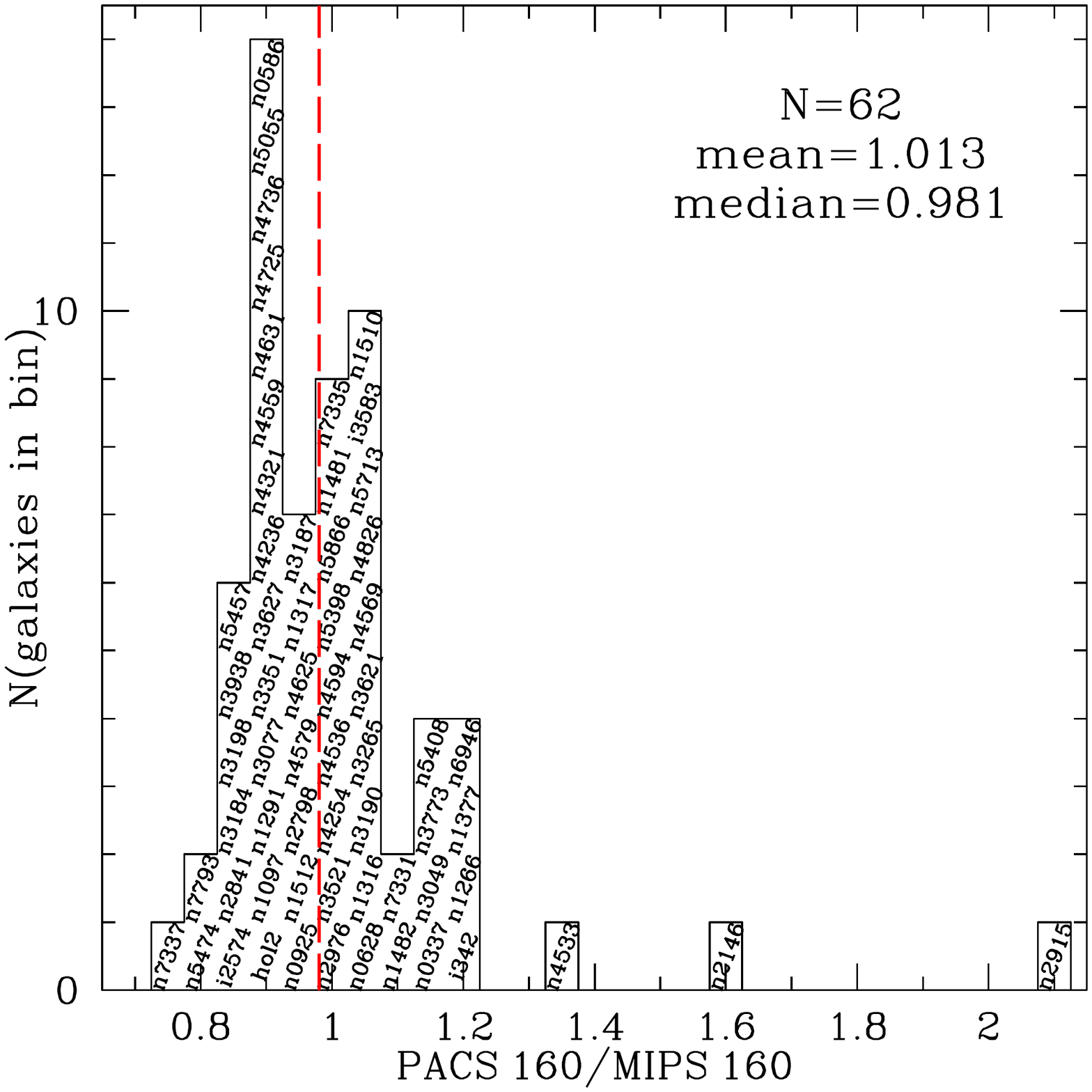} 
\caption{PACS/MIPS global photometry for the KF62 sample
  (see Table \ref{tab:subsamples}).
  Eight galaxies where dust was not reliably detected
  have been excluded (see text).  Dashed lines show medians.
  PACS and MIPS photometry typically differs by $\sim$20\% at $70\micron$, and
  $\sim$10\% at $160\micron$, except for outliers
  (NGC\,7337 and NGC\,1481 at $70\micron$;
  NGC\,2146 and NGC\,2915 at $160\micron$).
  PACS70 fluxes are systematically higher than MIPS70.
  \label{fig:PACS-MIPS}
  \btdnote{f2a.pdf, f2b.pdf}
}
\end{figure*}
\subsection{Post-Planck renormalization of DL07 dust masses and
starlight intensities}
\label{sec:renormalization}

\citet{Planck_DL07_2016} fitted the DL07 dust model to
all-sky maps in the {\it Planck} 
857, 545, 353, 217, 143, and 100 GHz
($350\micron$, $550\micron$, $850\micron$, $1.4\mm$, $2.1\mm$, and $3.0\mm$)
bands,
DIRBE $100\micron$, $140\micron$, and $240\micron$ bands,
IRAS $60\micron$ and $100\micron$ bands,
and the WISE $12\micron$ band,
to estimate the dust mass surface density for over 50 million
$1.7\arcmin\times1.7\arcmin$ pixels.
About 270,000 of these pixels contain spectroscopically-confirmed SDSS
quasars, which were used to estimate the correlation of quasar
reddening with the reddening predicted by the DL07 dust model.
It was discovered that the DL07 model tends to overpredict the
reddening by a factor
$\sim$2.
The {\it Panchromatic Hubble Andromeda Treasury} (PHAT) study of stars in 
M31 \citep{Dalcanton+Fouesneau+Hogg+etal_2015}
also found that the DL07 dust model,
if constrained to reproduce the observed infrared emission
\citep{Draine+Aniano+Krause+etal_2014}, overpredicted the reddening of
stars in M31 by a factor $\sim$2.
The SDSS quasars allow the bias factor to be estimated:
\citet{Planck_DL07_2016} found
that the bias appeared to depend on the value of $\Umin$:
\beq
\frac{E(B-V)_{\rm QSO}}{E(B-V)_{\rm DL07}} \approx 0.42 + 0.28\Umin
{\rm ~~for~~} 0.4\ltsim\Umin\ltsim1.0
~.
\eeq
If the reddening $E(B-V)$ has been overestimated, 
it is reasonable to suppose that
the dust mass/area has also been overestimated, by approximately the
same factor as the reddening.
Therefore, we will correct the
DL07 dust mass estimates by the same empirical correction factor as
for the reddening.

Because the KINGFISH sample includes some pixels with
high $\Umin$, we choose to limit the {\it Planck}-derived correction factor
for $\Umin>1$:
\beqa \label{eq:Mdcorrected}
\Sigma_{\Mdust,{\rm renorm},j} &=& C_{{\rm dust},j} 
\times \Sigma_{\Mdust,{\rm DL07},j}
\\
C_{{\rm dust},j} &=&
0.42 + 0.28\min(U_{{\rm min,j}},1.0)
~~~.
\eeqa

Because the dust models are required to reproduce the dust
luminosity, a reduction in the estimated amount of dust implies
a corresponding increase in the estimated starlight intensities.  
Thus we take, for pixel $j$:
\beq \label{eq:Ubar corrected}
    \bar{U}_{{\rm renorm},j} =
    \frac{1}{C_{{\rm dust},j}} \bar{U}_{{\rm DL07},j}
\eeq
and
\beq \label{eq:Umin corrected}
U_{{\rm min},{\rm renorm},j} \approx   
    \frac{1}{C_{{\rm dust},j}} U_{{\rm min},{\rm DL07},j}
~~~.
\eeq
All of the dust and starlight parameters ($M_{\rm d}$, $U_{\rm min}$) reported
below are ``renormalized'' values from Eq.\ (\ref{eq:Mdcorrected}) and
(\ref{eq:Umin corrected}).

\newtext{The above ``renormalization'' is required because observations
indicate that the far-infrared and submm opacity of interstellar dust 
per unit reddening is
somewhat larger than the DL07 model values.\footnote{%
  The empirical finding that $\Cdust$ depends on $\Umin$ suggests that
  the dust opacity may decline less rapidly with increasing $\lambda$ than
  assumed by DL07.}
We will find below (Table \ref{tab:dust}) 
that the global average $\langle \Cdust\rangle$ for the 62
galaxies ranges from $0.45$ to $0.69$, with median 
$\langle\Cdust\rangle=0.62$.  Thus the typical correction relative to the
DL07 model is a reduction in $\Mdust$ by a factor $\sim$0.62.

Because Eq.\ (\ref{eq:Mdcorrected}) has $\Cdust=0.70$ for 
$\Umin\geq1$, the estimated dust-to-gas ratio in regions with
$\Umin\geq1$ is reduced by a constant factor 0.70.
This applies to the study of 
\citet{Sandstrom+Leroy+Walter+etal_2013},
which was dominated by regions with $\langle U\rangle > 1$.
However, because
$\Cdust=constant$,
the CO-to-$\HH$ ratios found by \citet{Sandstrom+Leroy+Walter+etal_2013}
are unaffected by the renormlization.}

\subsection{Why both MIPS and PACS are needed
\label{sec:mipspacs}
}

As discussed by \citet{Aniano+Draine+Gordon+Sandstrom_2011} the
MIPS160 PSF cannot be convolved safely into any of the PSFs of the
remaining cameras.  
Therefore, if we wish to include MIPS160 photometry in the dust
modeling,
we must ``degrade'' all other images into the MIPS160 PSF. 

There are two reasons why we want to include MIPS160 even 
though PACS160 imaging is available.
First, using the larger PSF increases the signal/noise ratio for the
imaging, thereby allowing photometry to be extended to lower surface
brightness regions.
Secondly, 
there are significant and unexplained
discrepancies between PACS160
and MIPS160
photometry.
Similar discrepancies are found between PACS70 and MIPS70.

Figure \ref{fig:PACS-MIPS} shows histograms
of the global PACS70/ MIPS70 flux ratio (left panel), and the global
PACS160/ MIPS160 flux ratio (right panel) for each of the KF62
galaxies with reliable dust detections.
Each histogram shows the names of galaxies in
the bin; ``NGC'', ``DDO'', ``Holmberg'' and ``IC'' are abbreviated
to ``n'', ``d'', ``Hol'', and ``i'', respectively. 

The PACS70 and MIPS70 bandpasses differ slightly, as do the
PACS160 and MIPS160 bandpasses.  However, AD12 show that
for reasonable dust SEDs
the slight difference in bandpasses can explain differences in reported
fluxes of only $\ltsim 9\%$ at $70\micron$, and $\sim2\%$ at 160$\micron$,
whereas much larger PACS/MIPS discrepancies are often observed.

AD12 (their Appendix F) found that even when the
global photometry has PACS/MIPS $\approx 1$, the PACS and MIPS
images (with PACS convolved to the MIPS PSF)
can have local surface brightnesses
discrepant by factors as large as 1.5-2.0.  
Similar discrepancies were found when comparing PACS and MIPS
imaging of M31 \citep{Draine+Aniano+Krause+etal_2014}
and NGC\,4449 \citep{Calzetti+Wilson+Draine+etal_2018}.

Figure \ref{fig:PACS-MIPS} illustrates that even after summing over
the full galaxy mask, PACS70 and MIPS70 often disagree by more than
a factor 1.2, and sometimes up to a factor 1.4.
The median ratio is 1.17.

PACS160 and MIPS160 are generally in better agreement, but often
have discrepancies larger than 10\%.
There are two outliers in Figure \ref{fig:PACS-MIPS}: NGC\,2146
(PACS160/MIPS160=1.6) and NGC\,2915
(PACS160/ MIPS160=2.3).
The high value of PACS160/MIPS160
for NGC\,2146 may be the result of sublinear response of MIPS160
on the very bright nucleus of NGC\,2146.
The case of NGC\,2915 is unclear -- the peak surface brightness is modest.
Perhaps the background has been oversubtracted in the MIPS160 image,
or undersubtracted in the PACS160 image.

Because it is usually unclear why PACS and MIPS disagree (the discrepancies
are too large to be attributed to differences in bandpasses),
we consider that both PACS and MIPS
photometry should be included if we wish to estimate the dust
parameters with the best accuracy available.  AD12 also found that,
for a given camera set, dust parameter estimates do not change
significantly when using a broader PSF, therefore modeling at MIPS160 PSF does
not significantly alter the dust parameter estimates.  We consider our
``gold standard'' (i.e., the PSF and camera combination that
gives the most accurate dust parameter estimates) 
to be resolved (i.e., multipixel) modeling done
using the MIPS160 PSF,
using photometry from all of the IRAC, MIPS, PACS, and SPIRE cameras.

\newpage
\section{\label{sec:results}Results}
For each galaxy in the KF62 sample,
Table \ref{tab:dust} presents the global dust parameters estimated for the
``gold standard'' modeling, 
including
information characterizing the intensity of the starlight heating the
dust in each galaxy.  The modeling was done at MIPS160 PSF, using all
the cameras available; we also give results of
modeling at S250 resolution.

The given quantities are obtained by summing
or averaging over the resolved maps using Equations
(\ref{eq:mean_mdust}-\ref{eq:mean_fpdr}).
The dust masses listed in Table \ref{tab:dust} we obtained using the
DL07 model, but then ``renormalized'' following 
Equation (\ref{eq:Mdcorrected}).
The renormalization factor $C_\dust$ depends on $\Umin$, and therefore
varies from pixel to pixel.
The overall renormalization factor
\beq
\langle \Cdust\rangle \equiv
\frac{\sum_{j=1}^N \Sigma_{{\rm Md,renorm},j}}{\sum_{j=1}^N \Sigma_{{\rm Md,DL07},j}}
=
\frac{\sum_{j=1}^N C_{{\rm dust},j} \Sigma_{{\rm Md,DL07},j}}
{\sum_{j=1}^N \Sigma_{{\rm Md,DL07},j}}
\eeq
for each galaxy
is given in Table \ref{tab:dust},
for both M160 and S250 resolution.
Henceforth, $M_\dust$, $\Umin$, and $\Ubar$ will refer to the renormalized
values of these quantities 
[see Eq.\ \ref{eq:Mdcorrected}-\ref{eq:Umin corrected}].

\subsection{One Example: NGC\,5457 = M\,101}

To illustrate the quality of the data and the modeling results for
the KINGFISH galaxies, we choose 
the large, nearly face-on spiral NGC\,5457 (M\,101) as an example.
As for all our galaxies, the dust mass, PAH abundance, 
and starlight heating parameters are adjusted separately for each pixel.

\begin{figure}
\begin{center}
\includegraphics[angle=0,width=8.0cm,
                 clip=true,trim=0.8cm 5.6cm 0.8cm 2.6cm]
{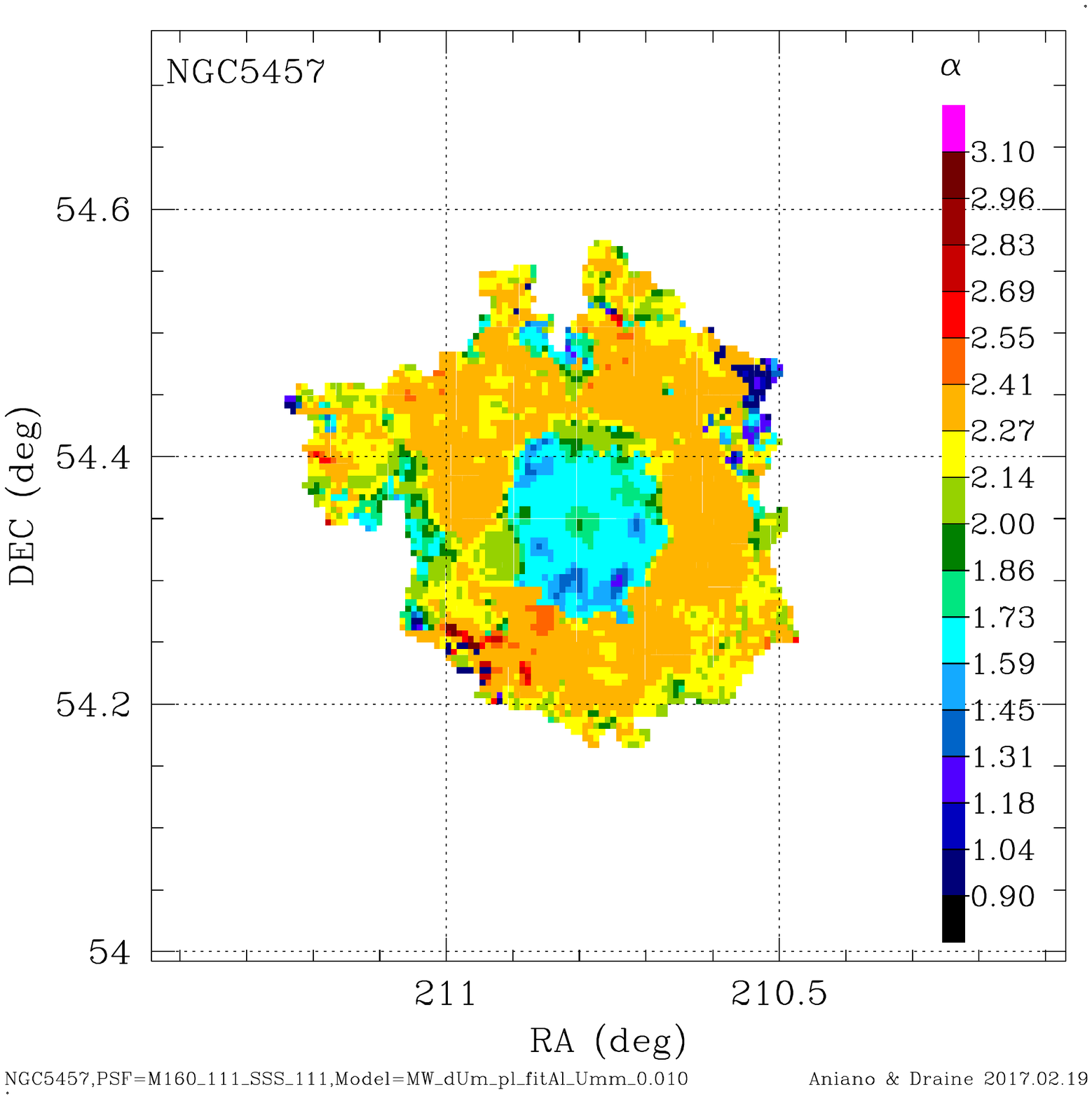} 
\includegraphics[angle=0,width=8.0cm,
                 clip=true,trim=0.8cm 5.6cm 0.8cm 2.6cm]
{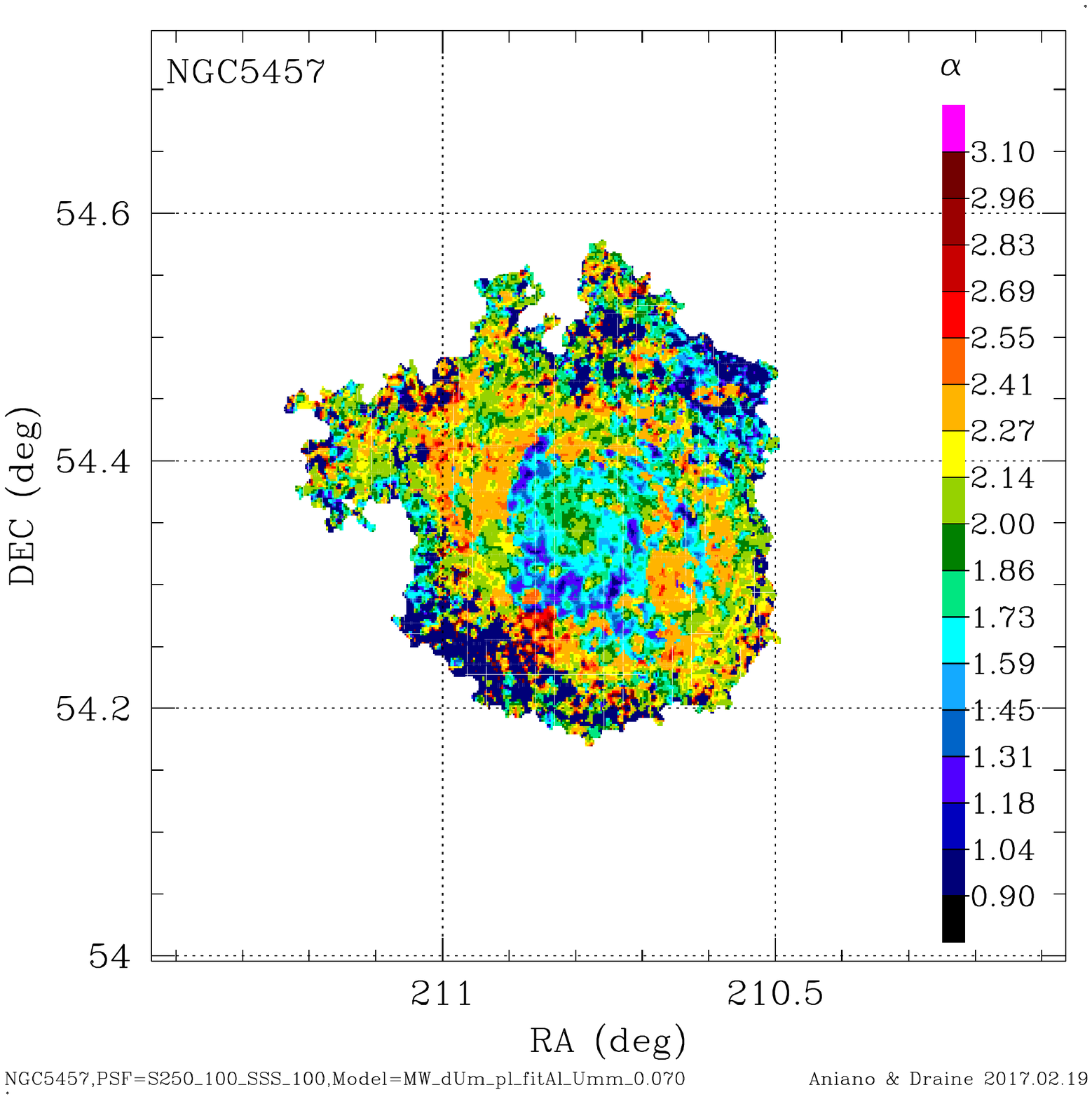} 
\caption{\label{fig:alphamaps} \footnotesize
Starlight heating parameter $\alpha$ for NGC\,5457\,=\,M101,
for modeling at M160 resolution (top) and at S250 resolution (bottom), for the
``galaxy mask'' defined by $\SigLd>\SigLdmin=0.67\Lsol\pc^{-2}$.
At M160 resolution, where all cameras are used to constrain the model,
$\alpha$ is azimuthally coherent but with a radial gradient:
$\alpha\approx1.7$ in the center, and $\alpha\approx 2.3$ in the outer regions.
The S250 map is noisier, because not all of the cameras can be used, and
the signal-to-noise ratio of the bands that can be used is reduced.
}
\end{center}
\end{figure}

The parameter $\alpha$ characterizes
the distribution of starlight intensities heating dust within a pixel
(see Eq.\ \ref{eq:dMdU}).
Figure \ref{fig:alphamaps} shows 
maps of the best-fit $\alpha$ values for the
M160 and S250 resolution modeling.
At M160 resolution, $\alpha$
is azimuthally coherent but has a notable 
radial gradient, with $\alpha\approx 1.7$
in the center, and $\alpha\approx 2.3$ beyond galactocentric radius
$\sim$$6\kpc$.  While the variation in best-fit $\alpha$ is apparent, these
values are all close to $\alpha=2$, the case where there is equal power
per unit $\log U$.
At S250 resolution, the signal-to-noise ratios are lower, and
the S350, S500, and M160 cameras are not used; 
the $\alpha$ map for the S250 resolution modeling
shows more pixel-scale variations, 
but with a radial trend similar to the M160 resolution modeling.

\begin{figure}
\begin{center}
\includegraphics[angle=0,width=8.0cm,
                 clip=true,trim=0.8cm 5.6cm 0.8cm 2.6cm]
{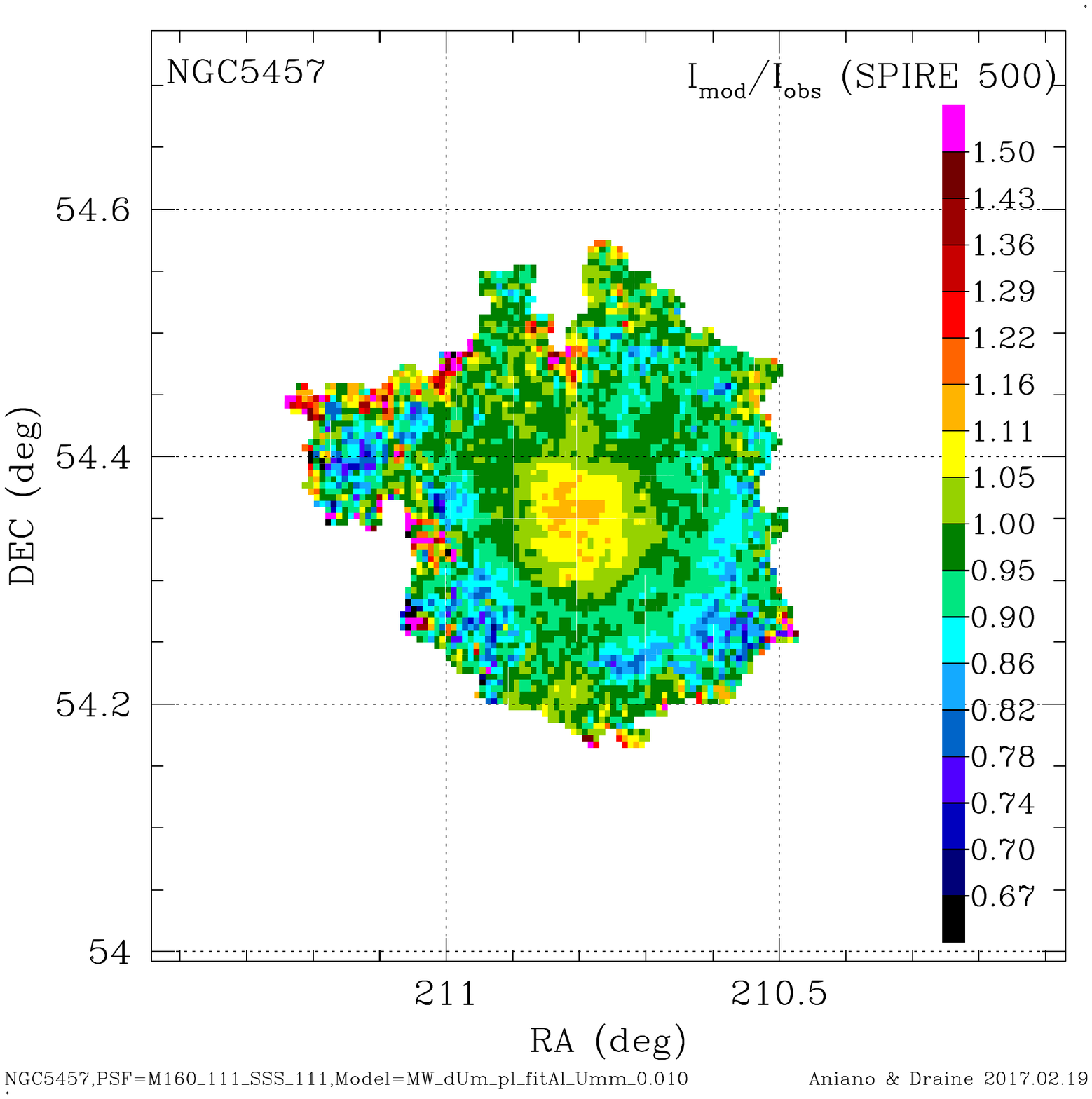}
\includegraphics[angle=0,width=8.0cm,
                 clip=true,trim=0.8cm 5.6cm 0.8cm 2.6cm]
{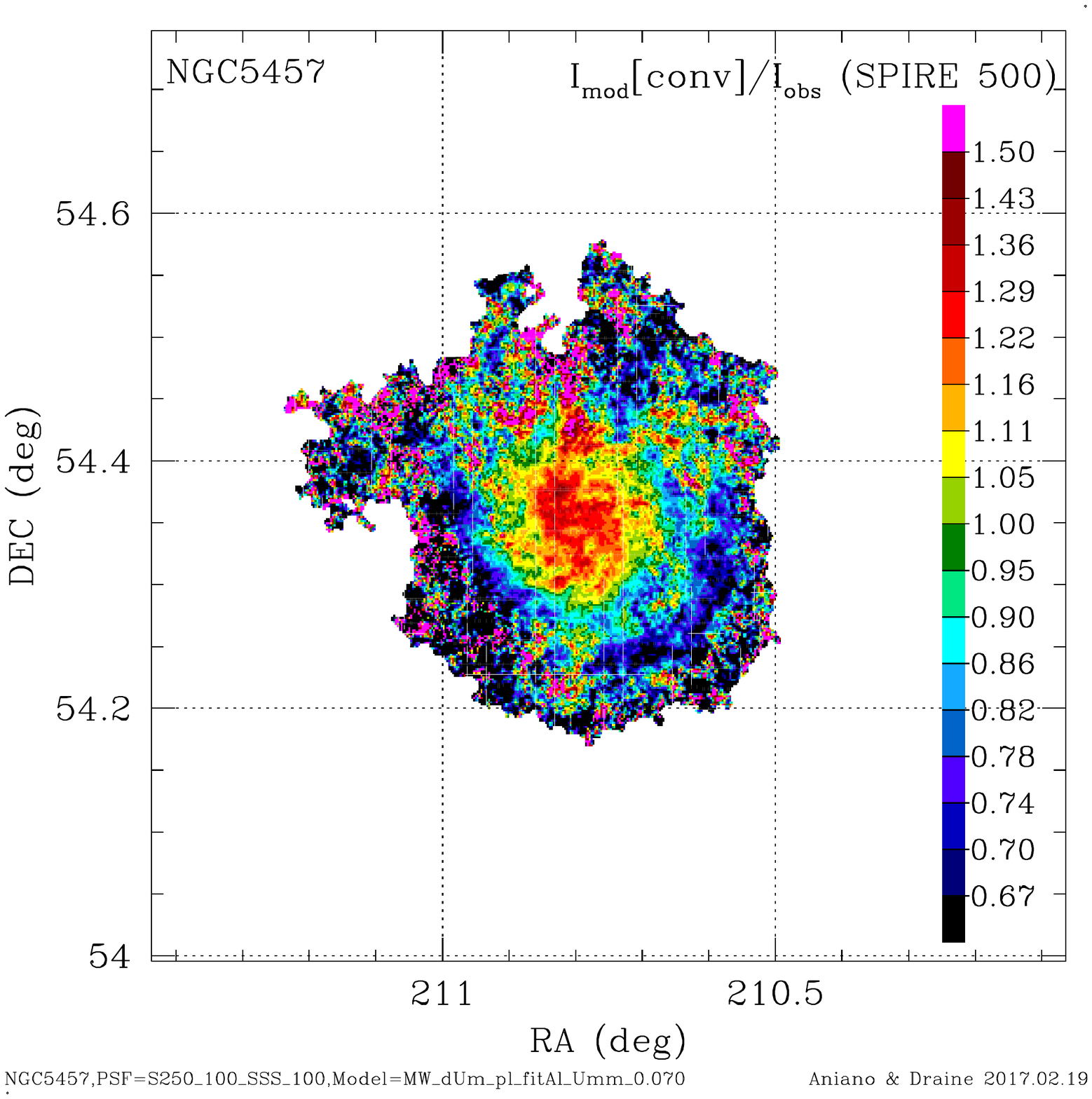}
\caption{\label{fig:NGC5457_S500fit} \footnotesize
$I_\nu^{\rm model}/I_\nu^{\rm obs}$ for NGC\,5457, 
for modeling at M160 resolution (top) and S250 resolution (bottom).  
At M160 resolution, the model reproduces the SPIRE500 observations to
within $\sim$$15\%$, with a clear radial gradient in model/observation,
suggesting a systematic change in the dust opacity with changing
metallicity (see text).
At S250 resolution, no data longward of $250\micron$ are used to 
constrain the model; the predicted
500$\micron$ intensity (after convolving to the SPIRE500 PSF)
agrees with observations to within $\sim$$25\%$.
A radial gradient is again seen.}
\end{center}
\end{figure}

\newtext{In general, the DL07 model successfully
reproduces the resolved SEDs in M\,101.
Figure \ref{fig:NGC5457_S500fit} compares the model $500\micron$ 
surface brightness with observations.  
The upper panel shows modeling at M160 resolution (the 
observed SPIRE500 intensity
is used as a model constraint).
The DL07 model is generally within $\pm10\%$ of the 
observed SPIRE500 intensity, 
except at the outer edges of the mask where the signal/noise is low.  
The model appears to fall short 
by $\sim$10\% in the outer regions 
(galactocentric radius $\sim$15\,kpc=0.13$^\circ$), 
where the metallicity has dropped to
$12+\log_{10}({\rm O/H})\approx 8.25$ \citep{Li+Bresolin+Kennicutt_2013}.
This could indicate that the frequency dependence of the dust opacity
becomes less steep as the metallicity drops -- consistent with the SED of the
SMC \citep{Israel+Wall+Raban+etal_2010,
           Bot+Ysard+Paradis+etal_2010,
           Planck_LMC_SMC_2011,
           Draine+Hensley_2012},
and with evidence for a submm excess in galaxies with metallicities
$12+\log_{10}({\rm O/H})\leq 8.3$ \citep{Remy-Ruyer+Madden+Galliano+etal_2013}.

The lower panel of Figure \ref{fig:NGC5457_S500fit}
compares modeling at S250 resolution 
(no data longward of $250\micron$ used to constrain the model)
with the SPIRE500 observations.
In the bright spiral arms, the $500\micron$ intensity  
is overpredicted by $\sim$25\%.
Once again we see a radial gradient: the model overpredicts SPIRE500
in the central regions, and underpredicts SPIRE500 at 
$R\gtsim 8\kpc=0.07^\circ$.
In the outer regions the fit is poorer,
presumably due to the low signal/noise ratio at S250 resolution.}

Figure \ref{fig:NGC5457_intext} shows maps of dust and starlight heating
parameters for M\,101. 
There are 2 sets of figures; the first set
(rows 1 and 2)
corresponds to modeling done at M160 resolution, using data from all
(IRAC, MIPS, PACS, and SPIRE) cameras, i.e., ``gold standard''
modeling, and the second
(rows 3 and 4)
to modeling done at S250 resolution, using
IRAC, MIPS24, PACS, and SPIRE250 cameras.  
This latter modeling is able
to resolve smaller scale structures in the galaxies, but is overall less
reliable, 
particularly in the outer regions where the surface brightness is lower 
and dust is cooler.

Because of the proximity of M\,101 ($D=6.7\Mpc$),
the spiral structure is visible
even at M160 resolution.
At M160 resolution 38.8\,arcsec FWHM),
the dust luminosity/area ranges from
the surface brightness 
$\SigLdmin=0.67\Lsol\pc^{-2}$ defining the boundary of
the galaxy mask to
a peak $\SigLd=10^{2.5}\Lsol\pc^{-2}$ $\sim8.5\kpc$ ESE of the center,
at the position of the
giant \ion{H}{2} region NGC\,5461 
\citep[see, e.g.,][]{Esteban+Bresolin+Peimbert+etal_2009}.

At S250 resolution the peak at NGC\,5461 has a dust/luminosity/area
$\SigLd=10^{3.2}\Lsol\pc^{-2}$
[corresponding to a dust luminosity $L_\dust=6\times10^7\Lsol$ in a
single $195\times195\pc^2$ S250 map pixel].
Thus at S250 resolution, we are able to measure the IR emission from the
dust over a dynamic range of $\sim$2000 in $\SigLd$.

\renewcommand{\galname}{NGC5457}
\begin{figure*}
\epsscale{1.05}
\plotone{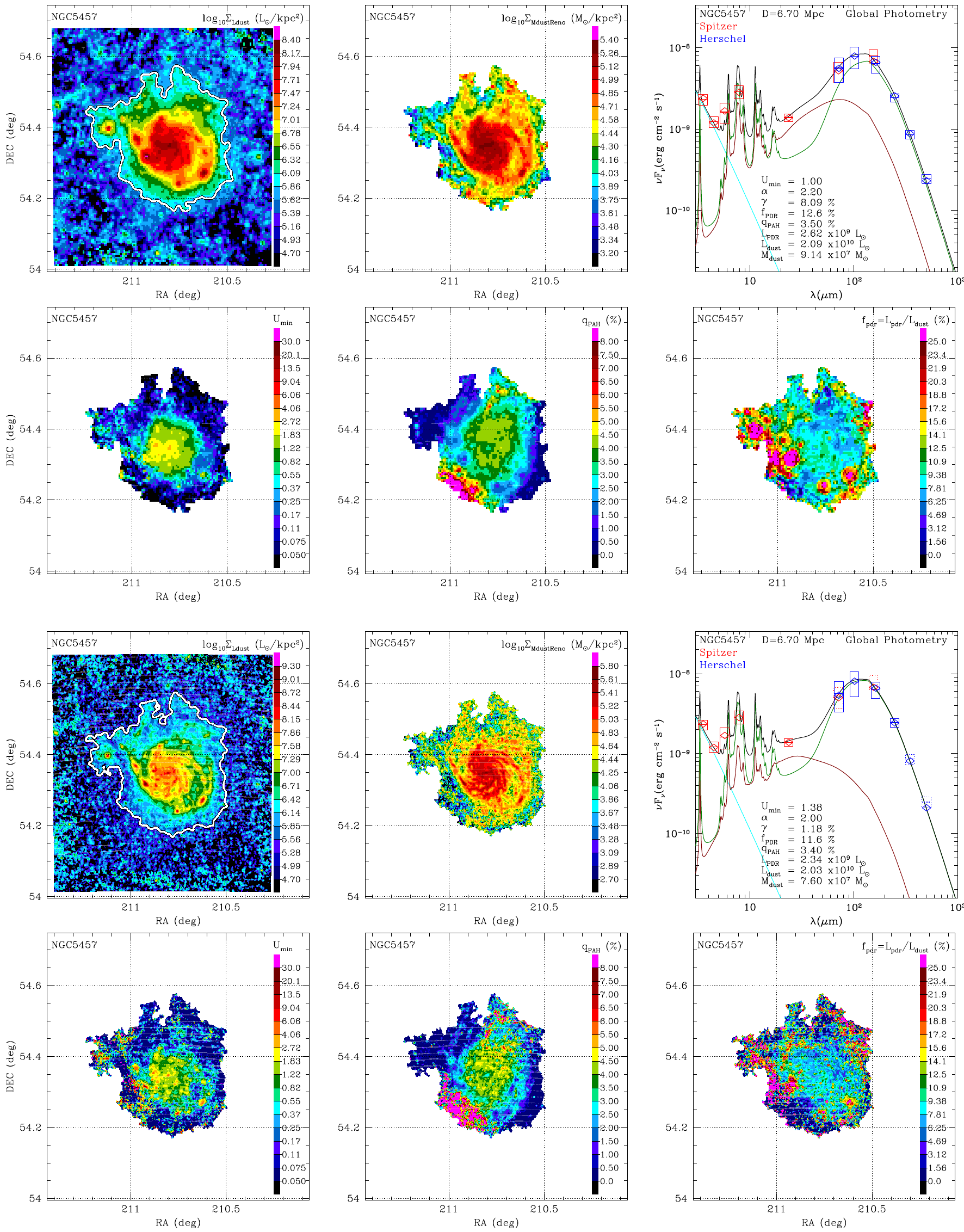}

\caption{\galname\,=\,M101: 
model results at M160 PSF (rows 1 and 2) and at S250 PSF (rows 3 and 4).
Dust luminosity per area $\SigLd$ (column 1, rows 1 and 3) is shown for entire 
field, with adopted galaxy mask boundary in white.
Dust mass per area $\SigMd$ (column 2, rows 1 and 3) is after renormalization 
(see text).
$U_{\rm min,DL07}$, $\qpah$ and $\fPDR$ are shown in rows 2 and 4.
The global SED (column 3, rows 1 and 3) is shown for single-pixel modeling, with
contributions from dust heated by $\Umin$ (green), 
dust heated by $U>\Umin$ (red) and starlight (cyan);
\newtext{values of $\Umin$ and $\Mdust$ in the figure label are
for the DL07 model before renormalization.}
{\it Herschel} (blue rectangles) and {\it Spitzer} (red rectangles) 
photometry is shown.
Diamonds show the band-convolved flux for the model.
Horizontal extent of rectangles and diamonds is an arbitrary
$\pm10\%$ wavelength range.
Vertical extent of photometry rectangles is $\pm1\sigma$.
\label{fig:NGC5457_intext}}
\end{figure*}

Maps of dust surface density $\SigMd$ 
are also shown for both the M160 and S250 modeling.
At both M160 and S250
resolution $\SigMd$ has a peak at the extranuclear
luminosity peak.
At S250 resolution
we estimate a peak dust surface density $5\times10^5\Msol\kpc^{-2}$,
corresponding to $\Mdust=2\times10^4\Msol$ of dust in a single
S250 map pixel.  

Maps of the starlight modeling parameter $U_{\rm min,DL07}$ 
are also shown at both M160
and S250 resolution.
In M101, $U_{\rm min,DL07}$ ranges from values as high as 30 
(the largest value permitted by our modeling) 
to values as low as $\sim0.07$ in the outer parts of the galaxy.
The highest values of $\Umin=30$ arise in the S250 modeling,
with high values of $\Umin$ 
appearing in a fraction of pixels in low surface brightness regions
to the east of the center.
The high $\Umin$ values found in these regions
using S250 resolution data 
are probably unphysical, arising as the
result of low S/N data: an upward fluctuation in PACS70 (or a
downward fluctuation in SPIRE250) can drive
the fitting to a high $\Umin$ value.  
Within $\sim5\kpc$ of the center, with higher surface brightnesses, 
we generally find {\bf $0.5\ltsim\Umin\ltsim4$}.
And in the M160 modeling, we do not obtain very high values of $\Umin$ even
in the low surface brightness outer regions.

Maps of $\qpah$ are also shown at both M160 and S250 resolution.
The modeling finds a very high value of $\qpah$ along the
SSE edge of the galaxy;
this is seen in both the M160 and S250 modeling of
an extended region approximately $12\kpc$ 
SSE of the center.
The high estimates for $\qpah$ could arise from errors in the
IRAC 5.6 and 8$\micron$ photometry,
probably due to errors in background subtraction.

Maps of $\fpdr$ -- the fraction of the dust luminosity that is contributed by
dust heated by starlight with $U>100$ -- 
are shown at both M160 and S250 resolution.
High values of $\fpdr$ are seen at many of the positions 
where $\SigLd$ peaks, which
is consistent with the idea that these are regions with active star formation,
with some fraction of the dust exposed to intense radiation fields in or near OB
associations.  However, we also see high $\fpdr$ values in some of the lowest
surface brightness regions near the edge of the galaxy mask -- 
this is presumably
an indication that photometric errors and errors in background
subtraction are leading 
to overestimation of 24 or 70$\micron$
emission relative to the total dust luminosity.  Thus our derived values of
$\fpdr$ appear to be unreliable in the lowest surface brightness regions.

We also show the global SED for M101, extracted from the galaxy mask.
In the upper right panel, the rectangular symbols show the measured fluxes
$\pm1\sigma$ for the 7 Spitzer cameras and the 6 Herschel cameras.
At 70$\micron$ and $160\micron$ both red and blue rectangles are shown, 
with the MIPS and PACS photometry.
Also shown is a {\it single-pixel} DL07 model, 
where the DL07 model is fitted to the
global photometry.  
The diamonds show the model fluxes for each of the instrumental
bandpasses.  In the case of M101, the model (with 6 adjustable parameters
-- $\Lstar,\Mdust,\qpah,\Umin,\gamma,\alpha$) is
consistent with the photometry at 11 independent wavelengths ($3.6\micron$ to
$500\micron$).
In 
row 3 column 3
we show a single-pixel
DL07 model fitted to only the photometry that is used for
the S250 modeling (i.e.,
MIPS70, MIPS160, SPIRE350, or SPIRE500 are {\it not} used 
when adjusting the model parameters).
The dashed rectangles show these unused measurements; 
we see that for M\,101 the single-pixel model
does quite well at {\it predicting} the fluxes at 350 and 500$\micron$, 
with only a $1\sigma$ underprediction
even at $500\micron$.
The single-pixel global fit parameters are given in the SED plots.

Table \ref{tab:M101 dust} compares 
total dust mass estimates for M101.
Column 2 reports the dust mass estimated from the DL07
model at either M160 or S250 resolution, after summing
the dust model over the galaxy mask.
Because we opted to use the same $\SigLdmin$ for the S250 and M160
modeling, the galaxy masks for the two cases are essentially the same.
Column 3 reports the result of fitting a DL07 model to
the global photometry -- this is referred to as ``single pixel'' modeling.
In columns 4 and 5 we show the multipixel or single pixel dust masses
after renormalizing following Equation \ref{eq:Mdcorrected}.

Multipixel vs.\ single-pixel modeling is of course expected to produce
different estimates because the models are nonlinear.
One notes in Table \ref{tab:M101 dust} that the discrepancies 
between the multipixel and single pixel mass estimators are
reduced when going from the original DL07 model to the renormalized model.
It is not
clear why this is the case, but this is a welcome result.

\begin{table}
\footnotesize
\caption{Dust mass estimates for NGC\,5457=M\,101}
\label{tab:M101 dust}
\begin{tabular}{|c | c c | c c|c|}
\hline
&\multicolumn{4}{c|}{$M_\dust/10^7\Msol$}&\\
PSF & \multicolumn{2}{c|}{DL07} & \multicolumn{2}{c|}{renorm.\ DL07} & \!\!multipix\!\!\\
           & \!\!multipixel\!\!\! & \!\!\!single pixel\!\! &
             \!\!multipixel\!\!\! & \!\!\!single pixel\!\! & 
$\langle \Cdust\rangle$ \\
\hline
\!M160\!   & $12.7\pm 0.5$ & $9.14\pm0.32$ & $6.97\pm 0.20$ & $6.40\pm0.89$ 
           & 0.549 \\
\!S250\!   & $12.1\pm 2.9$ & $7.60\pm0.10$ & $6.88\pm 1.34$ & $5.32\pm0.28$ 
           & 0.569 \\
\hline
\end{tabular}
\end{table}

\subsection{Full KINGFISH Sample}

Dust is detected reliably for every galaxy in the KF62 sample.
Selected images for each of these galaxies
are given in
Appendix \ref{app:maps}
(Figures 17.1-17.62, following the
scheme used for M\,101 in Fig.\ \ref{fig:NGC5457_intext}.
This is only a fraction of the maps and images that are available
online -- see Appendix \ref{app:online data} for a description of
the data set.

The ``galaxy mask'' for each galaxy is shown
for both the M160 PSF and the S250 PSF.
As for M101, we have opted to use the same $\SigLdmin$ for both
the M160 and S250 modeling, hence the M160 and S250 $\SigLdmin$-based
galaxy masks are
nearly identical for each galaxy, except for the 8 where dust emission
is so weak that we treat them
as nondetections (see Appendix \ref{app:upper limits}).
The flux densities $F_\nu$ measured by \Spitzer\ and \Herschel\ within the 
M160 and S250 galaxy mask for each galaxy 
have been given above in Tables \ref{tab:spitzer} and \ref{tab:herschel}.
The model-derived parameters for the dust and starlight are given in
Table \ref{tab:dust}.
$\alpha$ is not included in Table \ref{tab:dust} because there is no natural way to define a ``mean'' $\alpha$ for multipixel modeling.
The uncertainties listed for the
parameters are based on repeating the fitting procedure
with the ``observed'' fluxes obtained by Monte-Carlo sampling
from Gaussian distributions with means and widths given by the original
observed values and uncertainty estimates
(see discussion in Appendix E of AD12).
Systematic errors associated with the DL07 model itself have not been
estimated.

Figures
17.1-17.62
have twelve panels in all,
with the top six panels showing results of modeling with the M160 PSF,
and the lower six panels repeating this for the S250 PSF.
For each PSF, the top row shows maps of the
dust luminosity surface density $\Sigma_{\Ldust}$ (upper left) and 
modeled dust surface density $\Sigma_{M_\dust}$ (upper center), 
and the model SED (upper right).
The lower row shows maps of the starlight intensity parameter
$\Umin$ (left), 
the PAH abundance parameter $\qpah$ (center), 
and the PDR fraction $f_\PDR$ (right), all
restricted to the ``galaxy mask''.

In the SED plot, the observed global photometry is represented by
rectangular boxes [{\it Spitzer} (IRAC and MIPS) in red; {\it Herschel}
(PACS and SPIRE) in blue].  The vertical extent of each box
shows the $\pm1\sigma$ uncertainty in the photometry for each band.  
The black line is the total DL07 model spectrum,
and its different components are represented by three colors.
The cyan line is the stellar contribution, the dark red line is the
emission from dust heated by the power-law $U$ distribution, and the
dark green line is emission from dust heated by $U=\Umin$.
The DL07 model used in this SED plot is a single-pixel model,
which tries to reproduce the global photometry treating
the entire galaxy as a single pixel.  These ``single pixel'' models
generally do a good job at reproducing the global photometry.
Multipixel models, where the photometry in every pixel is fit
independently, have many more adjustable parameters, and naturally
do an even better job of reproducing the global photometry 
after summing
over all the pixels in the galaxy mask.  
It is reasonable to presume that models that do a better job of
reproducing the photometry will also be preferred for dust mass estimation.

The dust mass surface densities and dust luminosities per unit area
range over three orders of magnitude in our brightest galaxies.
Figures
17.1-17.62 illustrate that the
DL07 model does a satisfactory job of fitting the SEDs.
Although each pixel is modeled independently of its neighbors, it is
noteworthy that the dust parameters are smoothly varying over the
confines of the galaxy, except for the low surface brightness outer
regions at S250 resolution, where the S/N in individual pixels may
become low enough that certain dust and starlight parameters, such
as $\SigMd$ and $\Umin$, become somewhat noisy.  

\subsection{Special Cases}

\subsubsection{NGC\,1404}

The E1 galaxy NGC\,1404 is faint at infrared wavelengths, and 
there is a foreground star $\sim180\arcsec$
to the SSE.
However, NGC\,1404 is
unambiguously
detected in IRAC bands 1-4,
and by MIPS24.  At $\lambda\geq70\micron$ there appears to be excess emission
at the position of NGC\,1404,
but the surface brightness is low and the possibility that the emission is
from a foreground or background source cannot be excluded
\citep{Dale+Aniano+Engelbracht+etal_2012}.
Using the procedure described in Appendix \ref{app:upper limits}, we find a
$3\sigma$ upper bound $M_{\rm dust} < 2.0\times10^6\Msol$ for
NGC\,1404.

NGC\,1404 has an estimated
stellar mass $M_\star\approx 8\times10^{10}\Msol$, and evolved stars
are probably injecting dust at a rate $\dot{M}_{\rm dust}\approx 
0.005\times M_\star/10^{10}\yr = 0.04\Msol\yr^{-1}$.\footnote{%
   A rough estimate, 
   assuming $\sim$1$\Msol$ stars each lose $\sim$0.5$\Msol$ of envelope before
   becoming white dwarfs, and that $\sim$1\% of the envelope mass consists of
   dust.}
Thus the observed
dust mass upper limit
would be consistent with a dust lifetime $\Mdust/\dot{M}_{\rm dust}
\ltsim 5\times10^7\yr$.
If the ISM in NGC\,1404 has a temperature $T\approx10^7\K$,
silicate or carbonaceous dust grains would be eroded by thermal
sputtering at a rate 
$da/dt \approx -1\times10^{-10}\cm\yr^{-1}(\nH/\cm^{-3})$
\citep{Draine+Salpeter_1979a}, for a grain lifetime
$a/|da/dt|=10^7(a/0.1\micron)(0.01\cm^{-3}/\nH)\yr$.
The dust mass upper limit would thus be consistent with erosion of 
$a\approx0.1\micron$ grains by
sputtering in a $T\approx10^7\K$ ISM with $\nH\gtsim 0.002\cm^{-3}$.

\subsubsection{NGC\,1377 \label{sec:NGC1377}}
  NGC\,1377 has a compact dusty core, with an extremely high
  far-infrared surface brightness.
  The infrared spectrum \citep{Roussel+Helou+Smith+etal_2006} shows that it is
  optically thick at $\lambda \ltsim 20\micron$.
  Weak PAH emission is detected, but because of the uncertain infrared
  extinction it is not possible to reliably estimate the PAH abundance
  parameter $\qpah$.  The dust mass estimates should also be regarded
  as uncertain because of the unusual nature of the interstellar medium
  in this galaxy.

\begin{table*}
\scriptsize
  \begin{center}
    \caption{\label{tab:dust}
             Dust Model Parameters for S250 and M160 Galaxy Masks}
    \begin{tabular}{| c c | c c c c c c | c |}
      \hline
      Galaxy & mask & $M_{\rm dust}(10^6\Msol)^a$& $q_{\rm PAH}(\%)$& $\langle U_{\rm min}\rangle^a$& $\langle\bar{U}\rangle^a$& $f_{\rm PDR}(\%)$& $\langle C_{\rm dust}\rangle$& $L_{\rm dust}(10^9\Lsol)$\\
\hline
Hol2     & S250 & $0.12\pm0.05$         & $0.95\pm0.67$         & $ 2.4\pm 4.6$         & $ 3.1\pm 3.7$         & $15.3\pm4.1$          & 0.500 & $0.061\pm0.009$       \\
" & M160 & $0.134\pm0.033$       & $0.68\pm0.34$         & $ 1.0\pm 1.6$         & $ 2.7\pm 0.9$         & $18.1\pm3.4$          & 0.491 & 
$0.058\pm0.006$       \\
IC342    & S250 & $ 41.\pm  8.$         & $ 4.4\pm 0.8$         & $ 1.9\pm 1.3$         & $ 2.5\pm 1.2$         & $12.9\pm2.4$          & 0.614 & $16.5\pm2.3$          \\
" & M160 & $ 35.\pm  5.$         & $4.25\pm0.26$         & $ 2.2\pm 0.6$         & $ 2.7\pm 0.5$         & $12.8\pm3.7$          & 0.640 & 
$16.0\pm0.8$          \\
IC2574   & S250 & $1.08\pm0.23$         & $0.48\pm0.07$         & $0.56\pm0.57$         & $0.95\pm0.50$         & $11.8\pm3.7$          & 0.469 & $0.169\pm0.025$       \\
" & M160 & $1.17\pm0.34$         & $0.44\pm0.27$         & $0.22\pm0.21$         & $0.85\pm0.45$         & $12.5\pm1.0$          & 0.452 & 
$0.163\pm0.015$       \\
NGC0337  & S250 & $10.6\pm2.2$          & $2.05\pm0.25$         & $ 6.3\pm 1.8$         & $ 7.0\pm 1.9$         & $11.0\pm2.3$          & 0.685 & $12.1\pm1.3$          \\
" & M160 & $12.2\pm1.0$          & $ 2.4\pm 0.6$         & $ 5.0\pm 0.6$         & $ 5.4\pm 0.6$         & $13.6\pm0.6$          & 0.678 & 
$10.8\pm0.6$          \\
NGC0628  & S250 & $20.5\pm1.1$          & $ 3.5\pm 0.7$         & $1.96\pm0.44$         & $2.16\pm0.34$         & $11.5\pm4.2$          & 0.613 & $ 7.3\pm 1.0$         \\
" & M160 & $18.7\pm1.0$          & $ 3.6\pm 0.7$         & $2.16\pm0.21$         & $2.30\pm0.27$         & $11.4\pm2.1$          & 0.622 & 
$7.08\pm0.18$         \\
NGC0925  & S250 & $ 16.\pm  5.$         & $2.59\pm0.36$         & $ 1.2\pm 1.3$         & $ 1.5\pm 1.3$         & $ 8.5\pm 2.5$         & 0.525 & $3.88\pm0.36$         \\
" & M160 & $17.1\pm2.7$          & $2.65\pm0.47$         & $0.56\pm0.60$         & $1.32\pm0.37$         & $ 9.3\pm 1.2$         & 0.481 & 
$3.721\pm0.027$       \\
NGC1097  & S250 & $ 93.\pm 31.$         & $ 3.7\pm 0.9$         & $ 2.4\pm 1.3$         & $ 2.9\pm 1.2$         & $ 16.\pm  5.$         & 0.549 & $ 44.\pm  7.$         \\
" & M160 & $ 65.\pm  7.$         & $ 3.2\pm 0.9$         & $ 3.4\pm 1.0$         & $ 4.0\pm 0.9$         & $16.6\pm3.4$          & 0.622 & 
$42.6\pm2.5$          \\
NGC1266  & S250 & $13.2\pm1.9$          & $0.60\pm0.16$         & $ 13.\pm  5.$         & $ 13.\pm  5.$         & $13.6\pm3.9$          & 0.576 & $27.5\pm4.7$          \\
" & M160 & $ 9.6\pm 1.0$         & $0.61\pm0.44$         & $ 17.\pm  6.$         & $15.7\pm2.6$          & $ 15.\pm  5.$         & 0.635 & 
$24.7\pm2.3$          \\
NGC1291  & S250 & $ 25.\pm  7.$         & $ 2.6\pm 0.8$         & $0.54\pm0.49$         & $0.64\pm0.57$         & $ 8.4\pm 0.6$         & 0.470 & $ 2.7\pm 0.7$         \\
" & M160 & $16.0\pm4.4$          & $ 2.4\pm 1.2$         & $0.97\pm0.27$         & $1.04\pm0.25$         & $ 7.6\pm 3.3$         & 0.516 & 
$2.72\pm0.09$         \\
NGC1316  & S250 & $12.0\pm2.9$          & $ 1.8\pm 1.0$         & $ 2.8\pm 1.8$         & $ 3.2\pm 1.8$         & $10.2\pm2.7$          & 0.612 & $6.22\pm0.45$         \\
" & M160 & $ 8.7\pm 1.3$         & $ 1.9\pm 1.3$         & $ 3.8\pm 0.7$         & $ 4.2\pm 0.6$         & $10.5\pm3.1$          & 0.660 & 
$5.90\pm0.31$         \\
NGC1377  & S250 & $ 2.5\pm 0.6$         & $0.68\pm0.15$         & $ 17.\pm  7.$         & $ 34.\pm 15.$         & $56.7\pm3.7$          & 0.580 & $13.8\pm1.3$          \\
" & M160 & $1.48\pm0.22$         & $0.73\pm0.80$         & $25.1\pm3.8$          & $ 53.\pm  8.$         & $55.9\pm1.3$          & 0.673 & 
$12.81\pm0.13$        \\
NGC1482  & S250 & $20.4\pm2.0$          & $3.05\pm0.38$         & $13.8\pm2.2$          & $14.7\pm2.4$          & $12.2\pm1.1$          & 0.663 & $ 49.\pm  6.$         \\
" & M160 & $16.6\pm2.1$          & $ 3.3\pm 0.5$         & $15.5\pm3.4$          & $16.9\pm3.4$          & $11.7\pm3.5$          & 0.686 & 
$46.4\pm4.8$          \\
NGC1512  & S250 & $10.1\pm1.3$          & $ 3.4\pm 0.7$         & $1.63\pm0.44$         & $1.80\pm0.41$         & $ 7.5\pm 1.5$         & 0.584 & $2.99\pm0.32$         \\
" & M160 & $12.4\pm1.2$          & $ 3.3\pm 0.5$         & $1.48\pm0.35$         & $1.50\pm0.26$         & $ 7.5\pm 1.8$         & 0.553 & 
$3.06\pm0.16$         \\
NGC2146  & S250 & $ 53.\pm 13.$         & $ 4.1\pm 0.7$         & $14.6\pm4.5$          & $15.3\pm3.5$          & $12.6\pm3.1$          & 0.665 & $135.\pm  12.$        \\
" & M160 & $40.2\pm2.2$          & $ 4.2\pm 0.7$         & $17.9\pm4.4$          & $20.6\pm1.7$          & $11.8\pm3.1$          & 0.692 & 
$137.\pm   7.$        \\
NGC2798  & S250 & $15.8\pm2.1$          & $1.91\pm0.41$         & $14.7\pm3.4$          & $13.9\pm2.8$          & $14.5\pm2.1$          & 0.604 & $36.0\pm3.9$          \\
" & M160 & $11.4\pm0.6$          & $2.19\pm0.28$         & $16.9\pm2.1$          & $17.99\pm0.44$        & $13.6\pm2.0$          & 0.683 & 
$33.4\pm1.2$          \\
NGC2841  & S250 & $ 68.\pm  9.$         & $ 3.4\pm 1.1$         & $0.84\pm0.19$         & $0.92\pm0.15$         & $ 6.5\pm 3.2$         & 0.540 & $10.37\pm0.36$        \\
" & M160 & $53.0\pm4.0$          & $3.40\pm0.48$         & $1.11\pm0.11$         & $1.15\pm0.11$         & $ 6.6\pm 2.3$         & 0.584 & 
$10.03\pm0.20$        \\
NGC2915  & S250 & $0.052\pm0.029$       & $ 1.3\pm 0.6$         & $ 3.7\pm 2.0$         & $ 4.5\pm 1.7$         & $10.1\pm1.9$          & 0.588 & $0.0377\pm0.0046$     \\
" & M160 & $0.031\pm0.010$       & $ 1.4\pm 0.7$         & $ 6.0\pm 2.2$         & $ 6.6\pm 1.9$         & $11.1\pm2.0$          & 0.676 & 
$0.0332\pm0.0023$     \\
NGC2976  & S250 & $1.76\pm0.31$         & $2.93\pm0.20$         & $ 2.6\pm 0.6$         & $ 2.8\pm 0.5$         & $10.6\pm1.2$          & 0.611 & $0.800\pm0.034$       \\
" & M160 & $1.77\pm0.11$         & $ 3.2\pm 1.0$         & $ 1.9\pm 0.5$         & $2.67\pm0.31$         & $11.4\pm2.2$          & 0.661 & 
$0.778\pm0.049$       \\
NGC3049  & S250 & $ 6.6\pm 2.5$         & $1.81\pm0.21$         & $ 2.1\pm 4.5$         & $ 3.6\pm 4.7$         & $27.4\pm3.0$          & 0.542 & $3.93\pm0.32$         \\
" & M160 & $ 9.5\pm 0.8$         & $ 2.1\pm 0.6$         & $0.28\pm0.52$         & $2.09\pm0.19$         & $33.6\pm1.5$          & 0.472 & 
$3.24\pm0.11$         \\
NGC3077  & S250 & $1.83\pm0.33$         & $ 3.5\pm 0.6$         & $ 2.2\pm 1.1$         & $ 2.6\pm 0.9$         & $13.9\pm1.4$          & 0.506 & $0.79\pm0.11$         \\
" & M160 & $1.51\pm0.15$         & $ 3.1\pm 0.8$         & $2.69\pm0.11$         & $3.03\pm0.33$         & $14.1\pm2.1$          & 0.529 & 
$0.754\pm0.020$       \\
NGC3184  & S250 & $35.7\pm4.1$          & $ 3.8\pm 0.8$         & $1.45\pm0.28$         & $1.58\pm0.14$         & $ 9.8\pm 1.5$         & 0.581 & $ 9.4\pm 0.8$         \\
" & M160 & $30.2\pm2.7$          & $ 3.8\pm 0.8$         & $1.72\pm0.21$         & $1.82\pm0.24$         & $ 9.6\pm 3.4$         & 0.612 & 
$9.08\pm0.33$         \\
NGC3190  & S250 & $21.6\pm3.8$          & $3.00\pm0.26$         & $1.77\pm0.48$         & $1.87\pm0.31$         & $ 4.0\pm 0.5$         & 0.588 & $6.64\pm0.24$         \\
" & M160 & $14.8\pm2.0$          & $ 2.8\pm 0.5$         & $ 2.4\pm 0.6$         & $2.55\pm0.48$         & $ 3.8\pm 1.3$         & 0.681 & 
$6.19\pm0.33$         \\
NGC3198  & S250 & $26.8\pm4.0$          & $2.90\pm0.38$         & $ 1.4\pm 0.6$         & $1.82\pm0.47$         & $14.6\pm1.8$          & 0.581 & $8.02\pm0.14$         \\
" & M160 & $28.4\pm2.8$          & $ 3.1\pm 0.8$         & $1.49\pm0.31$         & $1.65\pm0.39$         & $15.0\pm3.0$          & 0.555 & 
$7.71\pm0.28$         \\
NGC3265  & S250 & $ 2.4\pm 1.1$         & $1.73\pm0.47$         & $ 5.2\pm 9.3$         & $ 7.1\pm 9.7$         & $ 27.\pm  7.$         & 0.542 & $2.74\pm0.10$         \\
" & M160 & $ 2.0\pm 0.9$         & $2.30\pm0.35$         & $ 4.6\pm 2.4$         & $ 7.6\pm 2.7$         & $30.1\pm3.3$          & 0.576 & 
$2.49\pm0.17$         \\
NGC3351  & S250 & $ 24.\pm  6.$         & $ 3.9\pm 1.5$         & $ 1.4\pm 0.6$         & $ 1.9\pm 0.5$         & $ 17.\pm  6.$         & 0.538 & $ 7.4\pm 0.9$         \\
" & M160 & $14.9\pm1.0$          & $ 3.2\pm 0.6$         & $ 2.3\pm 0.6$         & $2.85\pm0.45$         & $17.3\pm3.8$          & 0.623 & 
$6.99\pm0.45$         \\
NGC3521  & S250 & $ 89.\pm 12.$         & $4.16\pm0.35$         & $2.21\pm0.29$         & $2.34\pm0.40$         & $ 8.9\pm 2.0$         & 0.586 & $34.5\pm0.7$          \\
" & M160 & $ 69.\pm  8.$         & $4.17\pm0.16$         & $ 2.8\pm 0.6$         & $2.96\pm0.39$         & $ 8.7\pm 4.4$         & 0.648 & 
$33.7\pm1.3$          \\
NGC3621  & S250 & $ 20.\pm  6.$         & $4.18\pm0.38$         & $ 2.0\pm 0.8$         & $ 2.2\pm 0.6$         & $10.8\pm1.1$          & 0.569 & $7.39\pm0.41$         \\
" & M160 & $19.5\pm3.3$          & $ 4.8\pm 0.7$         & $ 2.5\pm 0.7$         & $2.23\pm0.38$         & $10.7\pm2.8$          & 0.567 & 
$7.28\pm0.25$         \\
NGC3627  & S250 & $ 38.\pm  7.$         & $ 3.3\pm 0.7$         & $ 3.9\pm 1.8$         & $ 4.3\pm 1.9$         & $12.0\pm1.5$          & 0.645 & $27.0\pm2.7$          \\
" & M160 & $30.1\pm1.5$          & $3.23\pm0.32$         & $4.85\pm0.47$         & $5.43\pm0.40$         & $11.7\pm1.9$          & 0.691 & 
$26.9\pm0.8$          \\
NGC3773  & S250 & $1.12\pm0.49$         & $1.27\pm0.35$         & $ 2.6\pm 4.1$         & $ 3.4\pm 4.0$         & $21.3\pm4.3$          & 0.519 & $0.63\pm0.10$         \\
" & M160 & $0.62\pm0.07$         & $1.90\pm0.43$         & $ 4.0\pm 1.5$         & $ 5.5\pm 1.0$         & $23.8\pm3.6$          & 0.631 & 
$0.563\pm0.023$       \\
NGC3938  & S250 & $ 52.\pm 14.$         & $3.71\pm0.35$         & $ 1.8\pm 0.7$         & $ 2.0\pm 0.8$         & $ 9.7\pm 3.4$         & 0.572 & $16.9\pm1.4$          \\
" & M160 & $41.5\pm2.5$          & $ 3.7\pm 0.7$         & $2.26\pm0.30$         & $2.41\pm0.31$         & $ 9.3\pm 2.2$         & 0.627 & 
$16.5\pm1.0$          \\
NGC4236  & S250 & $2.12\pm0.45$         & $0.94\pm0.32$         & $0.64\pm0.76$         & $ 1.1\pm 0.7$         & $14.4\pm4.3$          & 0.494 & $0.384\pm0.043$       \\
" & M160 & $ 2.6\pm 1.4$         & $0.86\pm0.39$         & $0.23\pm0.10$         & $0.85\pm0.29$         & $15.5\pm1.7$          & 0.449 & 
$0.367\pm0.038$       \\

      \hline
      \multicolumn{9}{l}{$a$ Renormalized as described in \S\ref{sec:renormalization}.}
    \end{tabular}
  \end{center}
\btdnote{uses tab\_dust\_guts1.tex created by tablemaker\_v9}
\end{table*}
\addtocounter{table}{-1}
\begin{table*}
  \scriptsize
  \begin{center}
    \caption{Dust Model Parameters, contd.}
\label{tab:DL07 model params}
    \begin{tabular}{| c c | c c c c c c | c |}
      \hline
      Galaxy & mask& $M_{\rm dust}(10^6\Msol)^a$& $q_{\rm PAH}(\%)$& $\langle U_{\rm min}\rangle^a$& $\langle\bar{U}\rangle^a$& $f_{\rm PDR}(\%)$& $\langle C_{\rm dust}\rangle$& $L_{\rm dust}(10^9\Lsol)$ \\
\hline
NGC4254  & S250 & $ 66.\pm 16.$         & $ 4.3\pm 1.1$         & $ 3.4\pm 1.0$         & $ 3.6\pm 1.1$         & $11.8\pm3.0$          & 0.631 & $39.0\pm3.3$          \\
" & M160 & $51.5\pm3.6$          & $ 4.1\pm 0.8$         & $4.03\pm0.42$         & $4.42\pm0.33$         & $11.6\pm1.9$          & 0.685 & $37.6\pm0.8$          \\
NGC4321  & S250 & $ 83.\pm 13.$         & $ 4.1\pm 1.7$         & $2.08\pm0.40$         & $2.26\pm0.50$         & $10.4\pm3.7$          & 0.619 & $31.1\pm1.7$          \\
" & M160 & $63.5\pm3.4$          & $ 3.7\pm 1.2$         & $2.67\pm0.20$         & $2.92\pm0.23$         & $10.2\pm1.7$          & 0.681 & $30.6\pm1.6$          \\
NGC4536  & S250 & $ 32.\pm 10.$         & $ 2.9\pm 1.2$         & $ 3.6\pm 1.8$         & $ 4.3\pm 1.5$         & $ 18.\pm  8.$         & 0.578 & $22.6\pm3.1$          \\
" & M160 & $27.1\pm4.4$          & $ 3.4\pm 0.5$         & $ 4.4\pm 1.3$         & $ 4.7\pm 0.7$         & $19.8\pm3.2$          & 0.591 & $20.8\pm0.5$          \\
NGC4559  & S250 & $7.71\pm0.26$         & $2.47\pm0.42$         & $ 2.1\pm 0.8$         & $2.16\pm0.22$         & $ 9.5\pm 2.4$         & 0.579 & $2.73\pm0.15$         \\
" & M160 & $ 8.5\pm 1.2$         & $ 2.7\pm 0.8$         & $ 1.9\pm 0.9$         & $1.89\pm0.43$         & $10.1\pm2.1$          & 0.541 & $2.65\pm0.15$         \\
NGC4569  & S250 & $12.9\pm4.0$          & $ 3.7\pm 0.6$         & $ 1.9\pm 0.6$         & $ 2.4\pm 0.7$         & $14.2\pm1.6$          & 0.635 & $5.14\pm0.25$         \\
" & M160 & $ 9.2\pm 0.8$         & $ 3.4\pm 0.9$         & $2.63\pm0.36$         & $3.12\pm0.33$         & $13.9\pm1.6$          & 0.694 & $4.71\pm0.40$         \\
NGC4579  & S250 & $ 39.\pm  7.$         & $ 3.1\pm 0.9$         & $1.55\pm0.26$         & $1.65\pm0.28$         & $ 7.9\pm 2.8$         & 0.611 & $10.58\pm0.32$        \\
" & M160 & $29.0\pm1.0$          & $ 2.9\pm 0.9$         & $1.88\pm0.07$         & $2.066\pm0.046$       & $ 9.0\pm 1.6$         & 0.675 & $9.83\pm0.31$         \\
NGC4594  & S250 & $16.4\pm3.6$          & $ 2.6\pm 1.4$         & $1.06\pm0.26$         & $1.15\pm0.24$         & $ 6.2\pm 3.5$         & 0.574 & $3.10\pm0.16$         \\
" & M160 & $13.3\pm1.0$          & $ 2.6\pm 0.7$         & $1.26\pm0.11$         & $1.34\pm0.14$         & $ 6.6\pm 0.9$         & 0.619 & $2.94\pm0.08$         \\
NGC4625  & S250 & $1.47\pm0.31$         & $ 3.7\pm 0.9$         & $ 2.1\pm 0.5$         & $2.23\pm0.49$         & $ 9.2\pm 2.8$         & 0.590 & $0.542\pm0.025$       \\
" & M160 & $1.18\pm0.21$         & $ 4.2\pm 0.9$         & $ 2.7\pm 0.8$         & $ 2.6\pm 1.0$         & $ 9.0\pm 3.7$         & 0.625 & $0.512\pm0.021$       \\
NGC4631  & S250 & $28.4\pm4.1$          & $ 3.0\pm 0.5$         & $ 4.4\pm 1.3$         & $ 4.6\pm 1.4$         & $10.9\pm1.8$          & 0.669 & $21.6\pm1.4$          \\
" & M160 & $31.1\pm2.9$          & $2.90\pm0.23$         & $ 4.9\pm 0.6$         & $ 4.3\pm 0.6$         & $10.9\pm1.0$          & 0.631 & $22.0\pm0.8$          \\
NGC4725  & S250 & $ 42.\pm  9.$         & $3.68\pm0.49$         & $0.90\pm0.27$         & $0.99\pm0.23$         & $ 5.7\pm 1.9$         & 0.541 & $ 6.9\pm 0.7$         \\
" & M160 & $37.4\pm3.7$          & $ 3.6\pm 0.8$         & $1.09\pm0.16$         & $1.09\pm0.19$         & $ 5.4\pm 2.6$         & 0.566 & $6.72\pm0.16$         \\
NGC4736  & S250 & $ 8.5\pm 0.6$         & $ 4.2\pm 0.6$         & $ 4.0\pm 0.8$         & $ 4.3\pm 0.6$         & $ 7.0\pm 2.6$         & 0.552 & $5.99\pm0.34$         \\
" & M160 & $ 5.4\pm 0.5$         & $3.70\pm0.50$         & $ 6.6\pm 1.7$         & $ 6.5\pm 1.6$         & $ 6.7\pm 1.8$         & 0.642 & $ 5.8\pm 0.7$         \\
NGC4826  & S250 & $ 4.0\pm 1.0$         & $2.59\pm0.39$         & $ 5.5\pm 2.1$         & $ 5.8\pm 1.3$         & $ 6.5\pm 3.4$         & 0.665 & $3.76\pm0.16$         \\
" & M160 & $3.46\pm0.30$         & $2.57\pm0.16$         & $ 6.5\pm 1.3$         & $ 6.5\pm 0.9$         & $ 6.4\pm 3.6$         & 0.673 & $3.67\pm0.33$         \\
NGC5055  & S250 & $ 63.\pm 14.$         & $ 4.2\pm 0.8$         & $1.70\pm0.46$         & $1.79\pm0.40$         & $ 7.8\pm 1.3$         & 0.594 & $18.70\pm0.44$        \\
" & M160 & $49.0\pm3.0$          & $3.89\pm0.23$         & $2.12\pm0.17$         & $2.28\pm0.12$         & $ 7.8\pm 3.8$         & 0.659 & $18.4\pm0.6$          \\
NGC5398  & S250 & $0.69\pm0.16$         & $2.25\pm0.37$         & $ 1.4\pm 0.5$         & $ 3.1\pm 0.8$         & $28.0\pm2.1$          & 0.528 & $0.352\pm0.013$       \\
" & M160 & $0.75\pm0.22$         & $ 1.8\pm 0.7$         & $0.61\pm1.42$         & $ 2.4\pm 1.6$         & $33.0\pm4.8$          & 0.488 & $0.297\pm0.042$       \\
NGC5408  & S250 & $0.049\pm0.007$       & $0.14\pm0.28$         & $ 8.5\pm 3.8$         & $ 21.\pm  6.$         & $ 37.\pm  5.$         & 0.627 & $0.172\pm0.016$       \\
" & M160 & $0.047\pm0.017$       & $0.14\pm0.64$         & $ 10.\pm  6.$         & $ 20.\pm  7.$         & $ 39.\pm  5.$         & 0.661 & $0.154\pm0.009$       \\
NGC5457  & S250 & $ 69.\pm 13.$         & $ 3.0\pm 0.6$         & $1.58\pm0.47$         & $1.79\pm0.42$         & $12.4\pm2.2$          & 0.569 & $20.3\pm1.2$          \\
" & M160 & $69.7\pm2.0$          & $ 3.0\pm 0.9$         & $1.84\pm0.35$         & $1.77\pm0.10$         & $12.7\pm1.8$          & 0.550 & $20.3\pm0.6$          \\
NGC5474  & S250 & $ 1.5\pm 1.2$         & $ 1.7\pm 0.8$         & $ 1.5\pm 1.0$         & $ 1.8\pm 0.9$         & $ 6.9\pm 2.1$         & 0.552 & $0.466\pm0.043$       \\
" & M160 & $1.81\pm0.29$         & $2.10\pm0.05$         & $0.40\pm0.10$         & $1.49\pm0.27$         & $ 7.7\pm 1.2$         & 0.489 & $0.442\pm0.020$       \\
NGC5713  & S250 & $ 28.\pm  6.$         & $2.81\pm0.41$         & $ 6.8\pm 2.0$         & $ 7.6\pm 2.1$         & $17.5\pm4.0$          & 0.639 & $34.7\pm3.8$          \\
" & M160 & $28.1\pm2.4$          & $2.74\pm0.27$         & $ 7.3\pm 1.9$         & $ 7.2\pm 1.1$         & $18.3\pm3.3$          & 0.610 & $33.3\pm1.9$          \\
NGC5866  & S250 & $ 9.4\pm 0.8$         & $ 1.7\pm 0.9$         & $3.36\pm0.31$         & $2.92\pm0.33$         & $ 1.3\pm 1.0$         & 0.600 & $4.48\pm0.21$         \\
" & M160 & $ 6.0\pm 0.5$         & $ 1.6\pm 0.8$         & $4.63\pm0.41$         & $4.54\pm0.34$         & $0.32\pm0.61$         & 0.692 & $4.46\pm0.08$         \\
NGC6946  & S250 & $ 57.\pm  6.$         & $ 3.5\pm 0.8$         & $ 3.4\pm 0.6$         & $ 3.8\pm 0.6$         & $13.5\pm2.9$          & 0.658 & $35.7\pm3.2$          \\
" & M160 & $47.1\pm4.1$          & $ 3.5\pm 0.7$         & $ 3.8\pm 0.6$         & $ 4.4\pm 0.6$         & $14.3\pm2.9$          & 0.688 & $34.1\pm2.0$          \\
NGC7331  & S250 & $116.\pm  32.$        & $ 4.2\pm 0.9$         & $ 2.3\pm 1.0$         & $ 2.4\pm 1.0$         & $ 7.6\pm 4.1$         & 0.619 & $45.7\pm4.0$          \\
" & M160 & $ 97.\pm  9.$         & $3.95\pm0.36$         & $2.79\pm0.44$         & $2.73\pm0.42$         & $ 7.1\pm 1.8$         & 0.641 & $43.7\pm2.9$          \\
NGC7793  & S250 & $ 5.5\pm 1.3$         & $ 3.0\pm 0.9$         & $ 1.9\pm 0.8$         & $ 1.9\pm 0.7$         & $ 8.1\pm 0.9$         & 0.585 & $1.76\pm0.13$         \\
" & M160 & $ 6.0\pm 0.8$         & $ 3.2\pm 0.6$         & $ 2.0\pm 0.9$         & $ 1.8\pm 0.5$         & $ 7.9\pm 2.5$         & 0.545 & $1.78\pm0.08$         \\
\hline
EIC3583  & S250 & $1.75\pm0.08$         & $2.12\pm0.13$         & $1.53\pm0.46$         & $1.91\pm0.32$         & $ 8.3\pm 1.7$         & 0.569 & $0.548\pm0.041$       \\
" & M160 & $1.53\pm0.23$         & $ 2.3\pm 0.6$         & $1.82\pm0.42$         & $2.05\pm0.38$         & $ 9.3\pm 0.9$         & 0.523 & $0.514\pm0.040$       \\
ENGC0586 & S250 & $ 4.0\pm 1.0$         & $3.873\pm0.039$       & $0.89\pm0.62$         & $ 1.0\pm 0.7$         & $10.9\pm2.2$          & 0.549 & $0.67\pm0.10$         \\
" & M160 & $2.31\pm0.23$         & $3.78\pm0.12$         & $1.50\pm0.31$         & $1.66\pm0.18$         & $ 8.6\pm 4.2$         & 0.639 & $0.633\pm0.022$       \\
ENGC1317 & S250 & $11.3\pm1.5$          & $ 3.0\pm 0.7$         & $ 4.1\pm 1.5$         & $ 3.7\pm 0.9$         & $ 4.2\pm 0.8$         & 0.596 & $ 6.9\pm 0.6$         \\
" & M160 & $ 7.6\pm 0.7$         & $3.13\pm0.28$         & $ 5.0\pm 1.0$         & $ 5.1\pm 0.9$         & $ 3.9\pm 2.2$         & 0.687 & $6.34\pm0.38$         \\
ENGC1481 & S250 & $0.94\pm0.15$         & $ 1.5\pm 1.4$         & $ 5.9\pm 3.7$         & $ 6.0\pm 2.1$         & $16.1\pm2.8$          & 0.567 & $0.93\pm0.11$         \\
" & M160 & $0.95\pm0.24$         & $ 2.3\pm 1.7$         & $ 3.6\pm 3.2$         & $ 5.9\pm 2.4$         & $18.6\pm4.0$          & 0.578 & $0.93\pm0.10$         \\
ENGC1510 & S250 & $0.244\pm0.048$       & $0.60\pm0.35$         & $ 4.0\pm 2.9$         & $ 8.5\pm 3.4$         & $35.1\pm3.8$          & 0.569 & $0.339\pm0.011$       \\
" & M160 & $1.16\pm0.20$         & $0.70\pm0.26$         & $0.47\pm0.12$         & $1.94\pm0.37$         & $34.9\pm1.6$          & 0.465 & $0.367\pm0.014$       \\
ENGC3187 & S250 & $ 7.5\pm 1.8$         & $1.80\pm0.48$         & $ 1.2\pm 1.0$         & $ 1.4\pm 1.0$         & $11.6\pm1.8$          & 0.541 & $1.75\pm0.21$         \\
" & M160 & $ 9.1\pm 0.9$         & $2.92\pm0.30$         & $0.56\pm0.55$         & $1.12\pm0.26$         & $11.4\pm3.5$          & 0.476 & $1.67\pm0.13$         \\
ENGC4533 & S250 & $1.49\pm0.27$         & $ 2.4\pm 0.5$         & $ 1.5\pm 0.8$         & $ 1.6\pm 0.7$         & $ 6.4\pm 2.5$         & 0.538 & $0.393\pm0.048$       \\
" & M160 & $1.183\pm0.043$       & $ 2.3\pm 0.5$         & $ 1.5\pm 0.5$         & $1.83\pm0.16$         & $ 8.2\pm 1.5$         & 0.534 & $0.355\pm0.021$       \\
ENGC7335 & S250 & $ 54.\pm  7.$         & $ 1.6\pm 0.5$         & $1.35\pm0.42$         & $1.49\pm0.36$         & $ 5.6\pm 2.1$         & 0.573 & $13.1\pm1.6$          \\
" & M160 & $38.2\pm2.9$          & $ 2.1\pm 0.5$         & $ 1.9\pm 0.5$         & $ 2.0\pm 0.5$         & $ 6.7\pm 3.6$         & 0.608 & $12.3\pm1.3$          \\
ENGC7337 & S250 & $ 66.\pm 30.$         & $3.51\pm0.13$         & $0.59\pm0.27$         & $0.68\pm0.23$         & $ 8.6\pm 1.5$         & 0.490 & $ 7.4\pm 0.7$         \\
" & M160 & $ 30.\pm  6.$         & $ 3.1\pm 1.2$         & $1.25\pm0.31$         & $1.45\pm0.40$         & $ 7.3\pm 5.4$         & 0.589 & $ 7.0\pm 0.9$         \\

      \hline
      \multicolumn{9}{l}{$a$ Renormalized as described in \S\ref{sec:renormalization}.}
    \end{tabular}
  \end{center}
\btdnote{uses tab\_dust\_guts2.tex created by tablemaker\_v9}
\end{table*}

\subsection{Gold-Standard DL07 fit results for KINGFISH galaxies}
\label{sec:goldstandard}

\begin{figure*}  
\centering
\includegraphics[width=6.0cm,clip=true,trim=0.5cm 5.0cm 0.5cm 2.5cm]
                {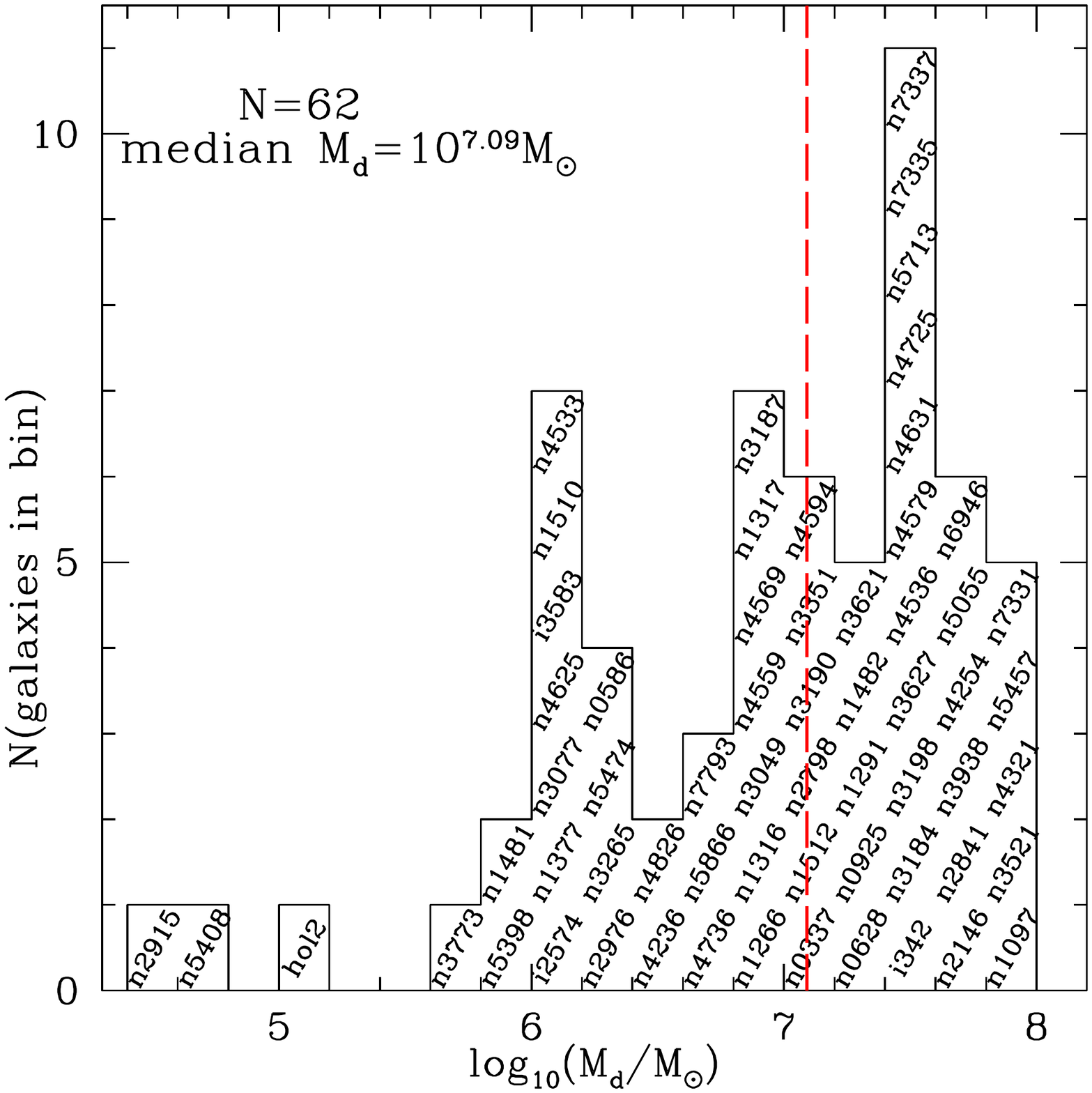} 
\includegraphics[width=6.0cm,clip=true,trim=0.5cm 5.0cm 0.5cm 2.5cm]
                {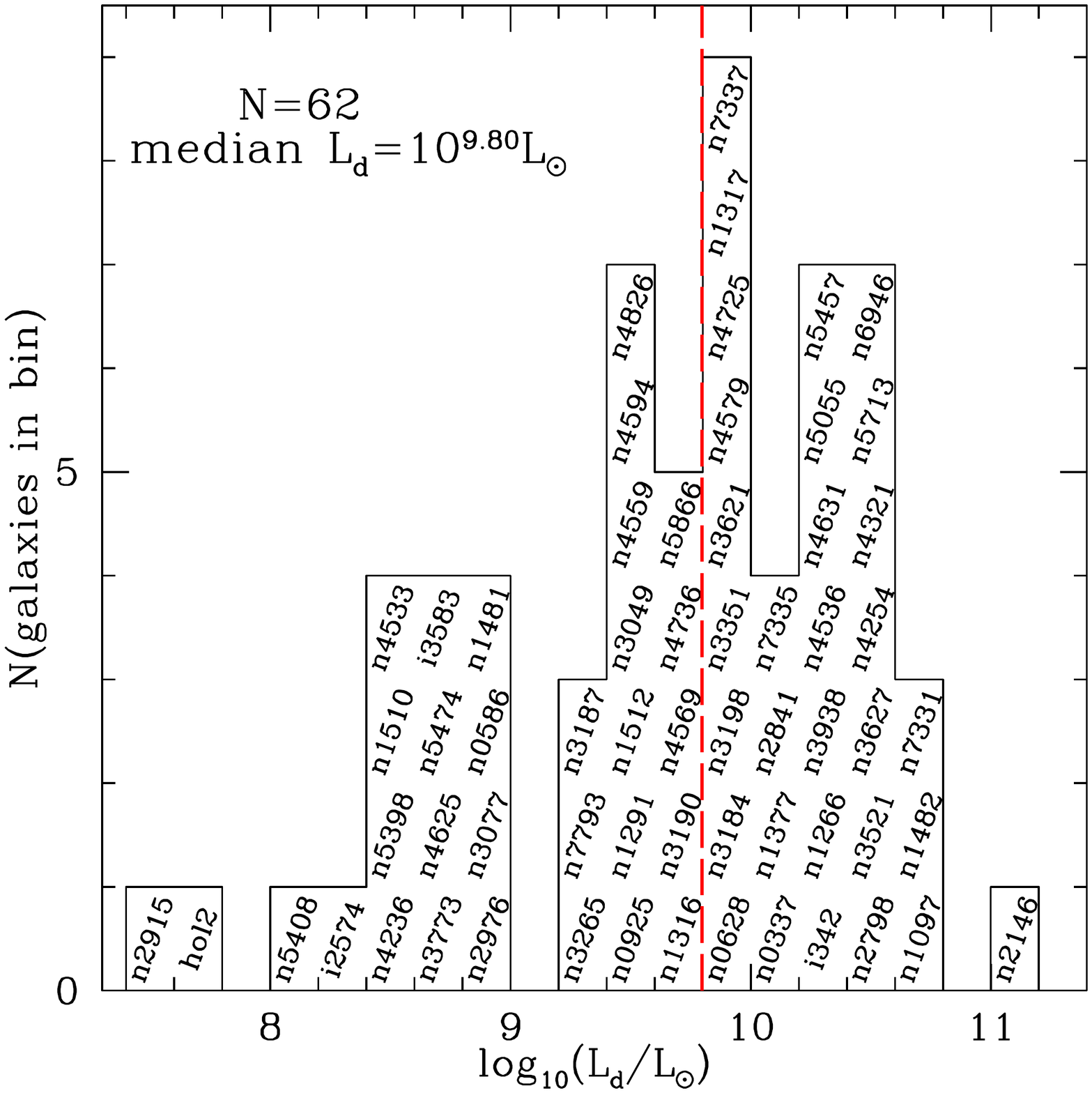} 
\\
\includegraphics[width=6.0cm,clip=true,trim=0.5cm 5.0cm 0.5cm 2.5cm]
                {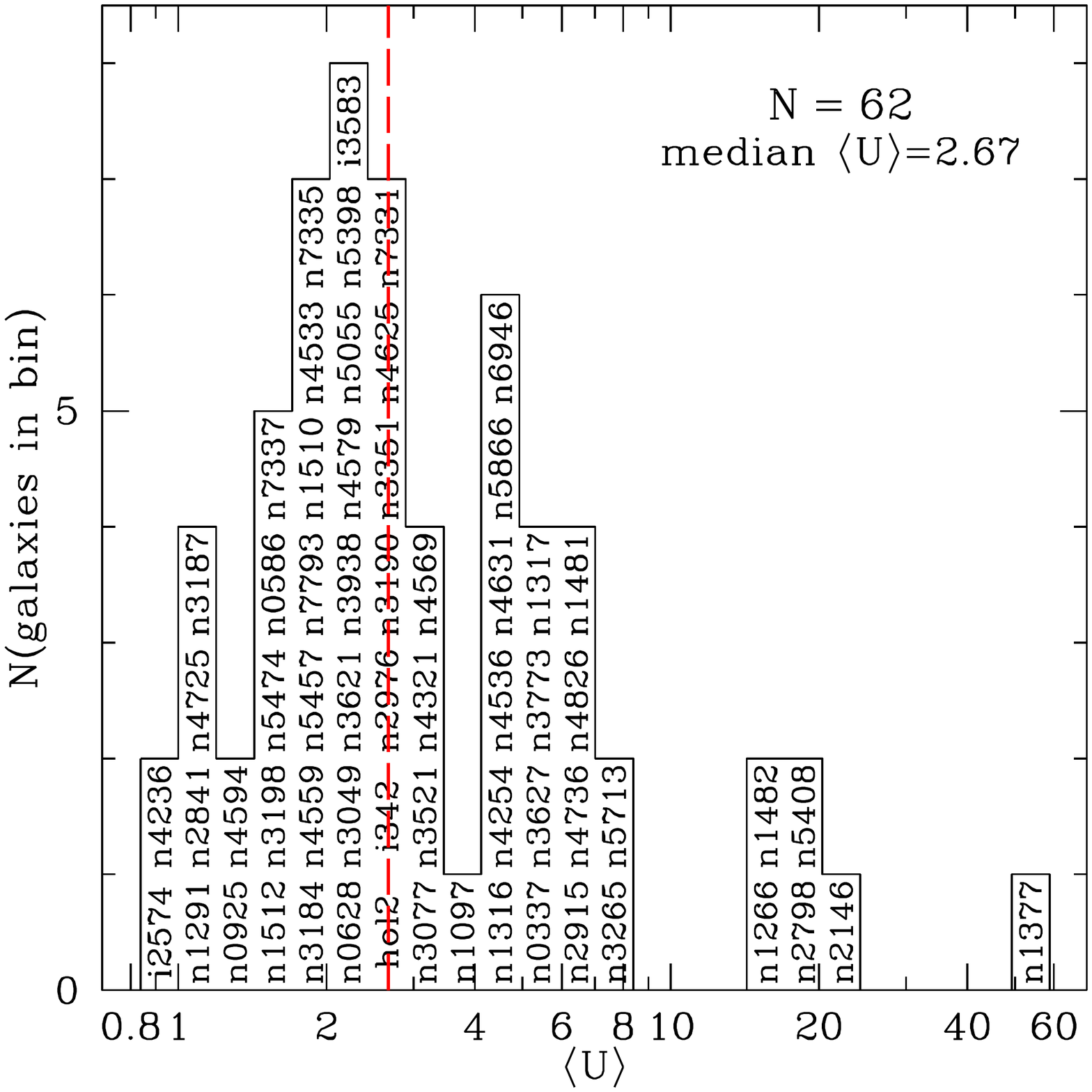} 
\includegraphics[width=6.0cm,clip=true,trim=0.5cm 5.0cm 0.5cm 2.5cm]
                {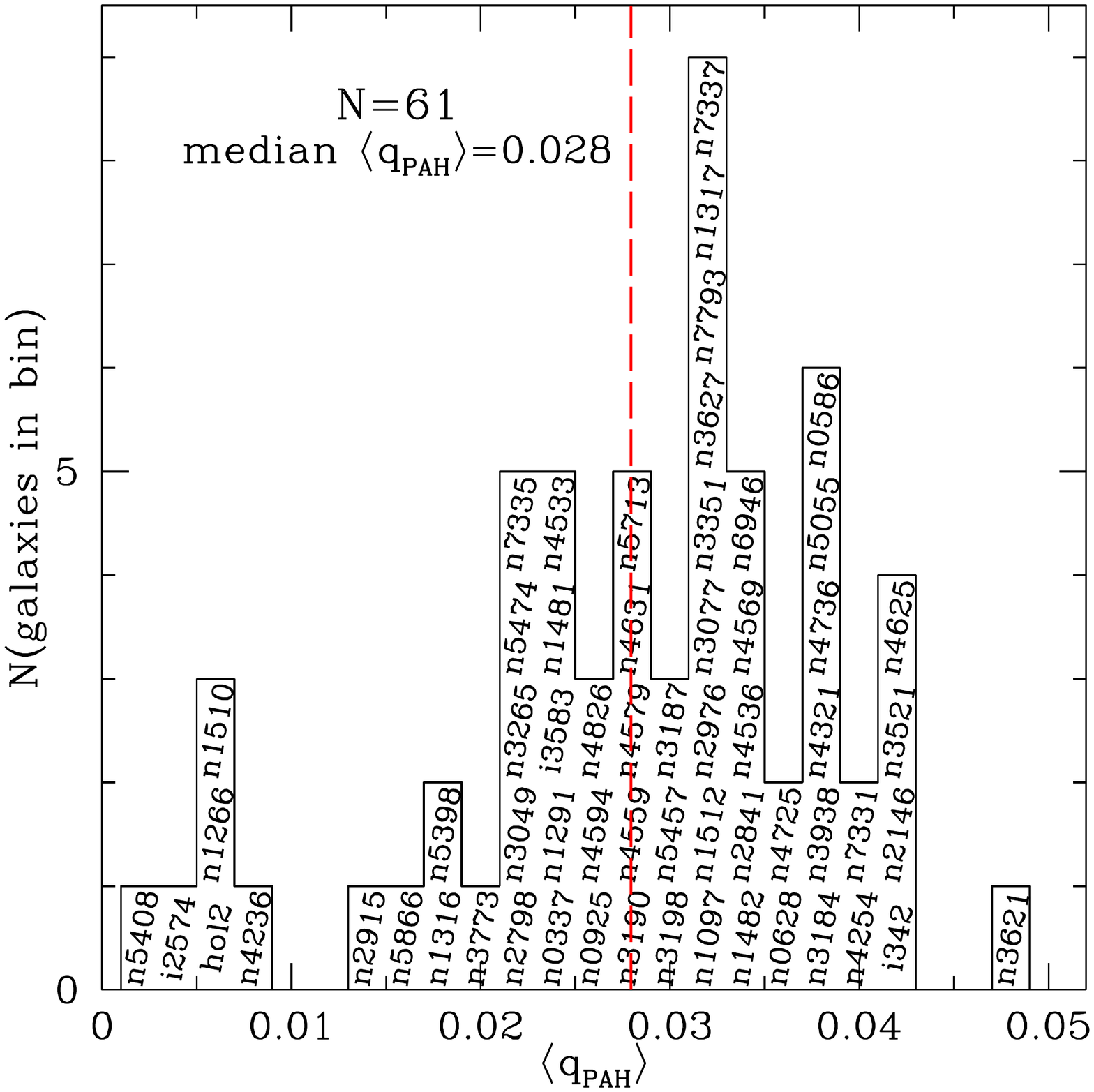} 
\\
\includegraphics[width=6.0cm,clip=true,trim=0.5cm 5.0cm 0.5cm 2.5cm]
                {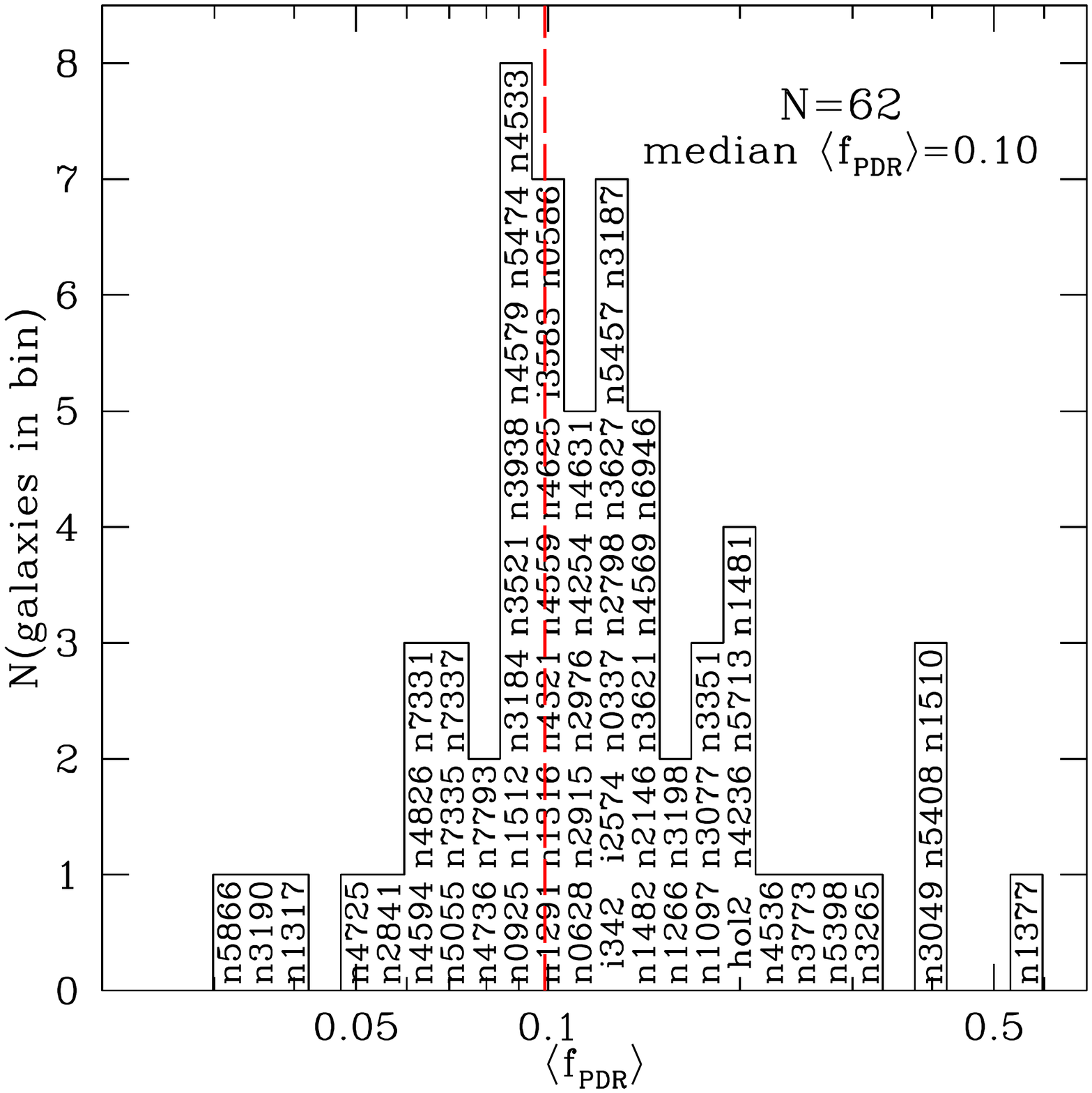} 
\includegraphics[width=6.0cm,clip=true,trim=0.5cm 5.0cm 0.5cm 2.5cm]
                {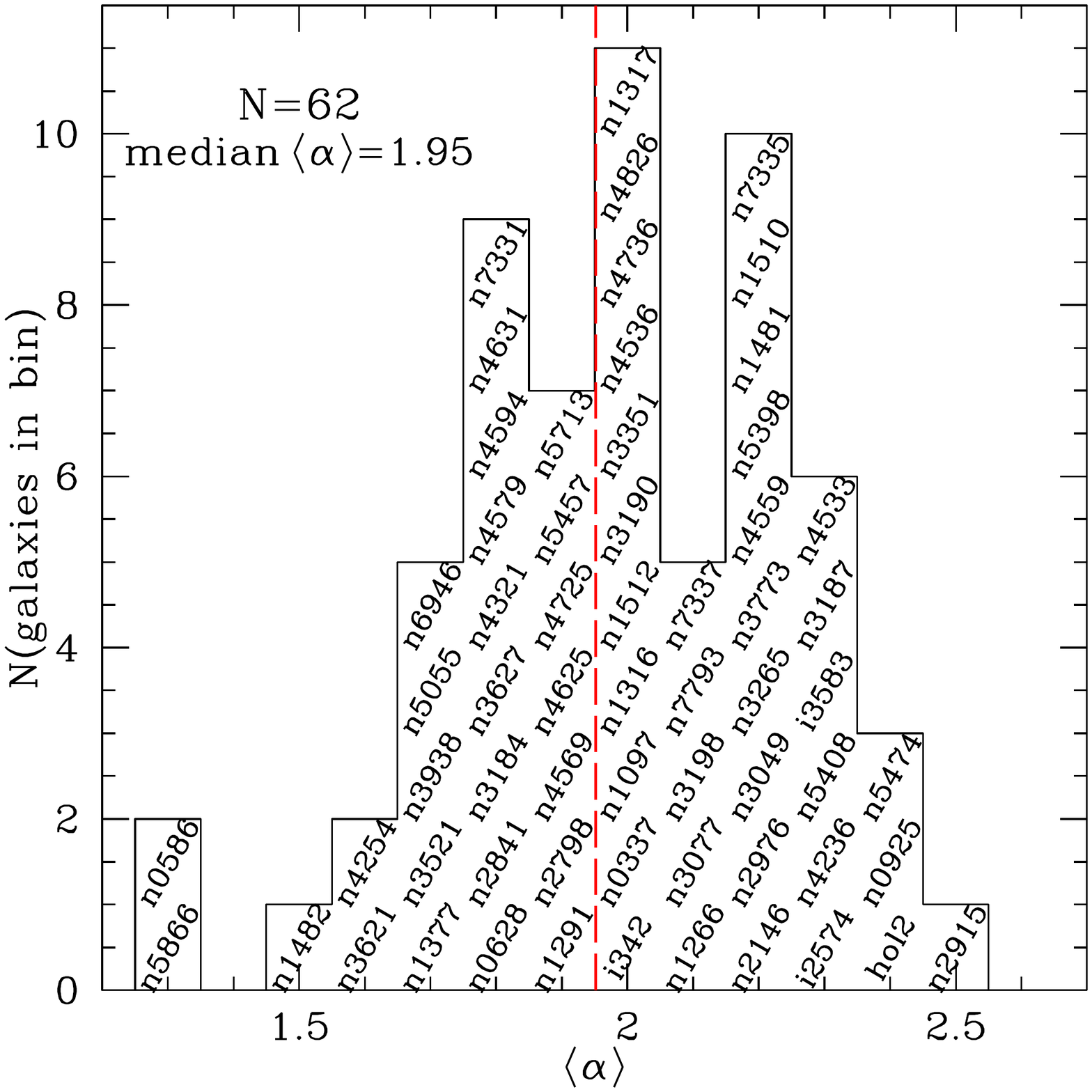} 
\\
\caption{\label{fig:dustparams}
         Distributions of dust and starlight parameters for
         the KF62 sample.
         NGC\,1377 is omitted from
         the $q_{\rm PAH}$ histogram (see text).
         $M_\dust$ and $\langle U\rangle$ are renormalized values.
         \btdnote{fhist\_Md.pdf, fhist\_Ld.pdf, fhist\_Ubar.pdf,
         fhist\_qpah.pdf, fhist\_fpdf.pdf, fhist\_alpha.pdf}
}
\end{figure*}

Figure \ref{fig:dustparams} shows the dust parameter distributions for the 61
KINGFISH galaxies plus 9 ``extras''.
The dust parameters shown are the result of
the ``gold standard'' modeling -- multipixel modeling for
each galaxy using the M160 PSF and data from all cameras.  

The first row
shows the distributions of $\Mdust$ (left column) and \lumdust\ 
(right column) for the KF62 galaxies.
The second
row shows the distributions of
$\langle U\rangle$ (left column) and $\qPAH$ right column),
and the bottom row shows the distributions of $\langle\fPDR\rangle$ 
(left column) and $\langle\alpha\rangle$ (right
column).
In these histograms, the dust masses $M_\dust$ and $\langle U\rangle$ are
renormalized,
as discussed in Section \ref{sec:renormalization}.

Figure \ref{fig:dustparams} illustrates the large
region in the model parameter space spanned by the
KINGFISH 
sample, allowing us to probe the dust properties in a variety of ISM conditions.
The total dust mass 
and dust luminosity found in the galaxies spans almost 4
decades:
$10^{4.5}\leq \,M_{\rm d} / \Msol\, \leq 10^{8.0}$ and
$10^{7.5}\leq \,L_{\rm d} / \Lsol\, \leq 10^{11.1}$, 
from the blue dwarf
NGC\,2915 ($L_\dust=3.3\times10^7\Lsol$)
to the luminous starburst galaxy NGC2146 ($L_\dust=1.4\times10^{11}\Lsol$).

\begin{table*}
  \scriptsize
  \begin{center}
     \caption{\label{tab:dust-to-gas}
              Gas Masses for M160 Resolution Galaxy Mask and Dust/Gas Ratio}
     \begin{tabular}{| c | c c c c |}
        \hline
        Galaxy& $M({\rm H\,I})$$^a$& $M({\rm H}_2)$$^a$& $M_{\rm dust}$$^b$& $M_{\rm dust}/M_{\rm H}$$^a$\\
& ($10^9\Msol$) & ($10^9\Msol$) & ($10^6\Msol$) &\\ \hline
DDO053   & $0.0382\pm0.0038$     & \nodata               & $<  0.21$             & $<0.0060$             \\
DDO154   & $0.128\pm0.013$       & $<0.0010$             & $<  0.61$             & $<0.0053$             \\
DDO165   & $0.069\pm0.007$       & \nodata               & $<  0.52$             & $<0.0085$             \\
Hol1     & $0.0366\pm0.0037$     & $<0.00077$            & $<  0.67$             & $< 0.020$             \\
Hol2     & $0.232\pm0.023$       & $<0.0016$             & $0.112\pm0.006$       & $<0.00057$            \\
IC342    & $14.3\pm4.3$          & \nodata               & $41.28\pm0.21$        & $<0.0041$             \\
IC2574   & $0.87\pm0.09$         & $0.0084\pm0.0014$     & $0.782\pm0.019$       & $0.00089\pm0.00012$   \\
M81dwB   & $0.0079\pm0.0008$     & $<0.00057$            & $< 0.081$             & $< 0.012$             \\
NGC0337  & $4.13\pm0.41$         & $0.389\pm0.042$       & $15.67\pm0.32$        & $0.0035\pm0.0005$     \\
NGC0584  & $0.157\pm0.047$       & \nodata               & $<  1.59$             & $< 0.014$             \\
NGC0628  & $2.30\pm0.23$         & $1.38\pm0.14$         & $21.44\pm0.08$        & $0.0058\pm0.0012$     \\
NGC0855  & $0.115\pm0.034$       & \nodata               & $<  1.02$             & $< 0.013$             \\
NGC0925  & $4.45\pm0.45$         & $0.256\pm0.027$       & $13.19\pm0.17$        & $0.0028\pm0.0004$     \\
NGC1097  & $ 4.4\pm 1.3$         & \nodata               & $73.3\pm0.7$          & $< 0.024$             \\
NGC1266  & $>0.0095$             & \nodata               & $12.14\pm0.29$        & $<  1.31$             \\
NGC1291  & $1.32\pm0.39$         & \nodata               & $13.63\pm0.23$        & $< 0.015$             \\
NGC1482  & $0.67\pm0.20$         & \nodata               & $22.5\pm0.6$          & $< 0.050$             \\
NGC1512  & $ 2.9\pm 0.9$         & \nodata               & $12.10\pm0.13$        & $<0.0060$             \\
NGC2146  & $ 3.3\pm 1.0$         & $ 8.5\pm 0.9$         & $55.9\pm0.6$          & $0.0047\pm0.0014$     \\
NGC2798  & $0.96\pm0.10$         & $2.52\pm0.25$         & $15.95\pm0.44$        & $0.0046\pm0.0014$     \\
NGC2841  & $ 6.6\pm 0.7$         & $0.89\pm0.09$         & $48.42\pm0.30$        & $0.0064\pm0.0009$     \\
NGC2915  & $0.35\pm0.11$         & \nodata               & $0.039\pm0.006$       & $<0.00018$            \\
NGC2976  & $0.142\pm0.014$       & $0.073\pm0.007$       & $1.948\pm0.031$       & $0.0090\pm0.0019$     \\
NGC3049  & $1.01\pm0.10$         & $0.144\pm0.018$       & $6.84\pm0.21$         & $0.0059\pm0.0010$     \\
NGC3077  & $0.429\pm0.043$       & $0.0161\pm0.0017$     & $1.425\pm0.018$       & $0.0032\pm0.0004$     \\
NGC3184  & $3.68\pm0.37$         & $1.98\pm0.20$         & $31.91\pm0.27$        & $0.0056\pm0.0011$     \\
NGC3190  & $0.46\pm0.14$         & $0.058\pm0.013$       & $18.2\pm0.5$          & $0.035\pm0.012$       \\
NGC3198  & $ 8.1\pm 0.8$         & $0.62\pm0.06$         & $26.71\pm0.30$        & $0.0031\pm0.0004$     \\
NGC3265  & $0.17\pm0.05$         & \nodata               & $2.00\pm0.24$         & $< 0.018$             \\
NGC3351  & $1.08\pm0.11$         & $1.02\pm0.10$         & $16.34\pm0.15$        & $0.0078\pm0.0018$     \\
NGC3521  & $10.4\pm1.0$          & $4.18\pm0.42$         & $82.6\pm0.9$          & $0.0057\pm0.0011$     \\
NGC3621  & $ 5.2\pm 0.5$         & \nodata               & $20.75\pm0.27$        & $<0.0045$             \\
NGC3627  & $1.12\pm0.11$         & $3.02\pm0.30$         & $42.22\pm0.20$        & $0.0102\pm0.0029$     \\
NGC3773  & $0.109\pm0.033$       & \nodata               & $0.689\pm0.024$       & $<0.0093$             \\
NGC3938  & $ 5.1\pm 0.5$         & $2.51\pm0.25$         & $48.42\pm0.32$        & $0.0063\pm0.0012$     \\
NGC4236  & $1.91\pm0.19$         & $0.0028\pm0.0010$     & $1.77\pm0.06$         & $0.00092\pm0.00012$   \\
NGC4254  & $4.93\pm0.49$         & $ 7.2\pm 0.7$         & $69.99\pm0.42$        & $0.0058\pm0.0015$     \\
NGC4321  & $3.38\pm0.34$         & $ 6.9\pm 0.7$         & $81.16\pm0.39$        & $0.0079\pm0.0021$     \\
NGC4536  & $4.62\pm0.46$         & $1.86\pm0.19$         & $29.5\pm0.6$          & $0.0045\pm0.0009$     \\
NGC4559  & $3.16\pm0.32$         & $0.051\pm0.006$       & $7.90\pm0.14$         & $0.0025\pm0.0003$     \\
NGC4569  & $0.248\pm0.025$       & $1.30\pm0.13$         & $11.88\pm0.16$        & $0.0077\pm0.0024$     \\
NGC4579  & $0.73\pm0.07$         & $2.42\pm0.24$         & $32.95\pm0.11$        & $0.0105\pm0.0030$     \\
NGC4594  & $0.20\pm0.06$         & $0.043\pm0.007$       & $12.57\pm0.11$        & $0.023\pm0.008$      $^c$ \\
NGC4625  & $0.241\pm0.024$       & $0.0285\pm0.0039$     & $1.365\pm0.046$       & $0.0051\pm0.0008$     \\
NGC4631  & $ 7.7\pm 0.8$         & $1.62\pm0.16$         & $36.84\pm0.30$        & $0.0040\pm0.0006$     \\
NGC4725  & $3.51\pm0.35$         & $0.67\pm0.07$         & $34.40\pm0.20$        & $0.0082\pm0.0013$     \\
NGC4736  & $0.50\pm0.05$         & $0.59\pm0.06$         & $6.676\pm0.034$       & $0.0061\pm0.0015$     \\
NGC4826  & $0.112\pm0.011$       & $0.66\pm0.07$         & $4.66\pm0.06$         & $0.0060\pm0.0019$     \\
NGC5055  & $3.70\pm0.37$         & $3.35\pm0.33$         & $57.54\pm0.27$        & $0.0082\pm0.0018$     \\
NGC5398  & $0.26\pm0.08$         & \nodata               & $0.593\pm0.038$       & $<0.0035$             \\
NGC5408  & $0.24\pm0.07$         & \nodata               & $0.0603\pm0.0049$     & $<0.00039$            \\
NGC5457  & $11.5\pm1.1$          & $2.58\pm0.26$         & $69.15\pm0.17$        & $0.0049\pm0.0008$     \\
NGC5474  & $0.62\pm0.06$         & $<0.0049$             & $1.429\pm0.035$       & $<0.0026$             \\
NGC5713  & $3.24\pm0.32$         & $4.00\pm0.40$         & $34.1\pm0.6$          & $0.0047\pm0.0012$     \\
NGC6946  & $3.52\pm0.35$         & $ 6.6\pm 0.7$         & $63.46\pm0.30$        & $0.0063\pm0.0017$     \\
NGC7331  & $10.7\pm1.1$          & $ 5.2\pm 0.5$         & $116.\pm   1.$        & $0.0073\pm0.0014$     \\
NGC7793  & $0.99\pm0.10$         & \nodata               & $5.75\pm0.05$         & $<0.0065$             \\

        \hline
     \multicolumn{5}{l}{$^a$ He is not included.}\\
     \multicolumn{5}{l}{$^b$ Renormalized as described in \S\ref{sec:renormalization}.}\\
     \multicolumn{5}{l}{$^c$ $M_\Ha$ includes $M=2.9\times10^8\Msol$ of hot gas \citep{Li+Jones+Forman+etal_2011}}\\
     \end{tabular}
   \end{center}
\btdnote{1. uses tab\_mdmh\_guts.tex created by tablemaker\_v9.
         2. Why do NGC3190 and  NGC4594 have such high 
            $\Mdust/M_\Ha$??}
\end{table*}

The mean value of the starlight heating parameter $\Umin$
also presents wide variations across the galaxy sample.
$\langle\Umin\rangle$ spans the range 
$0.1\leq\,\langle\Umin\rangle\, \leq \, 8.5$
(these values of $\Umin$ are for the DL07 model without renormalization). 
The PAH mass fraction $\qpah$
also shows wide variation, from 0.005 to 0.045, with median
$\qpah=0.027$.
The mean fraction of the dust luminosity coming from dust
heated by high-intensity radiation fields, $\langle\fPDR\rangle$,
typically ranges from 0.05 to 0.20.  

There are 4 KINGFISH galaxies 
where the fitted DL07 dust models have very
high values of $\langle\fPDR\rangle\,> 0.30$:
NGC\,1316 = Fornax A (SAB0),
NGC\,3049 (SBab),
NGC\,3265 (E),
NGC\,5408 (IBm),
and the ``extra'' galaxy
NGC\,1510 (SA0).  
NGC\,1316 = Fornax A has a central AGN/LINER spectrum, 
and NGC\,3265 has an emission-line
nuclear region \citep{Dellenbusch+Gallagher+Knezek+Noble_2008}, 
NGC\,1510 hosts a strong central starburst, and NGC\,3049 and NGC\,5408
are often classified as starburst galaxies.  
Thus the high $\langle\fpdr\rangle$ values for these
galaxies may be indicative of concentrated star formation or nuclear activity.

Finally, the mean
power-law exponent $\langle\alpha\rangle$ spans $1.5 \leq\,
\langle\alpha\rangle\,\leq 2.5$.
Allowing $\alpha$ to vary does improve the quality of the fit to the
observed SED, but in most cases the fit quality does not suffer
greatly if $\alpha$ is held fixed at $\alpha=2$, reducing the number
of free parameters.
Recall that $\alpha=2$ corresponds to equal amounts
of dust power per logarithmic interval in starlight intensity $U$.

Compared with the ``gold standard'',
modeling using PSFs smaller than M160, and hence having fewer
cameras available,
can affect the derived dust and starlight parameters.
As the PSF shrinks, data are provided by fewer cameras, the
wavelength coverage shrinks as
the PSF is reduced below S500, 
and the photometry becomes noisier because it is being
smoothed over smaller PSFs.  Above we have compared two cases:
modeling with the M160 PSF, versus modeling with the S250 PSF, but additional
comparisons are made
in Appendix \ref{app:dependence on psf}.
Here we simply note some trends.
In general
$M_\dust$ is
fairly robust: the S250 modeling typically overestimates $M_\dust$
by $\sim$25\%, but agrees with the ``gold standard'' to within a factor $1.5$
for over 75\% of the galaxies (see Fig.\ \ref{fig:multipix s250 vs m160}).
$\qpah$ estimates are also robust, with typical changes of less than
15\%.
Longer wavelength coverage (SPIRE350 and  SPIRE500) gives more
reliable dust estimates.
Even comparing resolved and global modeling of dust properties
can give different results;
although most parameters are consistent to within a few percent,
global modeling can
underestimate $M_{\rm d}$ by
as much as 35\%, as for NGC~1481 and NGC~3077
(see Figure \ref{fig:multipix vs singlepix}).
This is because the resolved models can have ``cold'' 
regions with low $\Umin$ values
that contribute to dust mass estimates but do not emerge in the global results
\citep[e.g.][]{Galliano+Hony+Bernard+etal_2011,
               Galametz+Kennicutt+Albrecht+etal_2012}.

\subsection{Dependence of global PAH fraction on metallicity}
\label{sec:pahglobaloh}

\begin{figure}
\begin{center}
\includegraphics[angle=0,width=\linewidth,
                 clip=true,trim=0.0cm 5.0cm 0.0cm 2.5cm]
                {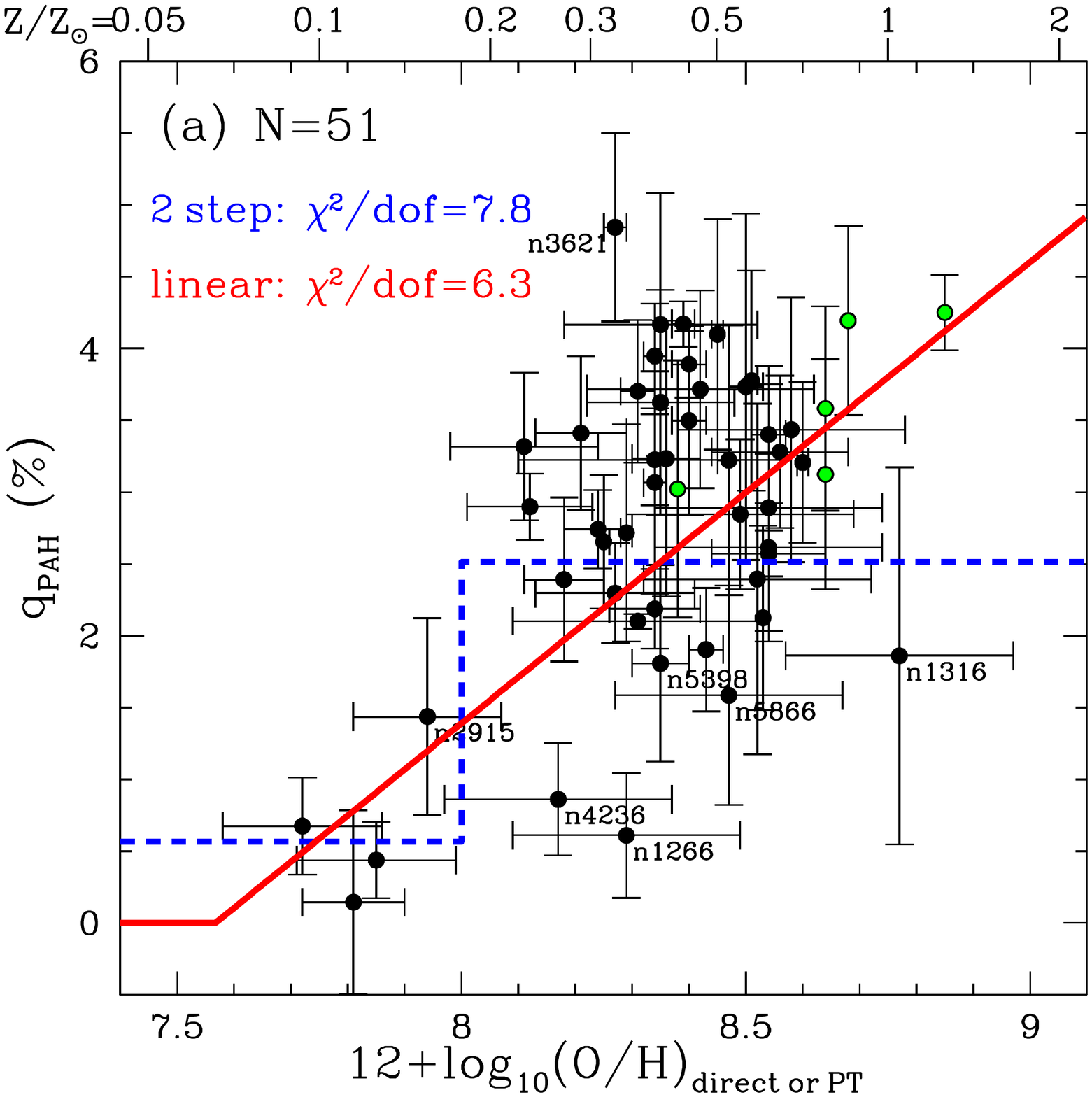} 
\includegraphics[angle=0,width=\linewidth,
                 clip=true,trim=0.0cm 5.0cm 0.0cm 2.5cm]
                {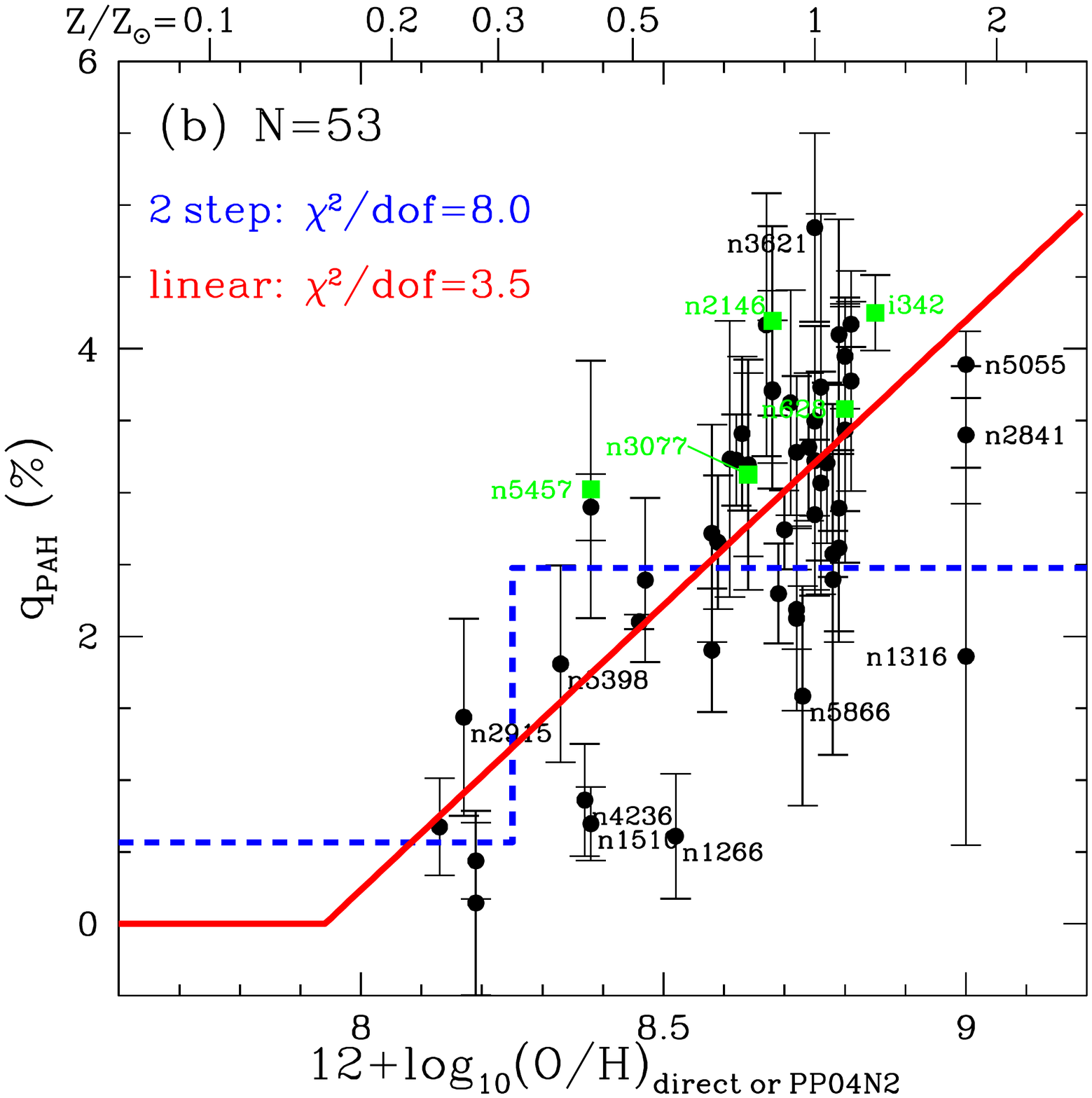} 
\vspace*{-0.3cm}
\caption{ 
  (a) PAH abundance parameter $\qpah$ versus oxygen abundance
  (direct or PT) for 51 galaxies (see text).
Two $\qpah$ estimators are shown: one is a step function, and the other
is linear above a threshold value.
Selected galaxies have been labeled.
The step function and linear estimators have similar
$\chi^2/{\rm dof}=$7.8 and 6.3.
(b) Same but for PP04N2 oxygen abundance, now for 52
galaxies (a PP04N2 oxygen abundance is available for NGC\,1512).
The 5 galaxies where O/H has been determined by
``direct'' methods are shown in green.
The linear fit of $\qpah$ vs. \logoh\ (Eq.\ \ref{eq:pah_vs_oh})
gives an improved fit,
with $\chi^2/{\rm dof}=3.5$, and a threshold (O/H) $\approx$
0.15(O/H)$_\odot$.  
\btdnote{fqpah\_pt.pdf and fqpah\_pp04n2.pdf}
}
\label{fig:qpah}
\end{center}
\end{figure}

Figure \ref{fig:qpah}a shows $\qpah$ vs.\ 
$\log({\rm O/H})$
for 51 galaxies
using direct determinations of (O/H) where available (5 galaxies),
and PT estimates otherwise.
19 galaxies have been omitted: 8 dust nondetections
have been excluded, 
NGC1377
(a dense starburst with a core that is
optically thick at 8$\micron$ -- see Section \ref{sec:NGC1377}),
plus 10 galaxies for which we have no PT estimate for O/H.
The oxygen abundance in these galaxies ranges over more than a factor of 10, and
$\qpah$ shows a clear tendency to increase with increasing O/H,
although there is considerable scatter.
The observed behavior can be approximated by a step function, with an
abrupt increase in $\qpah$ when \logohpt\ rises above $\sim$8.0
Alternatively, $\qpah$ can be approximated by a linear dependence
on $\log({\rm O/H})$.  
Best-fit step function and linear function are
shown in Fig.\ \ref{fig:qpah}a, with $\chi^2$ per degree of freedom
of 8.0 and 6.6, respectively.

Figure \ref{fig:qpah}b shows $\qpah$ vs.\ the PP04N2 estimate for metallicity.
Again, we show both step functions and a linear dependence on
$\log_{10}({\rm O/H})$.
In this case, the function linear in  $\log({\rm O/H})_{\rm PP04N2}$
gives a much better fit to the data:
\beq \label{eq:pah_vs_oh}
\qpah \approx 0.0396 \left[(12+\log_{10}({\rm O/H})_{\rm PP04N2})-7.94\right]
\eeq
(for \logohpp$>$7.94).
This fit, with 53-2=51 degrees of freedom (dof), has $\chi^2$/dof$=3.5$:
the PP04N2 metallicity is evidently a {\it much} better predictor of $\qpah$
than is the PT metallicity.\footnote{%
For some galaxies we use ``direct'' metallicities
rather than the PT or PP04N2 weak-line estimates,
but $\chi^2$ is dominated by
    the 51 galaxies where we use PP04N2 instead of the PT
    metallicity estimate.
}
This strongly suggests
that the PP04N2 metallicities are more tightly related to the
properties of the ISM -- including metallicity --
that regulate the balance between PAH formation and
destruction.

The observed tendency for $\qpah$ to increase with increasing
metallicity is consistent with many previous studies.
The connection between PAH abundance and metallicity was first noted in
ground-based spectroscopy by \citet{Roche+Aitken+Smith+Ward_1991}, and
further investigated using ISO data
\citep{Boselli+Lequeux+Sauvage+etal_1998,
       Sturm+Lutz+Tran+etal_2000,
       Madden_2000}.
\citet{Hunt+Bianchi+Maiolino_2005,Hunt+Thuan+Izotov+Sauvage_2010}
found PAH emission to be
weak in low-metallicity blue compact dwarf galaxies.
\citet{Engelbracht+Gordon+Rieke+etal_2005} used IRAC and MIPS24
photometry to show that there was an abrupt drop in the 
8$\micron$/24$\micron$
flux ratio when the metallicity dropped below 8.2,
interpreting this as due to a sharp drop in the abundance of PAHs that
normally dominate the emission at 8$\micron$.
\citet{Draine+Dale+Bendo+etal_2007} estimated $\qpah$ for 61 SINGS galaxies,
using the DL07 model
with IRAC and MIPS
photometry, and found a similar result: a sharp increase
in $\qpah$ when \logohpt\ rises above $\sim$8.2.

Nevertheless, there are outliers in Fig.\ \ref{fig:qpah}b.
The SB0 galaxy NGC\,1266 has 
$\qpah=0.70\%$, unusually low for a galaxy with
\logohpp=8.51.
The {\it Spitzer} and {\it Herschel} 
photometry of NGC\,1266 (see Fig.\ 17.8) appears
to be reliable.  
Because the optical spectrum of NGC\,1266 is AGN-dominated, 
the metallicity is not based on emission lines, and is therefore
highly uncertain.  \citet{Moustakas+Kennicutt+Tremonti+etal_2010} estimated
the metallicity from an assumed luminosity-metallicity relation.
The resulting \logohpp=8.51 is consistent with the
stellar mass-metallicity relation \citep{Andrews+Martini_2013}.
Perhaps
the PAH abundance in this galaxy has been 
suppressed by phenomena associated with the active galactic nucleus (AGN)
that is driving a molecular outflow characterized by shocked gas
\citep{Alatalo+Blitz+Young+etal_2011,Pellegrini+Smith+Wolfire+etal_2013,
Alatalo+Lacy+Lanz+etal_2015}.

The SAB0 galaxy NGC\,1316 = Fornax A
is another outlier.  The dust emission is weak
relative to the starlight, making the $\qpah$ estimate uncertain.
In addition, the starlight heating the dust is likely from an old population,
similar to the bulge of M31, and our estimate of $\qpah$ (based on single-photon heating by starlight assumed to have the solar neighborhood spectrum)
would then be biased low.
The estimate for $\qpah$ in the center of M31 increases by almost a factor of
two when calculated using the correct starlight spectrum
\citep{Draine+Aniano+Krause+etal_2014}, and a similar correction might bring
$\qpah$ for NGC\,1316 closer to the general trend in Figure \ref{fig:qpah}b.
In addition, the high metallicity estimated for NGC\,1316 may
be influenced by the AGN contribution to the emission line spectrum.

In Figure \ref{fig:qpah} it is striking that the bulk of the galaxies with
\logohpp$>8.3$ have $\qpah$ in the 1.5--5\% range.
Evidently the physical processes responsible for formation and destruction
of PAHs in normal star-forming 
galaxies tend to maintain PAH abundances near $3\%$ provided
that the metallicity $Z/Z_\odot\gtsim 0.3$.
From Eq.\ (\ref{eq:pah_vs_oh}) it appears that
there is a threshold metallicity for PAH formation:
$\qpah\approx0$ for $12+\log_{10}({\rm O/H})_{\rm PP04N2}\ltsim
7.94$, or $Z/Z_\odot\ltsim 0.15$.

\begin{figure}
\begin{center}
\includegraphics[angle=0,width=\linewidth,
                 clip=true,trim=0.0cm 5.0cm 0.0cm 2.5cm]
                {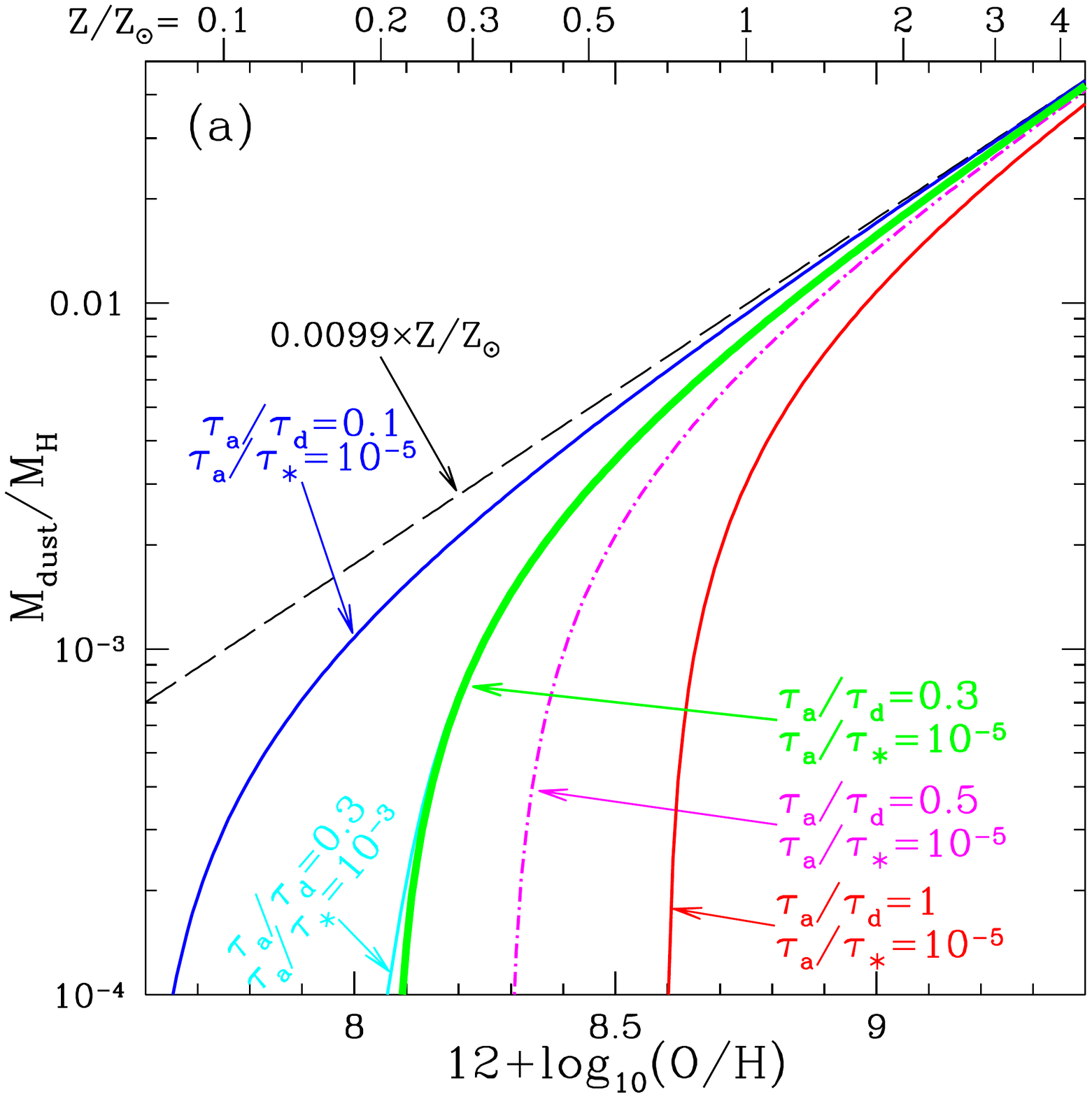} 
\includegraphics[angle=0,width=\linewidth,
                 clip=true,trim=0.0cm 5.0cm 0.0cm 2.5cm]
                {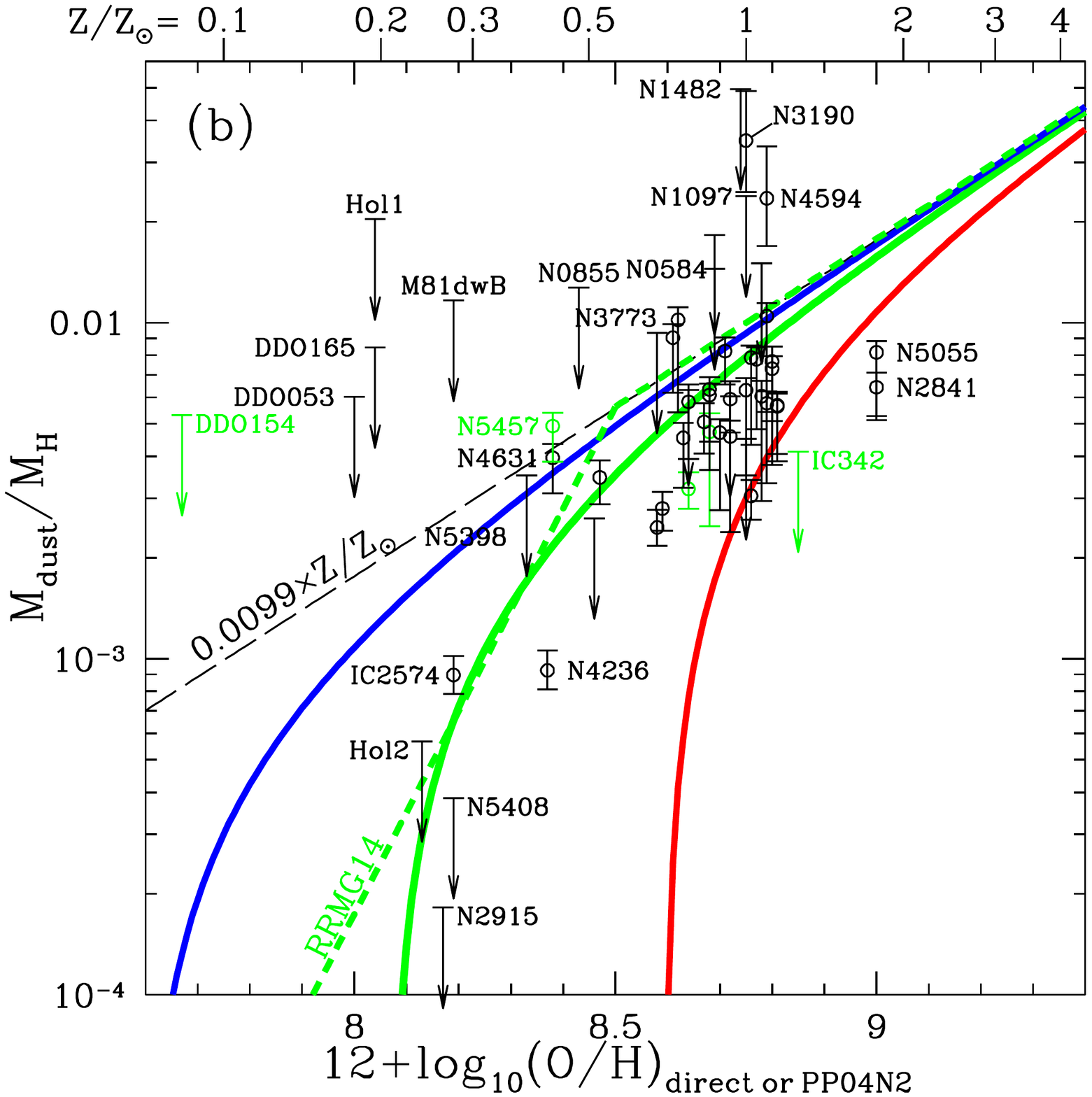} 
\vspace*{-0.3cm}
\caption{(a) Dust/H mass ratio vs. metallicity
         for the toy model (Eq.\ \ref{eq:toy model solution})
         for selected values of $\tau_a/\tau_d$ and $\tau_a/\tau_\star$
         (see text).
         (b) Measured $\Mdust/\MH$ (for the M160 resolution galaxy mask),
         versus \logoh\ (see text) for 
         57 KINGFISH galaxies.
         Selected galaxies have been labeled.
         Arrows indicate upper limits for the 16 galaxies lacking CO data.
         The green curve is Eq.\ (\ref{eq:toy model solution})
         for $\tau_a/\tau_d=0.3$ and $\tau_a/\tau_\star \ltsim 10^{-3}$;
         the blue and red curves are for $\tau_a/\tau_d=0.1$ and $1$.
         The green dashed line is the broken power-law fit from
         \citet{Remy-Ruyer+Madden+Galliano+etal_2014}.
         }
\label{fig:mdmh}
\end{center}
\end{figure}

\subsection{Dependence of global dust-to-gas ratio on metallicity}
\label{sec:globaloh}
\subsubsection{Theoretical Expectations}
\label{subsec:theory}

The abundance of dust in the ISM is the result of
competition between processes that form dust (dust formation in
stellar outflows, and dust growth in the ISM)
and processes  that return material to the gas phase
(e.g., sputtering in hot gas, and vaporization in high-speed
grain-grain collisions).
In the Milky Way and other star-forming galaxies
with near-solar metallicity, accretion of atoms onto grains is
rapid in the
cool, dense phases of the ISM, and the balance between grain growth
and grain destruction maintains a large fraction of the refractory
elements in grains.
Most of the dust in the Milky Way must have been grown in the ISM
-- there is simply no other way to understand the observed extreme 
depletions of
elements like Si, Al, Ca, Ti, and Fe in the diffuse ISM
\citep[see, e.g.,][]{Draine_1990a,Weingartner+Draine_1999,Draine_2009b}.

The black dashed line in Fig.\ \ref{fig:mdmh}a shows
the expected dependence of $\Mdust/\MH$ on O/H if all galaxies had
heavy element abundances proportional to solar abundances, and
the same depletion pattern as measured in the well-studied cloud toward the
nearby star $\zeta$Oph; in this cloud the refractory elements 
(e.g., Mg, Si, Fe...) 
are almost completely incorporated into grains, and
we infer a dust/H mass ratio 0.0099
\citep[see Table 23.1 of][]{Draine_2011a}.
For this scenario, we then expect
\beq \label{eq:mdmh vs o/h}
\left(\frac{\Mdust}{\MH}\right) = 0.0099 \left(\frac{Z}{Z_\odot}\right)
~~~,
\eeq
where we take $Z/Z_\odot=1$ for \logoh=8.72
\citep[][corrected for diffusion]{Asplund+Grevesse+Sauval+Scott_2009}.
However, in the overall ISM, $\Mdust/\MH$ will fall below this limiting value,
because of dust destruction processes.

A simple toy model can illustrate the competition between 
formation and destruction processes.
\citep[Similar models have been discussed by, e.g.,][]{%
   Edmunds_2001,
   Mattsson+Andersen+Munkhammar_2012,
   Asano+Takeuchi+Hirashita+Inoue_2013}.

Let $Z_m$ be the fraction of the ISM mass in ``refractory'' elements
($Z_m\approx0.007\times(Z/Z_\odot)=0.007\times10^{(12+\log_{10}({\rm O/H})-8.75)}$),
and $Z_d$ be the fraction of the ISM mass in dust grains made of these
refractory elements ($\Mdust/\MH=1.4Z_d$).
Clearly $Z_d<Z_m$, since some of the refractory elements are in the
gas phase.

Destruction and grain growth in the ISM both
contribute to the rate of change of $Z_d$.
We also include a term representing injection of solid grains into the ISM
from stellar sources (AGB stars, supernovae, etc.).
The rate of change of $Z_d$ is given by
\beq \label{eq:dust evolution}
\dot{Z}_d = 
- \frac{Z_d}{\tau_d} 
+ \frac{(Z_d/0.007)}{\tau_a}\left(Z_m-Z_d\right) 
+ \frac{Z_m}{\tau_\star}
~~~.
\eeq
The first term $-Z_d/\tau_d$ is the rate of dust destruction:
$\tau_d$ is the lifetime of solid material in the ISM against 
destructive processes that
return material to the gas phase.
The destruction rate $\tau_d^{-1}$ is a mass-weighted average
over the dust in the multiphase ISM.
Studies of the effects of supernova blastwaves in the local ISM
suggest timescales $\tau_d\approx 4\times10^8\yr$
\citep[see discussion in, e.g.,][]{Draine_2009b}.
Realistic estimation of $\tau_d$
requires a detailed dynamic multiphase model
of the ISM \citep[e.g.,][]{Zhukovska+Dobbs+Jenkins+Klessen_2016}.
The appropriate value of $\tau_d$ will obviously vary with
galactocentric radius within a galaxy, and from galaxy to galaxy.

The term in Eq.\ (\ref{eq:dust evolution}) representing grain
growth is proportional to $Z_d(Z_m-Z_d)$ because it depends on
grain surface area
($\propto Z_d$, for a fixed distribution of grain sizes)
and on the gas-phase abundance
of condensible elements ($\propto (Z_m-Z_d)$).
$(Z_d/0.007)\tau_a^{-1}$ is the probability per unit time that a refractory
atom in the gas phase will collide with and stick to a grain.

The last term, $Z_m/\tau_\star$, represents injection of
dust into the ISM from stellar sources, such
as cool AGB stars, planetary nebulae, and core-collapse supernovae.  
This term will obviously depend on the stellar populations.
Here, for illustration, 
we take the injection rate to be proportional to
the metallicity $Z_m$.
For galaxies of interest here, 
this injection term is small compared to the other terms
in Eq.\ (\ref{eq:dust evolution}), and the precise form adopted in
Eq.\ (\ref{eq:dust evolution}) is not critical.

If the shortest of the time scales 
$\{\tau_a,\tau_d\}$ is short compared to the $\sim$10$^9\yr$ timescale for
galactic chemical evolution, and
$\tau_\star \gg \{\tau_a,\tau_d\}$, we can
neglect time-dependence of the metallicity $Z_m$.
The toy model will approach
a quasi steady-state solution with $\dot{Z}_d\approx0$:
\beq \label{eq:toy model solution}
Z_d = \frac{1}{2}\left(Z_m - \frac{0.007}{\tau_d/\tau_a}\right)
+\frac{1}{2}\left[\left(Z_m - \frac{0.007}{\tau_d/\tau_a}\right)^2
+\frac{0.028}{\tau_\star/\tau_a}Z_m\right]^{1/2}
~~~.
\eeq
This solution for $Z_d$ depends only on $Z_m$ 
and on ratios of time scales, $\tau_a/\tau_d$ and $\tau_\star/\tau_d$.
Eq.\ (\ref{eq:toy model solution})
for $Z_d$ is plotted in Fig.\ \ref{fig:mdmh}a for
several choices of the ratios
$\tau_a/\tau_d$ and $\tau_a/\tau_\star$.
Note that for all of our examples we take 
$\tau_\star^{-1} \ll \tau_a^{-1}$: dust formation in stellar outflows is
secondary to dust growth in the ISM (i.e., only a small fraction of
interstellar dust is ``stardust'').
For large values of $Z_m$, all models approach 
the upper limit $Z_d/Z_m=1$
(long-dashed line in Fig.\ \ref{fig:mdmh}b).

Models of interest have $\tau_a < \tau_d$, so that for near-solar abundances,
accretion is faster than destruction, and a solar-metallicity 
ISM can maintain a large
fraction of the refractory elements in dust (i.e.,
$Z_d/Z_m\gtsim 0.5$).
However, for sufficiently low O/H, 
accretion rates become slow, resulting in low values of $Z_d/Z_m$.

\subsubsection{Observations}
\label{subsec:observations}

Using dust mass estimates based on modeling the infrared emission,
radial variations in dust-to-gas ratios (DGRs)
were found for galaxies in the SINGS sample
\citep{Munoz-Mateos+Gil_de_Paz+Boissier+etal_2009}
and for M101 \citep{Vilchez+Relano+Kennicutt+etal_2019}.
The dust-to-metals ratio was
approximately constant for KK04
metallicities \logoh$_{\rm KK}\geq9.0$, 
but for \logoh$_{\rm KK}\leq8.8$ the dust-to-metals
ratio appeared to decline with decreasing metallicity.
\citet{Chiang+Sandstrom+Chastenet+etal_2018} found variations in the
dust-to-metals ratio in M101, which they related to both
variations in metallicity and $\HH$ fraction.
\citet{DeCia+Ledoux+Mattsson+etal_2016} found similar behavior in
a sample that included 55 damped Lyman alpha systems (DLAs), where
dust abundances were inferred from depletions of Si, and metallicities
from [Zn/Fe].
It appears that 
as metallicity decreases below a certain threshold
(e.g., \logoh$_{\rm KK}<8.8$),
an increasing fraction of refractory
elements (Mg, Si, Fe, ...) remains in the gas phase.

Dust-to-gas ratios (DGRs) for the KF57 sample
(see Table \ref{tab:subsamples})
are plotted against O/H in Fig.\ \ref{fig:mdmh}b,
with dust masses estimated from our model, gas masses 
taken from Table \ref{tab:dust-to-gas}, and
the
PP04N2 estimate for O/H.
14 galaxies have detections of both dust and \ion{H}{1}, but
were either not observed or not detected in CO,
resulting in DGR upper limits.
An additional 7 galaxies
were detected in \ion{H}{1} but not in dust,
resulting in DGR upper limits.

Figure \ref{fig:mdmh}b shows a 
clear dependence of dust/gas ratio on metallicity.
With some exceptions, 
the observed  dust/H mass ratios for the KF57 sample
are in broad agreement with
the toy model 
(Eq.\ \ref{eq:mdmh vs o/h})
for $0.1\ltsim\tau_a/\tau_d\ltsim 1$, with $\tau_a/\tau_d=0.3$ (green curve
in Fig.\ \ref{fig:mdmh}b) providing a reasonable fit to the main
trend in $\Mdust/\MH$ vs O/H.

We do not expect all galaxies to be characterized by a
single value of $\tau_a/\tau_d$.
Allowing for reasonable variation of
$\tau_a/\tau_d$ from galaxy to galaxy
(ranging from $\tau_a/\tau_d=1$ for the red curve to
$\tau_a/\tau_d=0.1$ for the blue curve)
can accommodate almost all of the measured values.
However, there are some notable exceptions:
\begin{itemize}
\item NGC\,1482 (type SA0), 
This galaxy with near-solar O/H 
has a measured dust/H mass ratio
several times
larger than the ``upper limit'' $0.0099(Z/Z_\odot)$
(although NGC\,1482 is missing CO measurements).
It is notable that
the ISM appears to have been subject to
unusual activity.
NGC\,1482 shows evidence
of a galactic-scale ``superwind'':
the X-ray morphology shows a striking ``hour-glass'' shape emerging 
from the plane of the disk
\citep{Strickland+Heckman+Colbert+etal_2004,Vagshette+Pandge+Pandey+etal_2012}.
Interestingly, this galaxy is completely missing \ion{H}{1} in its 
central region, with
atomic gas only found in two blobs $\sim$2\,kpc distant from its center,
roughly at the confines of the X-ray emission \citep{Hota+Saikia_2005}.
CO observations of NGC\,1482 are needed.
If NGC\,1482 were found to have
$M(\HH)+M({\rm H}\,{\rm II})\approx1.3\times10^9\Msol$,
the $\Mdust/\MH$ ratio would be normal for its metallicity.

\item
NGC\,4594 (M104 ``Sombrero'', type SAb) also has
near-solar O/H, but a dust/H mass ratio several times larger than the
expected upper limit $0.0099(Z/Z_\odot)$.
NGC\,4594 has diffuse X-ray emission, 
suggesting the presence of a galactic-scale outflow
\citep{Li+Jones+Forman+etal_2011}.
\citet{Li+Jones+Forman+etal_2011} estimate the hot gas to have a 
temperature $T\approx 6\times10^6\K$ and total mass
$M_{\rm hot}\approx2.9\times10^8\Msol$.
Adding this to the \citet{Bajaja+vanderBurg+Faber+etal_1984} value for
\ion{H}{1},
and the H$_2$ mass estimated with a standard $X_{\rm CO}$ factor,
we find $M_\Ha=6.0\times10^8\Msol$, and
$M_\dust/M_\Ha=0.023$,
about a factor of 2.5 above the ratio expected for
metallicity $Z/Z_\odot\approx 1$.  
The gas in the hot phase, with a density
$\nH\approx0.1\cm^{-3}$, has a cooling time 
$\tau\approx5\times10^7\yr$ \citep{Li+Jones+Forman+etal_2011}.
Some of the hot gas may
have cooled
down to $\sim$$10^4\K$, perhaps making an additional
contribution to the total gas mass present in NGC\,4594.
We suggest that NGC\,4594 may 
contain a substantial mass of diffuse \ion{H}{2}
at $\sim$10$^4\K$ that
has not yet been detected.

Gravity, radiation pressure, and inertia can all lead to velocity
differences between gas and dust, allowing the two to separate.
However, because dust is generally well-coupled to the gas by both gas drag
and the Lorentz force on charged grains, 
gas-grain ``slip'' velocities are generally small
\citep[e.g.,][]{Weingartner+Draine_2001b}, and
scenarios where gas is removed but dust is left behind are not viable
unless the gas flows are slow enough that the small
gas-grain ``slip'' velocities suffice to prevent the dust grains from leaving
the galaxy.
Even if gas is stripped or lost in an outflow, we expect the 
metallicity in the remaining gas [and therefore the upper bound 
Eq.\ (\ref{eq:mdmh vs o/h}) 
on the dust/mass ratio] 
to be unaffected.
If NGC\,1482 and NGC\,4594 truly have high dust/gas ratios, then
this would appear to require a mechanism
for concentrating the dust in part of the gas, and removing the
dust-poor gas via an outflow or stripping.
Alternatively, perhaps the dust/gas ratio is actually normal, but the
dust mass has been overestimated because the dust material 
for some reason has a far-infrared/submm 
opacity that is significantly 
larger than found in normal star-forming galaxies.
The elevated dust/gas mass ratios in NGC\,1482 and NGC\,4594 
require further study.

\smallskip

\item NGC\,2841 (Type SAb) and NGC\,5055 (M63, Type SAbc):
These two 
galaxies have much {\it lower} dust/gas ratios than would be expected
given their high estimated metallicities (\logohpp$=9.31$: $Z/Z_\odot=3.6$).
The photometry for these galaxies is reliable, and the models reproduce
the SED out to 500$\micron$.  
However, it seems 
likely that the metallicities given in Table \ref{tab:geom} are overestimated.
\citet[][]{Moustakas+Kennicutt+Tremonti+etal_2010} 
with KK04 found that NGC\,2841 and NGC\,5055 have 
\logoh\ significantly greater than 9, $\sim$0.2\,dex higher 
than the values found for the 
central regions in the same galaxies by
\citet{Pilyugin+Grebel+Kniazev_2014}.
\citet{Pilyugin+Grebel+Kniazev_2014} also deduce 
strong metallicity gradients in these two galaxies implying that 
at $0.5\,R_{25}$, the characteristic \logoh$\sim$8.6. 
Such metallicities, better representing the average over the galactic disk,
would be consistent with 
the observed dust/gas ratios for these two galaxies.
\end{itemize}

In Fig.\ \ref{fig:mdmh}b we also show the broken power-law
empirical trend found
by \citet{Remy-Ruyer+Madden+Galliano+etal_2014}, with
$\Mdust/\MH \propto ({\rm O/H})$ for O/H above a critical value,\footnote{%
    Using PT metallicities, \citet{Remy-Ruyer+Madden+Galliano+etal_2014}
    estimated this critical metallicity to be \logoh=8.02.  Here we
    adjust the critical value to 8.42 to allow for the systematic
    offset of $\sim$0.4 between PT and PP04N2 metallicities 
    at low O/H (see Fig.\ \ref{fig:oxyoxy}).
    }
but with $\Mdust/\MH \propto ({\rm O/H})^{3.02}$ for lower values of (O/H).
This empirical result
is seen to fall close to our toy model with $\tau_a/\tau_d=0.3$.

\begin{figure}
\begin{center}
  \includegraphics[angle=0,width=\linewidth]
  {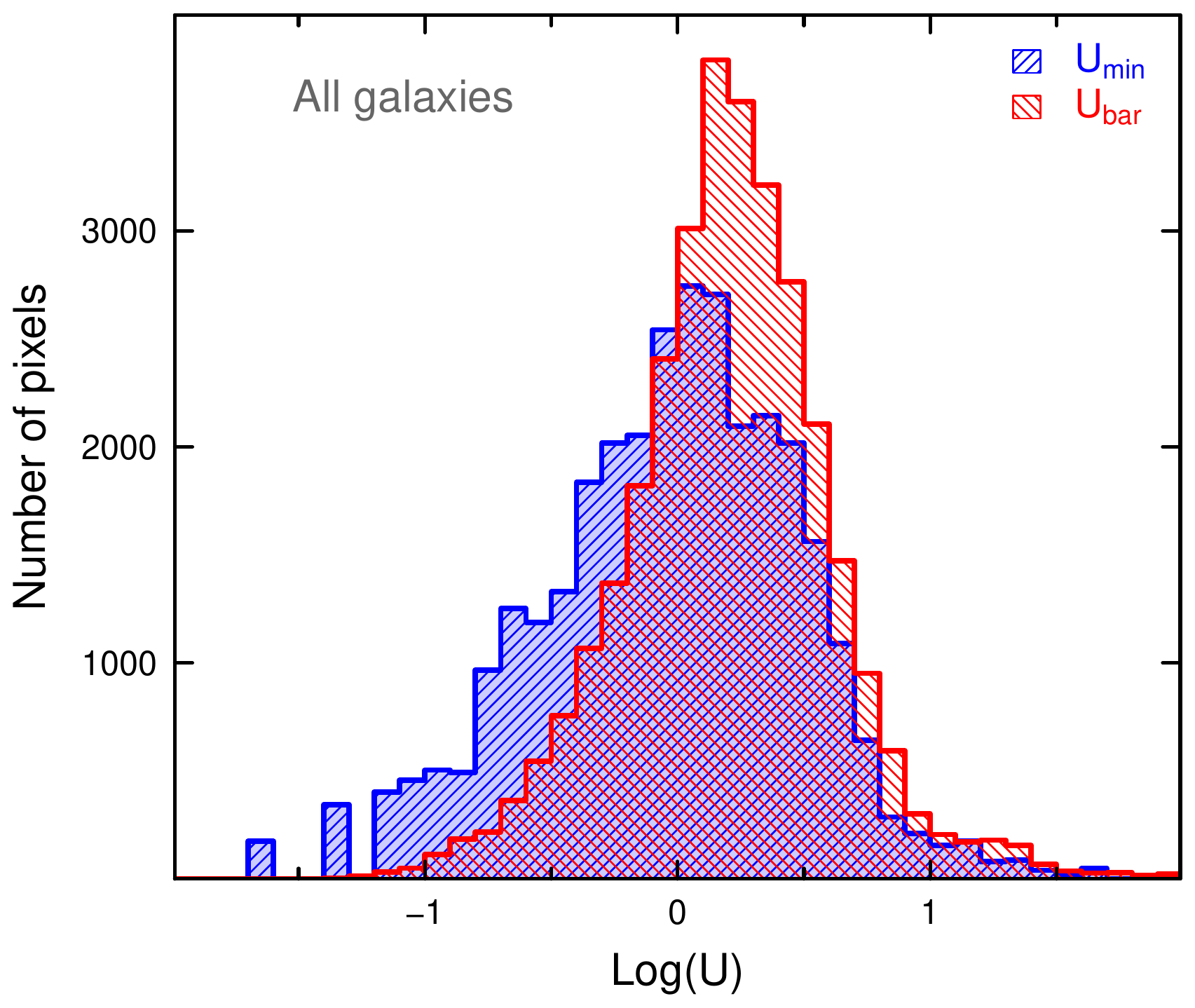}
  \caption{Distributions of $\Umin$ and $\Ubar$
    (U$_{\rm bar}$) for all galaxies.
    $\Umin$ and $\Ubar$ are the
    renormalized values (see
Eq.\ \ref{eq:Mdcorrected}-\ref{eq:Umin corrected}).
Both distributions are fairly broad; 
for a given pixel, $\Ubar-\Umin$ may be smaller than or 
comparable to $\Umin$ given that $\fpdr$ is usually modest.
Thus, it is not surprising that the two peaks are similar.
}
\label{fig:histograms_uminubar}
\end{center}
\end{figure}

\begin{figure*}
\begin{center}
  \includegraphics[angle=0,width=0.8\linewidth]
                  {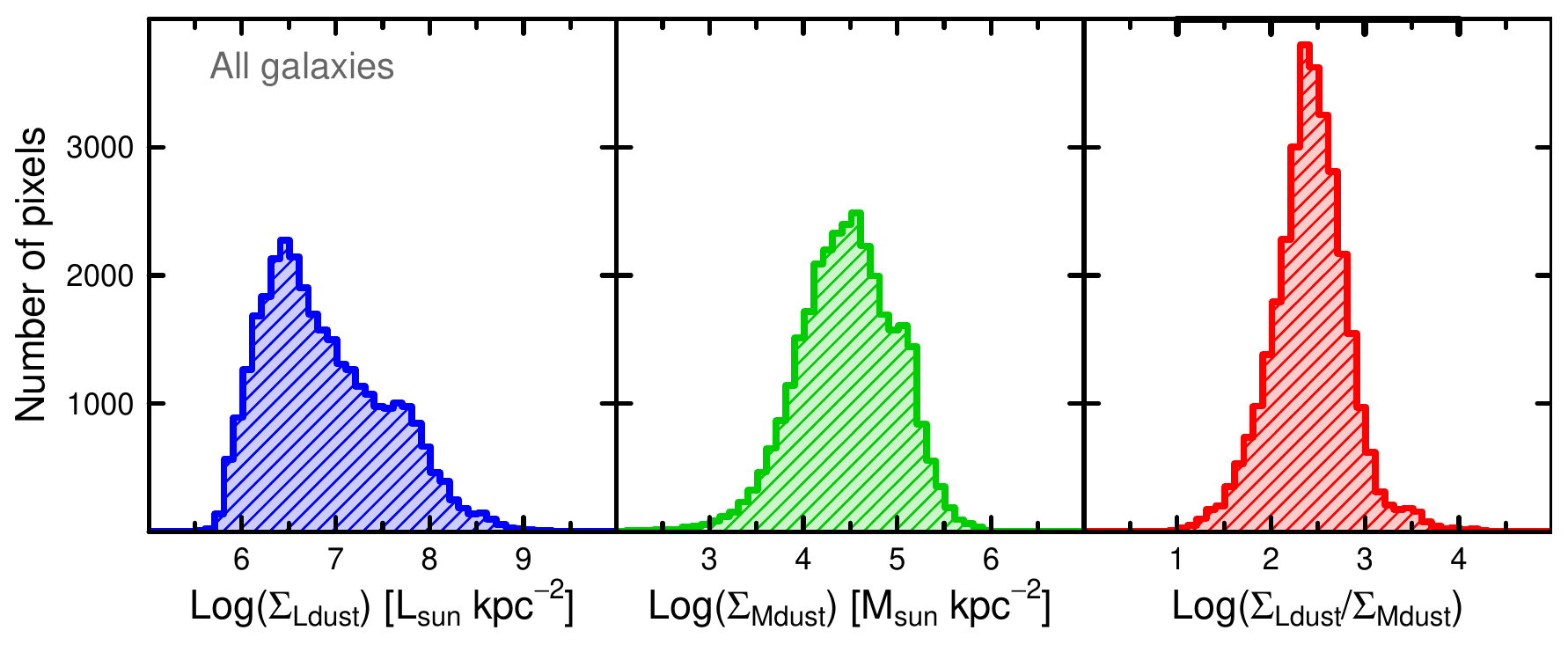}
\caption{Distributions of $\SigLd$ (left panel), 
$\SigMd$ (middle),
and $\SigLd/\SigMd$ (right) for all galaxies.
The cutoffs at low $\SigLd$ and low $\SigMd$ are due to limitations in
sensitivity.
The total dust luminosity $\Ldust$ is contributed mainly by higher
surface brightness pixels, with $\SigLd\approx10^8\Lsol\kpc^{-2}$.
The total mass is contributed mainly by pixels with
$\SigMd\approx 10^{5.2}\Msol\kpc^{-2}$, corresponding to extinction
$A_V\approx 1\,$mag.
The right panel shows that most of the dust has $\Ldust/\Mdust\approx 150\Lsol/\Msol$, corresponding to a heating rate $\Ubar\approx 1$.
}
\label{fig:histograms_ldustmdust}
\end{center}
\end{figure*}

\begin{figure}
\begin{center}
  \includegraphics[angle=0,width=\linewidth]
                  {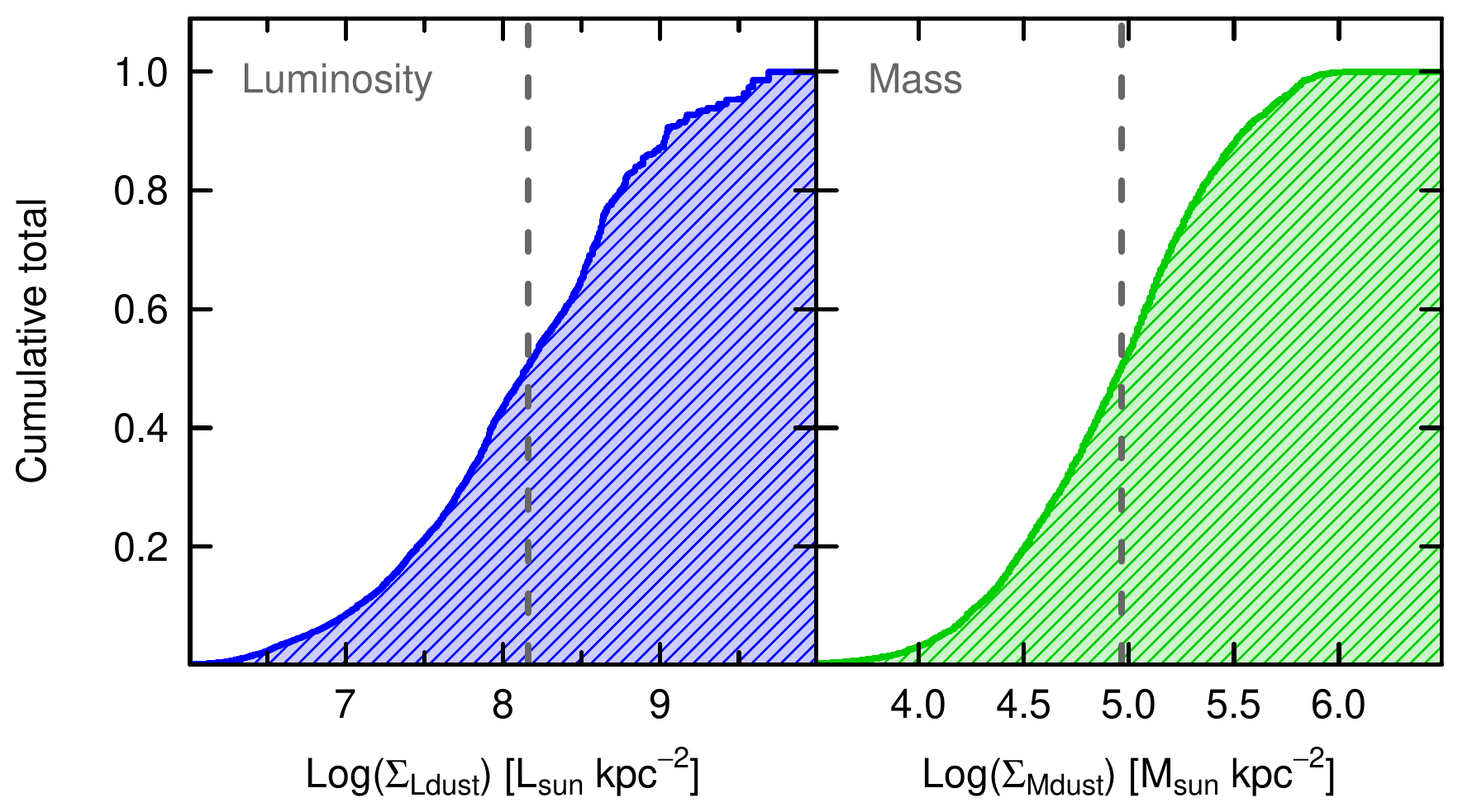}
\caption{Cumulative distributions of 
dust luminosity $\Ldust$ (left panel) and dust mass $\Mdust$ (right
panel) for all galaxies.
The vertical dashed lines show 
the surface brightness $\SigLd$ and surface density $\SigMd$
above and below which provides 50\% of the total {\bf dust} luminosity 
and {\bf dust} mass,
respectively: 
$\SigLd=10^{8.2}\Lsol\kpc^{-2}$;
and
$\SigMd=10^{5.1}\Msol\kpc^{-2}$. 
The regions with 
$\SigLd>10^{8.2}\Lsol\kpc^{-2}$
comprise $\sim$3\% of the pixels, and those with 
$\SigMd>10^{5.1}\Msol\kpc^{-2}$
$\sim$22\% of the pixels.
}
\label{fig:cumulative_ldustmdust}
\end{center}
\end{figure}

\begin{figure*}
\begin{center}
  \includegraphics[angle=0,width=0.8\linewidth]
                  {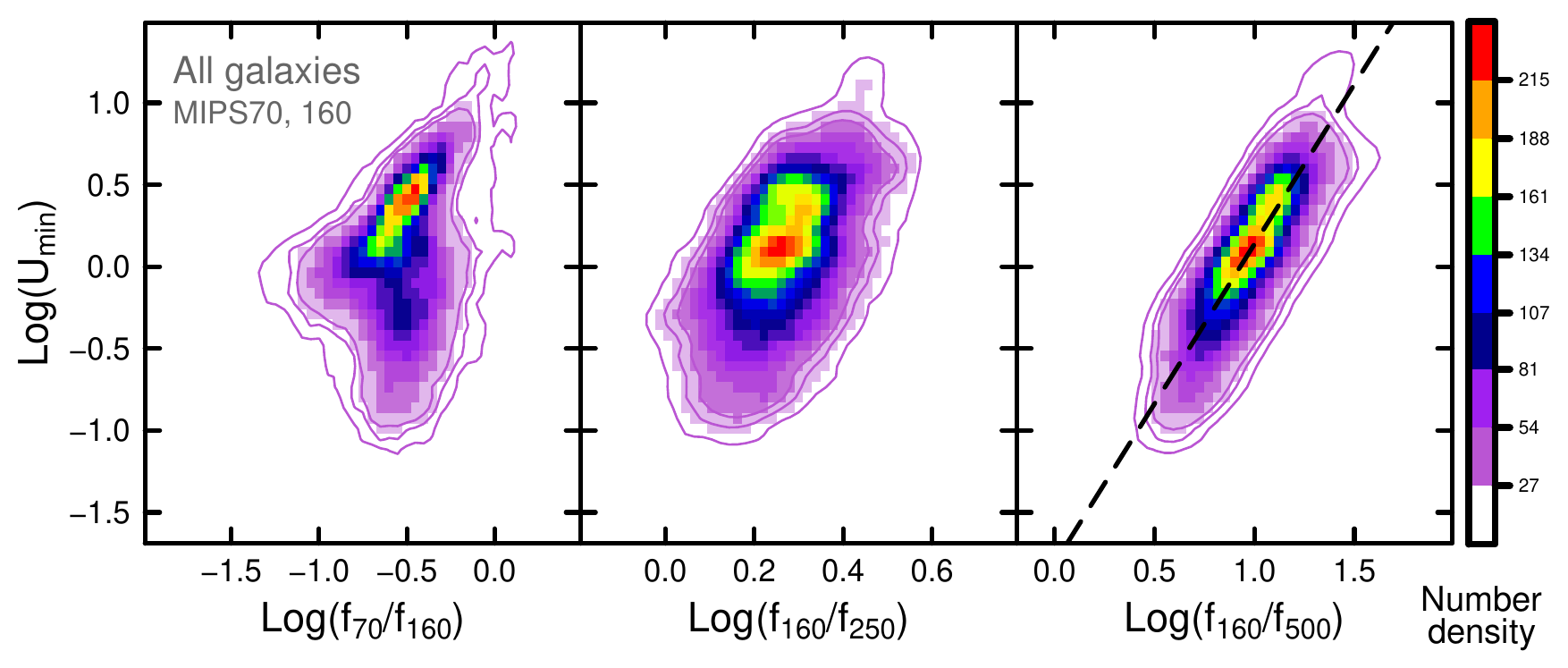}
\caption{Minimum starlight heating intensity $\Umin$ vs.\ 
$f_{70}/f_{160}$ (left panel),
$f_{160}/f_{250}$ (middle), and 
$f_{160}/f_{500}$ (right) for all galaxies.
The 70 and 160$\micron$ flux densities$f_{70}$ and $f_{160}$ are
from MIPS only (see text).
The color coding corresponds to number density of pixels 
as shown by the rightmost color table.
The left panel shows
that $f_{70}/f_{160}$ is not 
a good indicator of $\Umin$, because $f_{70}$ is sensitive to
both single-photon heating and the
emission from
dust exposed to starlight intensities $U>\Umin$.
The middle panel with $f_{160}/f_{250}$ 
avoids using $f_{70}$, 
but the wavelength range is insufficient to
adequately
sample $\Umin$ and a luminosity-weighted dust temperature.
Instead, the 
right panel shows the 
tight correlation between $\Umin$
and $f_{160}/f_{500}$
(the dashed line is Eq.\ \ref{eq:umin}),
illustrating the close relationship between the minimum heating intensity
and the coolest dust.
\btdnote{NB: limited to pixels with $\Sigma_L>2\times10^6$}
\btdnote{statistics/RenoFewerDL07\_UminvsPhotOnly160250WithWedge\_AllGalaxies\_Lim2e6-crop.pdf}
}
\label{fig:umin_fluxratios}
\end{center}
\end{figure*}
\begin{figure*}
\begin{center}
  \includegraphics[angle=0,width=0.8\linewidth]
                  {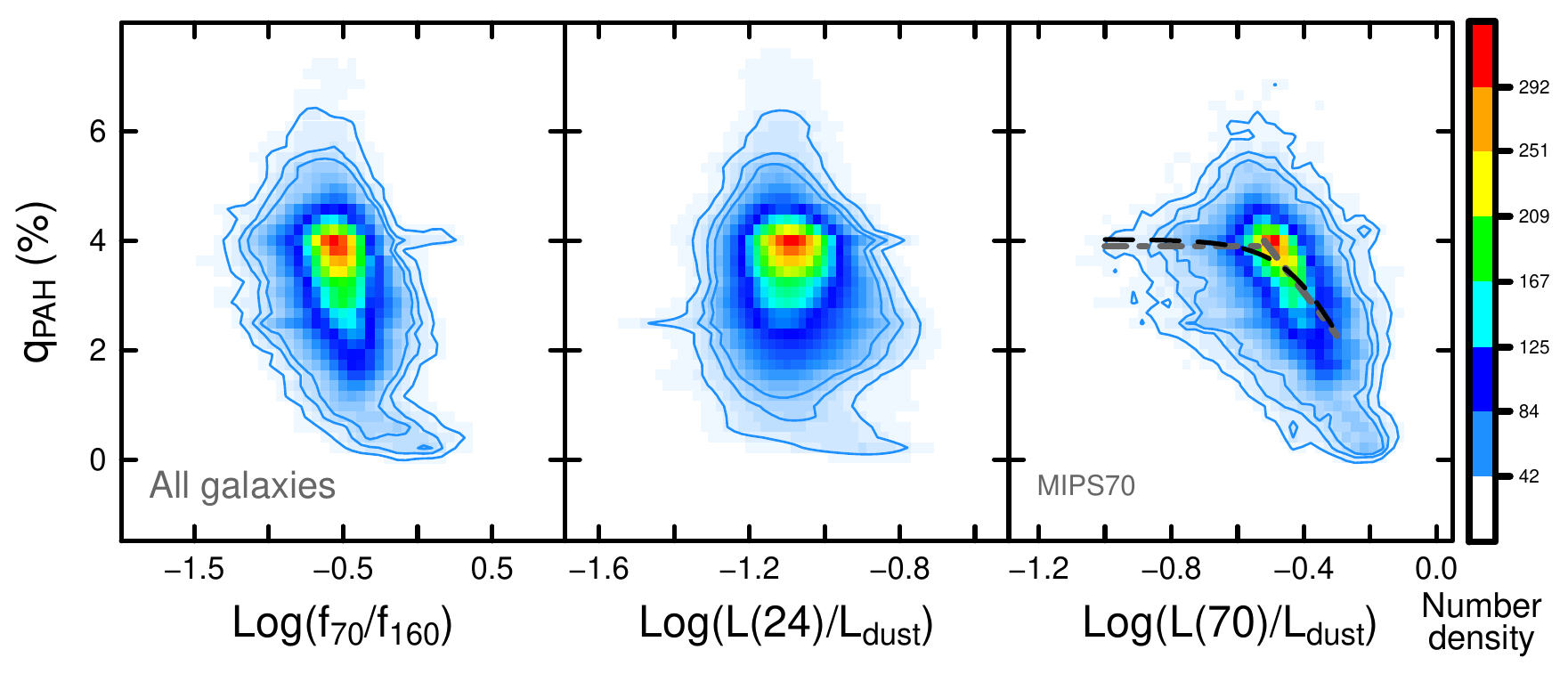}
\caption{PAH fraction $\qpah$ versus 
$f_{70}/f_{160}$ (left panel), 
\nulummipsshort/\lumdust\ (middle) and 
\nulummipsmid/\lumdust\ (right) for all galaxies.
The 70 and 160$\micron$ flux densities$f_{70}$ and $f_{160}$ are
from MIPS only (see text).
The color coding corresponds to number density of pixels as shown by the rightmost color table.
The trend in the right panel for $\qpah$ to decrease with increasing 
\nulummipsshort/\lumdust\
reflects the power of \nulummipsmid/\lumdust\ 
to trace $\qpah$.
The (black) long-dashed line represents the best-fit function given in 
Eq.\ (\ref{eq:qpahl70})
with rms residuals of $\sim$1.2\% on $\qpah$;
similar residuals are given by the 
(grey) dashed-dotted line, 
a broken power-law fit as given in Eq.\ (\ref{eq:qpahbpl}).
}
\label{fig:qpah_lums}
\end{center}
\end{figure*}

\subsection{Resolved trends of DL07 parameters}
\label{sec:resolveddl07}

Using data at M160 resolution,
the synergy of \hers\ and \spit\ for the KINGFISH sample 
enables an assessment of dust properties on kpc scales in nearby galaxies
(the FWHM of the M160 PSF, 38\farcs8, 
corresponds to 1.86\,kpc at the median KINGFISH
sample distance of 9.9\,Mpc).
The number of M160 $18\arcsec\!\times\!18\arcsec$ pixels 
in each galaxy ranges from 20 for the smallest galaxies
(M81\,dwB, NGC\,584) to $>$4000 pixels for the largest ones
(NGC\,5457\,=\,M\,101 and IC\,342);
the resolved sample as a whole, including the nine ``extra'' galaxies,
comprises $>$32\,000 
pixels with well-defined dust parameters
and photometry.

Figure \ref{fig:histograms_uminubar} shows how the starlight
intensity parameters $U_{\rm min,DL07}$ and $\overline{U}_{\rm DL07}$ 
are distributed over the
$\sim$32\,000 galaxy mask pixels 
(i.e., $\SigLd>\SigLdmin$) where we are able to estimate the
dust and starlight parameters.  Half of the pixels have
$\Ubar<1$,
and half of the pixels have 
$\Umin<1$.
The $\bar{U}$ distribution for the KF62 sample
(Figure \ref{fig:histograms_uminubar}) is similar to that for Local Group
galaxies
\citep{Utomo+Chiang+Leroy+etal_2019}.
  
The distributions of dust luminosity densities $\SigLd$ and 
dust mass densities $\SigMd$ 
are displayed in Fig.\ \ref{fig:histograms_ldustmdust}.
The $\SigLd$ distribution peaks toward fainter
$\SigLd$, increasing
down to the lowest values of $\SigLd\approx 10^6\Lsol\kpc^{-2}$ 
allowed by the luminosity surface density cutoff $\SigLdmin$ defining the 
``galaxy mask'' for each galaxy.\footnote{Because our
$\SigLdmin$ cutoff varies from galaxy to galaxy, ranging from
$10^{5.6}\Lsol\kpc^{-2}$ for DDO\,165 to $10^{7.4}\Lsol\kpc^{-2}$
for NGC2146, the pixel histogram has a 
broad peak near $\sim10^{6.3}\Lsol\kpc^{-2}$.}
While the pixel histogram peaks at faint $\SigLd \approx 10^6\Lsol\kpc^{-2}$, 
the infrared luminosity is dominated by the bright pixels with
$\SigLd\approx10^8\Lsol\kpc^{-2}$.
The distribution of dust surface densities $\SigMd$ peaks near 
$10^5\Msol\kpc^{-2}$,
which corresponds to $A_V\approx 0.7\,$mag,
and $\sim$90\% of the dust mass is contributed by pixels with
$\SigMd\gtsim 10^{4.75}\Msol\kpc^{-2}$, or $A_V\gtsim 0.4\,$mag.
The distribution of the light-to-mass ratio $\SigLd/\SigMd$ is shown in
the right 
panel of Fig.\ \ref{fig:histograms_ldustmdust}.
This is of course equivalent to the distribution of $\Ubar$.  The
histogram peaks at $\Ldust/\Mdust\approx 150\Lsol/\Msol$, corresponding to
a starlight heating rate parameter $U\approx1$.

The dust light and mass surface densities that most contribute to the
total dust budget are more clearly seen in 
Fig.\ \ref{fig:cumulative_ldustmdust}, 
where we show the cumulative distributions of
dust luminosity $L_\dust$ and dust mass $M_\dust$
plotted against $\SigLd$ and $\SigMd$, respectively.

The vertical dashed lines in Fig.\ \ref{fig:cumulative_ldustmdust} show 
the surface-density 
thresholds that provide 50\% of the total:
$\SigLd=10^{8.2}\Lsol\kpc^{-2}$ and 
$\SigMd=10^{5.1}\Msol\kpc^{-2}$.
Regions with dust light and mass surface 
densities greater than these values 
comprise only a small fraction of the total;
from Fig.\ \ref{fig:histograms_ldustmdust} we see that
50\% of the 
dust light comes from only $\sim$3\% of the 
(brightest) 
pixels, and 50\% of 
the total dust mass from
$\sim$22\% of the (densest) pixels.

In what follows
we have applied a limit in dust surface brightness
$\SigLd \geq 2 \times 10^6\Lsol\kpc^{-2}$;
thus the low signal-to-noise faint outer regions of the 
sample galaxies 
(where estimates of parameters such 
as $\SigMd$ and $\qpah$ may become unreliable)
will not be considered.
As seen above, such regions contribute very little to either the light budget or
the mass budget of the dust over the sample as a whole.
Applying such a cut ensures that the plotted DL07 parameters 
(and the photometric
quantities) will be as accurate as possible, given the constraints of the data;
the total number of 
$18\arcsec\!\times\!18\arcsec$ pixels in the sample is reduced to $\sim$25\,500.

We now investigate the IR observational 
signatures associated with dust heating ($\Umin$).
Figure \ref{fig:umin_fluxratios} shows 
$\Umin$ for all galaxies plotted versus MIPS and SPIRE flux density ratios,
$f_{70}/f_{160}$, $f_{70}/f_{250}$, and $f_{160}/f_{500}$. 
Because of the unexplained discrepancies between
MIPS and PACS photometry (see Figure \ref{fig:PACS-MIPS}), 
we have elected to use only MIPS photometry for $f_{70}$ and
$f_{160}$.
The left panel shows that the flux ratio $f_{70}/f_{160}$ 
is not a very good predictor of $\Umin$.  This is because when
$\Umin\ltsim 1$, the 70\,$\micron$ emission 
has an appreciable contribution from
(1) single-photon
heating of small grains, and (2)
dust in regions with high starlight intensities
(assuming $\gamma>0$, which is almost always the case).
The flux ratio $f_{160}/f_{250}$, shown in the middle panel, ameliorates
the potential domination of the emission by small-grain stochastic heating, 
but the wavelength ratio of the two fluxes is insufficient to reliably
sample $\Umin$;
a small range in flux ratio corresponds to as much as an order of
magnitude change in $\Umin$. 
However, the right panel shows that the $f_{160}/f_{500}$ flux ratio correlates
quite well with $\Umin$ because 
the emission at both $160\,\micron$ 
and $500\,\micron$ is dominated by the larger grains heated by starlight
intensities near $\Umin$. 
Because 160\,$\micron$ is not in the Rayleigh-Jeans
limit for the grain temperatures in these galaxies,
the $f_{160}/f_{500}$ ratio is sensitive to large-grain temperature, and hence
to starlight heating rate.
The best-fit correlation,
obtained with median clipping and a ``robust'' regression algorithm,
effective for minimizing the effects of outliers
\citep{R_Core_Team_2014},
is given by:
\begin{equation}
\log_{10} (\Umin)=(-1.81\pm0.01) + (1.95\pm0.01)\log_{10}\left(\frac{f_{160}}{f_{500}}\right).
\label{eq:umin}
\end{equation}
This relation predicts $\Umin$ to within
0.21\,dex {\bf (rms)}
over a range of $\Umin$ of
more than two orders of magnitude.
Because the emission at these wavelengths is dominated completely
by large grains, this long-wavelength ratio predicts very well the minimum starlight heating intensity.

\begin{figure*}
\begin{center}
  \includegraphics[angle=0,width=0.8\linewidth]
                  {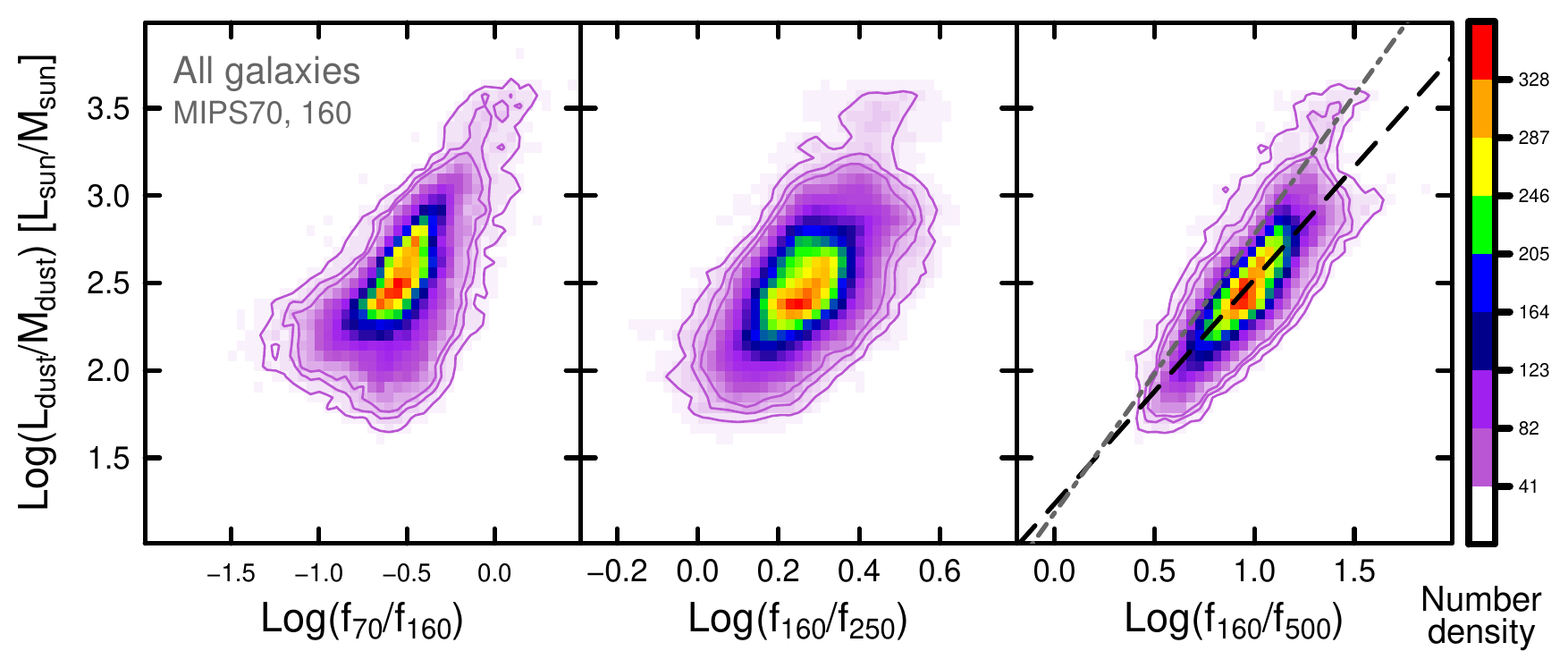}
\caption{Dust mass-to-light ratio \lumdust/\mdust\
vs.
$f_{70}/f_{160}$ (left panel), 
$f_{160}/f_{250}$ (middle), and
$f_{160}/f_{500}$ (right) for all galaxies.
The color coding corresponds to pixel number density as shown by the rightmost color table.
The best-fit (robust) regression for f$_{160}$/f$_{500}$ is shown as a 
(black) long-dashed line, 
and corresponds to rms residuals of 
$\sim$0.18\,dex (see Eq.\ \ref{eq:ldustmdust_160500}).
The (grey) dashed-dotted line is the analogous best-fit regression 
for only IC\,2574 and NGC\,2146 
(with 383 dof, see Fig.\ \ref{fig:ldustmdust_fluxratios3}).
}
\label{fig:ldustmdust_fluxratios}
\end{center}
\end{figure*}
\begin{figure*}
\begin{center}
  \includegraphics[angle=0,width=0.75\linewidth]
                  {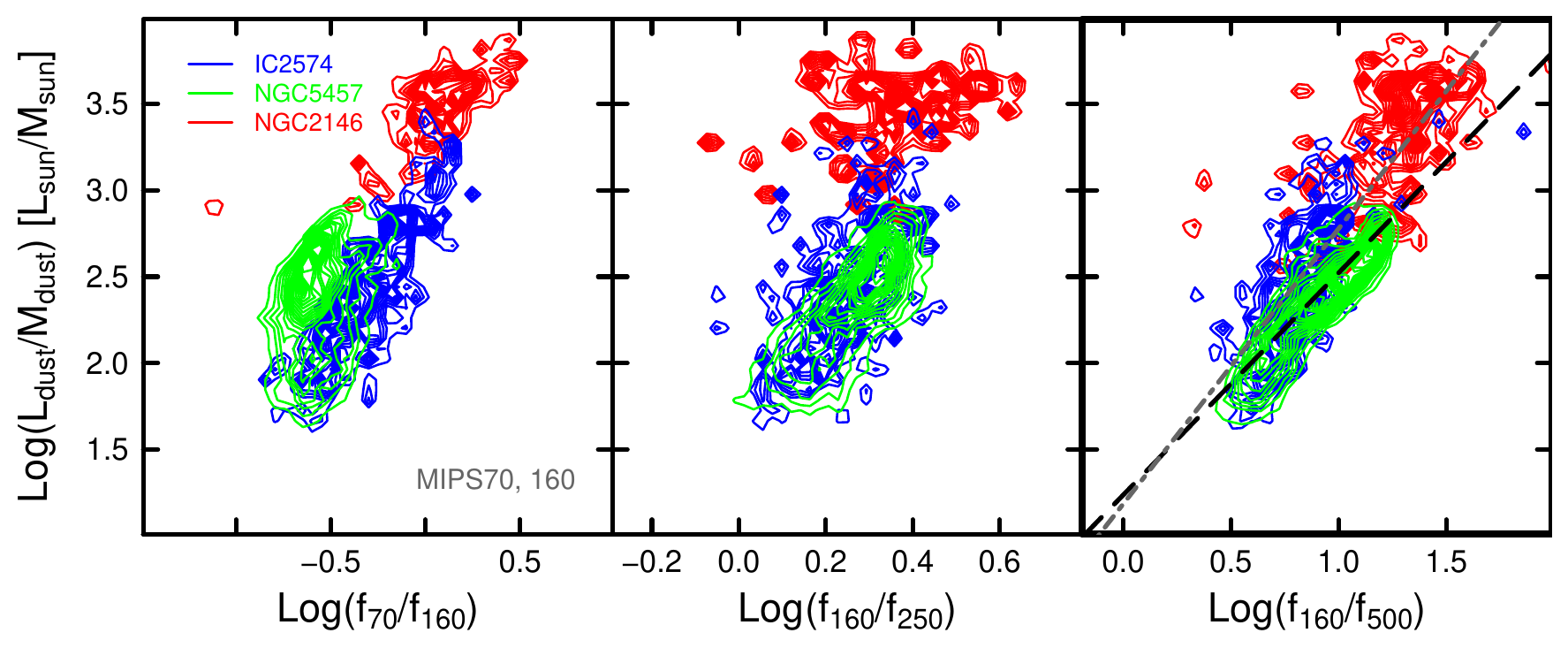}
\caption{Dust mass-to-light ratio \lumdust/\mdust\
plotted against
f$_{70}$/f$_{160}$ (left panel)
f$_{160}$/f$_{250}$ (middle), and
f$_{160}$/f$_{500}$ (right)
for three galaxies separately: IC\,2547, a low-metallicity dwarf;
NGC\,5457 (M\,101), a large grand-design spiral;
and NGC\,2146, a LIRG.
The contours reflect the individual galaxies 
(IC\,2547 blue, NGC\,5457 green, and NGC\,2146 red),
and correspond to pixel number densities.
In the right panel,
the (black) long-dashed line corresponds to the 
best-fit regression reported in Fig.\ \ref{fig:ldustmdust_fluxratios}
for the sample as a whole [see Eq.\ (\ref{eq:ldustmdust_160500})], and
the (grey) dashed-dotted line
corresponds to the analogous best-fit regression 
for only IC\,2574 and NGC\,2146 (383 dof).
\btdnote{limited to pixels with $\Sigma_L>2\times10^6$}
\btdnote{statistics/RenoFewerDL07\_LdustMdustvs3Phot160250\_3galaxies\_Lim2e6-crop.pdf}
}
\label{fig:ldustmdust_fluxratios3}
\end{center}
\end{figure*}

The PAH abundance parameter $\qpah$ varies from galaxy to galaxy,
as discussed in Section \ref{sec:pahglobaloh}, where it is apparent that
there is a correlation between $\qpah$ and the gas-phase metallicity O/H.
$\qpah$ also exhibits significant variations within individual galaxies,
as can be seen from the map of $\qpah$ in M101 
(see Fig.\ \ref{fig:NGC5457_intext}) as well as for other well-resolved
galaxies (see Figs.\
17.1-17.62).
If $\qpah$ is sensitive to metallicity, then we may expect radial
variations within galaxies, with $\qpah$ generally
declining with radius.
However, our $\qpah$ maps also exhibit substantial azimuthal variations,
suggesting that the PAH abundance responds to changes in environmental
conditions beyond metallicity alone.

In Fig.\ \ref{fig:qpah_lums}, we explore
-- using three different proxies for the starlight intensity -- whether  
$\qpah$ is affected by the intensity of the radiation field.
The left panel in Fig.\ \ref{fig:qpah_lums} 
indicates that $\qpah$ seems to be relatively independent
of variations in the  $f_{70}/f_{160}$ flux ratio.
The $f_{70}/f_{160}$ flux ratio is apparently not uniquely 
tracing the temperature of the larger grains; 
as seen in Fig.\ \ref{fig:umin_fluxratios}, 
and discussed below, this ratio begins to reflect $\Umin$,
and thus large-grain temperature,
only above a certain $\Umin$ threshold ($\Umin\gtsim 0.5$).
The middle panel, 
shows little correlation between $\qpah$ and 
$\nu L_\nu(24\micron)/L_{\rm dust}$,
but the right panel shows a stronger trend where $\qpah$ tends to fall significantly when 
$\nu L_\nu(70\micron)/L_{\rm dust}$ rises to the highest levels.
The lack of dependence on the $L(24)/\Ldust$ ratio 
(and the relatively small 0.5\,dex range in $L(24)/\Ldust$)
arises
because single-photon heating generally dominates at 24\,$\micron$;
big grains only get hot enough to radiate at 24\,$\micron$ when the radiation 
field is extremely intense.
Instead, at 70\,$\micron$,
single-photon heating makes a significant contribution only
for $\Umin\ltsim 0.5$. 
Thus $L(70)/\Ldust$ is a better indicator of warm large grains than $L(24)/\Ldust$,
and
it is these warm large grains that are the signature of 
high-intensity radiation fields that could be associated with PAH
destruction.

As seen in the right panel of Fig.\ \ref{fig:qpah_lums},
the PAH fraction appears to vary with $L(70)/L_{\rm dust}$ according
to the empirical relation
\beq
\qpah \approx \frac{0.0402}{1+15\,[L(70)/L_{\rm dust})]^{4.4}}
~~~.
\label{eq:qpahl70}
\eeq
\btdnote{need new fit params}
where the normalization constant 0.0402 corresponds to $\qpah\approx 4.0$\%;
the root-mean-square (rms) residual of this fit is 
1.2\% on $\qpah$.
A similar empirical fit is given by the broken power-law shown
by the grey dashed-dotted line in Fig.\ \ref{fig:qpah_lums}:
\begin{equation}
\qpah \approx\ \begin{cases}
0.039 & \text{if $\log_{10}(\frac{L70}{L_{\rm dust}}) \leq -0.52$} \\
0.3 - 7.0\log_{10}(\frac{L70}{L_{\rm dust}})  & \text{if $\log_{10}(\frac{L70}{L_{\rm dust}}) > -0.52$} \\
\end{cases}
\label{eq:qpahbpl}
\end{equation}
Such a trend
may reflect a tendency for PAH destruction to occur in
star-forming regions, where O stars supply high energy photons
that photodestroy PAHs, and a significant fraction of the dust is
exposed to starlight intensities high enough to elevate the 
$L(70)/L_{\rm dust}$ ratio.
Many studies have previously noted 
suppression of PAH emission in \ion{H}{2} regions
\citep[e.g.,][]{Giard+Bernard+Lacombe+etal_1994,
Helou+Roussel+Appleton+etal_2004,
Povich+Stone+Churchwell+etal_2007,
Relano+Kennicutt+Lisenfeld+etal_2016}.
In a detailed study of PAH abundances in the Magellanic Clouds,
\citet{Chastenet+Sandstrom+Chiang+etal_2019} show that $\qpah$ is reduced in
regions close to sources of H-ionizing radiation.

High values of $L(70)/\Ldust$ also occur in low-metallicity galaxies,
because of radiative transfer effects (hotter stars, less dust attenuation),
and is consistent with the tendency for lower $\qpah$ in a metal-poor ISM.

\begin{figure*}
\begin{center}
  \includegraphics[angle=0,width=0.8\linewidth]
                  {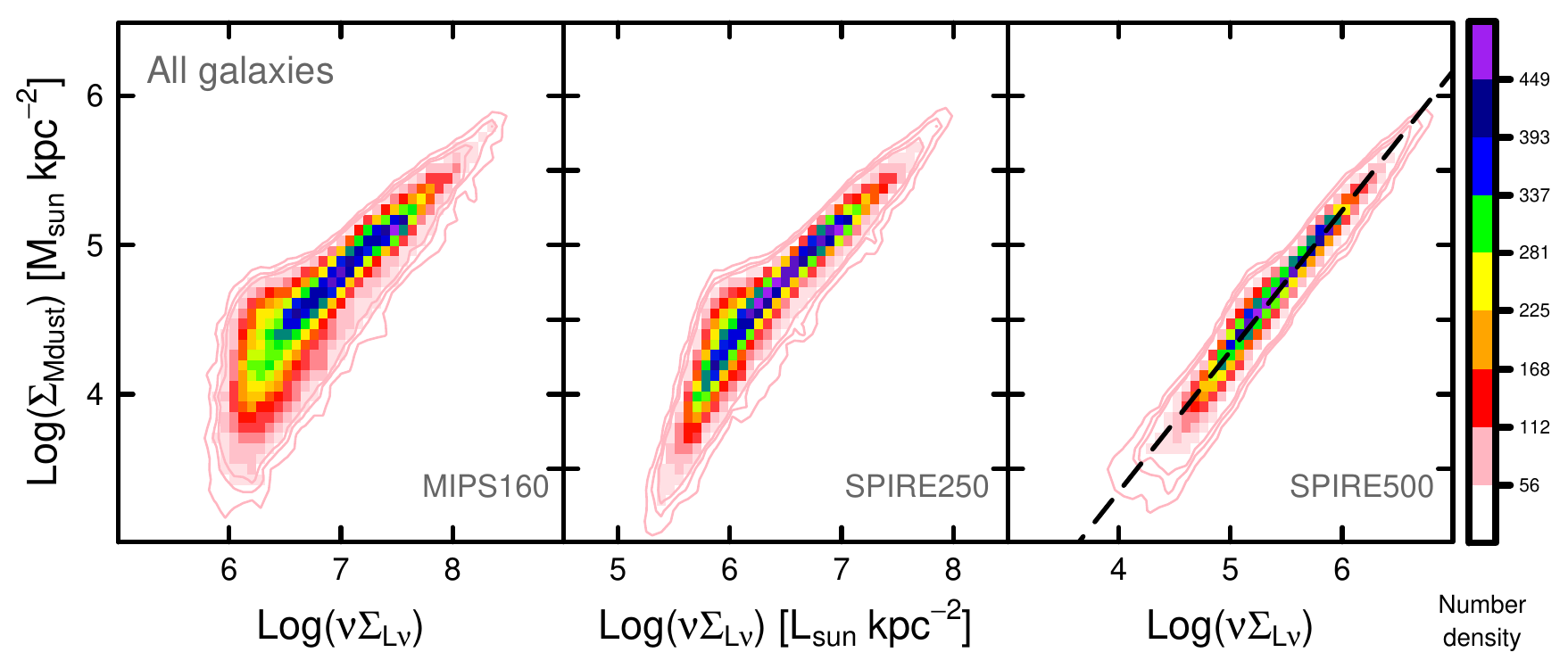}
\caption{Dust-mass surface density $\Sigma_{\rm Mdust}$ versus 
monochromatic surface brightness $\nu \Sigma_{L_\nu}=4\pi \nu I_\nu$
in the 
MIPS\,160\,\micron\ (left panel), 
SPIRE\,250\micron\  (middle), and 
SPIRE\,500\micron\ (right) bands,
for all galaxies.
The color coding corresponds to pixel number density as shown by the
rightmost color table.
The dashed line in the right panel (SPIRE500) shows the best-fit regression
relating dust mass surface density $\Sigma_{\rm Mdust}$
to 500\micron\ luminosity surface density $\nu \Sigma_{L\nu}(500\micron)$
(Eq.\ \ref{eq:mass500}).
See Sect. \ref{sec:resolveddl07_surface} for more details.
}
\label{fig:mdust_fluxratios}
\end{center}
\end{figure*}

\subsubsection{The resolved dust light-to-mass ratios}
\label{sec:resolveddl07_lightmass}

 
The dust light-to-mass ratio in galaxies and within galaxies, 
\lumdust/\mdust, 
should reflect the peak and spread of luminosity-weighted dust temperatures.
In the DL07 model, $\Ubar\propto\,$\lumdust/\mdust,
so $\Ubar$ also probes dust temperatures.
We would thus expect \lumdust/\mdust\ to depend on 
photometric flux ratios, as long as 
the two wavelengths in the flux ratios are sampling 
a sufficiently broad spectral range to be 
sensitive to large-grain temperature variations.
Figure \ref{fig:ldustmdust_fluxratios} illustrates 
the correlations in the resolved pixels of
all galaxies between $\SigLd/\SigMd$ 
(noted as \lumdust/\mdust\ in the ordinate axis label) and, 
as in Fig.\ \ref{fig:umin_fluxratios}, three flux density ratios, 
$f_{70}/f_{160}$, $f_{160}/f_{250}$, and $f_{160}/f_{500}$.
The longer the wavelength 
ratio (in this case 160\,\micron/500\,\micron),
the better that $\SigLd/\SigMd$ can be predicted from observations.
The right panel (black long-dashed line)
of Fig.\ \ref{fig:ldustmdust_fluxratios} shows the 
correlation with $f_{160}/f_{500}$ given by:
\begin{equation} \label{eq:ldustmdust_160500}
\log_{10}\left(\frac{\Ldust/\Mdust}{\Lsol/\Msol}\right)=
(1.24\pm0.01) + (1.28\pm0.01)
\log_{10}\left(\frac{f_{160}}{f_{500}}\right).
\end{equation}
This fit with $f_{160}/f_{500}$ has a rms deviation of 
0.16\,dex over $>$22\,000 degrees of freedom. 
The trend of $\Ldust/\Mdust$ with $f_{160}/f_{250}$ is much less reliable, 
so we have not shown any regression in the middle panel 
of Fig.\ \ref{fig:ldustmdust_fluxratios}.
Because of the limited wavelength lever arm for the $f_{160}/f_{250}$ flux ratio
(see also Fig.\ \ref{fig:umin_fluxratios}),
for a given {\bf $f_{160}/f_{250}$} ratio, 
$\Ldust/\Mdust$ can vary by a factor of 30 or more;
this makes it difficult to accurately determine the dust light-to-mass ratio
from $f_{160}/f_{250}$.
Nevertheless, if we know the dust luminosity \lumdust, 
and have a measure of a flux around the peak of dust emission
(e.g., $f_{160}$),
and one sufficiently far away and in the Rayleigh-Jeans regime
(e.g., $f_{500}$),
we can estimate the dust mass \mdust\ to 
within $\sim$50\%.

Fig.\ \ref{fig:ldustmdust_fluxratios3} shows the same quantities 
but separately for three galaxies representative of the extremes 
probed by the KINGFISH sample:
IC\,2574, a metal-poor dwarf; 
NGC\,5457 (M\,101), a face-on grand-design spiral;
and NGC\,2146, a luminous IR galaxy (LIRG).
For a 
flux density ratio with short$+$long wavelengths 
(e.g., $f_{70}/f_{160}$) 
the \lumdust/\mdust\ ratio within these galaxies can 
differ by up to an order of magnitude.
As has been seen in previous figures, 
because $f_{70}/f_{160}$ is sensitive to both single-photon heating and
the possible exposure of a small fraction of the dust to
starlight intensities $U>\Umin$,
the $f_{70}/f_{160}$ ratio does not strongly constrain the temperature
of the dust grains that dominate the total emission.
Instead, the longer-wavelength ratio (e.g., $f_{160}/f_{500}$) is a much better
indicator of large-grain temperature,
and consequently better
correlated with the dust light-to-mass ratio \lumdust/\mdust.
Two regressions are shown in
Figs.\ \ref{fig:ldustmdust_fluxratios}
and \ref{fig:ldustmdust_fluxratios3};
the (black) long-dashed line, described above 
[see Eq.\ (\ref{eq:ldustmdust_160500})], is for the entire sample.
The (grey) dashed-dotted one is the regression 
obtained for only IC\,2574 and NGC\,2146, and given by:
\begin{equation}
\log_{10}\!\left(\frac{\Ldust/\Mdust}{\Lsol/\Msol}\right)=
(1.18\pm0.05) + (1.59\pm0.04)
\,\log_{10}\!\left(\frac{f_{160}}{f_{500}}\right).
\label{eq:ldustmdust_160500_special}
\end{equation}
The regression for the entire sample is entirely 
consistent with NGC\,5457 (M\,101), but
not for IC\,2574 and NGC\,2146, 
which may be considered two ``extreme'' galaxies.
The overall radiation fields $\Ubar$ in IC\,2574 and NGC\,2146 are
higher (in the mean, by $\sim$40\% and a factor of 13, respectively) than that of NGC\,5457.
These more intense heating fields, possibly a signature of starbursts, 
result in a slightly steeper slope relating \lumdust/\mdust\ and
$f_{160}/f_{500}$ than in more quiescent 
environments such as the disk of NGC\,5457
(and most of the KINGFISH sample).

In the present model, the dust temperatures are determined by
the starlight intensity distribution within a pixel, which is characterized by
three parameters: $\Umin$, $\gamma$, and $\alpha$ 
(see Section \ref{sec:dustmodel}), and we need more than two bands
if we wish to determine the distribution of temperatures for the
emitting dust well enough to reliably estimate the mass of dust in the pixel.
On the other hand,
if flux ratios at longer wavelengths are considered 
(right panel of Fig.\ \ref{fig:ldustmdust_fluxratios3}),
there is much less variation within and between galaxies.
Such behavior was also seen with $\Umin$ in Fig.\ \ref{fig:umin_fluxratios},
and suggests that ratios at these longer wavelengths 
better trace \lumdust/\mdust,
because they provide better information about the
temperatures of the large grains that dominate the dust luminosity. 

\subsubsection{The resolved dust mass surface densities}
\label{sec:resolveddl07_surface}

Given the relative constancy of 
dust-to-metals ratios
for galaxies with metallicities \logoh$\ga$8.4
(see Fig.\ \ref{fig:mdmh}b),
the dust luminosity in the Rayleigh-Jeans (R-J) regime of the dust SED
has become a popular tracer of ISM mass 
\citep[e.g.,][]{Corbelli+Bianchi+Cortese+etal_2012,
                Eales+Smith+Auld+etal_2012,
                Groves+Schinnerer+Leroy+etal_2015,
                Scoville+Aussel+Sheth+etal_2014,
                Scoville+Sheth+Aussel+etal_2016,
                Scoville+Lee+VandenBout+etal_2017}.
This is an effective technique both locally and at high redshift because the R-J tail of the dust emission
probes optically-thin dust, and is relatively insensitive to dust temperature.
Here we explore whether this is also true for the spatially-resolved dust emission in the KINGFISH sample.
Figure \ref{fig:mdust_fluxratios} shows the dust mass surface density 
$\Sigma_{M_\dust}$ (estimated from the renormalized DL07 model)
plotted against 
monochromatic dust luminosity surface density
$\Sigma_{\nu L_\nu}=4\pi \nu I_\nu$,
in the MIPS 160\,\micron,
SPIRE 250\,\micron, and
SPIRE 500\,\micron\ bands.
It can be seen that at 160\,\micron, a wavelength that generally probes the 
dust emission peak, there is 
only a broad correlation with more than an order of magnitude dispersion
at low surface brightness.
As wavelength increases toward the SPIRE bands, the correlation improves,
and becomes very good at 500\,\micron,
similar to the trends found for KINGFISH global values by 
\citet{Groves+Schinnerer+Leroy+etal_2015}. 

The rightmost panel reports the best-fit correlation,
obtained with the robust regression algorithm: 
\beqa\nonumber \label{eq:mass500}
&\log_{10}\left(\frac{\Sigma_{M_{\rm dust}}}{\Msol\kpc^{-2}}\right) =
\quad\quad\quad\quad\quad\quad\quad\quad&
\\
&
(-0.42\pm0.01) + (0.942\pm0.001)
\log_{10}\left(\frac{\Sigma_{\nu L_\nu(500\mu{\rm m})}}{\Lsol\kpc^{-2}}\right)
.~~~~&
\eeqa
This fit gives an rms scatter $\sigma\,=\,$0.07\,dex on $\log_{10}$(\mdust) (with $\sim$25\,400 dof),
implying that dust mass surface densities can be inferred from 500\,\micron\ luminosity surface
densities to within
$\sim$20\%.
The slope is significantly sub-linear, over almost three decades 
of 500\,\micron\ luminosity surface
densities,
reflecting the tendency for dust to be somewhat warmer in pixels
where $\SigMd$ is high, presumably because these pixels are more likely
to harbor star-forming regions.
\citet{Groves+Schinnerer+Leroy+etal_2015} 
obtained a similar result globally for
inferring gas mass from $L_{500}$ for
all KINGFISH galaxies including dwarfs (stellar mass $\leq 10^9$\,\msun);
however, once \citet{Groves+Schinnerer+Leroy+etal_2015} 
considered only the more massive galaxies,
the slope steepened and became approximately linear.

The rms deviation of only 0.07 dex from Eq.\ (\ref{eq:mass500})
implies that one
can estimate $M_\dust$ more reliably from $L_\nu(500\micron)$ alone
than from the total dust luminosity $L_\dust$ and the ratio of 
two flux densities $L_\nu(160\micron)$ and $L_\nu(500\micron)$.
This is because obtaining $M_\dust$ from $L_\dust$ using
Eq.\ (\ref{eq:ldustmdust_160500}) in effect requires estimation
of $\langle T_\dust^{4+\beta}\rangle$,
whereas obtaining $M_\dust$ from $L_\nu(500\micron)$
from (\ref{eq:mass500}) (with an rms of 0.16\,dex)
requires estimating only $\langle T_\dust\rangle$, since
at 500$\micron$ the dust emission is in the Rayleigh-Jeans limit, with
$L_\nu(500\micron)\propto M_\dust\times\langle T_\dust\rangle$.
  
To estimate ISM mass from Eq.\ (\ref{eq:mass500}), 
the dust mass from Eq.\ (\ref{eq:mass500})
needs to be combined with a gas-to-dust ratio as discussed in Sect.\ \ref{sec:globaloh}.
However, this ratio depends on metallicity (see Fig.\ \ref{fig:mdmh}); thus
oxygen abundance needs to be incorporated to estimate gas mass for metal poor galaxies.
In any case,
Fig.\ \ref{fig:mdust_fluxratios} shows that the 
slope between dust mass and luminosity
is steeper, closer to unity, at {\it lower} surface
brightnesses, roughly independently of wavelength.
However, global integrated values of quantities such as long-wavelength IR luminosity are luminosity
weighted, thus sampling preferentially {\it higher} surface brightnesses.
Thus, our new result for resolved regions in KINGFISH galaxies is 
inconsistent with a strictly linear trend of
dust mass with long-wavelength IR luminosity.
Indeed, as noted above, a non-linear behavior would be expected since the dust in high $\SigMd$
pixels is, on average, somewhat warmer.

%
%

\section{Summary }
\label{sec:summary}

Dust modeling results for 70 galaxies (61 KINGFISH
galaxies, plus 9 additional galaxies present in the observed fields)
are presented here.
Dust is detected reliably in 62 galaxies, and upper limits
are reported for the remaining 8.
Tables \ref{tab:spitzer} and \ref{tab:herschel} 
report the global galaxy photometry, 
and the best-fit dust parameter estimates are given in Table
\ref{tab:dust}.
Dust parameter maps are displayed in
Figs.\
17.1-17.62.
The DL07 dust model successfully reproduces the dust SEDs over the
wide variety of environments present in the KINGFISH sample.

Long-wavelength imaging can be omitted in order to increase the
angular resolution of the modeling, but results become unreliable if
the long-wavelength coverage is insufficient.
For maximum reliability, we recommend using all cameras available,
including MIPS160 and SPIRE250, SPIRE350, and SPIRE500.
If better angular resolution is critical, the lowest-resolution cameras
(SPIRE500 and MIP160) can be left out, but estimates of dust mass
become unreliable unless at least SPIRE250 is included.
If SPIRE350, SPIRE500, and MIPS160 are not included, the DL07 model dust
masses can be low by as much as a factor 0.8, or high by as much as
a factor 2 (see Figs.\ \ref{fig:multipix md for different psfs});
the median factor is 1.25.
$\qPAH$ and $\fPDR$
estimates are fairly insensitive to the camera combination used, 
so they can be obtained reliably without $\lambda > 250\micron$ photometry,
provided that the signal/noise ratio is adequate.

Resolved (multipixel) and  global (single-pixel) 
modeling generate similar estimates of 
$\Mdust$, $\qpah$, and $\fPDR$
when all the \Spitzer\ and \Herschel\ cameras are employed.
The single-pixel modeling tends to slightly underestimate the
total dust mass $\Mdust$ by $\sim$13\% 
(see Fig.\ \ref{fig:multipix vs singlepix}).

Our analysis shows that $\qpah$, 
the fraction of the dust mass contributed by PAHs,
correlates much better with the PP04N2 estimate for O/H than for the PT
estimate, strongly suggesting that PP04N2 is a better strong-line
abundance estimator than the PT estimator.
We find that
$\qpah$ appears to increase monotonically with increasing
metallicity, with $\qpah$ varying linearly with $\log({\rm O/H})$ for
\logohpp$>7.94$ (see Fig.\ \ref{fig:qpah}b and Eq.\ \ref{eq:pah_vs_oh}).

For most star-forming galaxies with metallicity $Z\gtsim Z_\odot$,
the dust/gas ratio is close to the limiting value where nearly all of
the refractory elements are locked up in grains.
However, at lower metallicity, the dust/gas ratio is often well below
this limiting value, consistent with what is expected from a simple
toy model with accretion rate $\tau_a^{-1}\propto Z_d$
(see Fig.\ \ref{fig:mdmh}b).

The 
resolved regions in the KINGFISH galaxy sample 
show several trends
with $\Umin$, $\qpah$, and mass-to-light ratios for dust emission.
$\Umin$ can be estimated from long-wavelength flux ratios
(e.g., $f_{160}/f_{500}$) to within a factor of two over more than
two orders of magnitude in $\Umin$ [see Eq.\ (\ref{eq:umin})].
From the same flux ratio, and with a measurement of dust luminosity,
dust mass can be estimated to within $\sim$50\%
[see Eq.\ (\ref{eq:ldustmdust_160500})].
Despite a variation of $\ga$3 orders of magnitude in IR surface brightness,
for the adopted physical dust model
it is possible to estimate dust mass from IR luminosity at 500\,\micron\ to 
within $\sim$0.07\,dex), affording an accuracy of $\sim$20\%
[see Eq.\ (\ref{eq:mass500})].
There are of course systematic errors coming from the choice of
dust model, but these are difficult to estimate.
Estimating gas mass for metal-poor galaxies requires incorporating
metallicity, because of the metallicity dependence of dust-to-gas ratios.
Our formulations for inferring starlight heating intensity 
and dust mass from flux ratios
and integrated IR or monochromatic luminosities have been calibrated 
over $\gtsim$22\,000 independent regions in 62 galaxies,
spanning 
metal-poor dwarf irregulars to grand-design spiral disks
and actively star-forming LIRGs.
These calibrated prescriptions are designed with the
aim of facilitating comparison with high-redshift
galaxies, where frequently rest-frame $f_{160}$ and at
least one longer wavelength flux are available.

\acknowledgements

\newtext{We thank the referee for helpful comments.}
We are grateful to R.H. Lupton for availability of the SM graphics
program.
L.K.H. thanks Princeton University for kind hospitality during 
a very pleasant and productive visit,
and acknowledges funding from the INAF PRIN-SKA 2017 program 1.05.01.88.04.
This research was supported in part by JPL grants 1329088 and 1373687,
and by NSF grants AST-0406883, AST-1008570, and AST-1408723.
KS was supported in part by NSF grant AST-1615728 and NASA ADP grant
NNX17AF39G.

\facilities{
Spitzer Space Telescope;
Herschel Space Observatory;
Karl G.\ Jansky Very Large Array;
IRAM 30\,m telescope;
Westerbork Synthesis Radio Telescope;
Nobeyama Radio Observatory.}

\software{
CASA \citep{McMullin+Waters+Schiebel+etal_2007};
SINGS Fifth Data Delivery Pipeline; 
Local Volume Legacy Project Pipeline;
HIPE v11.1.0 \citep{Ott_2010}; 
Scanamorphos v24.0 \citep{Roussel_2013};
R \citep{R_Core_Team_2014};
SM.}

\bibliography{/u/draine/work/bib/btdrefs}

\appendix

\section{Resolved dust parameter maps for KINGFISH galaxies}
\label{app:maps}

As described in the text, each galaxy where we have a positive dust
detection has two figures:
the first (a) shows the model
done at MIPS160 resolution, 
using data from all cameras (IRAC, MIPS, PACS, and SPIRE) cameras.
This is our ``gold standard'' modeling.
The second (b) shows a model at
SPIRE250 resolution, using
IRAC, MIPS24, PACS, and SPIRE250 cameras (i.e., omitting MIPS70, MIPS160,
SPIRE350, SPIRE500).  
This latter modeling, while able
to resolve smaller scale structures in the galaxies, is overall less
reliable.

Figures 17.1-17.62
each have twelve panels.
For each of the resolutions.
the top row is a map of
dust luminosity surface density
$\Sigma_{\Ldust}$ (left), 
dust surface density
$\Sigma_{M_\dust}$ (center), 
and the model SED (right).  
The lower row shows the starlight intensity parameter
$U_{\rm min,DL07}$ (left), 
the PAH abundance parameter $\qpah$ (center), 
and the PDR fraction $f_\PDR$ (left).
The dust luminosity surface density $\Sigma_{\Ldust}$ is shown for
the full field, with
the white contour showing the minimum surface brightness $\SigLdmin$
below which we do not attempt to model the emission.
Maps of derived quantities
($\Sigma_{\Mdust}$, $\Umin$, $\qpah$, and $f_\PDR$) are limited to
the ``galaxy mask'' region with $\Sigma_{\Ldust}>\SigLdmin$.
In the SED plot, the observed photometry is represented by
rectangular boxes ({\it Spitzer} (IRAC, MIPS) in red; {\it Herschel}
(PACS, SPIRE) in blue) showing $\pm1\sigma$ uncertainties.
The black line is a single-pixel DL07 model
that seeks to reproduce the observed SED,
with different components shown.
\newtext{The values of $\Umin$ and $\Mdust$ in the label are for the
DL07 model before renormalization.}
The cyan line is the stellar contribution, the dark red line is the
emission from dust heated by the power-law $U$ distribution, and the
dark green line is emission from dust heated by $U=\Umin$.


\newtext{Figures 17.1-17.5 are shown below as examples.
This paper with a complete figure set is available at\\
\url{http://www.astro.princeton.edu/~draine/KFdust/KFdust_full.pdf}
}
\bigskip

\figsetstart
\figsetnum{17}

\figsettitle{Dust Maps for 62 Galaxies}


\renewcommand{\galname}{Hol2}
\figsetgrpstart
\figsetgrpnum{17.1}
\figsetgrptitle{\galname}
\figsetplot{\galname_panel.pdf}

\figsetgrpnote{\galname: Model results at M160 PSF (rows 1 and 2) 
  and at S250 PSF (rows 3 and 4).
  Dust luminosity per area $\SigLd$ (column 1, rows 1 and 3) 
  is shown for entire field, with
  adopted galaxy mask boundary in white.
  Dust mass per area $\SigMd$ (column 2, rows 1 and 3) 
  is after renormalization (see text).
  $U_{\rm min,DL07}$, $\qpah$ and $\fPDR$ are shown in rows 2 and 4.
  The global SED (column 3, rows 1 and 3) is shown for single-pixel modeling,
  with contributions from dust heated by $\Umin$ (green), 
  dust heated by $U>\Umin$ (red)
  and starlight (cyan); 
  \newtext{values of $\Umin$ and $\Mdust$ in the figure label are for the
  DL07 model before renormalization.}
  {\it Herschel} (blue rectangles) and {\it Spitzer} (red rectangles) 
  photometry is shown;
  vertical extent is $\pm1\sigma$.
  Diamonds show the band-convolved flux for the model.}
\figsetgrpend


\renewcommand{\galname}{IC342}
\figsetgrpstart
\figsetgrpnum{17.2}
\figsetgrptitle{\galname}
\figsetplot{\galname_panel.pdf}

\figsetgrpnote{As in Figure \ref{fig:Hol2}, but for \galname. 
The $\qpah$ map is truncated to the NW because $8\micron$ imaging was unavailable.}
\figsetgrpend


\renewcommand{\galname}{IC2574}
\figsetgrpstart
\figsetgrpnum{17.3}
\figsetgrptitle{\galname}
\figsetplot{\galname_panel.pdf}

\figsetgrpnote{As in Figure \ref{fig:Hol2}, but for \galname.}
\figsetgrpend


\renewcommand{\galname}{NGC0337}
\figsetgrpstart
\figsetgrpnum{17.4}
\figsetgrptitle{\galname}
\figsetplot{\galname_panel.pdf}

\figsetgrpnote{As in Figure \ref{fig:Hol2}, but for \galname.}
\figsetgrpend


\renewcommand{\galname}{NGC0628}
\figsetgrpstart
\figsetgrpnum{17.5}
\figsetgrptitle{\galname}
\figsetplot{\galname_panel.pdf}

\figsetgrpnote{As in Figure \ref{fig:Hol2}, but for NGC0628\,=\,M74.}
\figsetgrpend


\renewcommand{\galname}{NGC0925}
\figsetgrpstart
\figsetgrpnum{17.6}
\figsetgrptitle{\galname}
\figsetplot{\galname_panel.pdf}

\figsetgrpnote{As in Figure \ref{fig:Hol2}, but for \galname.}
\figsetgrpend


\renewcommand{\galname}{NGC1097}
\figsetgrpstart
\figsetgrpnum{17.7}
\figsetgrptitle{\galname}
\figsetplot{\galname_panel.pdf}

\figsetgrpnote{As in Figure \ref{fig:Hol2}, but for \galname.}
\figsetgrpend

\renewcommand{\galname}{NGC1266}
\figsetgrpstart
\figsetgrpnum{17.8}
\figsetgrptitle{\galname}
\figsetplot{\galname_panel.pdf}

\figsetgrpnote{As in Figure \ref{fig:Hol2}, but for \galname.}
\figsetgrpend

\renewcommand{\galname}{NGC1291}
\figsetgrpstart
\figsetgrpnum{17.9}
\figsetgrptitle{\galname}
\figsetplot{\galname_panel.pdf}

\figsetgrpnote{As in Figure \ref{fig:Hol2}, but for \galname.}
\figsetgrpend

\renewcommand{\galname}{NGC1316}
\figsetgrpstart
\figsetgrpnum{17.10}
\figsetgrptitle{\galname}
\figsetplot{\galname_panel.pdf}

\figsetgrpnote{As in Figure \ref{fig:Hol2}, but for NGC1316\,=\,Fornax A.}
\figsetgrpend

\renewcommand{\galname}{NGC1377}
\figsetgrpstart
\figsetgrpnum{17.11}
\figsetgrptitle{\galname}
\figsetplot{\galname_panel.pdf}

\figsetgrpnote{As in Figure \ref{fig:Hol2}, but for \galname.}
\figsetgrpend

\renewcommand{\galname}{NGC1482}
\figsetgrpstart
\figsetgrpnum{17.12}
\figsetgrptitle{\galname}
\figsetplot{\galname_panel.pdf}

\figsetgrpnote{As in Figure \ref{fig:Hol2}, but for NGC1482.
High values of $\qpah$ to NW and SE may be due to bleeding in the original 
IRAC8 image.}
\figsetgrpend

\renewcommand{\galname}{NGC1512}
\figsetgrpstart
\figsetgrpnum{17.13}
\figsetgrptitle{\galname}
\figsetplot{\galname_panel.pdf}

\figsetgrpnote{As in Figure \ref{fig:Hol2}, but for \galname.}
\figsetgrpend

\renewcommand{\galname}{NGC2146}
\figsetgrpstart
\figsetgrpnum{17.14}
\figsetgrptitle{\galname}
\figsetplot{\galname_panel.pdf}

\figsetgrpnote{As in Figure \ref{fig:Hol2}, but for NGC2146.
High values of $\qpah$ to NNE and SSW may be due to bleeding in the original IRAC8 image.}
\figsetgrpend

\renewcommand{\galname}{NGC2798}
\figsetgrpstart
\figsetgrpnum{17.15}
\figsetgrptitle{\galname}
\figsetplot{\galname_panel.pdf}

\figsetgrpnote{As in Figure \ref{fig:Hol2}, but for \galname.}
\figsetgrpend

\renewcommand{\galname}{NGC2841}
\figsetgrpstart
\figsetgrpnum{17.16}
\figsetgrptitle{\galname}
\figsetplot{\galname_panel.pdf}

\figsetgrpnote{As in Figure \ref{fig:Hol2}, but for \galname.}
\figsetgrpend

\renewcommand{\galname}{NGC2915}
\figsetgrpstart
\figsetgrpnum{17.17}
\figsetgrptitle{\galname}
\figsetplot{\galname_panel.pdf}

\figsetgrpnote{As in Figure \ref{fig:Hol2}, but for \galname. 
}
\figsetgrpend

\renewcommand{\galname}{NGC2976}
\figsetgrpstart
\figsetgrpnum{17.18}
\figsetgrptitle{\galname}
\figsetplot{\galname_panel.pdf}

\figsetgrpnote{As in Figure \ref{fig:Hol2}, but for \galname.}
\figsetgrpend

\renewcommand{\galname}{NGC3049}
\figsetgrpstart
\figsetgrpnum{17.19}
\figsetgrptitle{\galname}
\figsetplot{\galname_panel.pdf}

\figsetgrpnote{As in Figure \ref{fig:Hol2}, but for \galname.}
\figsetgrpend

\renewcommand{\galname}{NGC3077}
\figsetgrpstart
\figsetgrpnum{17.20}
\figsetgrptitle{\galname}
\figsetplot{\galname_panel.pdf}

\figsetgrpnote{As in Figure \ref{fig:Hol2}, but for \galname.}
\figsetgrpend

\renewcommand{\galname}{NGC3184}
\figsetgrpstart
\figsetgrpnum{17.21}
\figsetgrptitle{\galname}
\figsetplot{\galname_panel.pdf}

\figsetgrpnote{As in Figure \ref{fig:Hol2}, but for \galname.}
\figsetgrpend

\renewcommand{\galname}{NGC3190}
\figsetgrpstart
\figsetgrpnum{17.22}
\figsetgrptitle{\galname}
\figsetplot{\galname_panel.pdf}

\figsetgrpnote{As in Figure \ref{fig:Hol2}, but for \galname.}
\figsetgrpend

\renewcommand{\galname}{NGC3198}
\figsetgrpstart
\figsetgrpnum{17.23}
\figsetgrptitle{\galname}
\figsetplot{\galname_panel.pdf}

\figsetgrpnote{As in Figure \ref{fig:Hol2}, but for \galname.}
\figsetgrpend

\renewcommand{\galname}{NGC3265}
\figsetgrpstart
\figsetgrpnum{17.24}
\figsetgrptitle{\galname}
\figsetplot{\galname_panel.pdf}

\figsetgrpnote{As in Figure \ref{fig:Hol2}, but for \galname.}
\figsetgrpend

\renewcommand{\galname}{NGC3351}
\figsetgrpstart
\figsetgrpnum{17.25}
\figsetgrptitle{\galname}
\figsetplot{\galname_panel.pdf}

\figsetgrpnote{As in Figure \ref{fig:Hol2}, but for NGC3351\,=\,M95.
High values of $\qpah$ to NNE and SSW may be due to bleeding in 
original IRAC8 image.}
\figsetgrpend

\renewcommand{\galname}{NGC3521}
\figsetgrpstart
\figsetgrpnum{17.26}
\figsetgrptitle{\galname}
\figsetplot{\galname_panel.pdf}

\figsetgrpnote{As in Figure \ref{fig:Hol2}, but for \galname.}
\figsetgrpend

\renewcommand{\galname}{NGC3621}
\figsetgrpstart
\figsetgrpnum{17.27}
\figsetgrptitle{\galname}
\figsetplot{\galname_panel.pdf}

\figsetgrpnote{As in Figure \ref{fig:Hol2}, but for \galname.}
\figsetgrpend

\renewcommand{\galname}{NGC3627}
\figsetgrpstart
\figsetgrpnum{17.28}
\figsetgrptitle{\galname}
\figsetplot{\galname_panel.pdf}

\figsetgrpnote{As in Figure \ref{fig:Hol2}, but for NGC3625\,=\,M66.}
\figsetgrpend

\renewcommand{\galname}{NGC3773}
\figsetgrpstart
\figsetgrpnum{17.29}
\figsetgrptitle{\galname}
\figsetplot{\galname_panel.pdf}

\figsetgrpnote{As in Figure \ref{fig:Hol2}, but for \galname.}
\figsetgrpend

\renewcommand{\galname}{NGC3938}
\figsetgrpstart
\figsetgrpnum{17.30}
\figsetgrptitle{\galname}
\figsetplot{\galname_panel.pdf}

\figsetgrpnote{As in Figure \ref{fig:Hol2}, but for \galname.}
\figsetgrpend

\renewcommand{\galname}{NGC4236}
\figsetgrpstart
\figsetgrpnum{17.31}
\figsetgrptitle{\galname}
\figsetplot{\galname_panel.pdf}

\figsetgrpnote{As in Figure \ref{fig:Hol2}, but for \galname.}
\figsetgrpend

\renewcommand{\galname}{NGC4254}
\figsetgrpstart
\figsetgrpnum{17.32}
\figsetgrptitle{\galname}
\figsetplot{\galname_panel.pdf}

\figsetgrpnote{As in Figure \ref{fig:Hol2}, but for NGC4254\,=\,M99. 
}
\figsetgrpend

\renewcommand{\galname}{NGC4321}
\figsetgrpstart
\figsetgrpnum{17.33}
\figsetgrptitle{\galname}
\figsetplot{\galname_panel.pdf}

\figsetgrpnote{As in Figure \ref{fig:Hol2}, but for NGC4321\,=\,M100.}
\figsetgrpend

\renewcommand{\galname}{NGC4536}
\figsetgrpstart
\figsetgrpnum{17.34}
\figsetgrptitle{\galname}
\figsetplot{\galname_panel.pdf}

\figsetgrpnote{As in Figure \ref{fig:Hol2}, but for \galname.}
\figsetgrpend

\renewcommand{\galname}{NGC4559}
\figsetgrpstart
\figsetgrpnum{17.35}
\figsetgrptitle{\galname}
\figsetplot{\galname_panel.pdf}

\figsetgrpnote{As in Figure \ref{fig:Hol2}, but for \galname.}
\figsetgrpend

\renewcommand{\galname}{NGC4569}
\figsetgrpstart
\figsetgrpnum{17.36}
\figsetgrptitle{\galname}
\figsetplot{\galname_panel.pdf}

\figsetgrpnote{As in Figure \ref{fig:Hol2}, but for NGC4569\,=\,M90.}
\figsetgrpend

\renewcommand{\galname}{NGC4579}
\figsetgrpstart
\figsetgrpnum{17.37}
\figsetgrptitle{\galname}
\figsetplot{\galname_panel.pdf}

\figsetgrpnote{As in Figure \ref{fig:Hol2}, but for NGC4579\,=\,M58.}
\figsetgrpend

\renewcommand{\galname}{NGC4594}
\figsetgrpstart
\figsetgrpnum{17.38}
\figsetgrptitle{\galname}
\figsetplot{\galname_panel.pdf}

\figsetgrpnote{As in Figure \ref{fig:Hol2}, but for NGC4594\,=\,M104\,=\,Sombrero.}
\figsetgrpend

\renewcommand{\galname}{NGC4625}
\figsetgrpstart
\figsetgrpnum{17.39}
\figsetgrptitle{\galname}
\figsetplot{\galname_panel.pdf}

\figsetgrpnote{As in Figure \ref{fig:Hol2}, but for \galname.}
\figsetgrpend

\renewcommand{\galname}{NGC4631}
\figsetgrpstart
\figsetgrpnum{17.40}
\figsetgrptitle{\galname}
\figsetplot{\galname_panel.pdf}

\figsetgrpnote{As in Figure \ref{fig:Hol2}, but for \galname.}
\figsetgrpend

\renewcommand{\galname}{NGC4725}
\figsetgrpstart
\figsetgrpnum{17.41}
\figsetgrptitle{\galname}
\figsetplot{\galname_panel.pdf}

\figsetgrpnote{As in Figure \ref{fig:Hol2}, but for \galname.}
\figsetgrpend

\renewcommand{\galname}{NGC4736}
\figsetgrpstart
\figsetgrpnum{17.42}
\figsetgrptitle{\galname}
\figsetplot{\galname_panel.pdf}

\figsetgrpnote{As in Figure \ref{fig:Hol2}, but for NGC4736\,=\,M94.}
\figsetgrpend

\renewcommand{\galname}{NGC4826}
\figsetgrpstart
\figsetgrpnum{17.43}
\figsetgrptitle{\galname}
\figsetplot{\galname_panel.pdf}

\figsetgrpnote{As in Figure \ref{fig:Hol2}, but for NGC4826\,=\,M64.}
\figsetgrpend

\renewcommand{\galname}{NGC5055}
\figsetgrpstart
\figsetgrpnum{17.44}
\figsetgrptitle{\galname}
\figsetplot{\galname_panel.pdf}

\figsetgrpnote{As in Figure \ref{fig:Hol2}, but for NGC5055\,=\,M63.}
\figsetgrpend

\renewcommand{\galname}{NGC5398}
\figsetgrpstart
\figsetgrpnum{17.45}
\figsetgrptitle{\galname}
\figsetplot{\galname_panel.pdf}

\figsetgrpnote{As in Figure \ref{fig:Hol2}, but for \galname.}
\figsetgrpend

\renewcommand{\galname}{NGC5408}
\figsetgrpstart
\figsetgrpnum{17.46}
\figsetgrptitle{\galname}
\figsetplot{\galname_panel.pdf}

\figsetgrpnote{As in Figure \ref{fig:Hol2}, but for \galname.}
\figsetgrpend

\renewcommand{\galname}{NGC5457}
\figsetgrpstart
\figsetgrpnum{17.47}
\figsetgrptitle{\galname}
\figsetplot{\galname_panel.pdf}

\figsetgrpnote{As in Figure \ref{fig:Hol2}, but for NGC5457\,=M101\,=Pinwheel.}
\figsetgrpend

\renewcommand{\galname}{NGC5474}
\figsetgrpstart
\figsetgrpnum{17.48}
\figsetgrptitle{\galname}
\figsetplot{\galname_panel.pdf}

\figsetgrpnote{As in Figure \ref{fig:Hol2}, but for \galname.}
\figsetgrpend

\renewcommand{\galname}{NGC5713}
\figsetgrpstart
\figsetgrpnum{17.49}
\figsetgrptitle{\galname}
\figsetplot{\galname_panel.pdf}

\figsetgrpnote{As in Figure \ref{fig:Hol2}, but for NGC5713.
High values of $\qpah$ to NNE and SSW may be due to bleeding in the IRAC8 image.}
\figsetgrpend

\renewcommand{\galname}{NGC5866}
\figsetgrpstart
\figsetgrpnum{17.50}
\figsetgrptitle{\galname}
\figsetplot{\galname_panel.pdf}

\figsetgrpnote{As in Figure \ref{fig:Hol2}, but for NGC5866\,=\,M102.}
\figsetgrpend

\renewcommand{\galname}{NGC6946}
\figsetgrpstart
\figsetgrpnum{17.51}
\figsetgrptitle{\galname}
\figsetplot{\galname_panel.pdf}

\figsetgrpnote{As in Figure \ref{fig:Hol2}, but for \galname.}
\figsetgrpend

\renewcommand{\galname}{NGC7331}
\figsetgrpstart
\figsetgrpnum{17.52}
\figsetgrptitle{\galname}
\figsetplot{\galname_panel.pdf}

\figsetgrpnote{As in Figure \ref{fig:Hol2}, but for \galname.}
\figsetgrpend

\renewcommand{\galname}{NGC7793}
\figsetgrpstart
\figsetgrpnum{17.53}
\figsetgrptitle{\galname}
\figsetplot{\galname_panel.pdf}

\figsetgrpnote{As in Figure \ref{fig:Hol2}, but for \galname.}
\figsetgrpend

\renewcommand{\galname}{IC3583}
\figsetgrpstart
\figsetgrpnum{17.54}
\figsetgrptitle{\galname}
\figsetplot{\galname_panel.pdf}

\figsetgrpnote{As in Figure \ref{fig:Hol2}, but for \galname.}
\figsetgrpend

\renewcommand{\galname}{NGC0586}
\figsetgrpstart
\figsetgrpnum{17.55}
\figsetgrptitle{\galname}
\figsetplot{\galname_panel.pdf}

\figsetgrpnote{As in Figure \ref{fig:Hol2}, but for NGC0586.
$\qpah$ is shown only for pixels where IRAC8 data were available.
The SED at 3.6, 4.5, 6.0, 8.0$\micron$ includes only the flux from the 
small fraction of the galaxy with IRAC imaging.}
\figsetgrpend

\renewcommand{\galname}{NGC1317}
\figsetgrpstart
\figsetgrpnum{17.56}
\figsetgrptitle{\galname}
\figsetplot{\galname_panel.pdf}

\figsetgrpnote{As in Figure \ref{fig:Hol2}, but for \galname.}
\figsetgrpend

\renewcommand{\galname}{NGC1481}
\figsetgrpstart
\figsetgrpnum{17.57}
\figsetgrptitle{\galname}
\figsetplot{\galname_panel.pdf}

\figsetgrpnote{As in Figure \ref{fig:Hol2}, but for \galname.}
\figsetgrpend

\renewcommand{\galname}{NGC1510}
\figsetgrpstart
\figsetgrpnum{17.58}
\figsetgrptitle{\galname}
\figsetplot{\galname_panel.pdf}

\figsetgrpnote{As in Figure \ref{fig:Hol2}, but for \galname.}
\figsetgrpend

\renewcommand{\galname}{NGC3187}
\figsetgrpstart
\figsetgrpnum{17.59}
\figsetgrptitle{\galname}
\figsetplot{\galname_panel.pdf}

\figsetgrpnote{As in Figure \ref{fig:Hol2}, but for \galname.}
\figsetgrpend

\renewcommand{\galname}{NGC4533}
\figsetgrpstart
\figsetgrpnum{17.60}
\figsetgrptitle{\galname}
\figsetplot{\galname_panel.pdf}

\figsetgrpnote{As in Figure \ref{fig:Hol2}, but for \galname.}
\figsetgrpend

\renewcommand{\galname}{NGC7335}
\figsetgrpstart
\figsetgrpnum{17.61}
\figsetgrptitle{\galname}
\figsetplot{\galname_panel.pdf}

\figsetgrpnote{As in Figure \ref{fig:Hol2}, but for \galname.}
\figsetgrpend

\renewcommand{\galname}{NGC7337}
\figsetgrpstart
\figsetgrpnum{17.62}
\figsetgrptitle{\galname}
\figsetplot{\galname_panel.pdf}

\figsetgrpnote{As in Figure \ref{fig:Hol2}, but for \galname.}
\figsetgrpend


\figsetend

\renewcommand{\galname}{Hol2}
\begin{figure*}
\epsscale{1.10}
\plotone{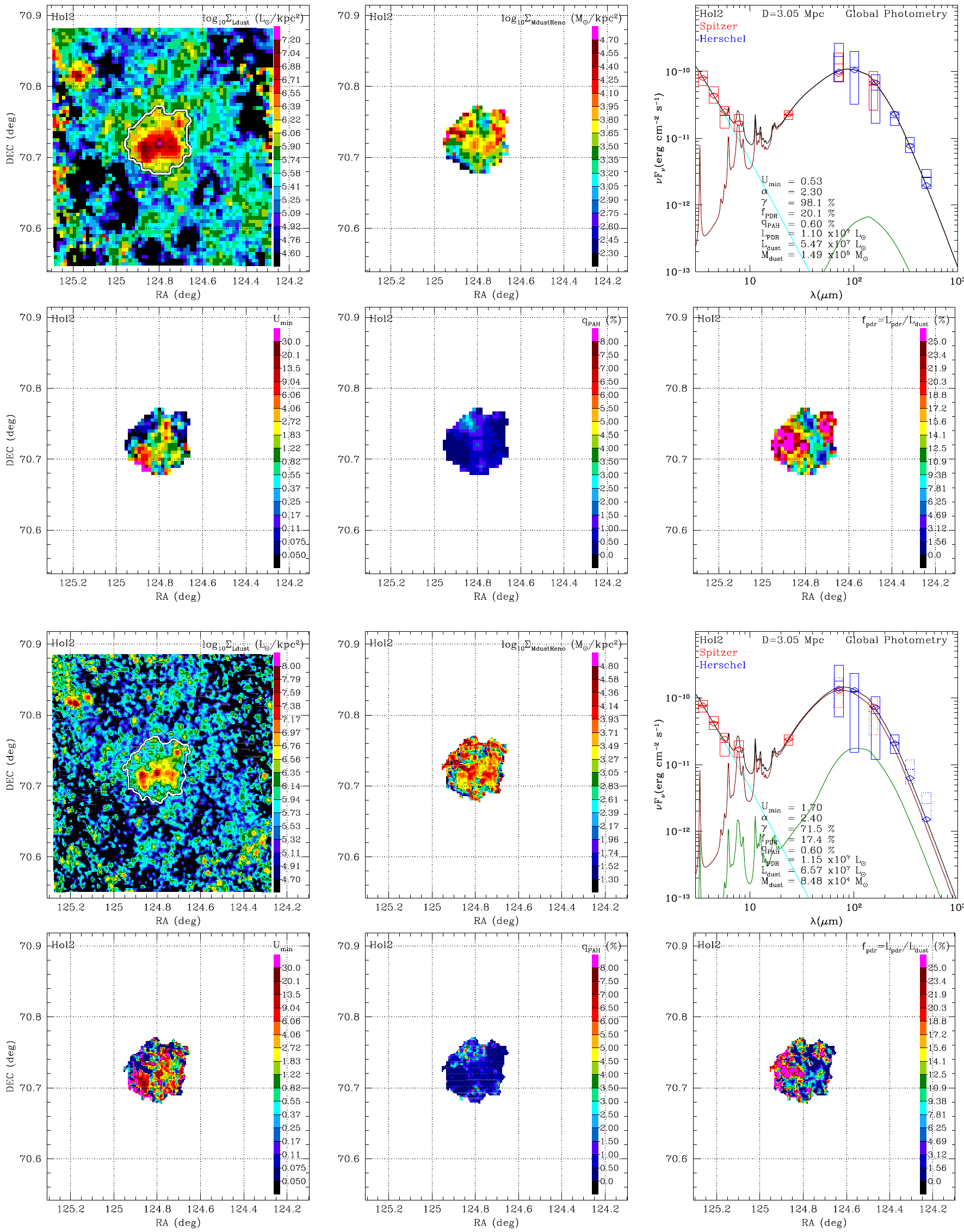}
\caption{\!\!\!{\bf 1} \galname: Model results at M160 PSF 
  (rows 1 and 2) and at S250 PSF (rows 3 and 4).
  Dust luminosity per area $\SigLd$ (column 1, rows 1 and 3) is shown 
  for entire field, with adopted galaxy mask boundary in white.
  Dust mass per area $\SigMd$ (column 2, rows 1 and 3) is after renormalization
  (see text).
  $U_{\rm min,DL07}$, $\qpah$ and $\fPDR$ are shown in rows 2 and 4.
  The global SED (column 3, rows 1 and 3) is shown for single-pixel modeling, with
  contributions from dust heated by $\Umin$ (green), dust heated by $U>\Umin$ (red)
  and starlight (cyan);
\newtext{values of $\Umin$ and $\Mdust$ in the figure label are for the
DL07 model before renormalization.} 
  {\it Herschel} (blue rectangles) and {\it Spitzer} (red rectangles) photometry is
  shown; vertical extent is $\pm1\sigma$.
  Diamonds show the band-convolved flux for the model.}
\end{figure*}

\renewcommand{\galname}{IC342}
\begin{figure*}
\addtocounter{figure}{-1}
\epsscale{1.10}
\plotone{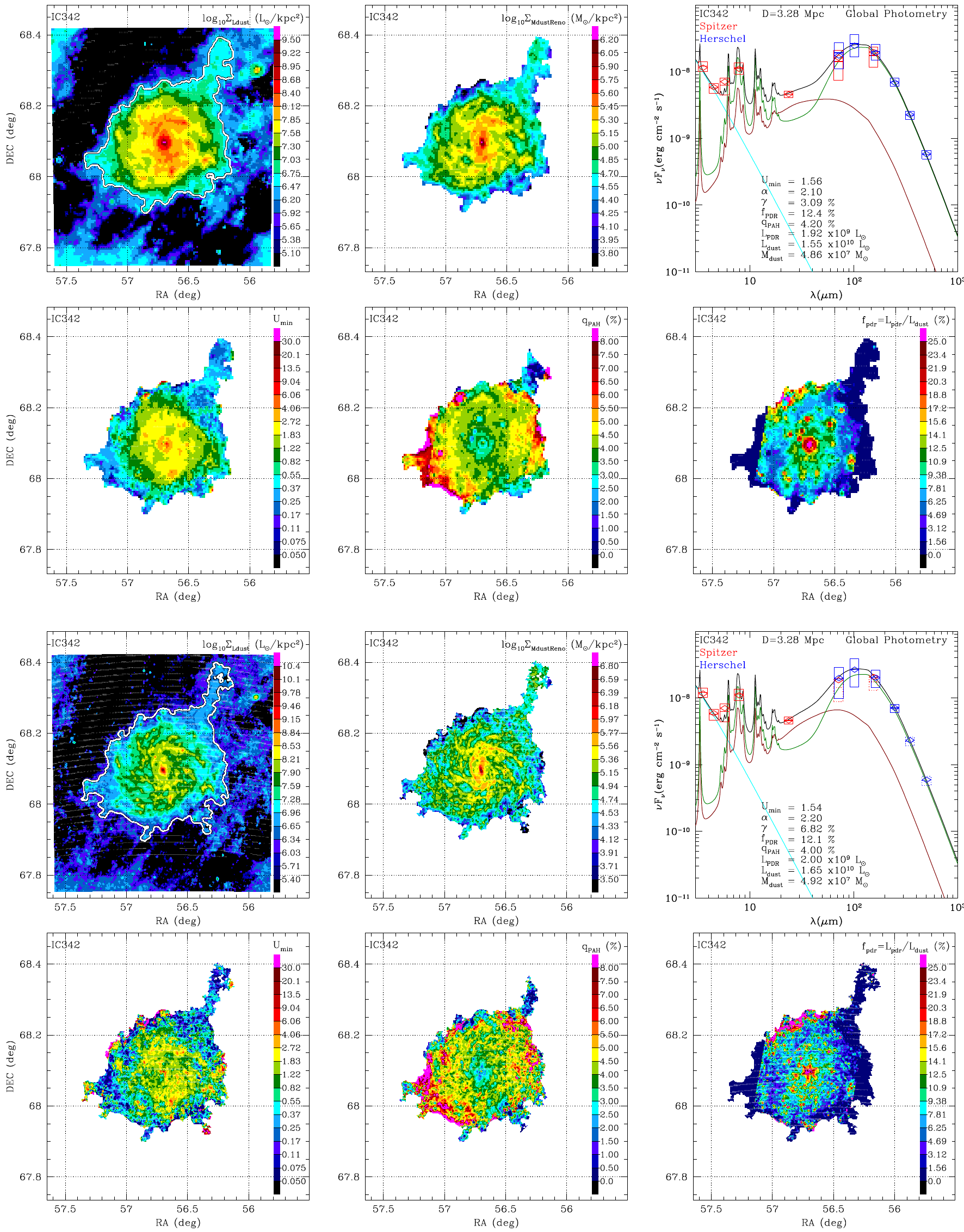}
\caption{\!\!\!{\bf 2} As in Figure 17.1, but for \galname.
  The $\qpah$ map is truncated to the NW because $8\micron$ imaging was unavailable.
  }
\end{figure*}

\renewcommand{\galname}{IC2574}
\begin{figure*}
\addtocounter{figure}{-1}
\epsscale{1.10}
\plotone{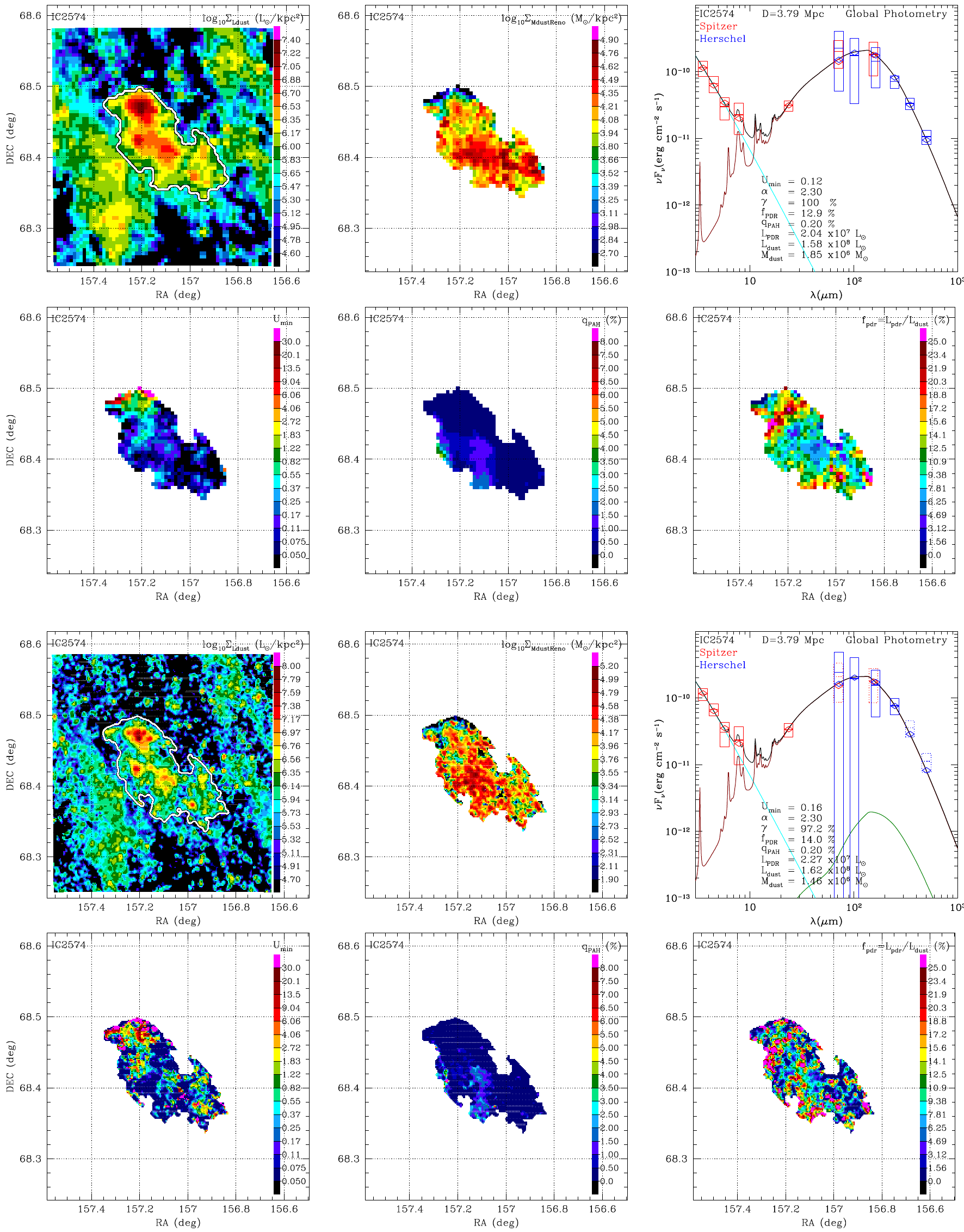}
\caption{\!\!\!{\bf 3} As in Figure 17.1, but for \galname.}
\end{figure*}

\renewcommand{\galname}{NGC0337}
\begin{figure*}
\addtocounter{figure}{-1}
\epsscale{1.10}
\plotone{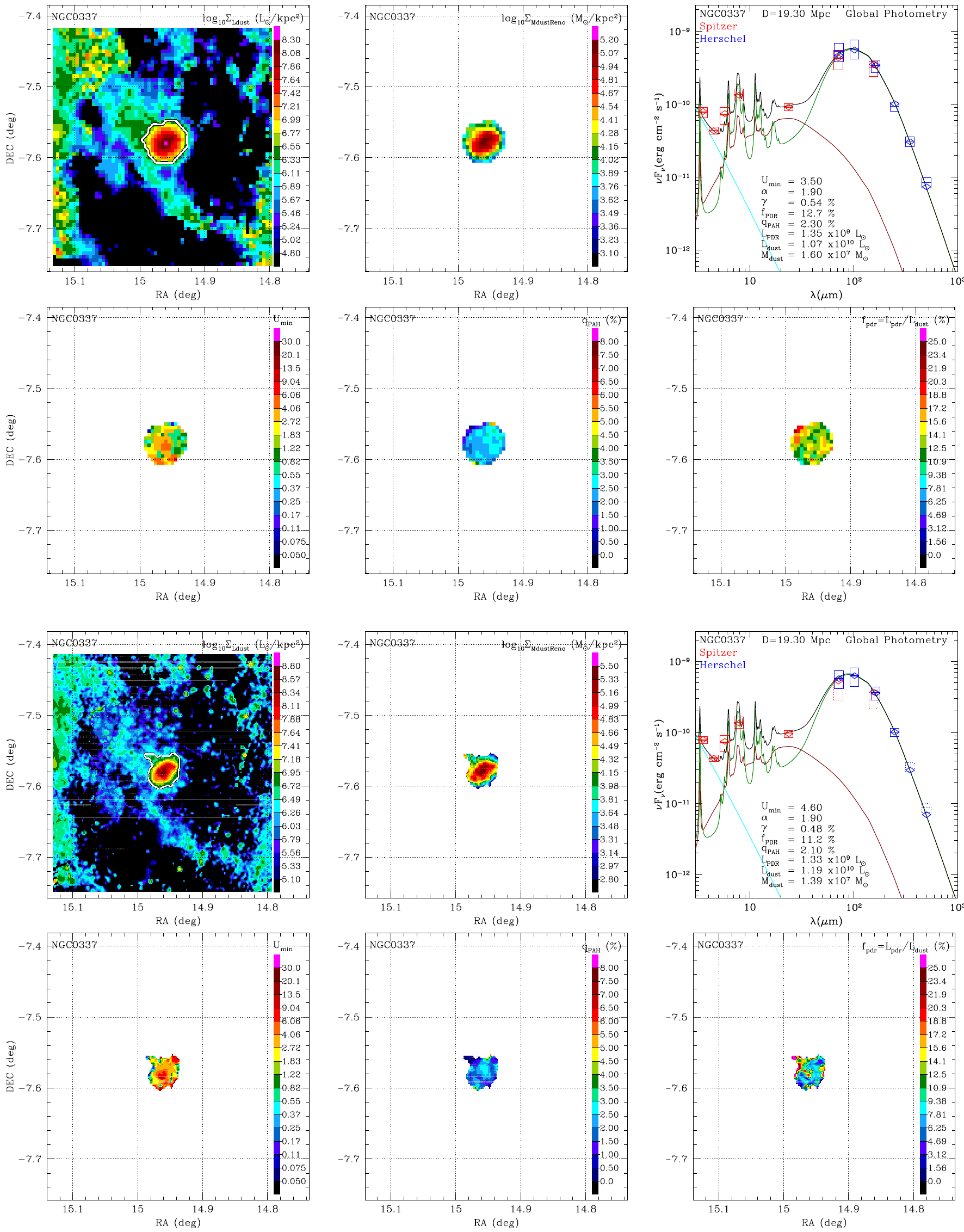}
\caption{\!\!\!{\bf 4} As in Figure 17.1, but for \galname.}
\end{figure*}

\renewcommand{\galname}{NGC0628}
\begin{figure*}
\addtocounter{figure}{-1}
\epsscale{1.10}
\plotone{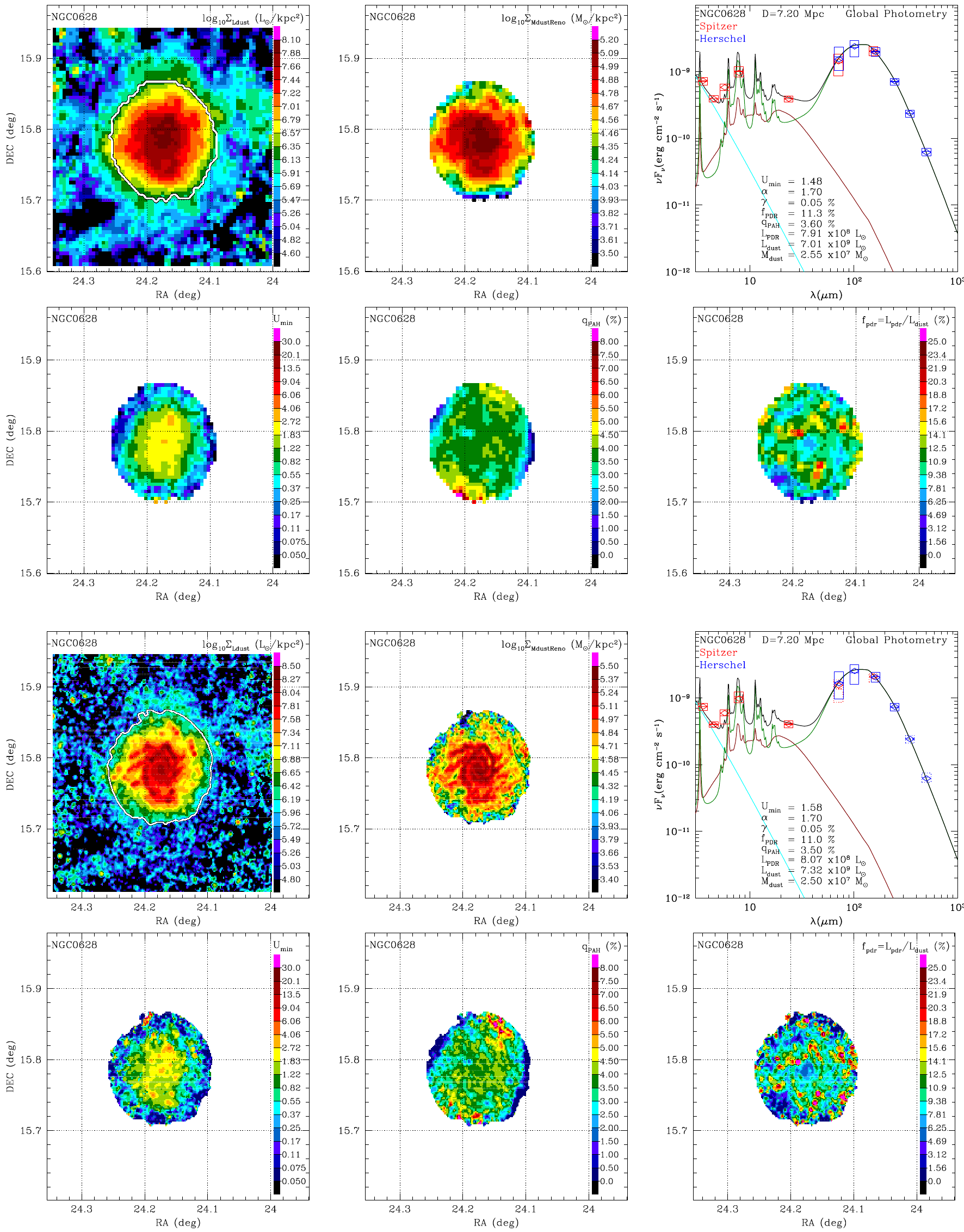}
\caption{\!\!\!{\bf 5} As in Figure 17.1, but for \galname\,=\,M74.}
\end{figure*}

\clearpage

\section{Cases with $3\sigma$ upper limits for dust mass}
\label{app:upper limits}
Eight galaxies in the KINGFISH sample – five dwarfs (DDO053, DDO154,
DDO165, Hol1, and M81dwB) and three ellipticals (NGC0584, NGC0855, and
NGC1404) yield upper limits on the dust mass from our modeling.  In
these cases, the signal from the galaxy in the far-IR is of comparable
magnitude to the contamination from background galaxies, leading to
uncertain dust mass measurements.

\newtext{For the 3 elliptical galaxies, we employ
a $\SigLd$-based mask that roughly
coincides with the optical galaxy.}
To obtain an upper limit on the dust mass in the case of these
non-detections, we randomly shift the \omittext{galaxy} mask around in the 
\newtext{M160 resolution} dust
mass image, avoiding overlap with the original mask, and remeasure the
total dust mass. 
We construct a distribution of these ``background''
dust mass measurements to obtain a mean and standard deviation.  In
all galaxies mentioned above, the mean of the ``background'' dust mass
values is positive, as would be expected due to the real signal at
far-IR wavelengths \omittext{due to confusion noise.}
\newtext{from unidentified background sources (``confusion noise'').}
The measured dust mass at
the expected galaxy location is within $\sim1\sigma$ of the mean.  In
the text and Table \ref{tab:Mdupperlimits} 
we provide the $3\sigma$ upper limit on the dust
mass generated with this procedure.

The definition of the galaxy mask itself is also potentially affected
by confusion noise.  In the case of the dwarf galaxies, we use the \ion{H}{1}
observations from THINGS and Little THINGS to create an alternative
galaxy mask, based on a cut at an \ion{H}{1} column density of 
$10^{20}\cm^{-2}$ from the \ion{H}{1} image convolved to M160 resolution.  The
\ion{H}{1}-based galaxy mask is typically somewhat larger than that defined by
the dust luminosity surface density cut.  We apply the same procedure
described above to obtain the background mean and standard deviation.

\begin{table}[h]
  \footnotesize
  \begin{center}
    \caption{\label{tab:Mdupperlimits} Dust Upper Limits}
    \begin{tabular}{l c c}
      \hline
      Galaxy  & $\Mdust (\Msol)$ & method\\
      \hline
      DDO053  & $<2.1\times10^5$ & \ion{H}{1} mask\\
      DDO154  & $<6.1\times10^5$ & \ion{H}{1} mask\\
      \omittext{Hol1}\newtext{DDO165}    & $<5.2\times10^5$ & \ion{H}{1} mask\\
      \omittext{Hol2}\newtext{Hol1}    & $<6.7\times10^5$ & \ion{H}{1} mask\\
      M81dwB  & $<8.1\times10^4$ & \ion{H}{1} mask\\
      NGC0584 & $<1.6\times10^6$ & $\SigLd$ mask\\
      NGC0855 & $<1.0\times10^6$ & $\SigLd$ mask\\
      NGC1404 & $<2.0\times10^6$ & $\SigLd$ mask\\
      \hline
    \end{tabular}\\
    \btdnote{data from maketables/kingfish\_dust\_nondet.dat}
  \end{center}
\end{table}
      
Table \ref{tab:Mdupperlimits} and Figures 18.1-18.8
provide the results of this procedure.  
\omittext{We report} \newtext{Table \ref{tab:Mdupperlimits} lists}
the $3\sigma$ upper limits for each galaxy using the \ion{H}{1}-based masks
for the dwarfs and the dust luminosity surface density masks (as
described in the text) for the ellipticals.  
\newtext{Histograms of the dust masses from the randomly shifted masks are shown in Figures 18.1-18.8.}
We note that the dust
mass limits from this procedure are expected to be very conservative.
Higher S/N could be obtained by a careful treatment of the integrated
photometry for each galaxy taking the confusion noise into account.

\newtext{Figure 18.1 (DDO053) and Figure 18.5 (NGC0584) are shown below
as examples.
This paper with a complete figure set is available at\\
\url{http://www.astro.princeton.edu/~draine/KFdust/KFdust_full.pdf}
}
\bigskip

\figsetstart
\figsetnum{18}
\figsettitle{8 Galaxies with Upper Limits for Dust Mass: 
DDO053, DDO154, DDO165, Hol1,
M81dwB, NGC0584, NGC0855, NGC1404}


\renewcommand{\galname}{DDO053}
\figsetgrpstart
\figsetgrpnum{18.1}
\figsetgrptitle{\galname}
\figsetplot{\galname_panel.pdf}

\figsetgrpnote{\galname: 
  Left: $\SigLd$ map;
  contours: $\SigLdmin=0.6\Lsol\pc^{-2}$. 
  Center: $\SigMd$ map \newtext{within the galaxy mask}. 
  Right: histogram of $\Mdust$ for \ion{H}{1}-based mask, shifted randomly.
  \newtext{Red dot-}dashed line: $\Mdust$ for mask centered on galaxy.
  68\% of random masks give $\Mdust$ between
  \omittext{dotted} \newtext{green dashed} lines.}
\figsetgrpend


\renewcommand{\galname}{DDO154}
\figsetgrpnum{18.2}
\figsetgrptitle{\galname}
\figsetplot{\galname_panel.pdf}
\figsetgrpnote{
  As in Fig.\ 18.1, but for \galname.
  Contours: $\SigLdmin=0.44\Lsol\pc^{-2}$.}
\figsetgrpend


\renewcommand{\galname}{DDO165}
\figsetgrpnum{18.3}
\figsetgrptitle{\galname}
\figsetplot{\galname_panel.pdf}
\figsetgrpnote{
  As in Fig.\ 18.1, but for \galname.
  Contours: $\SigLdmin=0.38\Lsol\pc^{-2}$.} 
\figsetgrpend


\renewcommand{\galname}{Hol1}
\figsetgrpnum{18.4}
\figsetgrptitle{\galname}
\figsetplot{\galname_panel.pdf}
\figsetgrpnote{
  As in Fig.\ 18.1, but for \galname.
  Contours: $\SigLdmin=0.48\Lsol\pc^{-2}$.}
\figsetgrpend


\renewcommand{\galname}{M81dwB}
\figsetgrpnum{18.5}
\figsetgrptitle{\galname}
\figsetplot{\galname_panel.pdf}
\figsetgrpnote{
  As in Fig.\ 18.1, but for \galname.
  Contours: $\SigLdmin=0.96\Lsol\pc^{-2}$.}
\figsetgrpend


\renewcommand{\galname}{NGC0584}
\figsetgrpnum{18.6}
\figsetgrptitle{\galname}
\figsetplot{\galname_panel.pdf}
\figsetgrpnote{
  As in Fig.\ 18.1, but for \galname.
  Contours: $\SigLdmin=1.4\Lsol\pc^{-2}$.
  Histogram: $\Mdust$ for $\SigLd$-based masks, shifted randomly.} 
\figsetgrpend


\renewcommand{\galname}{NGC0855}
\figsetgrpnum{18.7}
\figsetgrptitle{\galname}
\figsetplot{\galname_panel.pdf}
\figsetgrpnote{
  As in Fig.\ 18.1, but for \galname.
  Contours: $\SigLdmin=1.3\Lsol\pc^{-2}$.
  Histogram: $\Mdust$ for $\SigLd$-based masks, shifted randomly.}
\figsetgrpend

\renewcommand{\galname}{NGC1404}

\figsetgrpnum{18.8}
\figsetgrptitle{\galname}
\figsetplot{\galname_panel.pdf}
\figsetgrpnote{
  As in Fig.\ 18.1, but for \galname.
  Contours: $\SigLdmin=0.5\Lsol\pc^{-2}$.
  Histogram: $\Mdust$ for $\SigLd$-based masks, shifted randomly.}
\figsetgrpend


\figsetend

\renewcommand{\galname}{DDO053}
\begin{figure}[h]
\epsscale{1.10}
\plotone{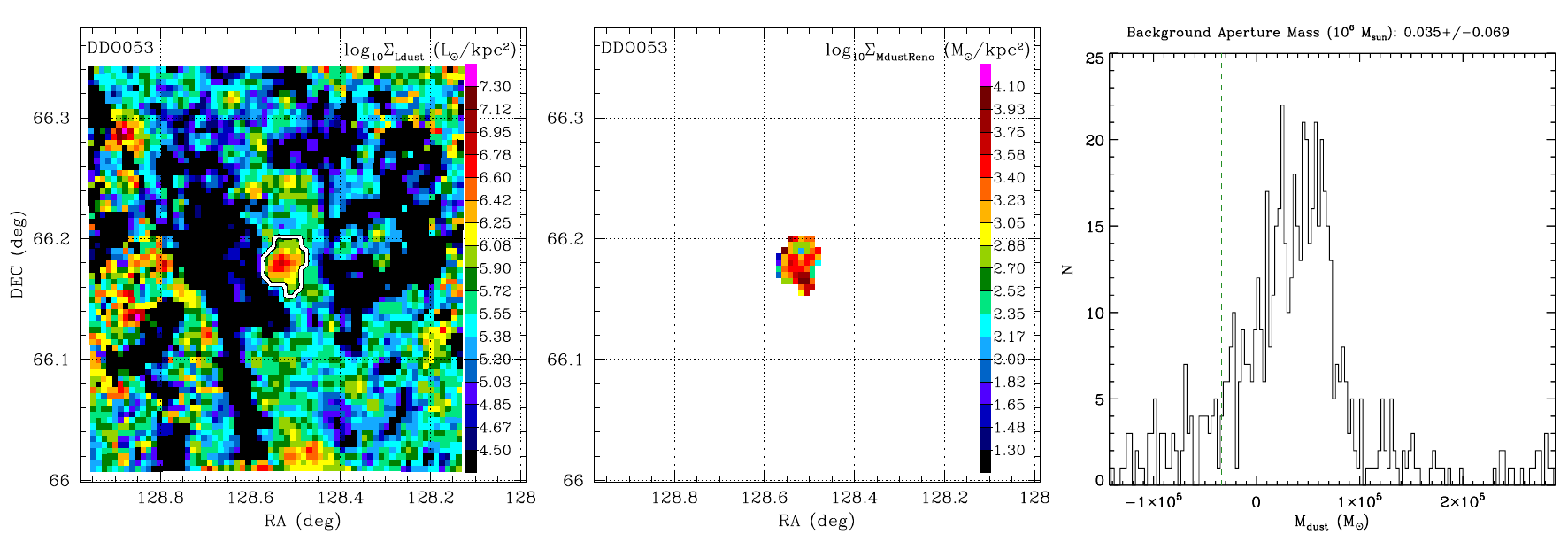}
\caption{\!\!\!{\bf 1} \galname:
  Left: $\SigLd$ map;
  contours: $\SigLdmin=0.6\Lsol\pc^{-2}$. 
  Center: $\SigMd$ map \newtext{within the galaxy mask}. 
  Right: histogram of $\Mdust$ for \ion{H}{1}-based mask, shifted randomly.
  \newtext{Red dot-}dashed line: $\Mdust$ for mask centered on galaxy.
  68\% of random masks give $\Mdust$ between
  \omittext{dotted} \newtext{green dashed} lines.}
\end{figure}

\renewcommand{\galname}{NGC0584}
\begin{figure}
\addtocounter{figure}{-1}
\epsscale{1.10}
\plotone{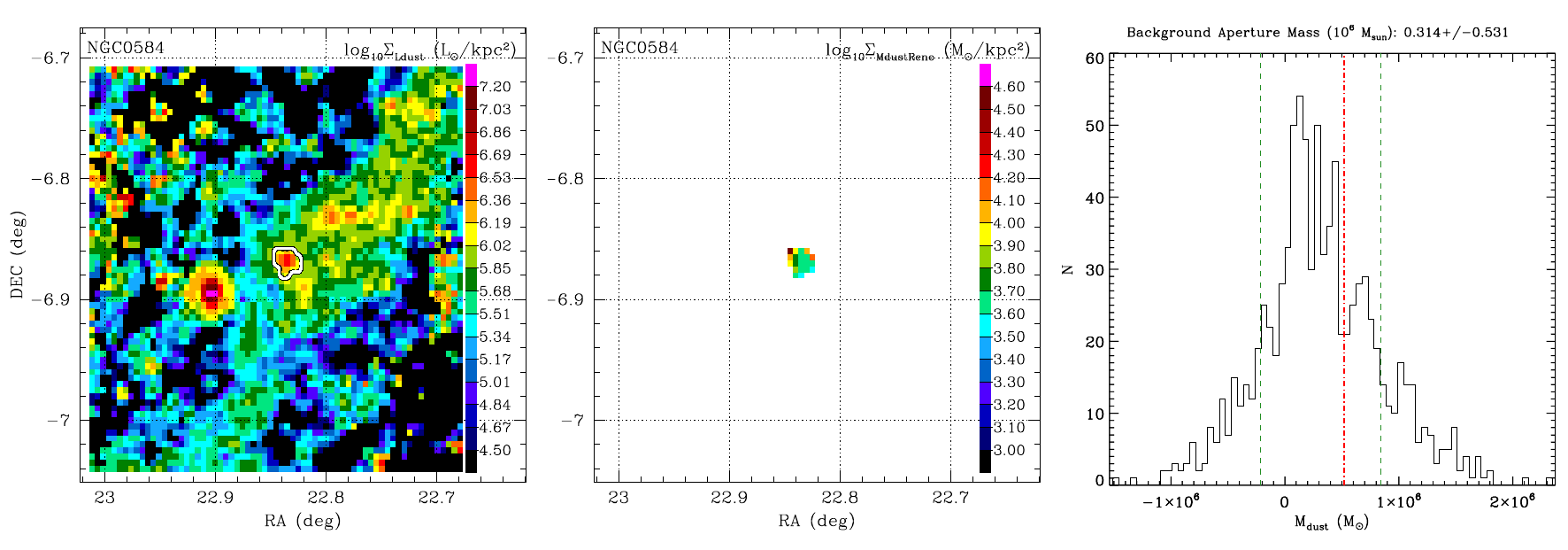}
\caption{\!\!\!{\bf 5} \galname:
  Left: $\SigLd$ map;  
  contours: $\SigLdmin=0.6\Lsol\pc^{-2}$.
  Center: $\SigMd$ map \newtext{within the galaxy mask}.
  Right: histogram of $\Mdust$ for $\SigLd$-based mask, shifted randomly.
  \newtext{Red dot-}dashed line: $\Mdust$ for mask centered on galaxy.
  68\% of random masks give $\Mdust$ between 
  \omittext{dotted} \newtext{green dashed} lines.}
\end{figure}


\newpage
\section{On-line KINGFISH data and dust models}
\label{app:online data}

The processed KINGFISH imaging and dust models are available online
at 
\url{http://www.astro.princeton.edu/~draine/KFdust/KFdustsite/}

Here we briefly describe the types of data that are available there.

For each of the 70 galaxies (61 KINGFISH galaxies + 9 ``extras'')
we provide results for resolved modeling at 4 different resolutions:
M160, S500, S350, and S250.
For each case, we use all compatible cameras (see Table \ref{tab:resolutions}).
FITS files of the following maps are provided:
\begin{itemize}
\item Dust mass surface density $\SigMd$ (renormalized).
\item Dust luminosity per unit projected area $\SigLd$.
\item PAH mass fraction $\qpah$.
\item $U_{\rm min,DL07}$ = minimum starlight intensity 
      parameter for the DL07 model.
      The renormalized $U_{\rm min}$ can be obtained from
      $U_{\rm min,DL07}$ using Eq.\ (\ref{eq:Umin corrected}).
\item $\bar{U}_{\rm DL07}$ = 
      mean starlight intensity parameter for the DL07 model.
      The renormalized $\Ubar$ can be obtained from $\bar{U}_{\rm DL07}$
      using Eq.\ (\ref{eq:Ubar corrected}).
\item $\fpdr$ = fraction of the total starlight heating of dust taking place in
      subregions where $U>10^2$.
\item global SED for the dust model.
\end{itemize}
For each case, the data in the FITS files are limited to the ``galaxy mask''
defined by $\SigLd>\SigLdmin$, where $\SigLdmin$ for each galaxy is given in
Table \ref{tab:geom}.

\section{Dependence of model results on PSF used and wavelength coverage}
\label{app:dependence on psf}

Obviously, one would like to model the dust emission with the best
angular resolution that is feasible.  Some of the cameras (e.g.,
PACS160) have small PSFs, which would seem to allow observations and
modeling with high angular resolution.  However, deciding to use
a small PSF means not being able to use data from cameras with
larger PSFs, which both reduces the amount of redundant data (e.g.,
MIPS70 and MIPS160) and limits the wavelength coverage by preventing
use of the longer wavelength cameras (e.g., SPIRE500).
In addition, use of a smaller PSF implies a lower
signal/noise ratio, which is a limiting factor in low surface
brightness regions.

Here we examine the degree to which derived dust and starlight
parameters are sensitive to the
choice of PSF.
We also compare the results obtained from the resolved modeling and those from global photometry.

\subsection{Comparison of modeling at S250
resolution with the gold standard (M160)}

\begin{figure*}  
\centering 
\includegraphics[width=6.0cm,clip=true,trim=0.5cm 5.0cm 0.5cm 2.5cm]{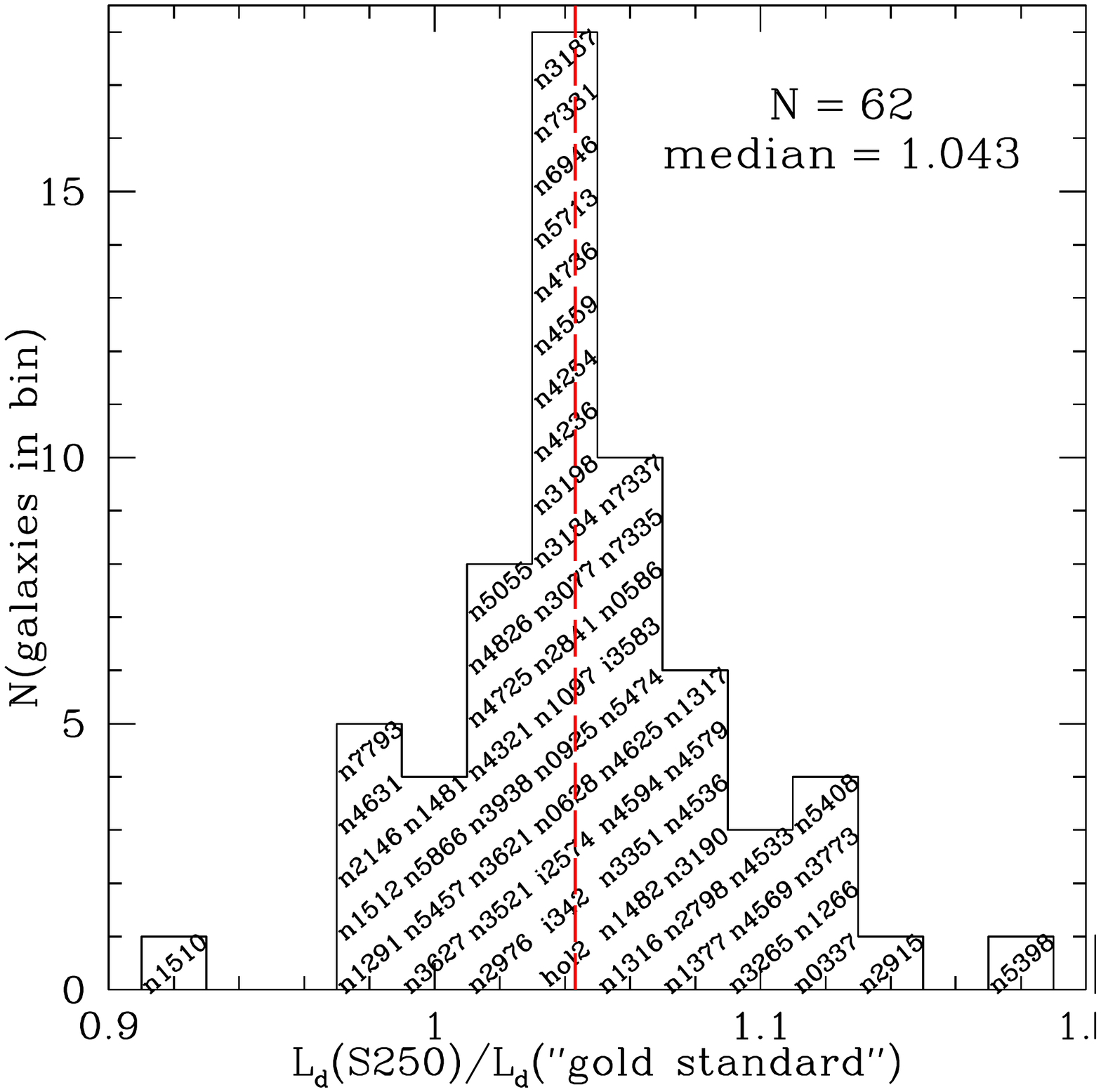} 
\includegraphics[width=6.0cm,clip=true,trim=0.5cm 5.0cm 0.5cm 2.5cm]{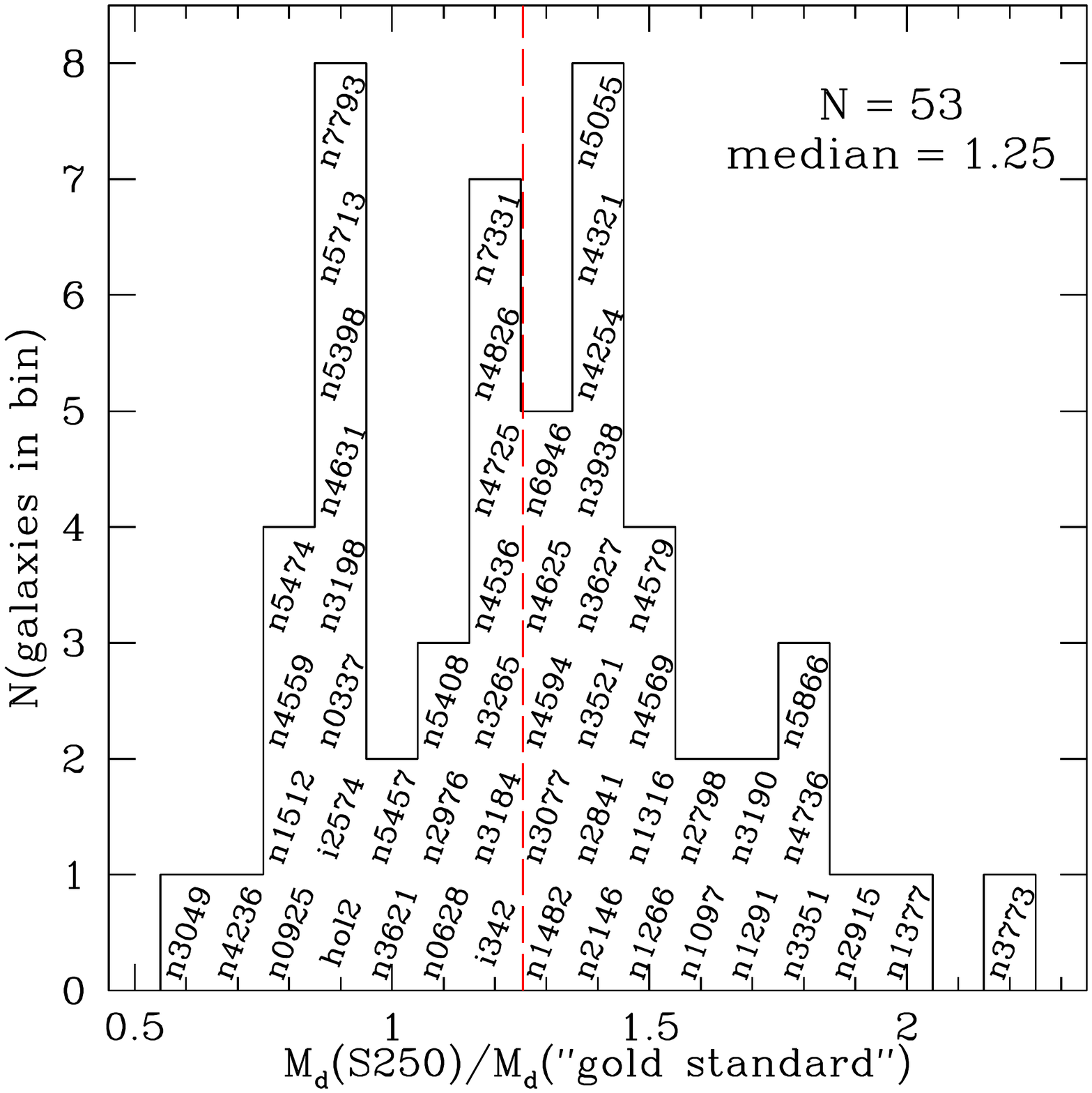} \\
\includegraphics[width=6.0cm,clip=true,trim=0.5cm 5.0cm 0.5cm 2.5cm]{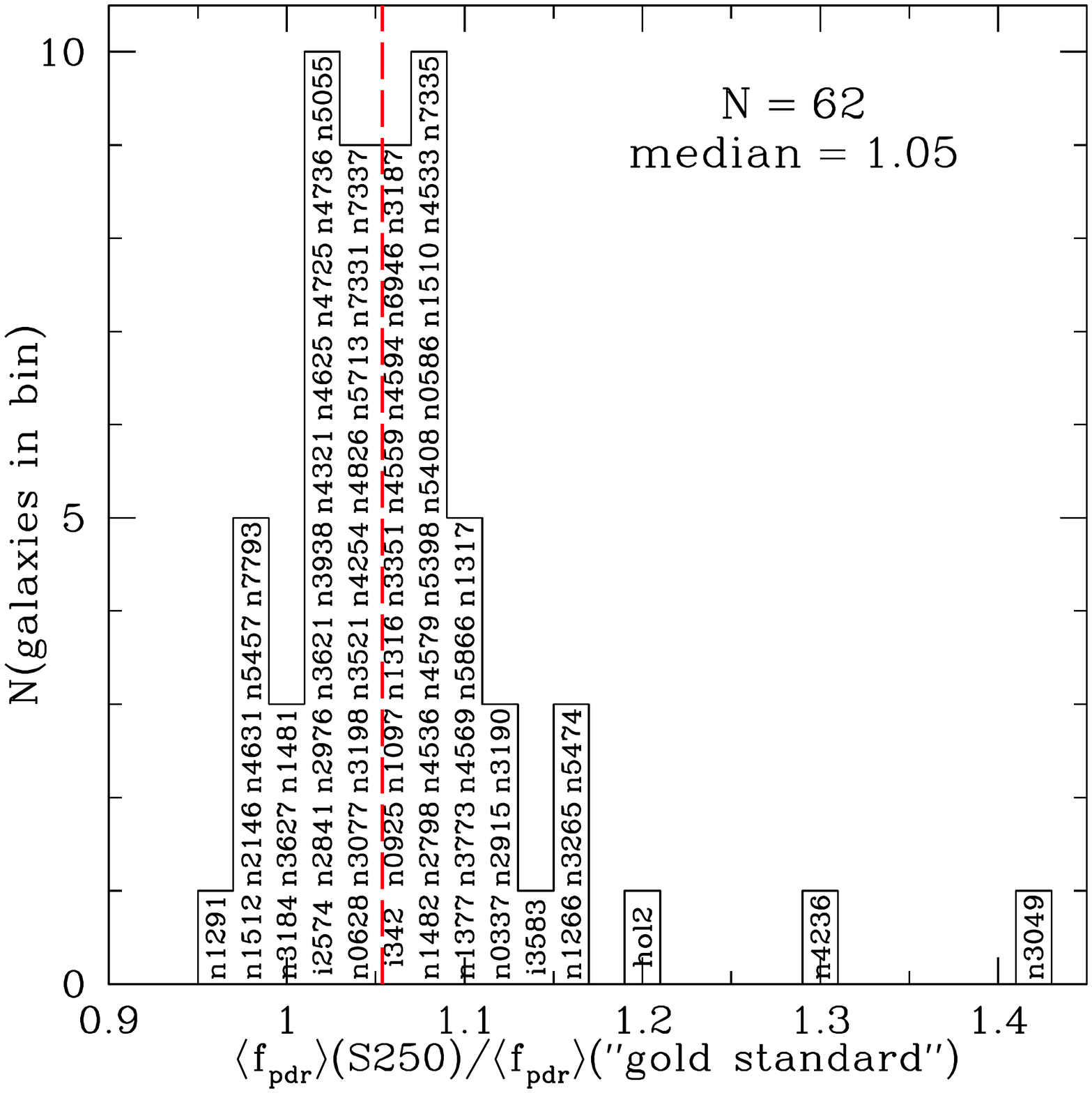} 
\includegraphics[width=6.0cm,clip=true,trim=0.5cm 5.0cm 0.5cm 2.5cm]{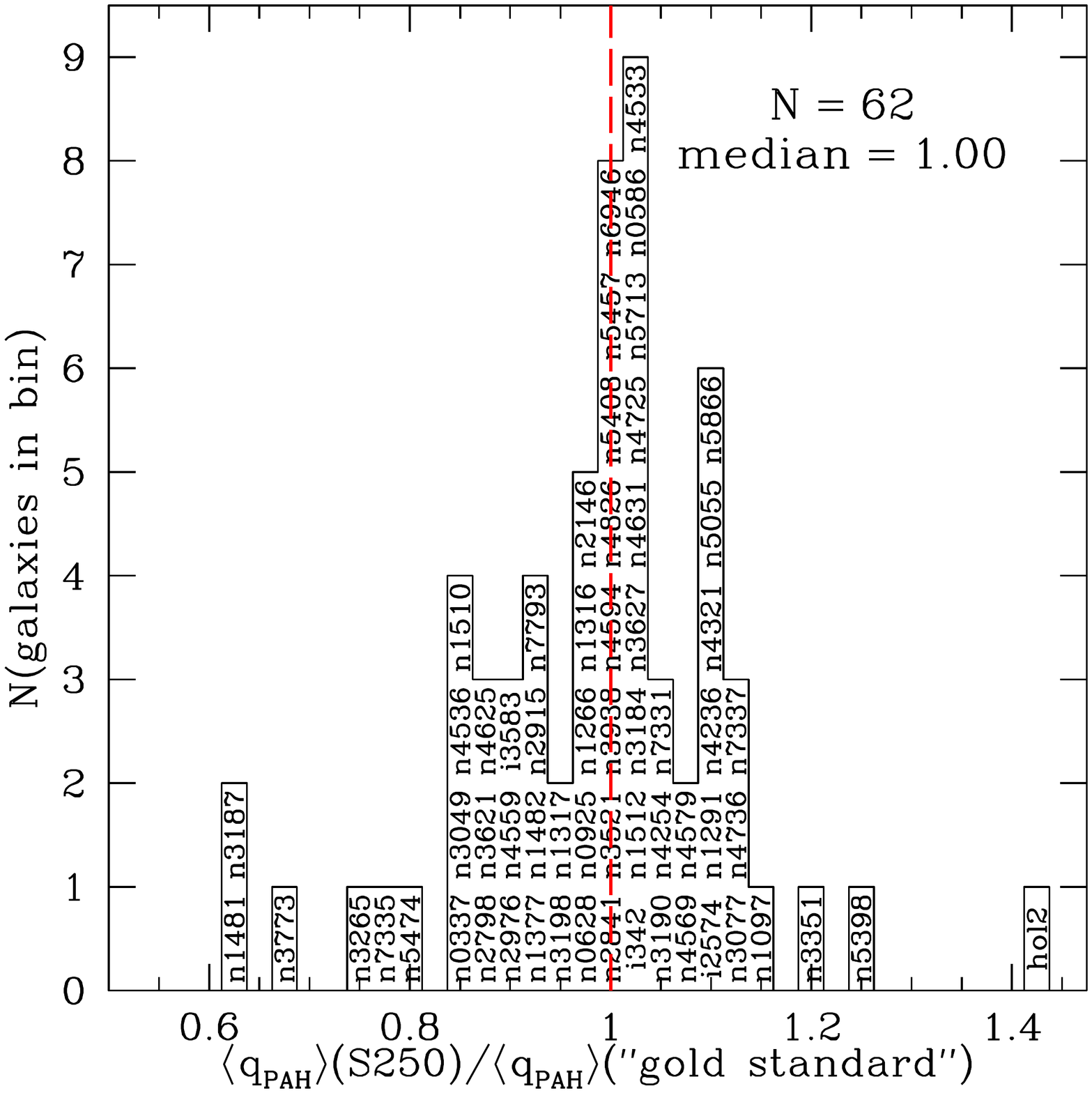} \\
\caption{Comparison of modeling with the S250 PSF to
results obtained with the M160 PSF (the ``gold standard'').
With S250 modeling, the dust mass tends to be slightly overestimated; the
median overestimation is $\sim$25\%.
\btdnote{fhist\_LdLd.pdf, fhist\_MdMd250.pdf, fhist\_fpdrfpdr.pdf, fhist\_qpahqpah.pdf}
\label{fig:multipix s250 vs m160}} 
\end{figure*}

Figure \ref{fig:multipix s250 vs m160} 
shows the comparison of the dust parameter estimates 
obtained from models using the S250 PSF (18.2\arcsec FWHM)
with parameters estimated from (our ``gold standard'') modeling
using all cameras (IRAC, MIPS, PACS, SPIRE) and the M160
PSF (39\arcsec FWHM).

The results for $L_\dust$ and $\Mdust$ at S250 resolution
appear to be quite robust: the median change in $L_\dust$ is only 5\%,
which may be due in part to calibration differences between
MIPS160 and PACS160, with MIPS160 data being used only in the M160 PSF
modeling.
$\Mdust$ shows more variation: with a median change of 25\%; this is likely
because loss of SPIRE350 and SPIRE500 may allow modeling at the S250 PSF
to include a bit more cool dust than is actually present. 
However, it is gratifying that the median change is only 25\%, indicating that
the DL07 model is relatively good at ``predicting'' $\lambda>300\micron$
emission using data shortward of $300\micron$.
However, in some cases the dust mass is overestimated by as much as
a factor of 2 (see Fig.\ \ref{fig:multipix md for different psfs}), 
and we therefore recommend using
M160 resolution modeling rather than the riskier S250 PSF.


\subsection{Dust mass estimates at different resolutions}

\begin{figure*}  
\centering 
\begin{tabular}{cc}  
\includegraphics[width=6.0cm,clip=true,trim=0.5cm 5.0cm 0.5cm 2.5cm]{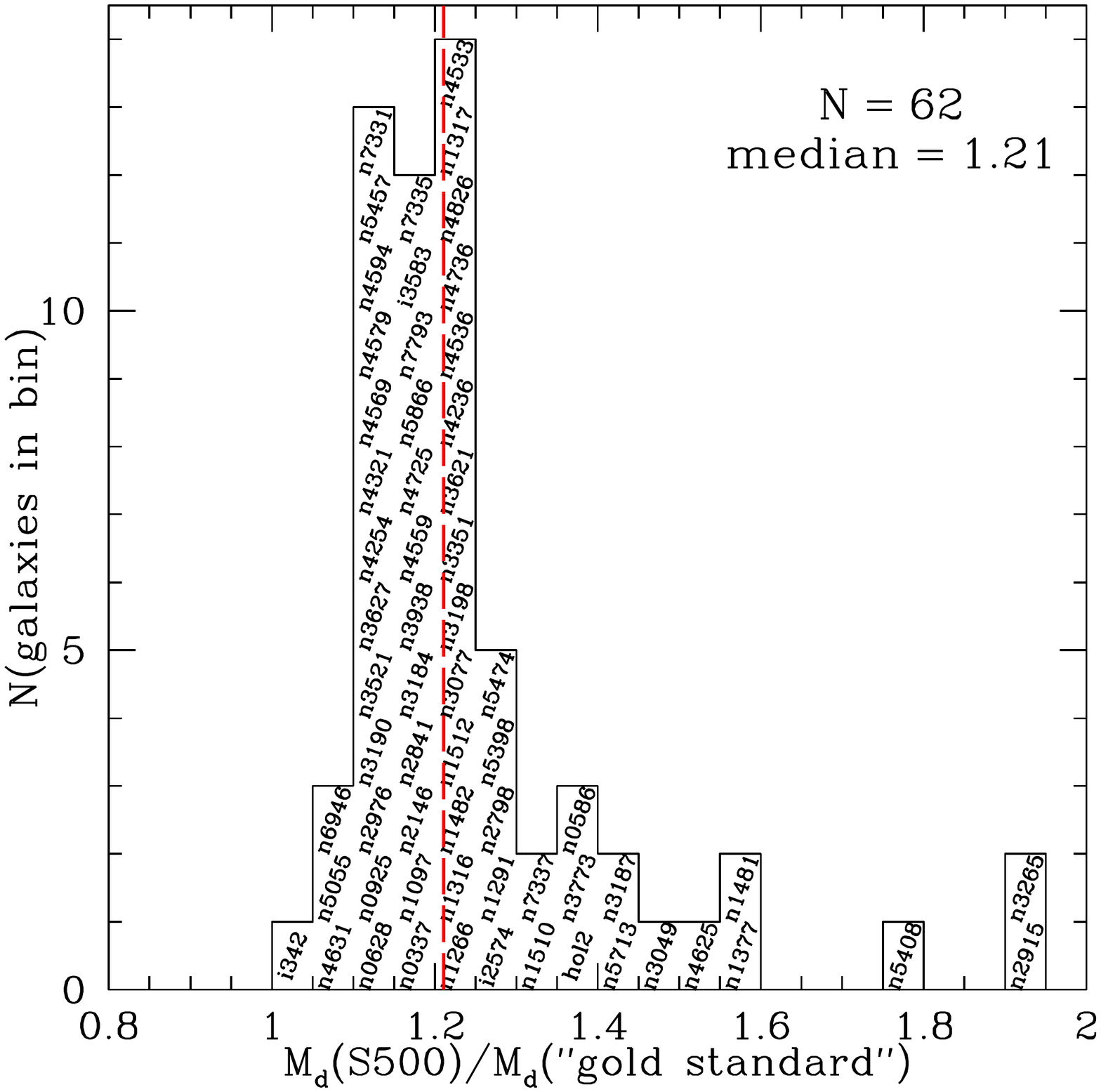} & 
\includegraphics[width=6.0cm,clip=true,trim=0.5cm 5.0cm 0.5cm 2.5cm]{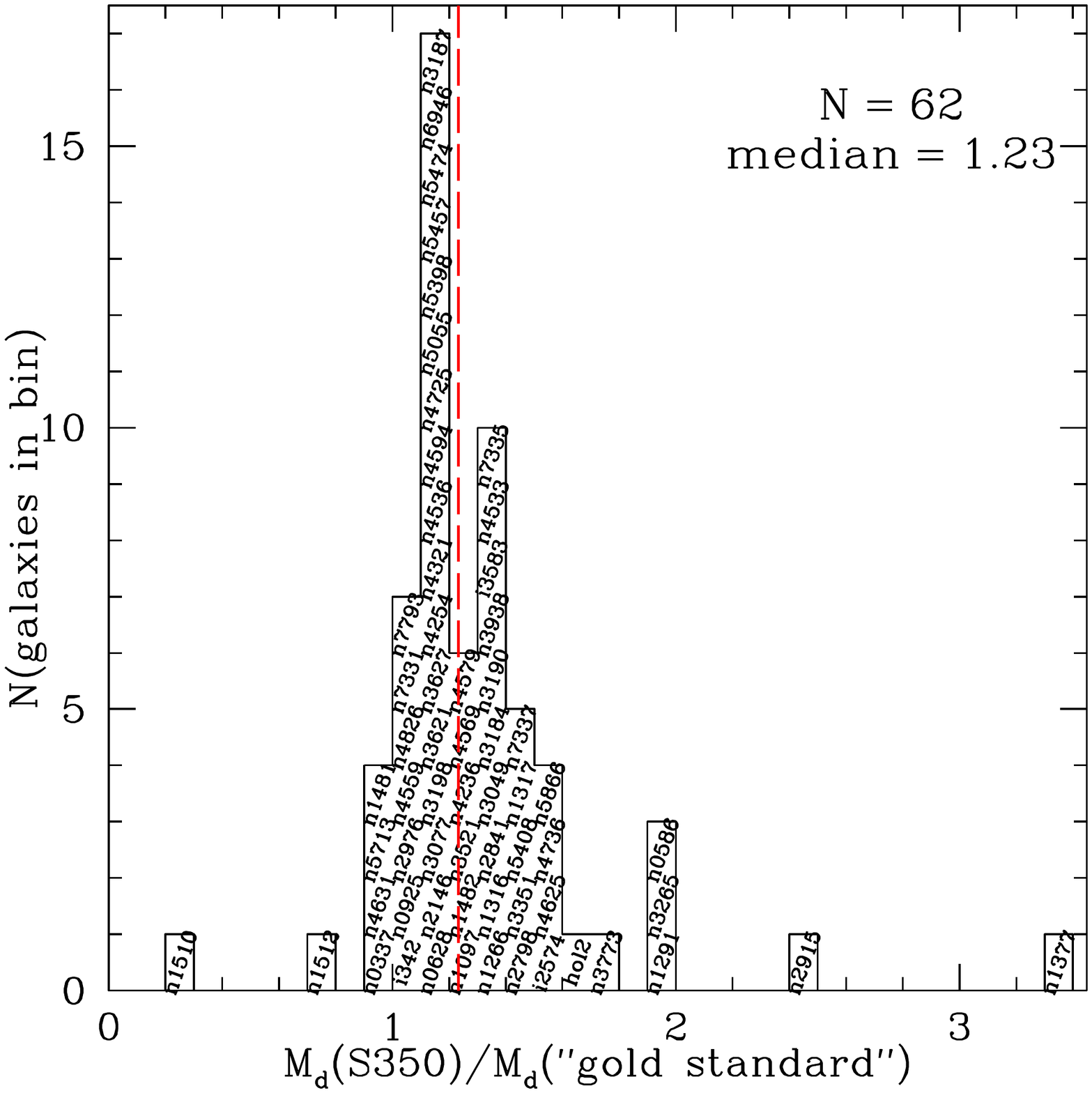} \\
\includegraphics[width=6.0cm,clip=true,trim=0.5cm 5.0cm 0.5cm 2.5cm]{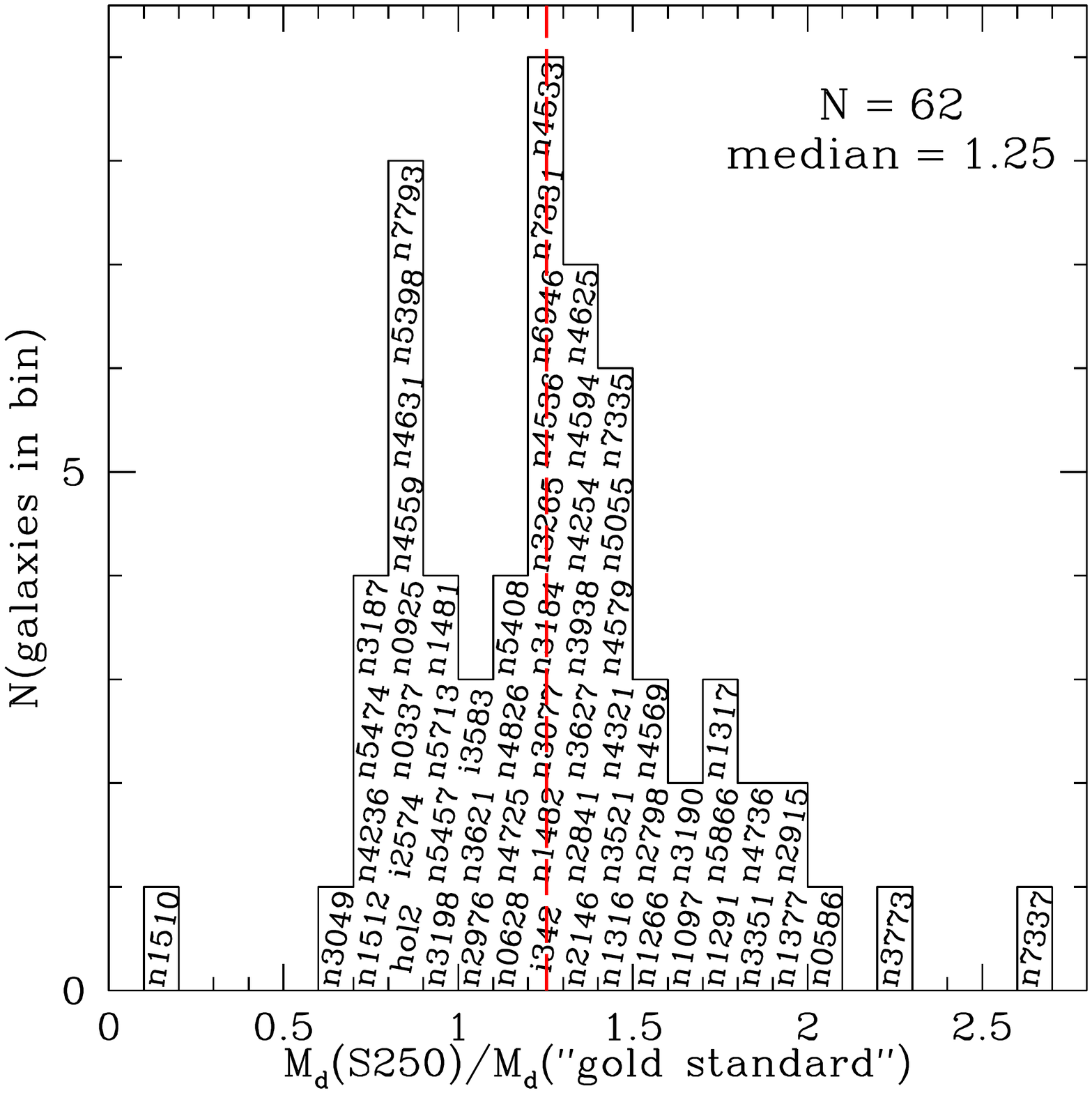} & 
\includegraphics[width=6.0cm,clip=true,trim=0.5cm 5.0cm 0.5cm 2.5cm]{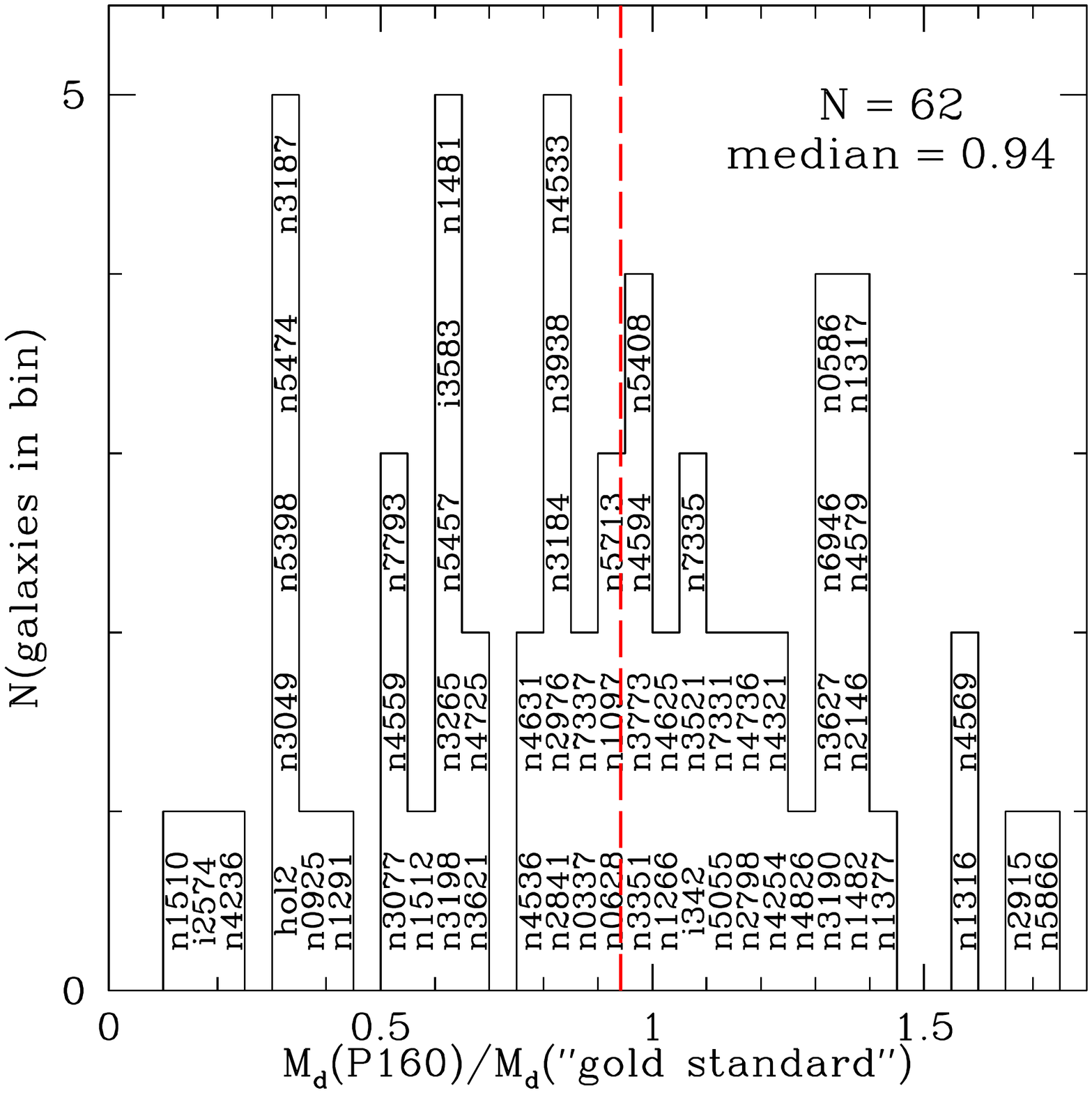} \\
\end{tabular}  
\caption{Comparison of estimates for dust mass $M_\dust$ obtained from
multipixel modeling with different PSFs and camera combinations.
The ``gold standard'' refers to modeling with the M160 PSF and all cameras.
Here we compare $M_\dust$ obtained with the S500, S350, S250, and P160
PSFs (see Table \ref{tab:resolutions}).
Because of the limited wavelength coverage, and lower signal/noise ratio,
for the P160 PSF
the dust mass can be in error by up to a large factor: 10/62 cases underestimate the dust mass by more than a factor of 2, and 2/62 cases by more than a factor of 5.
\btdnote{fhist\_MdMdP160.pdf, fhist\_MdMdS250.pdf,
  fhist\_MdMdS350.pdf, fhist\_MdMdS500.pdf}
\label{fig:multipix md for different psfs}} 
\end{figure*}

Figure \ref{fig:multipix md for different psfs} 
shows the comparison of the dust mass estimates for 4 different resolutions and camera combinations.
The compared resolutions are
\begin{itemize} 
\item M160 (the ``gold standard'') uses all the cameras (IRAC, MIPS, PACS,
SPIRE) at the
MIPS160 PSF; this is 
taken to be our best estimate for $\Mdust$
\item S500: IRAC, MIPS24, MIPS70, PACS, and SPIRE at the SPIRE500 PSF.
\item S350: IRAC, MIPS24, MIPS70, PACS, SPIRE250, and SPIRE350 
at the SPIRE350 PSF.
\item S250: IRAC, MIPS24, PACS, and SPIRE250 at the SPIRE250 PSF.
\item P160: IRAC, MIPS24, and PACS at the PACS160 PSF.
This is the riskiest PSF we are willing to consider.
\end{itemize}
For each resolution, Fig.\ \ref{fig:multipix md for different psfs} shows 
a histogram of the galactic total dust mass estimates 
divided by the gold standard estimate.

We observe that dust mass discrepancies can be large, with the errors and
bias increasing as fewer cameras are used, and long-wavelength data
are lost.
The S500 case (coverage out to 500$\micron$, a PSF that is not
much smaller than the M160 PSF, but no MIPS160 photometry) gives dust
mass estimates that are close to our gold standard estimate, with a median
ratio of 1.21.  However, there are a few outliers where $\Mdust$ appears
to be overestimated by as much a factor of 2.  These are all
galaxies with very weak dust emission and low signal/noise data, 
where loss of the data from one
camera (MIPS160) causes a significant change in the apparent SED.

The systematic bias in $\Mdust$ as well as the
scatter both increase as we move to smaller PSFs (S350, S250, P160).
At P160 resolution, fully 25\% of the cases have $\Mdust$
{\it under-}estimated by a factor of 2 or more.

On balance, it appears that modeling at S250 resolution is reasonable, although
slightly risky -- there is a significant chance that the dust mass
may be overestimated or underestimated
by a factor 1.5 or more.  S350 resolution is safer, and S500 even better.

\subsection{Modeling using only global photometry (single-pixel)}

The KINGFISH galaxies are close ($D<30\Mpc$) 
and large enough that they can be resolved using {\it Herschel} 
Space Telescope.
When studying galaxies at larger distances, only their global photometry 
may be available.
Here we compare our dust mass estimates using resolved imaging
and multipixel modeling with ``single-pixel'' modeling that makes
use of only the global SED.
We recall that the dust modeling is not a linear process, 
and differences in parameter estimates are to be expected.

\begin{figure*}  
\centering 
\includegraphics[width=10.0cm,clip=true,trim=0.5cm 5.0cm 0.5cm 2.5cm]{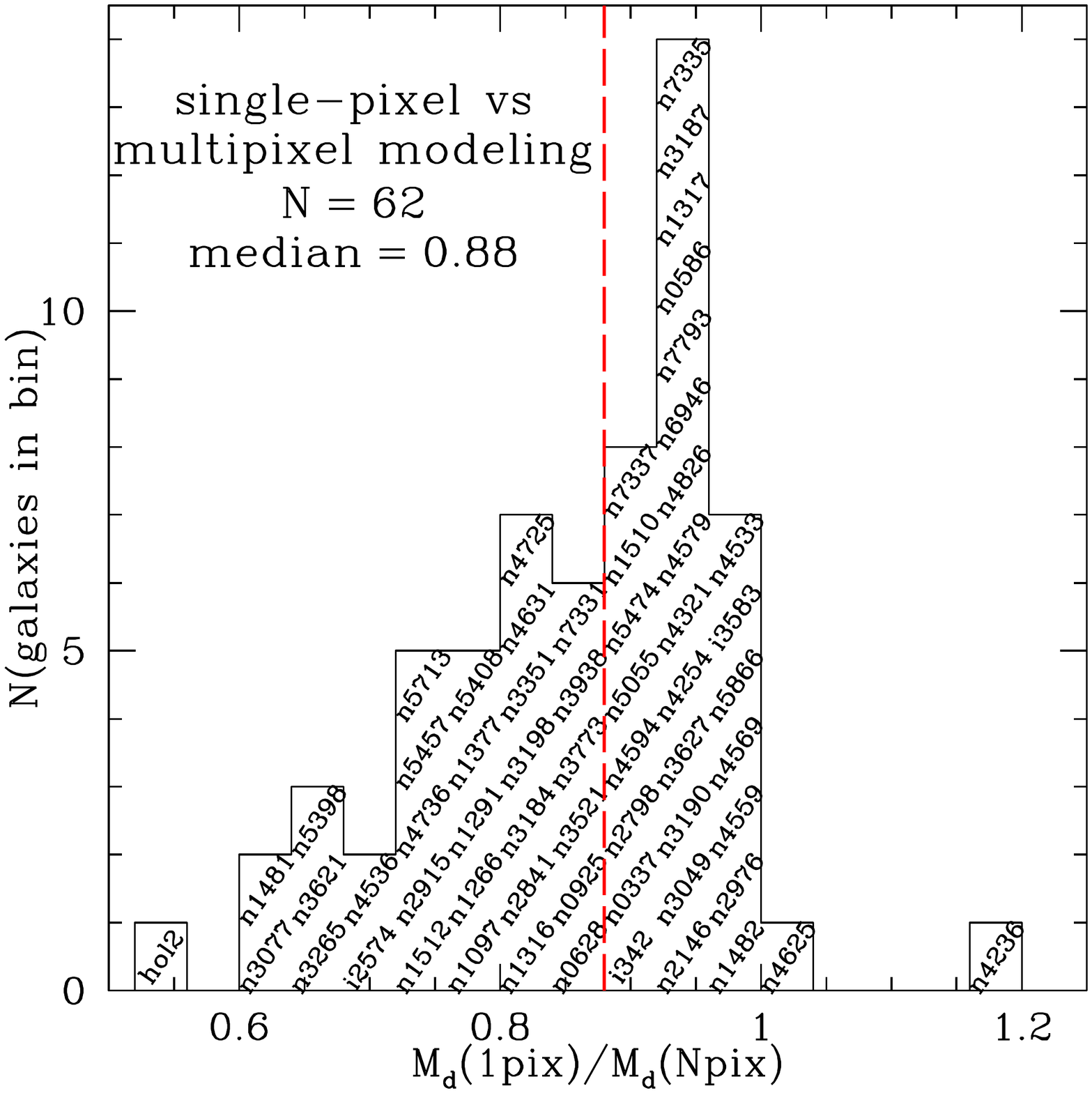}
\caption{Dust mass estimation using global (single-pixel) modeling versus
sum over resolved (multi-pixel) modeling at M160 resolution.
Single-pixel modeling estimates the total dust mass to within a factor
of 2 in the worst case (Holmberg II), but for $\sim$70\% of the cases
the single-pixel mass is within 25\% of the multipixel mass.
\btdnote{fhist\_Md\_Nvs1.pdf}
\label{fig:multipix vs singlepix}} 
\end{figure*}

Figure \ref{fig:multipix vs singlepix} shows the ratio of 
the dust model parameter estimates from fitting
global photometry (a ``single pixel'' model)
versus our ``gold standard'' multipixel modeling at 
M160 resolution, where each pixel is modeled separately.
In both cases we use all cameras: IRAC, MIPS, PACS and SPIRE.
We observe that $L_{{\rm d}}$ is very reproducible,
with the single-pixel luminosity estimate differing from the multipixel
result by only a few percent.

The dust mass estimate is probably most important, and is found to be
moderately robust: for 75\% of the cases, 
the single-pixel modeling
obtains a mass estimate within 25\% of the resolved multipixel analysis.
Thus dust mass estimates for unresolved distant galaxies should be reliable,
assuming only that the photometry covers a suitable range of
wavelengths (rest-frame wavelengths $50\ltsim \lambda \ltsim 300\micron$), with an
adequate signal/noise
ratio.

The 
$\langle\fPDR\rangle$ and $\langle\qpah\rangle$ estimates
are both quite robust, with the single pixel results agreeing 
with the multipixel analysis
to within $\sim$5\%
in most cases.

\end{document}